\documentclass[a4paper,fleqn,usenatbib]{mnras}

\usepackage{newtxtext,newtxmath}

\usepackage[T1]{fontenc}
\usepackage{ae,aecompl}

\usepackage{float}
\usepackage{graphicx}	
\usepackage{amsmath}	
\usepackage{graphicx}
\usepackage{mathtools}
\usepackage{txfonts}
\usepackage{pdflscape}
\usepackage{lscape}
\usepackage{float}
\usepackage{natbib}
\usepackage{threeparttable, tablefootnote}
\usepackage{microtype}
\usepackage{hyperref}

\title[Stellar substructures in the periphery of the MCs]{Stellar substructures in the periphery of the Magellanic Clouds with the VISTA Hemisphere Survey from the red clump and other tracers}

\author[D. El Youssoufi et al.]{Dalal El Youssoufi,$^{1}$\thanks{E-mail: delyoussoufi@aip.de (DE)} Maria-Rosa L. Cioni,$^{1}$ Cameron P. M. Bell,$^{1}$
\newauthor Richard de Grijs,$^{2,3,4}$ Martin A. T. Groenewegen,$^{5}$ Valentin D. Ivanov,$^{6}$ Gal Matijevi\u{c},$^{1}$ 
\newauthor Florian Niederhofer,$^{1}$ Joana M. Oliveira,$^{7}$ Vincenzo Ripepi,$^{8}$ Thomas Schmidt,$^{1}$  
\newauthor Smitha Subramanian,$^{9}$ Ning-Chen Sun,$^{10}$ and Jacco Th. van Loon$^{7}$
\\
$^{1}$Leibniz-Institut f\"ur Astrophysik Potsdam (AIP), An der Sternwarte 16, D-14482 Potsdam, Germany\\
$^{2}$Department of Physics and Astronomy, Macquarie University, Balaclava Road, Sydney, NSW 2109, Australia\\
$^{3}$Research Centre for Astronomy, Astrophysics and Astrophotonics, Macquarie University, Balaclava Road, Sydney, NSW 2109, Australia\\
$^{4}$International Space Science Institute--Beijing, 1 Nanertiao, Zhongguancun, Hai Dian District, Beijing 100190, China\\
$^{5}$Koninklijke Sterrenwacht van Belgi\"e, Ringlaan 3, B--1180 Brussels, Belgium \\ 
$^{6}$European Southern Observatory, Karl-Schwarzschild-Str. 2, D--85748 Garching bei M\"unchen, Germany\\
$^{7}$Lennard-Jones Laboratories, Keele University, ST5 5BG, UK\\
$^{8}$INAF -- Osservatorio Astronomico di Capodimonte, via Moiariello 16, I--80131, Naples, Italy\\
$^{9}$Indian Institute of Astrophysics, Koramangala II Block, Bangalore-34, India\\
$^{10}$Department of Physics and astronomy, University of Sheffield, Hicks Building, Hounsfield Road, Sheffield S3 7RH, UK\\
}

\date{Accepted 2021 April 12. Received 2021 March 20; in original form 2020 August 15}

\pubyear{2020}

\begin{document}
\label{firstpage}
\pagerange{\pageref{firstpage}--\pageref{lastpage}}
\maketitle

\begin{abstract}
We study the morphology of the stellar periphery of the Magellanic Clouds in search of substructure using near--infrared imaging data from the VISTA Hemisphere Survey (VHS). Based on the selection of different stellar populations using the ($J - K_\mathrm{s}, K_\mathrm{s}$) colour--magnitude diagram, we confirm the presence of substructures related to the interaction history of the Clouds and find new substructures on the eastern side of the LMC disc which may be owing to the influence of the Milky Way, and on the northern side of the SMC, which is probably associated to the ellipsoidal structure of the galaxy. We also study the luminosity function of red clump stars in the SMC and confirm the presence of a bi--modal distance distribution, in the form of a foreground population. We find that this bi--modality is still detectable in the eastern regions of the galaxy out to a 10$^\circ$ distance from its centre. Additionally, a background structure is detected in the North between 7$^\circ$ and 10$^\circ$ from the centre which might belong to the Counter Bridge, and a foreground structure is detected in the South between 6$^\circ$ and 8$^\circ$ from the centre which might be linked to the Old Bridge.
\end{abstract}




\section{Introduction}\label{section1}
The Large Magellanic Cloud (LMC) and the Small Magellanic Cloud (SMC) are the closest pair of interacting dwarf galaxies to the Milky Way (MW). Located at distances of 50 kpc (e.g.~\citealp{DeGrijs2014}) and 60 kpc (e.g.~\citealp{DeGrijs2015}), respectively, they represent important testbeds for fundamental astrophysical processes as they provide insights into resolved stellar populations, the cosmological distance scale, the interstellar medium, as well as galaxy interactions and morphology. The LMC is a barred Magellanic spiral galaxy, showing an asymmetric bar (e.g. \citealp{Zhao2000}), a few spiral arms (e.g. \mbox{\citealp{DeVaucouleurs1972,ElYoussoufi2019}}) and an inclined disc (e.g. \citealp{Cioni2001,VanderMarel2001a}) with a warp (e.g. \citealp{Salyk2002,Choi2018b}), while the SMC is an irregular dwarf galaxy characterised by a large line--of--sight depth (e.g.~\citealp{Subramanian2012,Jacyszyn-Dobrzeniecka2017,Scowcroft2016,Ripepi2017,Muraveva2018}). 
It has been found that stellar populations of different ages in the Magellanic Clouds (MCs) display different morphologies (e.g. \citealp{Cioni2000a,Nikolaev2000,Zaritsky2000,ElYoussoufi2019}). Additionally, the Magellanic system includes the Magellanic Bridge, a feature connecting the two Clouds that is most prominent in \ion{H}{I} gas \citep{Hindman1963} with young (e.g.~\citealp{Shapley1940,Harris2007,Skowron2014,Noel2015,Mackey2017}) as well as intermediate-age and old stars associated with it (e.g.~\citealp{Bagheri2013,Carrera2017,Jacyszyn-Dobrzeniecka2017}). It also includes the Magellanic Stream, a $\sim$200$^\circ$ long stream of gas trailing the Clouds in their orbit around the MW \citep{Mathewson1974} without an identified stellar component as well as the Leading Arm \citep{Putman1998}, a gaseous feature extending beyond the Galactic plane \citep{Nidever2010}, and across the Galactic disc \citep{McClure-Griffiths2008}.

Dynamical interactions have shaped the evolutionary history of the MCs. The LMC and SMC have been dynamically coupled since at least $\sim$2~Gyr ago and experienced a significant interaction $\sim$150~Myr ago (e.g.~\citealp{Diaz2012,Zivick2019}). Two mechanisms, tidal and ram pressure stripping, have been suggested to account for the creation of the Magellanic Stream and Bridge (e.g.~\citealp{Besla2007,Diaz2012, Salem2015}). Dynamical interactions can explain many of the diffuse stellar substructures, including arcs, clumps and overdensities, discovered in the periphery of the MCs. 

In the outskirts of the SMC (8$^\circ$ from its centre), \cite{Pieres2017} found a stellar overdensity, named SMCNOD, using imaging data from the Dark Energy Survey (DES; \citealp{Abbott2018}). Subsequently, imaging data from the MAGellanic SatelLITEs Survey (MagLiteS; \citealp{Drlica-Wagner2016}) was used to study SMCNOD finding that it lies at the same distance as the SMC and contains  mostly intermediate-age stars (6 Gyr old) with a small fraction of young stars (1 Gyr old). SMCNOD was most likely tidally stripped from the SMC. \cite{MartinezDelgado2019} revisited a shell-like overdensity in the outskirts of the SMC using the Survey of the MAgellanic Stellar History (SMASH; \citealp{Nidever2017}) data and found that it is solely composed of young stars ($\sim$ 150 Myr old). There is no evidence of it being of tidal origin and it may have formed in a recent star formation event resulting from an interaction with the LMC or the MW. \cite{Massana2020}, also using SMASH data, discovered a new faint feature located at ~14$^\circ$ from the centre of the SMC, possibly associated with a more distant structure.

In the periphery of the LMC, \cite{Mackey2016,Mackey2018} found numerous substructures in and around the galaxy, including an arc-like structure 13.5$^\circ$ North of the centre, of which the stellar populations are indistinguishable from those of the LMC disc. The stars in this feature were most likely stripped from the LMC due to either the tidal force of the MW or to an interaction with the SMC. \cite{Belokurov2019} used \textit{Gaia} DR2 to look for low surface brightness features in the periphery of the MCs and found two thin and long stellar streams in the northern and southern regions of the LMC. Numerical simulations indicate that the effect of the MW and SMC on the LMC are both important for the creation of tidal features. 
Using a combination of \textit{Gaia} data release \#1 (DR1; \citealp{Brown2016}) and Galaxy Evolution Explorer (GALEX, \citealp{Martin2005}) imaging data, \cite{Belokurov2017} demonstrated that the distribution of RR Lyrae stars traces a bridge joining the SMC from its trailing tidal tail to the LMC. However, the nature of this bridge is  still under debate as it has not been detected in OGLE \citep{Jacyszyn-Dobrzeniecka2017,Jacyszyn-Dobrzeniecka2020b} nor in \textit{Gaia} DR2 data \citep{Clementini2019}. \cite{Deason2017} found that Mira-like stars trace the LMC as far as 20$^\circ$ from its centre.

Established tracers of substructures within and around the Magellanic Clouds are red clump (RC) stars, low mass stars which have undergone the helium flash to ignite helium burning in their cores (e.g. \citealp{Girardi2016} and references therein). Owing to their constrained location in the colour--magnitude diagram (CMD), they are used as probes of stellar distances allowing the study of the three--dimensional (3D) structure and reddening of the MCs (e.g. \citealp{Subramanian2010,Subramanian2012,Subramanian2013,Tatton2013,Choi2018a,Choi2018b}). \cite{Nidever2013} discovered a strong distance bi--modality in the luminosity function of RC stars with one component at $\sim$67 kpc and a second component at $\sim$55 kpc in three eastern fields located 4$^\circ$ from the SMC centre. They suggested that the component found at $\sim$ 55 kpc is the tidally stripped stellar counterpart of the Magellanic Bridge, in good agreement with dynamical models (e.g. \citealp{Diaz2012}). Using the VISTA near-infrared $YJK_\mathrm{s}$ survey of the MCs system (VMC; \citealp{Cioni2011}), \cite{Subramanian2017} studied the luminosity function of RC stars in the inner 20 deg$^2$ region of the SMC. A bi--modality in distances to the RC stars was found in the eastern SMC, after having ruled out reddening, RC population effects, and line of sight depth effects as potential causes, it was interpreted as a foreground population $\approx$~11~kpc in front of the main body. This is most likely due to tidal stripping from the SMC during its most recent encounter 100 -- 300 Myr ago with the LMC. \cite{TattonThesis} and \cite{Tatton2020} showed how the luminosity function of the RC varies across all VMC tiles belonging to the SMC, while \cite{Omkumar2020} have investigated the RC bi--modality using \textit{Gaia} DR2 data. Additional evidence supporting the tidal stripping of SMC material during the most recent encounter comes in the form of accreted red giant branch stars belonging to the SMC and discovered in the LMC \citep{Olsen2011}, while \cite{Dobbie2014} found signatures of tidally stripped red giant branch stars in the outskirts of the SMC.

In this paper, we present a detailed analysis of the spatial distribution of different stellar populations in the periphery of the MCs using data from the VISTA Hemisphere Survey (VHS; \citealp{McMahon2013}). VHS enables us to obtain a comprehensive view of the morphology in the outskirts of the MCs over a continuous area, allowing us to confirm the presence of known substructures while also finding new ones. Furthermore, we also investigate the properties of the double RC feature, beyond the main body of the SMC, using a combination of VHS and VMC data. Identifying stellar substructures around the MCs using morphology and luminosity function studies is very important to understand the nature of the interactions between the MCs and will provide inputs for theoretical models. The paper is organised as follows, Section \ref{section2} describes the data set used in our study and our selection criteria, Section \ref{section3} focuses on the outer morphology of the MCs, while Section \ref{section4} studies the properties of the double RC feature. Finally, Section \ref{section5} presents a discussion and summary of our results. 

\begin{figure}
	\centering
	\includegraphics[scale=0.22]{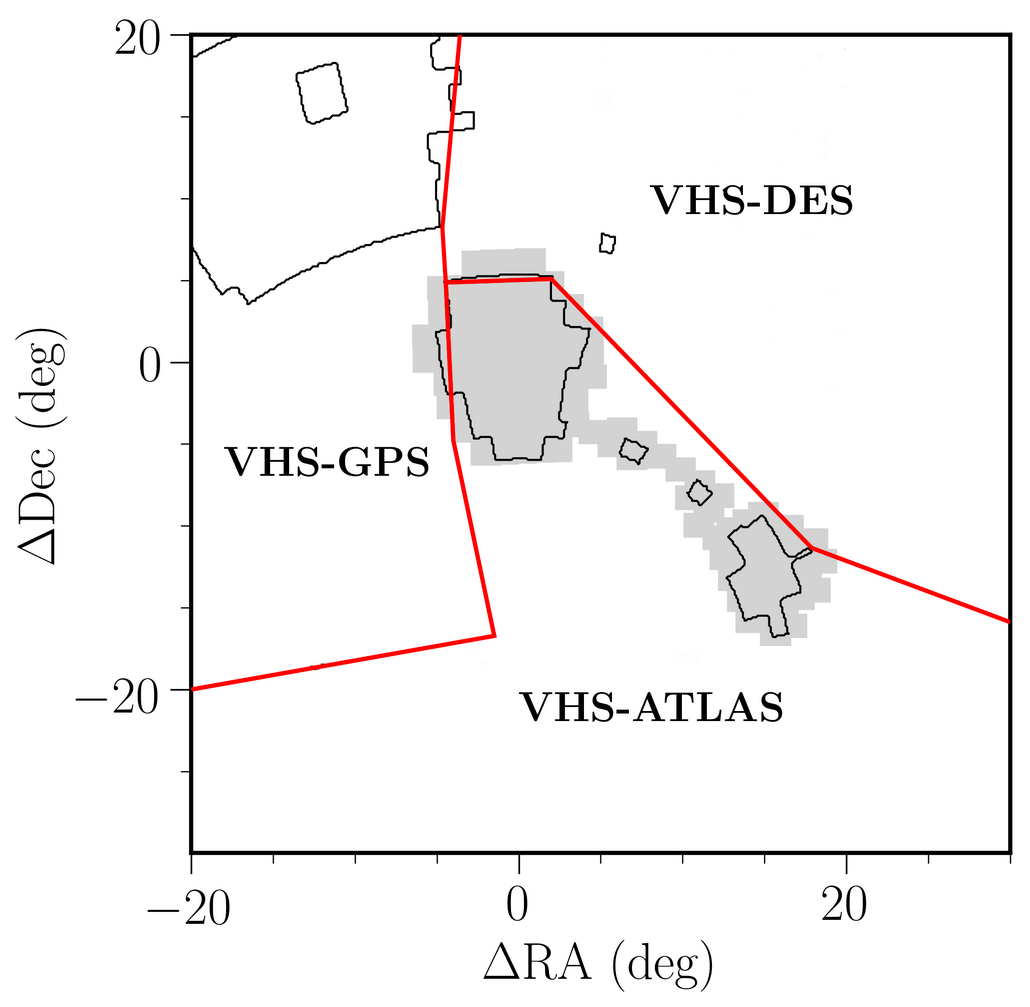}
	\caption{Schematic figure showcasing the footprints of the VHS and VMC surveys. The borders of the VHS sub--surveys are indicated in red and those of observational gaps in black, while the area covered by VMC tiles is displayed in light grey. The projection origin for the maps is set at (RA$_0$, Dec$_0$) = ($81.24^\circ$, $-69.73^\circ$) corresponding to the densest point in the LMC bar \protect\citep{DeVaucouleurs1972}.}
	\label{fig:surveys}
\end{figure}

\begin{figure}
	\includegraphics[scale=0.11]{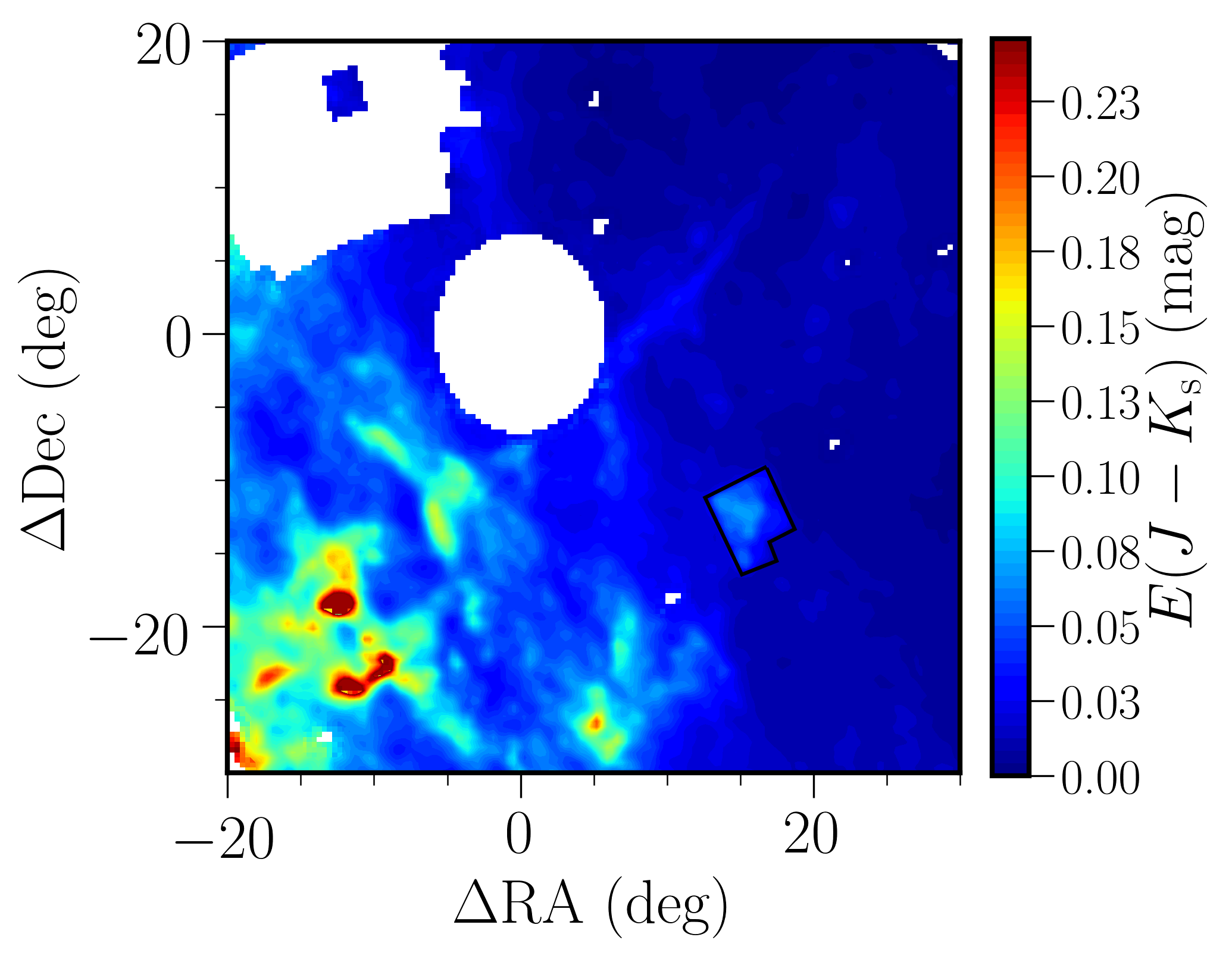}
	\caption{Distribution of reddening derived from the \protect\cite{Schlegel1998} dust map as well as from the star formation history study of the main body of the SMC (outlined in black) by \protect\cite{Rubele2018}  across the footprint analysed in this paper. The map is centred at (RA$_0$, Dec$_0$) = ($81.24^\circ$, $-69.73^\circ$) and the colour bar shows the variation of the $E$($J-K_\mathrm{s}$) colour excess. White patches refer to the central regions occupied by the LMC and to the incomplete VHS observations towards the NE, as well as to a few tiles in the outskirts of the MCs.}
	
	\label{fig:reddening}
\end{figure}
\begin{figure*}
	\centering
	\includegraphics[scale=0.09]{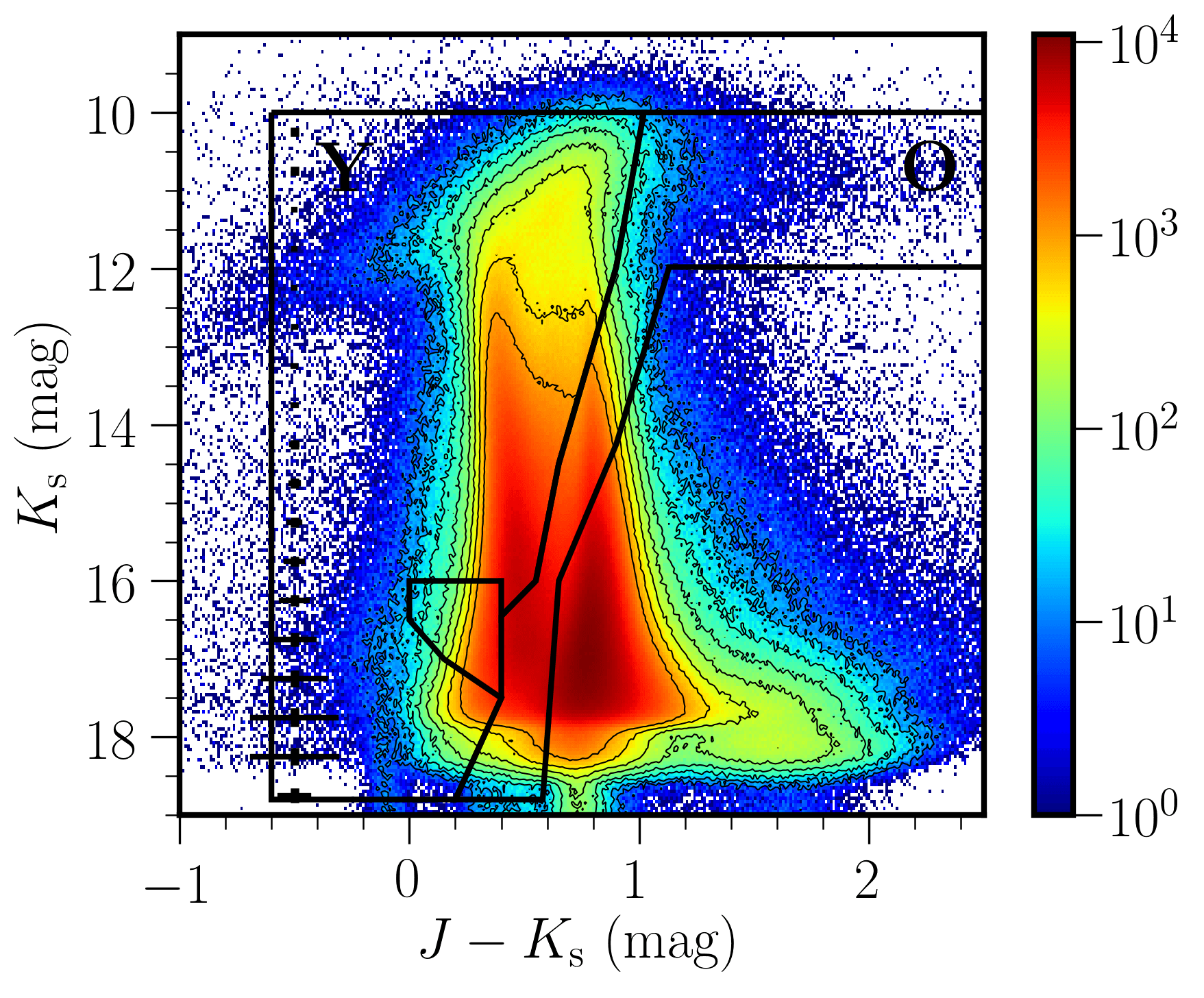}
	\includegraphics[scale=0.09]{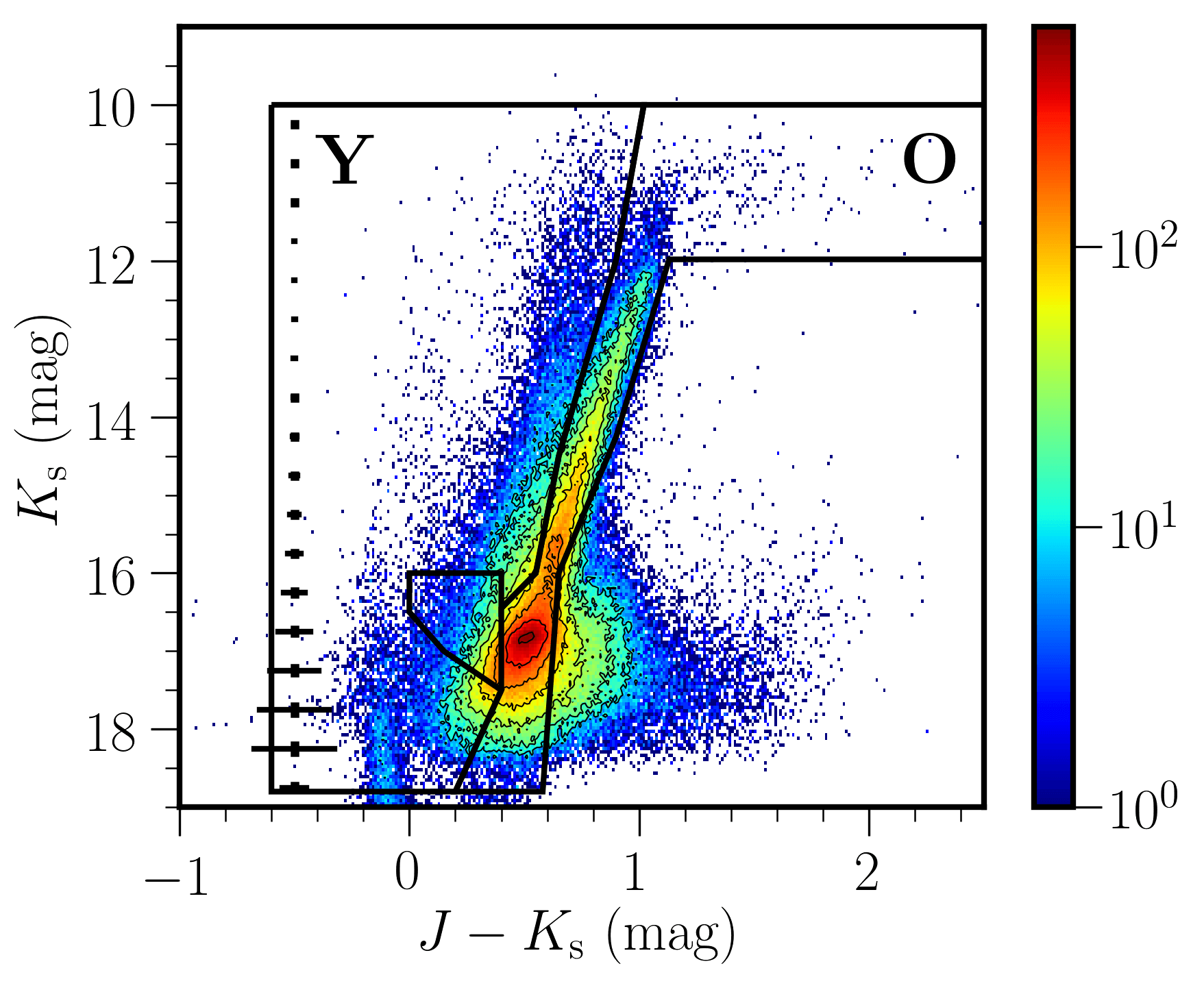}
	\includegraphics[scale=0.09]{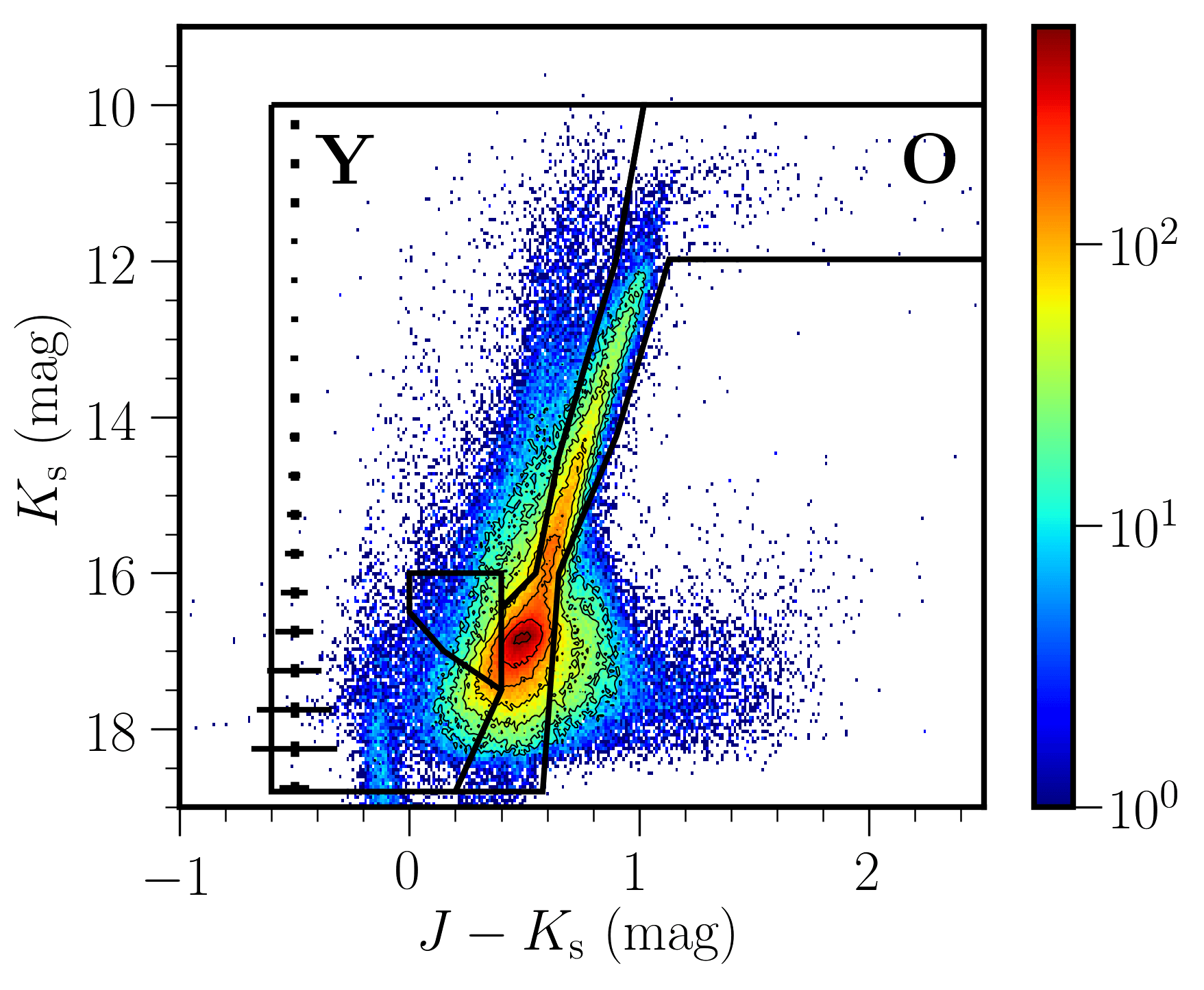}
	\caption{NIR ($J-K_\mathrm{s}$, $K_\mathrm{s}$) Hess diagrams of the stellar periphery of the MCs before the cross--match with \textit{Gaia} DR2 data (left), after the cross--match using the selection criteria to reduce the presence of MW stars (middle), and after correcting for reddening (right). The colour scale indicates the stellar density on a logarithmic scale whereas horizontal bars show the photometric uncertainties as a function of magnitude. Two different classes of objects are enclosed by black lines, Y referring to young stars and O referring to old stars. Stars belonging to the polygonal area ($16 \leq K_\mathrm{s} \leq 17.5$ mag) are not included in our study as they have mixed ages.}
	\label{fig:CMDS_Motphology}
\end{figure*}

\begin{figure}
	\centering
	\includegraphics[scale=0.1]{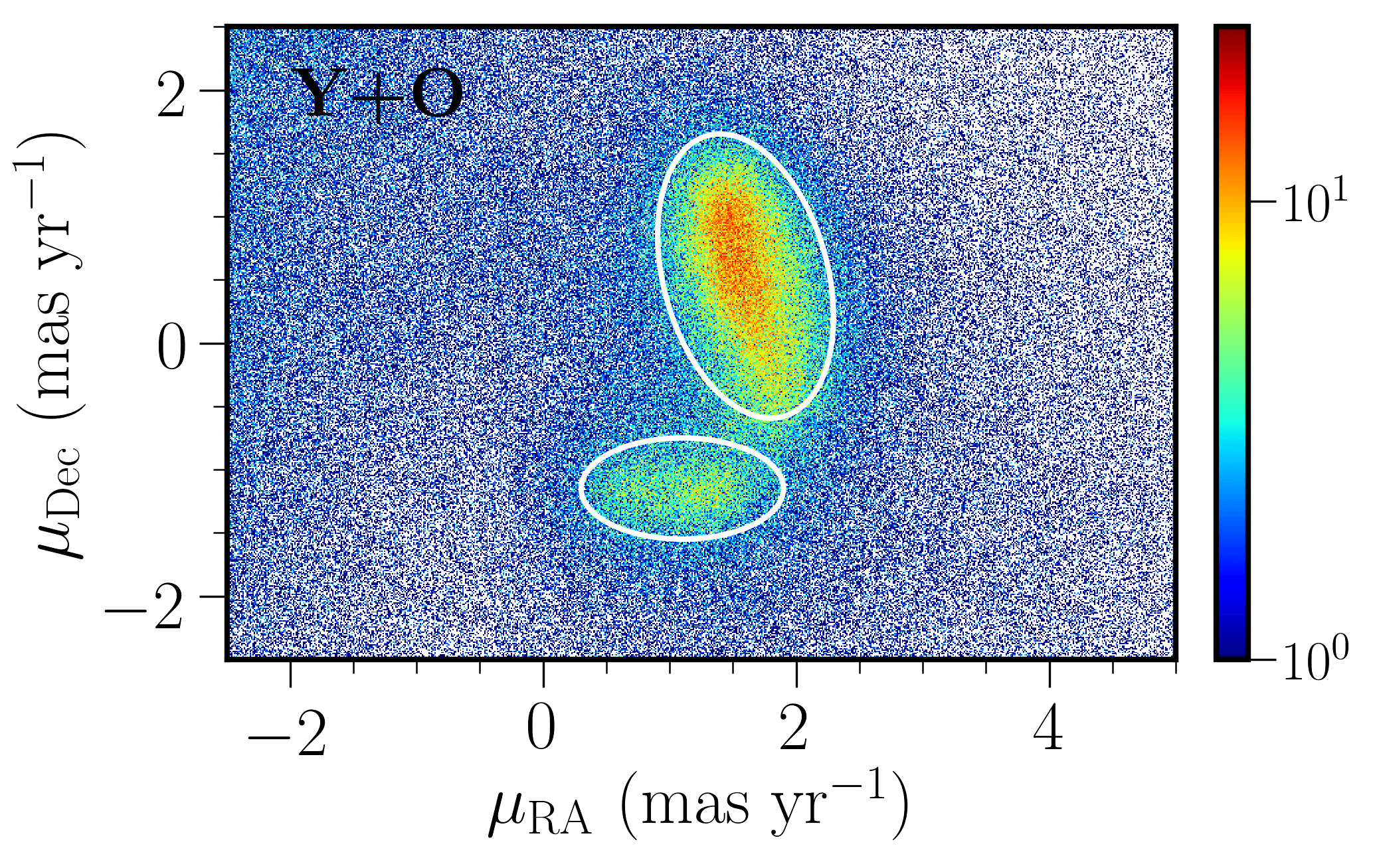}
	\caption{Stellar proper motions towards the direction of the MCs. Most of the objects with LMC and SMC proper motions are enclosed within the large and small ellipse, respectively, whereas many of the objects outside these ellipses belong to the MW.}
	\label{fig:PM_Motphology}
\end{figure}

\section{Observations and Data Selection}\label{section2}
\subsection{Data}
We used data from VHS, a near--infrared (NIR) survey, that covers the whole southern hemisphere when combined with other first--generation VISTA Public Surveys. VHS is carried out by the 4.1~m Visible and Infrared Survey Telescope for Astronomy (VISTA; \citealp{Sutherland2015}) and the VISTA infrared  camera (VIRCAM; \citealp{Dalton2006,Emerson2006}). It aims to cover 18,000 deg$^2$ with observations in at least two photometric bands, $J$ and $K_\mathrm{s}$ (central wavelengths: 1.25 $\mu$m and 2.15 $\mu$m, respectively). Median 5$\sigma$ point source limits of $J$ = 19.3 mag and $K_\mathrm{s}$~=~18.5 mag (in the Vega system) are achieved with a minimum exposure time of 60~s per waveband. 
VHS includes three programmes: VHS--GPS (Galactic Plane Survey), covering a region of $\approx$8200 deg$^2$ with exposure times of 60s in the $J$ and $K_\mathrm{s}$ bands;
VHS--DES, covering a region of $\approx$4500 deg$^2$ with exposure times of 120s in the $JHK_\mathrm{s}$ bands and VHS--ATLAS, covering a region of $\approx$5000 deg$^2$ evenly divided between the North and South Galactic caps with exposure times of 60s in the $YJHK_\mathrm{s}$ bands. VHS data from all three programmes border the MCs (Figure \ref{fig:surveys}). Our study makes use of all VHS observations obtained until 2017 September 30th.

We also make use of data from the VMC survey \citep{Cioni2011}, a deep NIR multi--epoch imaging survey of the MCs providing data in three photometric bands reaching 5$\sigma$ limits of $Y$ = 21.9, $J$ = 22, and $K_\mathrm{s}$ = 21.5 mag (in the Vega system). VMC observed the LMC (105 deg$^2$), SMC (42 deg$^2$), the Magellanic Bridge (21 deg$^2$) and the Magellanic Stream (3 deg$^2$). In this study we make use of observations in the SMC and Bridge components (Figure \ref{fig:surveys}). VHS data is combined with VMC data in order to fill observational gaps in the Magellanic Bridge (two small areas within it, see Figure \ref{fig:surveys}) and to study the luminosity function of RC stars with continuous coverage within 10 deg from the centre of the SMC. Our selection criteria for VMC sources are the same as those used for VHS sources (see below). Duplicate sources that are present in both surveys (e.g. in the outskirts of the LMC and SMC, as well as the Magellanic Bridge) were eliminated by keeping VMC detections.

The VISTA Data Flow System (VDFS; \citealp{Irwin2004}) was used for data reduction while the data were extracted from the VISTA Science archive (VSA\footnote{\url{http://horus.roe.ac.uk/vsa}}; \citealp{Cross2012}). The VHS data allow us to explore resolved stellar populations in the periphery of the MCs and to detect sources just below the RC. VHS showcases an improvement in depth over previous NIR surveys such as the Two Micron All Sky Survey (2MASS; \citealp{Skrutskie2006}) and the DEep Near Infrared Survey of the Southern Sky (DENIS; \citealp{Cioni2000a}), 
which have limiting magnitudes $\approx$4~mag brighter in $K_\mathrm{s}$. VHS is the deepest NIR survey covering the entire southern hemisphere to date, and although it is by design much shallower than its optical counterparts such as \textit{Gaia} DR2 and the DES \citep{Abbott2018}, VHS provides an essential contribution to the multi--wavelength view of the periphery of the MCs.

Figure \ref{fig:reddening} shows the line--of--sight Galactic dust reddening derived from the \cite{Schlegel1998} dust maps. In the Bar and Wing regions of the SMC, \cite{Schlegel1998} measurements can be unreliable due to contamination by SMC stars or unresolved temperature structure.
We combined the \cite{Schlegel1998} reddening map with reddening derived from the star formation history study of the main body of the SMC by \cite{Rubele2018}. \cite{Rubele2018} used VMC data across an area of 23.57  deg$^2$ with a bin size of 0.143 deg$^2$ and reconstructed the observed CMDs using stellar partial models, recovering the best--fitting extinction $A_{V}$ as an additional output of the star formation rate. The combination of the \cite{Schlegel1998} and \cite{Rubele2018} maps encompasses the entire area studied in this paper. For the \cite{Schlegel1998} map, $E$($B - V$) values were obtained using the python module \texttt{SFDMAP}\footnote{\url{https://github.com/kbarbary/sfdmap}} and converted to $E$($J-K_\mathrm{s}$) values using coefficients from \cite{Gonzalez-Fernandez2018}. These coefficients have been calculated assuming $R_V$ = 3.1 and account for the \cite{Schlafly2011} recalibration. For the \cite{Rubele2018} map, we converted $A_{V}$ to $E$($B - V$) values assuming $R_V$ = 3.1 and used the same coefficients to convert them to $E$($J-K_\mathrm{s}$) values. Despite the spatial distribution of reddening being generally low in the periphery of the MCs with $E$($J-K_\mathrm{s}$) $\approx$ 0.10~mag, particularly at the location of the substructures traced in this study, filamentary dust features with high reddening values are present near the Galactic disc. Moreover, the sharp borders (owing to the VISTA field--of--view) suggest that the level of reddening in the immediate outskirts of the SMC may be underestimated as we only account for the foreground reddening and not the intrinsic reddening of the SMC. In the analysis which follows, we therefore correct for reddening. However, the central regions of the LMC and SMC, including the region covered by the \cite{Rubele2018} reddening map, have been omitted from the morphological maps to enhance the distribution of stars in the outer regions.

\subsection{Selection of young and old stellar populations}\label{section22}
The selection of different stellar populations was based on ($J-K_\mathrm{s}$, $K_\mathrm{s}$) CMDs (see Figure \ref{fig:CMDS_Motphology}). We selected unique objects \texttt{(priOrSec$\leqslant$0 or priOrSec = frameSetID)} from the \textit{vhssource} table detected in both $J$ and $K_\mathrm{s}$ bands, classified as stars with at least a $70$\% probability \texttt{(flag mergedclass = $-1$ or mergedclass = $-2$)} and with photometric errors less than 0.2~mag. The \texttt{AperMag3}, which corresponds to the default point source aperture corrected magnitude (2$^{\prime\prime}$ aperture diameter), was retrieved for each source in each waveband. No selection criteria based on extraction quality flags \texttt{(flag ppErrbits)} were applied. VISTA magnitudes were transformed to the Vega system by adding $0.011$~mag to the $K_\mathrm{s}$ band while no corrections are required in the $J$ band \citep{Gonzalez-Fernandez2018}.

Stars belonging to the MCs were divided into two regions: region Y occupied by young stars, including tip of the main sequence and supergiant stars, and region O occupied by old stars, including red giant branch, asymptotic giant branch, and RC stars. These two regions result from the combination of CMD regions established by \cite{ElYoussoufi2019}: region Y comprises regions A, B, C, I, G, H and region O includes regions D, E, J, K, M. Region Y has been extended to $K_\mathrm{s}$ = 10~mag, instead of $K_\mathrm{s}$ = 11.98 mag and to $J-K_\mathrm{s}$ = $-$0.6~mag instead of  $J-K_\mathrm{s}$ = $-$0.2~mag, to maximise the number of stars after their cross--correlation with \textit{Gaia} DR2. Both regions Y and O have been limited to $K_\mathrm{s}$ = 18.8 mag instead of $K_\mathrm{s}$ = 19.8~mag, to reflect the sensitivity of the VHS observations. Due to the small number of stars in the periphery of the MCs, we combined several regions as opposed to using them separately. Studies of MC objects are prone to both foreground and background contamination. Most background galaxies were omitted from our selection by adopting the object classification flag \texttt{mergedclass} and avoiding the reddest region of the CMD ($J-K_\mathrm{s}$ > 1~mag and $K_\mathrm{s}$ > 13~mag). However, this region may contain reddened MCs sources and a minority of red giant branch stars scattered from the adjacent region of old stars. To reduce the presence of MW stars in our sample, we combined data from \textit{Gaia} DR2 with VHS and VMC data \footnote{The catalogues have been cross-matched by G. Matijevi\u{c} using his own software tools.}, retaining only sources with a maximum cross--matching distance of 1" and parallaxes$\leq$0.2~mas.

In addition, we applied a proper motion selection. We examined the stellar density of objects belonging to regions Y and O in proper motion space and determined two ellipses around the LMC and SMC proper motion distributions (Figure \ref{fig:PM_Motphology}). Throughout the paper, we refer to $\mu_\mathrm{RA}\mathrm{cos}$(Dec) as $\mu_\mathrm{RA}$. The LMC ellipse is centred at $\mu_\mathrm{RA}$~=~1.60~mas~yr$^{-1}$, $\mu_\mathrm{Dec}$~=~0.53~mas~yr$^{-1}$, has a semi-major axis of 1.35~mas~yr$^{-1}$, a semi-minor axis of 0.85~mas~yr$^{-1}$, and is rotated by 15$^\circ$, whereas the SMC ellipse is centred at $\mu_\mathrm{RA}$~=~1.10~mas~yr$^{-1}$ and $\mu_\mathrm{Dec}$~=~$-$1.15~mas~yr$^{-1}$, has a semi-major axis of 0.80~mas~yr$^{-1}$, a semi-minor axis of 0.40~mas~yr$^{-1}$ and no rotation. Objects belonging to these two ellipses constitute our morphology dataset which contains 312,748 sources, encompassing an area of $1800$ deg$^2$ around the MCs.

In order to assess the detection limit of the VHS survey, we examined the normalised number of sources in each sub--survey as a function of magnitude compared with the 2MASS and VMC surveys. We used circular areas of 0.5$^\circ$ radius with the following (RA, Dec) centres: 2MASS (32.9$^\circ$, $-$75.0$^\circ$), VMC (32.9$^\circ$, $-$75.0$^\circ$), VHS--ATLAS (43.5$^\circ$,$-$77.8$^\circ$), VHS--GPS (111.50$^\circ$,$-$75.66$^\circ$), and VHS--DES (42.00$^\circ$,$-$60.97$^\circ$). We found that the VHS programmes ATLAS, GPS, and DES have detection limits in $J$ of 19.84~mag, 19.55~ mag and 20.36~mag, respectively, while in $K_\mathrm{s}$ the detection limits are of 18.08~mag, 17.79~mag and 18.42~mag, respectively. Only sources detected in both $J$ and $K_\mathrm{s}$ with photometric uncertainties less than 0.2~mag were included. These values show that stellar populations brighter than and including the bulk of the RC are not significantly affected by incompleteness.

\begin{figure}
	\centering
	\includegraphics[scale=0.04]{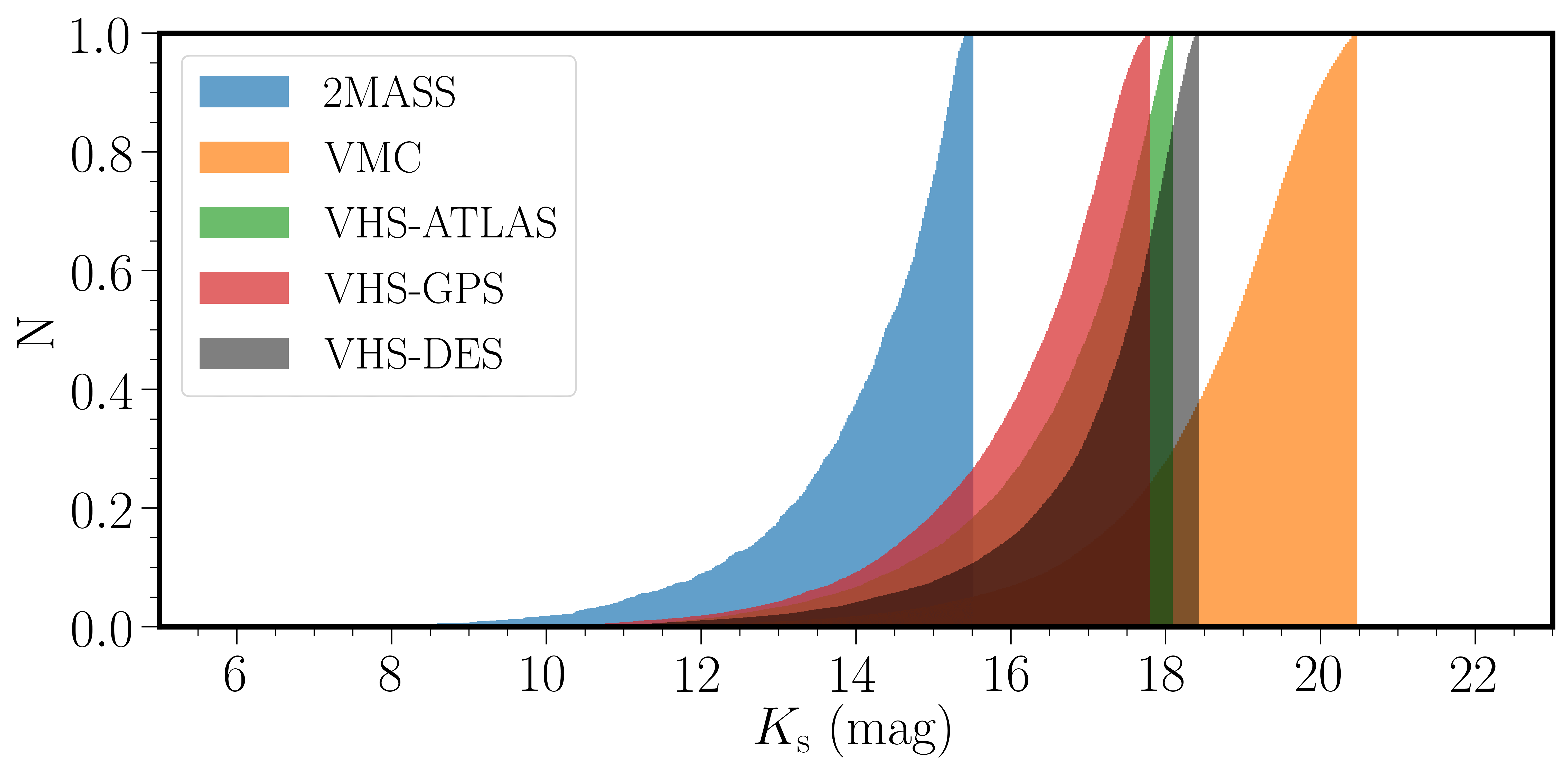}\\	
	\includegraphics[scale=0.04]{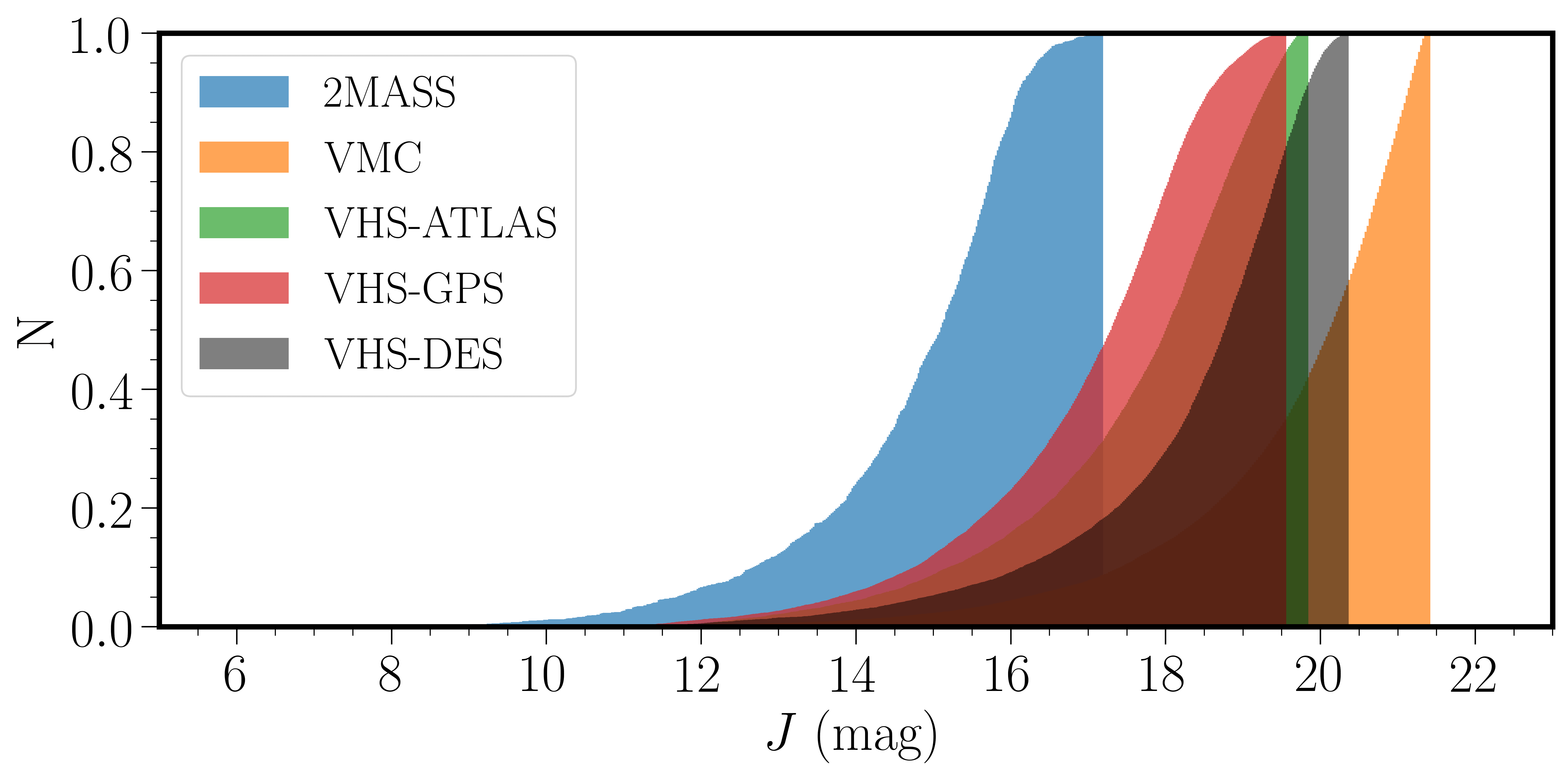}	
	\caption{Comparison between the detection limits of 2MASS, VMC, VHS-GPS, VHS-DES, and VHS-ATLAS in circular areas of 0.5$^\circ$ radius in the $J$ (left) and $K_\mathrm{s}$ (right) bands. The number of sources is normalised.}
	\label{fig:comp}
\end{figure}

\begin{figure*}
	\centering
	\includegraphics[scale=0.11]{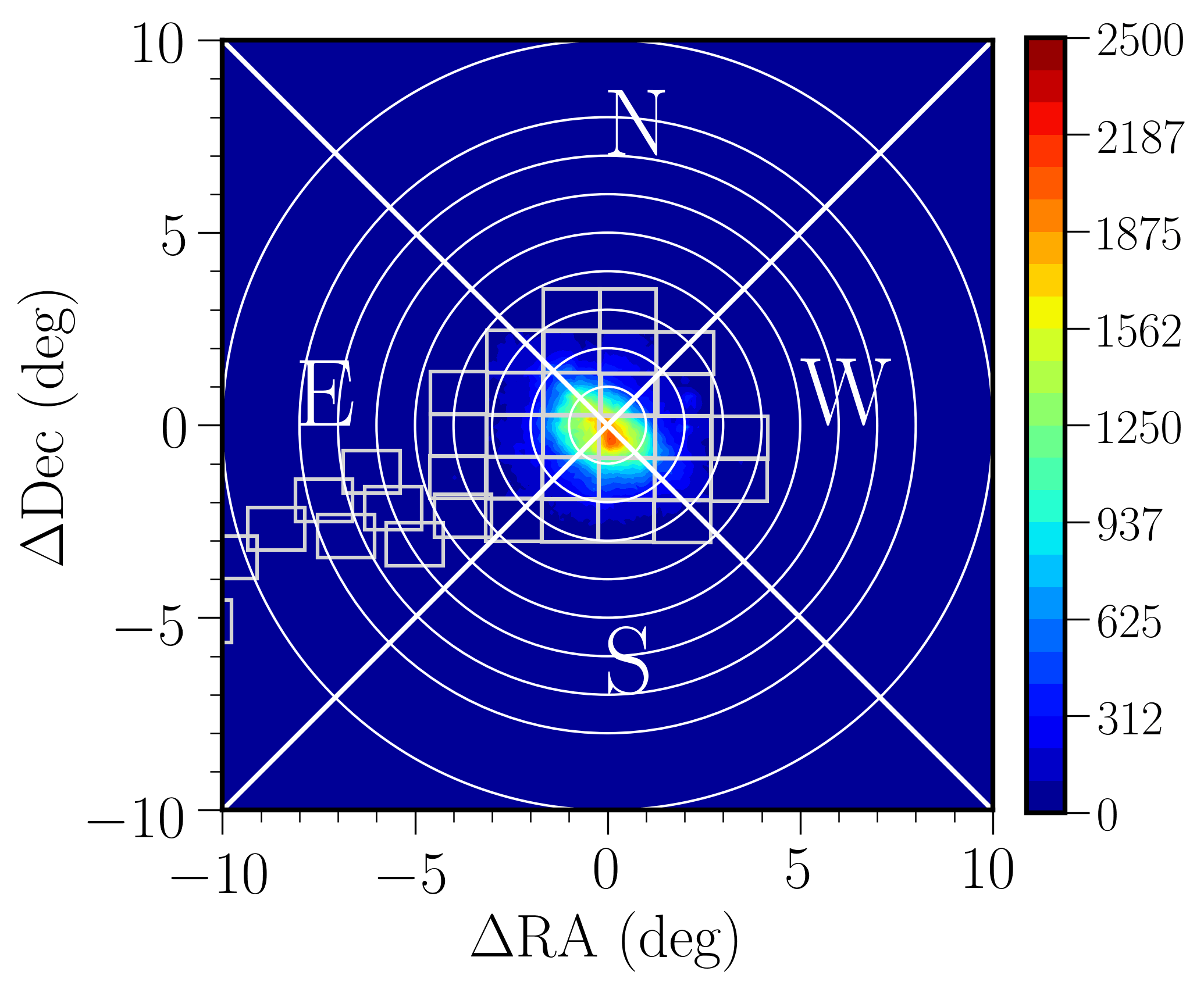}
	\includegraphics[scale=0.15]{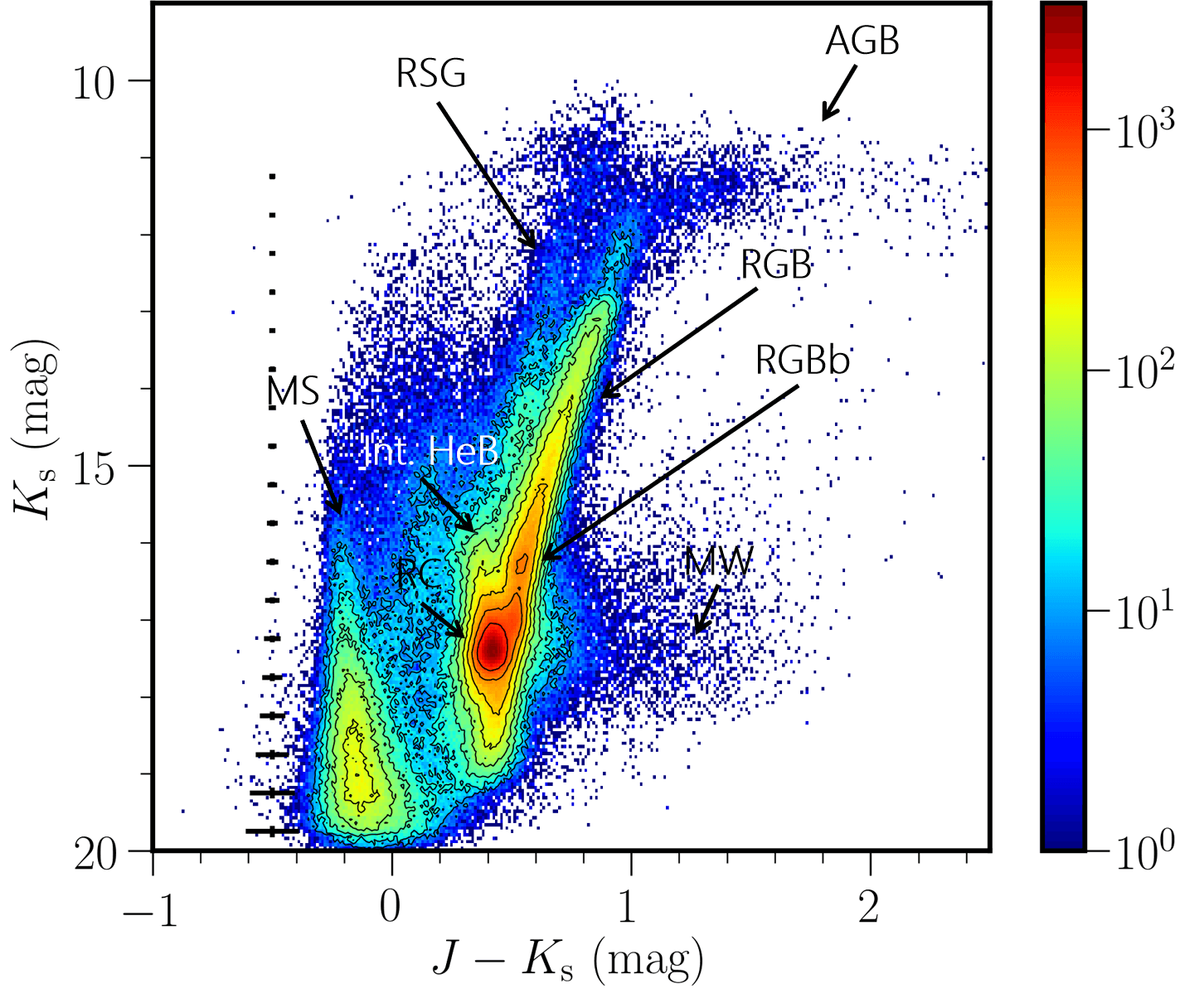}	
	\caption{(left) Spatial distribution of stars around the SMC and plotted annuli from 0$^\circ$ to 10$^\circ$ further divided into 4 regions: N, W, S, and E. VMC tiles are also shown. (right) NIR ($J-K_\mathrm{s}$, $K_\mathrm{s}$) Hess diagram (corrected for reddening) of all stars in the region shown in the right panel. Specific stellar populations are indicated as follows: asymptotic giant branch (AGB) stars, red supergiant (RSG) stars, tip of main sequence (MS) stars, intermediate--age Helium burning (Int.~HeB) stars as well as RGB, RC and MW stars while RGBb refers to the RGB bump.}
	\label{fig:DRC_1}
\end{figure*}

\begin{figure}
		\includegraphics[scale=0.1]{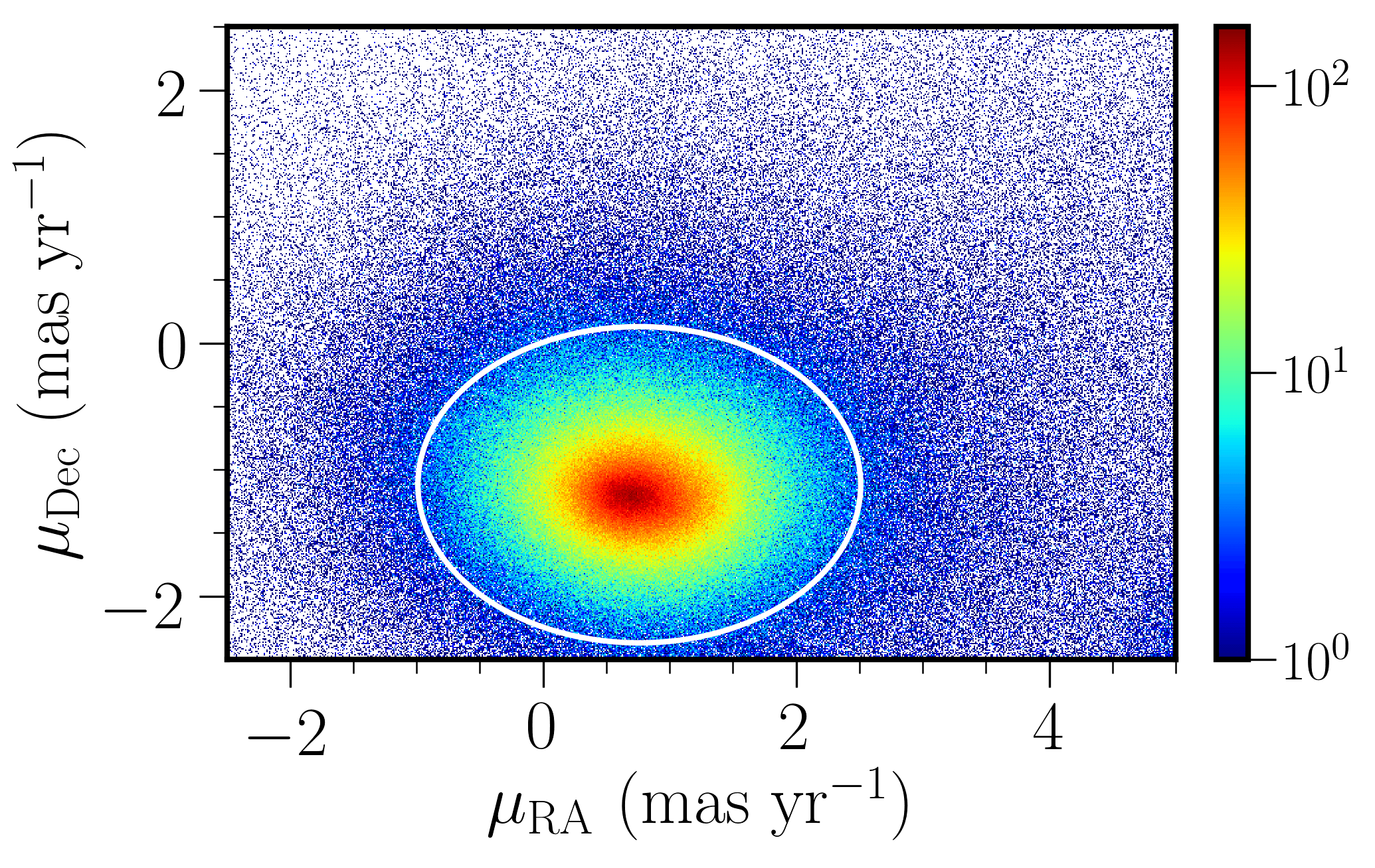}
		\caption{Stellar density of objects up to 10$^\circ$ from the SMC centre in proper motion space. The ellipse encloses a region with a reduced number of MW stars.}
		\label{fig:pmsmc}
\end{figure}

\subsection{Selection of RC stars around the SMC}\label{section23.png}
To study the distribution of RC stars in the outskirts of the SMC we proceeded as follows. We combined VHS with VMC observations of the SMC and Bridge which are within 10$^\circ$ from the SMC's optical centre. Due to the smaller number of stars in the periphery of the SMC than in the main body of the galaxy, we divided the SMC into circular annuli of radii 1$^\circ$, 2$^\circ$, 3$^\circ$, 4$^\circ$, 5$^\circ$, 6$^\circ$, 7$^\circ$, 8$^\circ$ and 10$^\circ$ from the centre of the galaxy and each annulus into four regions North (N), East (E), South (S) and West (W), see Figure \ref{fig:DRC_1} (left). In the ($J-K_\mathrm{s}$, $K_\mathrm{s}$) CMDs, specific stellar populations are indicated  (Figure \ref{fig:DRC_1} right). Foreground MW stars were removed following a similar procedure as that described above using a combination of parallax and proper motion cuts. For the proper motion selection, we examined the stellar density in proper motion space of sources up to 10$^\circ$ from the centre of the SMC and defined an ellipse with the following parameters: centre at 
$\mu_\mathrm{RA}$~=~0.76~mas~yr$^{-1}$ and $\mu_\mathrm{Dec}$~=~$-$1.12~mas~yr$^{-1}$, semi-major axis at 1.75~mas~yr$^{-1}$s and semi-minor axis at 1.25~mas~yr$^{-1}$ (Figure \ref{fig:pmsmc}). This ellipse is larger than the one defined in Section \ref{section22}, to distinguish between LMC and SMC stars, in order to include more RC stars in the outer regions of the SMC (not necessarily associated to the SMC itself) and at the same time minimise the influence of the MW. The small difference between the centres of the two proper motion SMC ellipses reflects the distribution of the type of stellar population dominating the inner and outer regions of the SMC. \cite{Zivick2020} showed that the centre of the proper motion distribution of red giant stars is clearly offset from where one would place the centre of the distribution of main sequence stars. The central region of the SMC is masked out in Figure \ref{fig:PM_Motphology} but it is included in Figure \ref{fig:pmsmc}.
In the ($J-K_\mathrm{s}$, $K_\mathrm{s}$) CMDs, we defined a box to encompass the full extent of RC stars, take into account the influence of RGB stars and minimise contamination from neighbouring stellar populations (Sect.~\ref{section4} and Fig.~\ref{fig:DRC}). The bright limit ($K_\mathrm{s}$ = 15~mag) was chosen to include stars beyond the RGB bump for better modelling of the RGB population, while the faint limit ($K_\mathrm{s}$ = 18.5~mag) represents the 5$\sigma$ point source limit in the $K_\mathrm{s}$ band. On the red side, neighbouring populations are RGB and MW stars. Therefore, the red borders of the box are defined as those of  K (RGB) and F (MW) from \cite{ElYoussoufi2019}. On the blue side, the main contaminants are horizontal branch stars, supergiant and subgiant stars. Beyond a radius of 4$^\circ$, photometric uncertainties and metallicity variations elongate the distribution of RC stars, shifting also RGB stars to bluer colours. Therefore, the blue limit for  $K_\mathrm{s}$$\leq$16.5~mag is the same as that of region I (red supergiants) from \cite{ElYoussoufi2019} to include the full width of the RGB, while for $K_\mathrm{s}$$\geq$16.5~mag, the limit is shifted by $\sigma_{J-K_\mathrm{s}}$ = 0.13~mag to include RC stars. RR Lyrae stars are generally confined to a compact region below the RC (0$\leq$$J-K_\mathrm{s}$$\leq$0.5~mag, 17.7$\leq$$K_\mathrm{s}$$\leq$19.4~mag). Therefore, they mostly affect the fainter end of the RGB.

\section{Stellar periphery of the Magellanic Clouds}\label{section3}
\begin{landscape}
\begin{figure}
	\centering
	\includegraphics[scale=0.09]{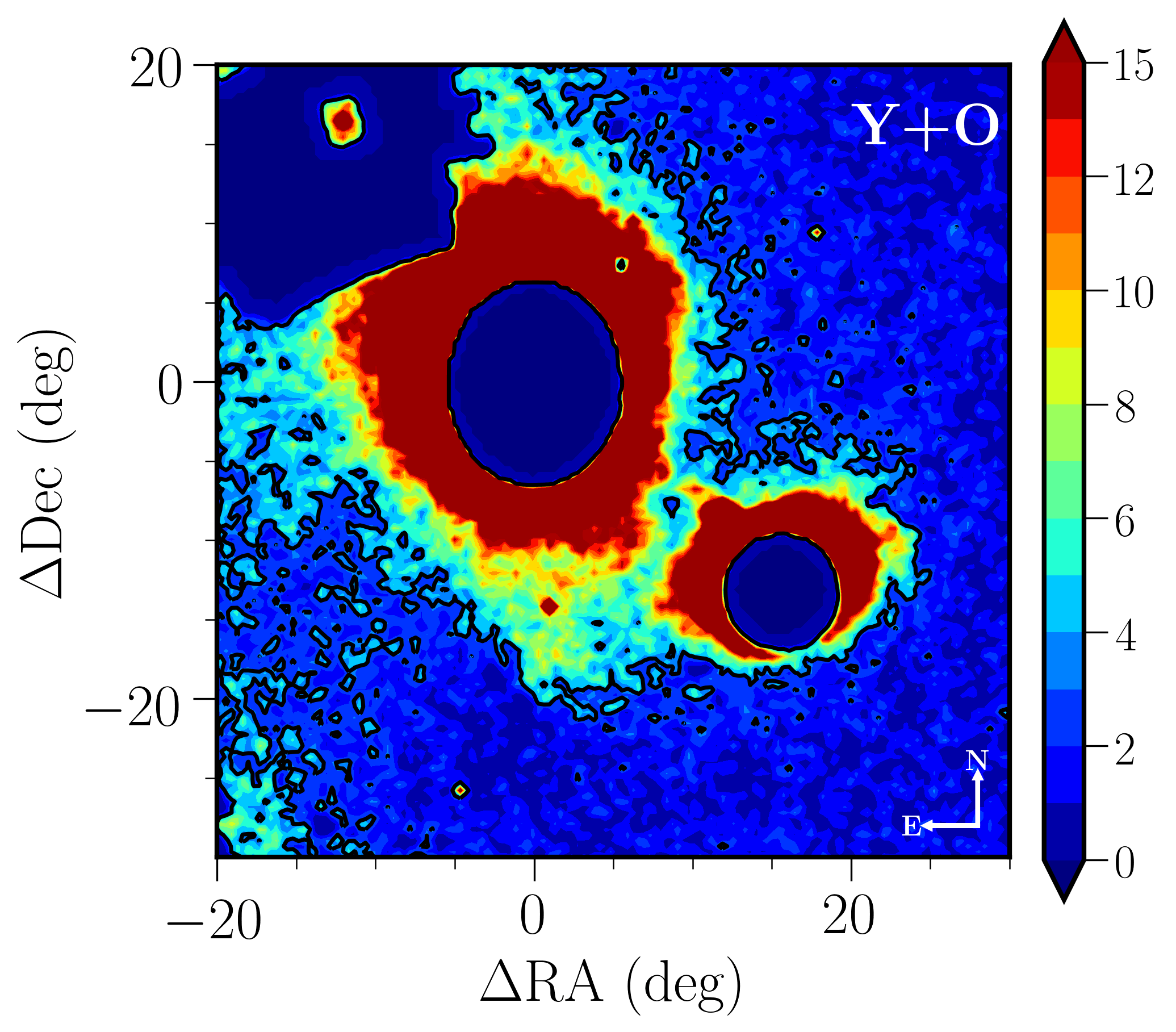}
	\includegraphics[scale=0.09]{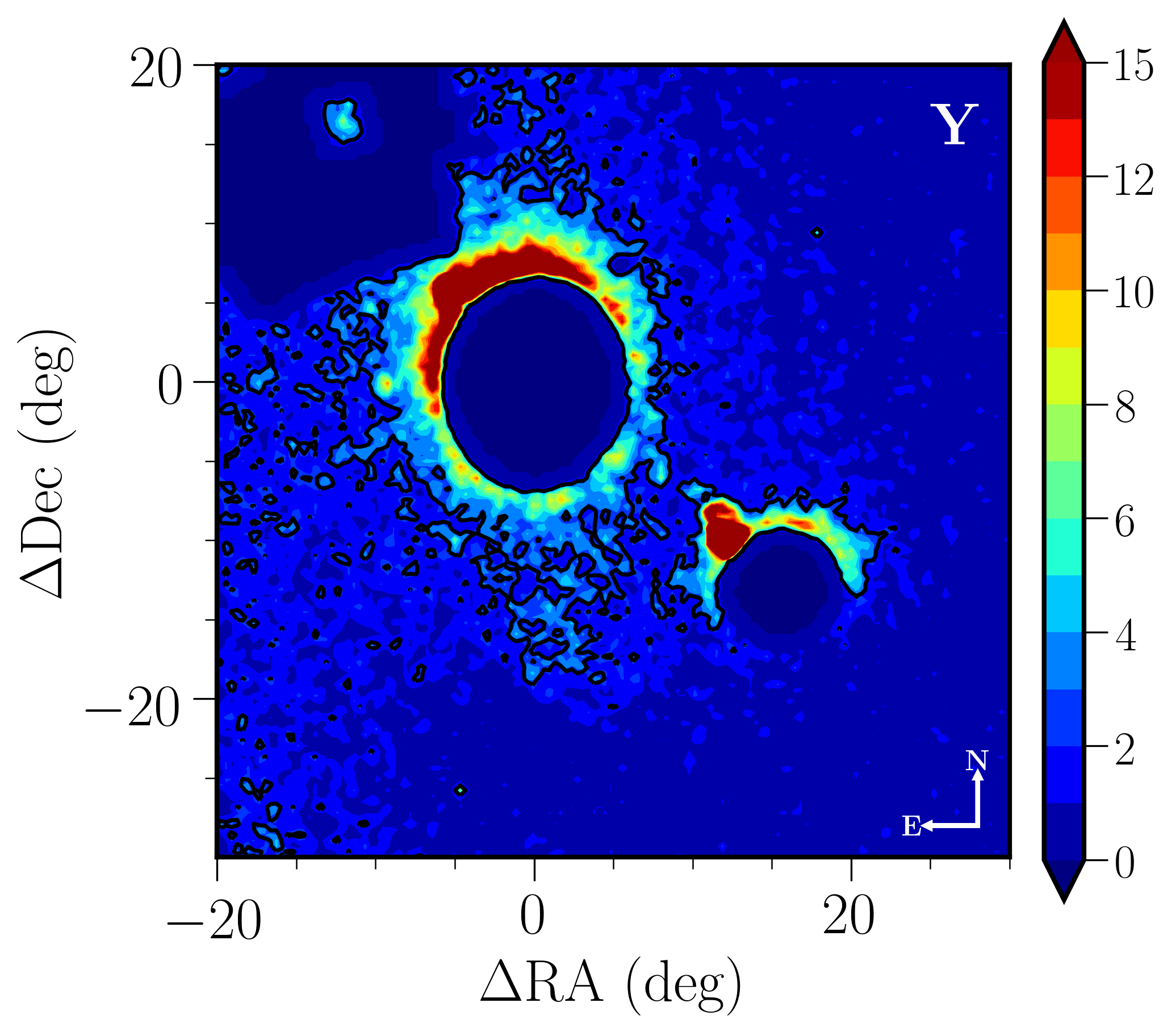}
	\includegraphics[scale=0.09]{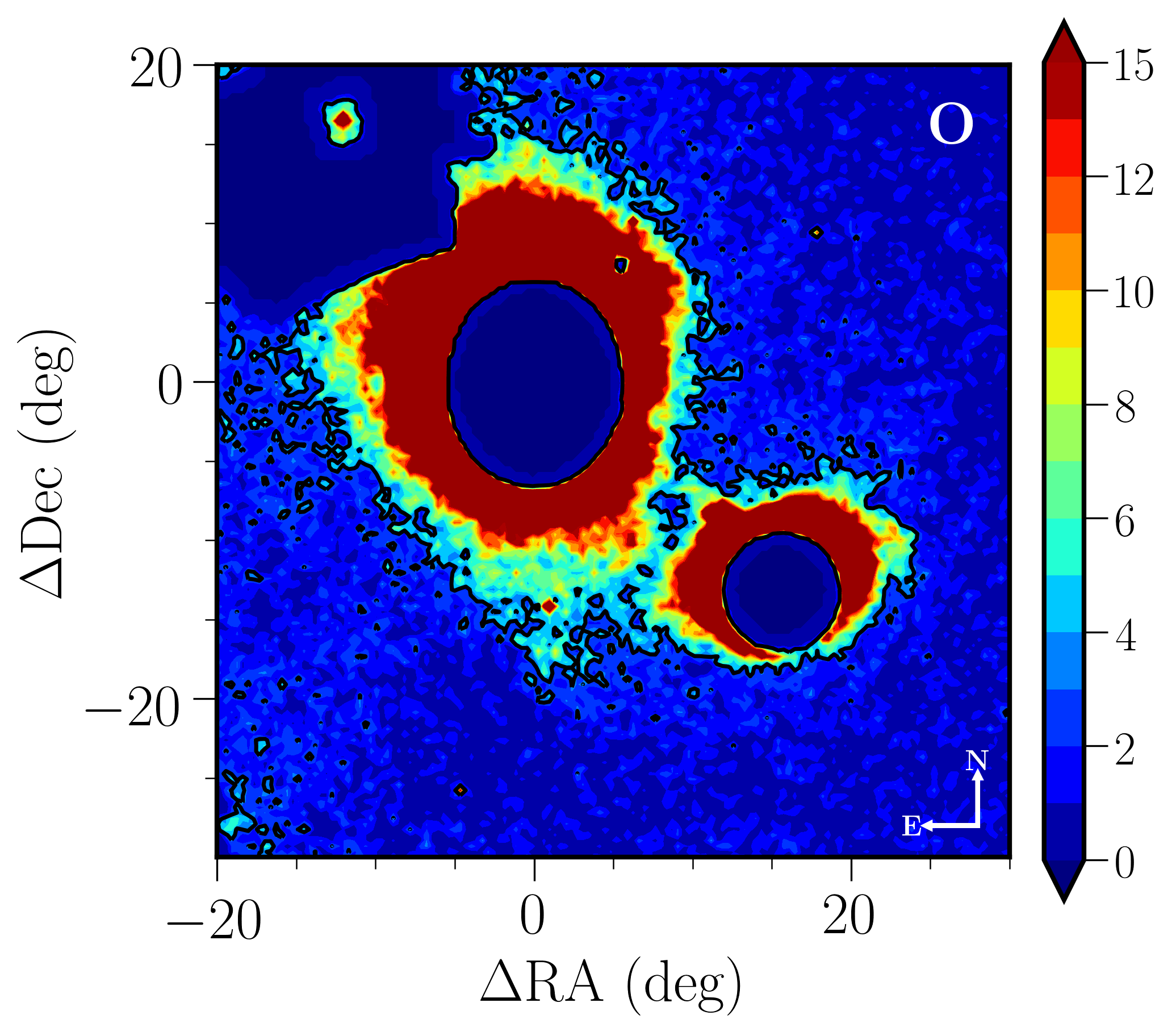}\\
	\includegraphics[scale=0.09]{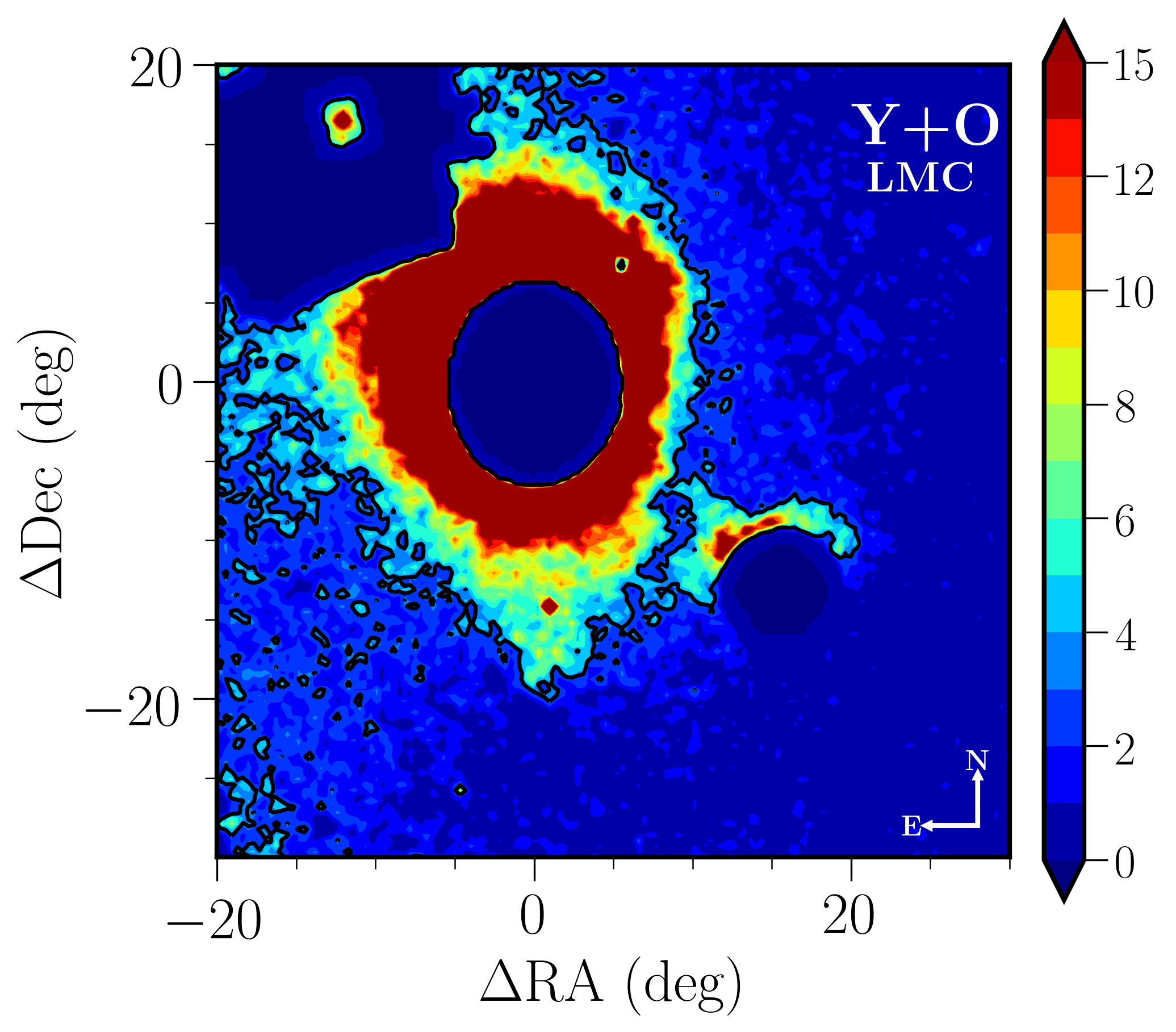}
	\includegraphics[scale=0.09]{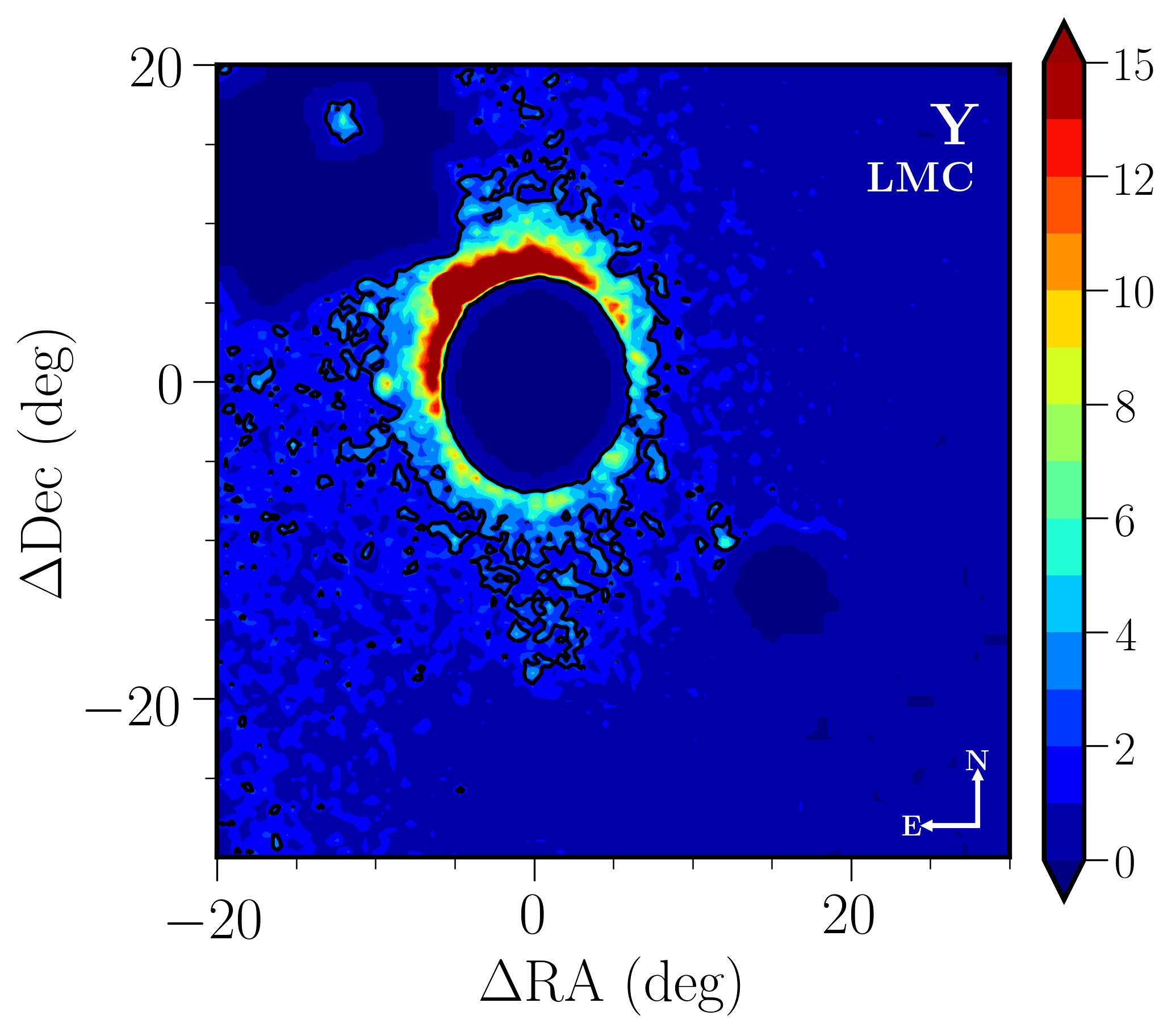}
	\includegraphics[scale=0.09]{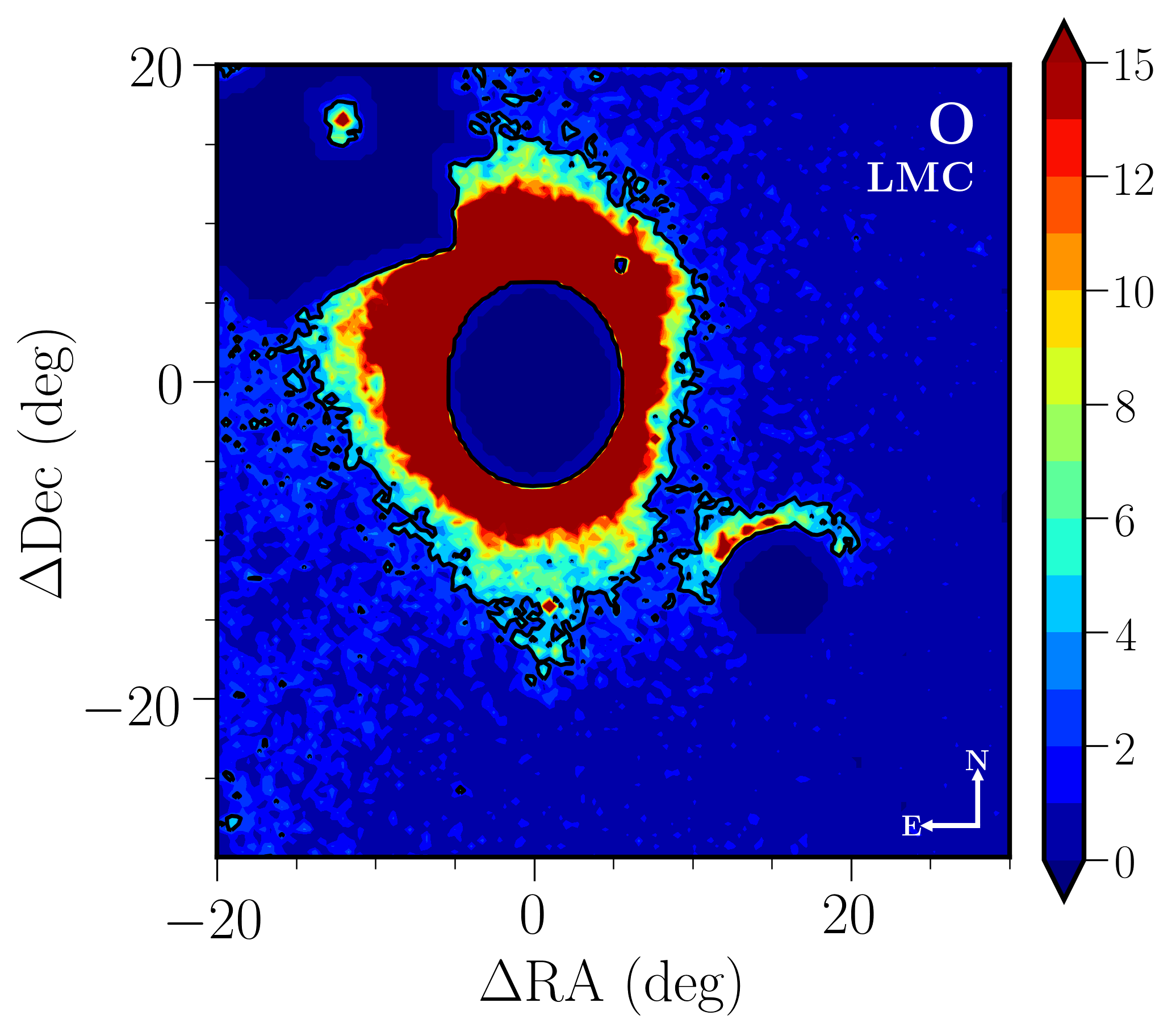}\\
	\includegraphics[scale=0.09]{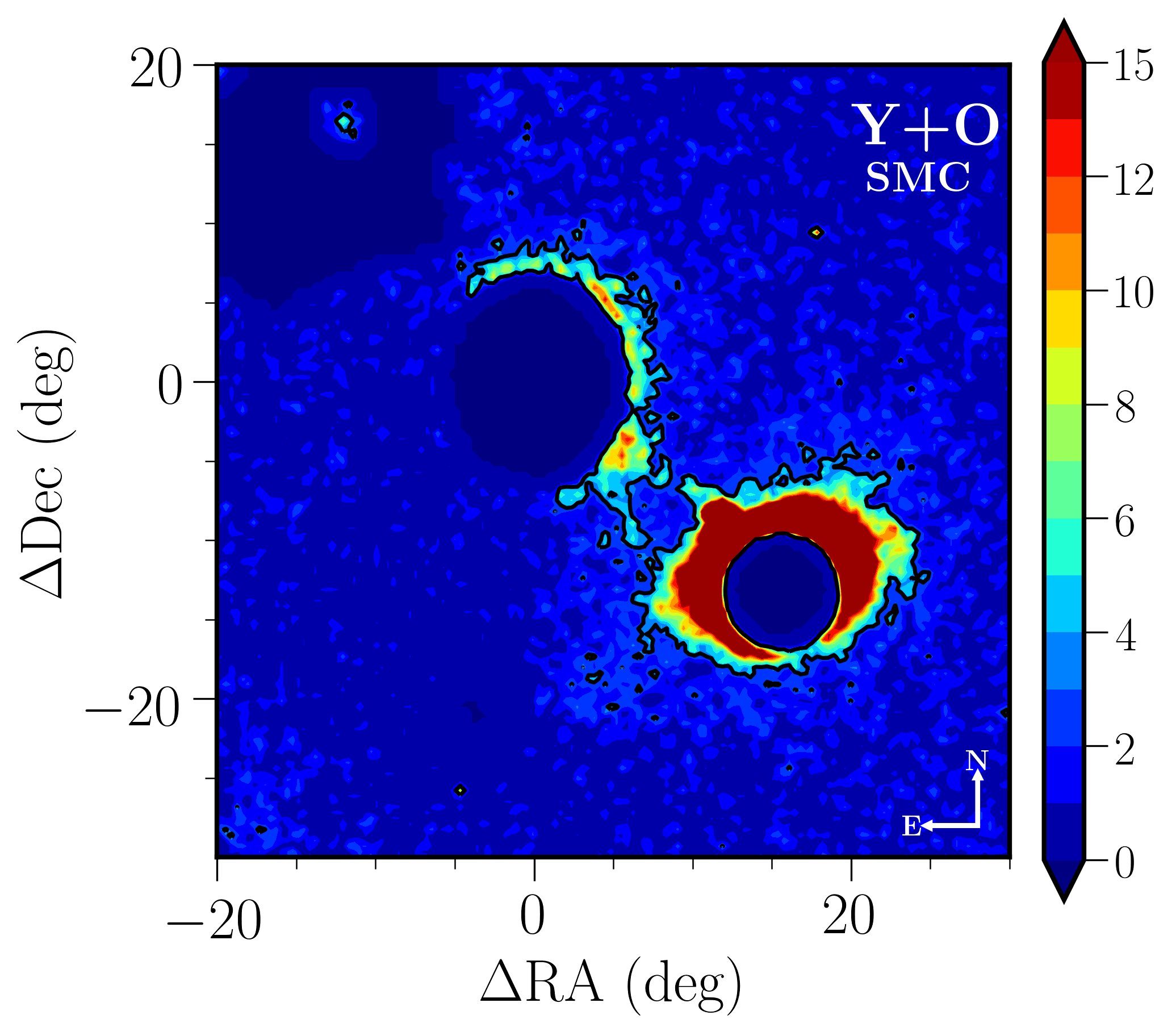}
	\includegraphics[scale=0.09]{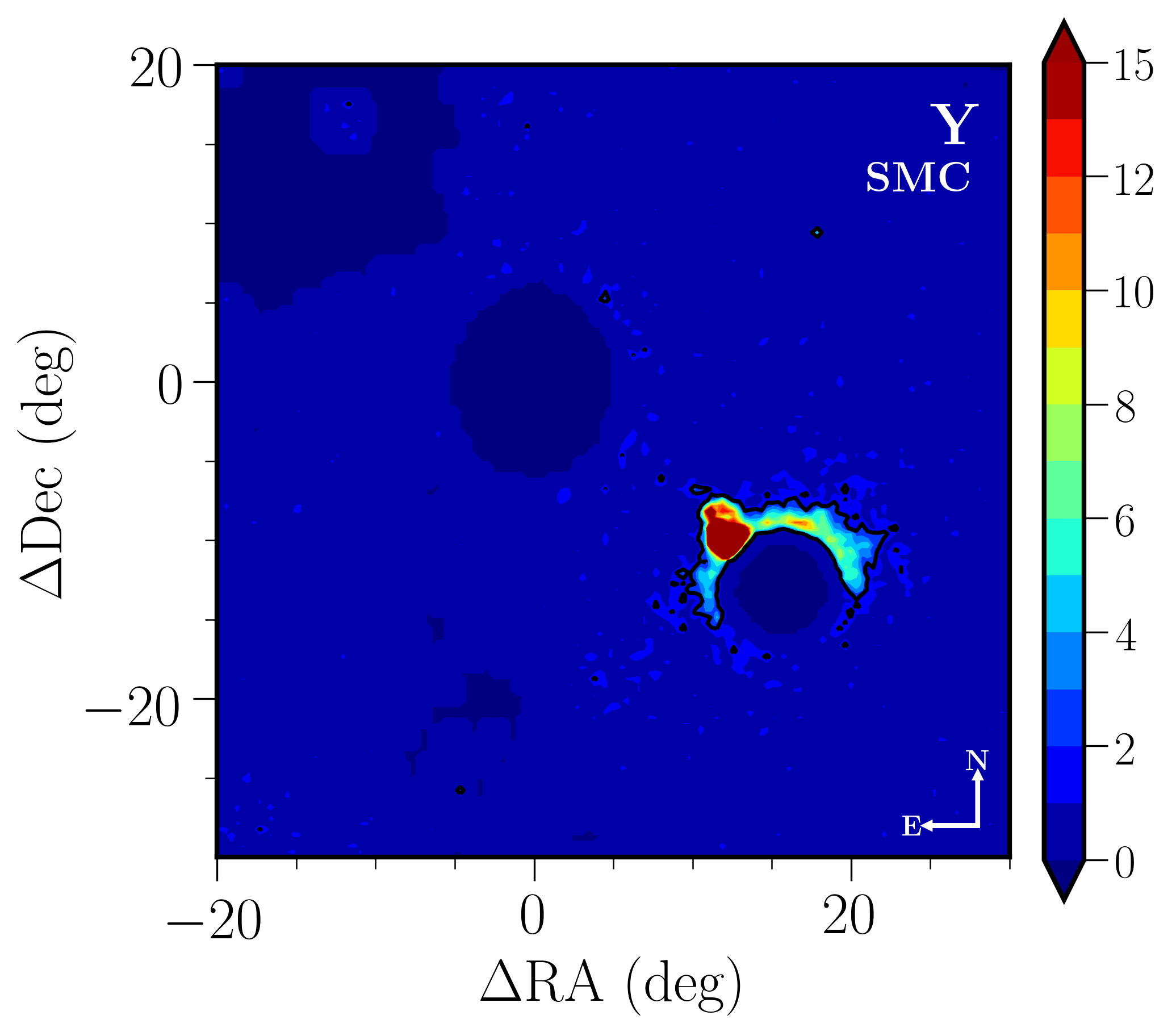}
	\includegraphics[scale=0.09]{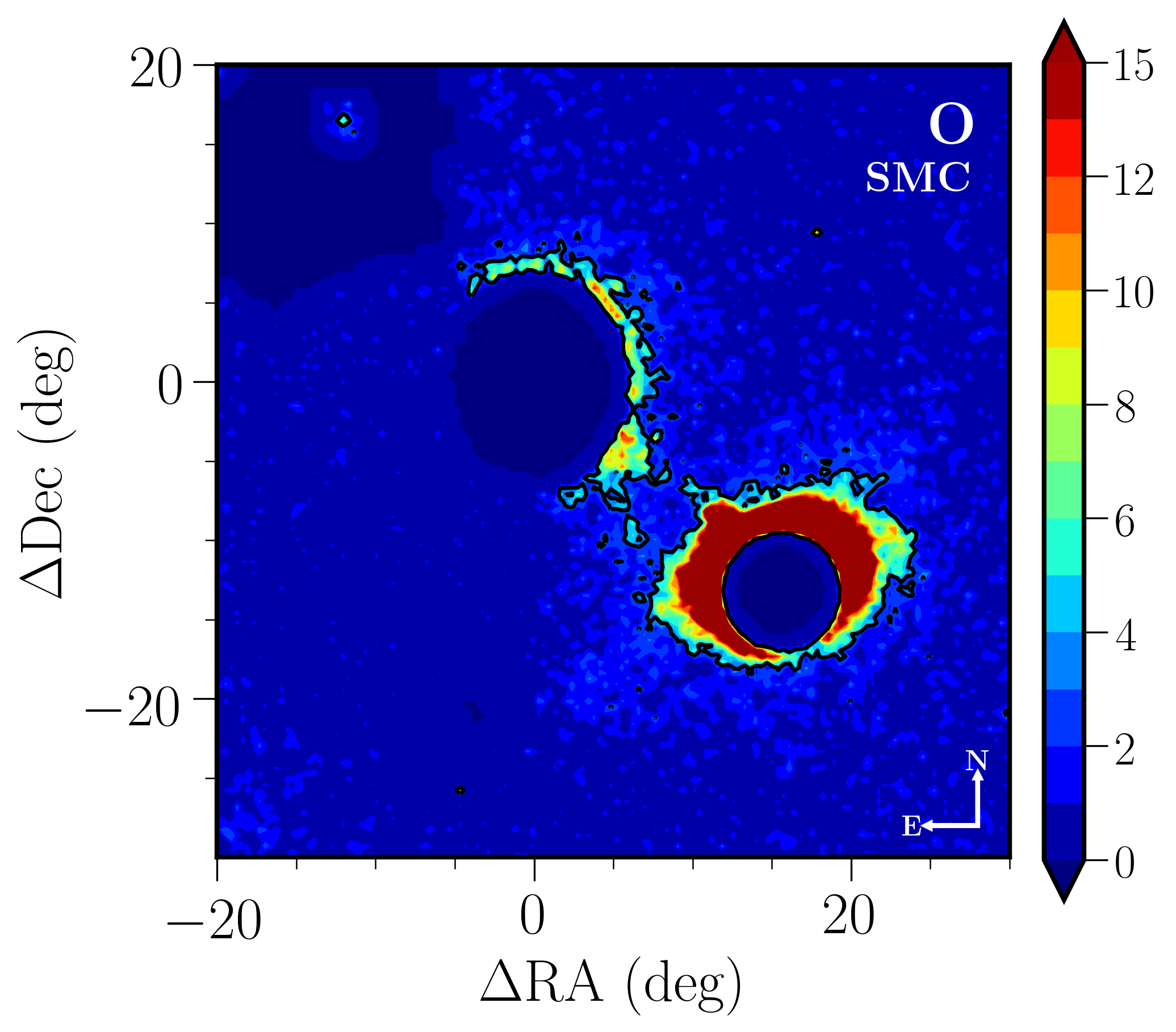}\\
	\caption{Morphology maps of young stars (Y; left) dominated by tip of MS and  supergiant stars, old stars (O; middle) dominated RGB and RC stars, young and old stars (Y+O; right) surrounding the MCs. The three rows show stars selected based on LMC or SMC proper motions (top), LMC proper motions (middle) and SMC proper motions (bottom). The bin size is $0.12$ deg$^2$ and the maps are centred at (RA$_0$, Dec$_0$) = ($81.24^\circ$, $-69.73^\circ$). The colour bars show the numbers of stars per bin. The central regions of the LMC and SMC have been masked out to enhance the distribution of stars in the outer regions.}
	\label{fig:Morphology}
\end{figure}
\end{landscape}

We used the NIR ($J-K_\mathrm{s}$, $K_\mathrm{s}$) CMD to divide the stellar populations into young and old stars. The VHS photometry reaches one magnitude below the RC. Figure \ref{fig:features} outlines the morphological features discussed in this work, while Figure \ref{fig:Morphology} shows morphology maps of the spatial density of young, old and young+old stars using LMC, SMC and LMC or SMC proper motion  selections (Section \ref{section22}). The bin size is 0.12~deg$^2$. The projection origin for the maps is set at (RA$_0$, Dec$_0$) = ($81.24^\circ$, $-69.73^\circ$) corresponding to the densest point in the LMC bar \citep{DeVaucouleurs1972}. The transformation of coordinates from angular to Cartesian was performed through a zenithal equidistant projection \citep{VanderMarel2001a}. The LMC globular clusters NGC 1841 ($\Delta$RA = $1^\circ$, $\Delta$Dec = $-14^\circ$), Reticulum ($\Delta$RA = $6^\circ$, $\Delta$Dec = $10^\circ$) and the MW globular cluster NGC 1261 ($\Delta$RA = $17.5^\circ$, $\Delta$Dec = $10^\circ$) appear as clear overdensities in the morphology maps. 

\subsection{Morphological features}
The morphology maps reveal striking substructures in the periphery of the MCs. The Northern Substructure 1 of the LMC detected by \cite{Mackey2016} ($\Delta$RA = $0^\circ$, $\Delta$Dec = $15^\circ$) is most prominent in the old  population extending to $\sim$ 20$^\circ$ from the LMC centre, it is present also in the map of the young population but to a lesser radial extent. The extension of this feature towards the NE is not visible due to incomplete VHS observations. 
We also detect two substructures on the eastern side of the LMC (Eastern Substructures 1 and 2), revealed at ($\Delta$RA = $-15^\circ$, $\Delta$Dec = $1^\circ$) and ($\Delta$RA = $-15^\circ$, $\Delta$Dec = $-6^\circ$), and emerging in the direction of the Galactic disc. We find that these substructures are present despite the strict proper motion cuts we applied (see Section \ref{section23}), suggesting they belong to the LMC rather than to the MW. In the young population and somewhat closer to the LMC disc, Eastern Substructure 1 delineates an approximate circle while Eastern Substructure 2 is similar to a thin stream. There is also a high-density clump between the two substructures at about ($\Delta$RA = $-10^\circ$, $\Delta$Dec = $0^\circ$). These features are less prominent in the old population. \cite{DeVaucouleurs1955} identified a circular feature in the eastern LMC disc at about ($\Delta$RA = $-14^\circ$, $\Delta$Dec = $2^\circ$) which most likely coincides with Eastern Substructure 1. Eastern Substructure 2 is a new discovery.

South of the LMC we find significant overdensities encompassing the locations of substructures previously identified by \cite{Mackey2018}: Southern Substructure 1 at ($\Delta$RA = $-2^\circ$, $\Delta$Dec = $-12.5^\circ$) and Southern Substructure 2 at ($\Delta$RA = $6^\circ$, $\Delta$Dec = $-10^\circ$). These substructures are most prominent in the old population. 
The LMC globular cluster NGC 1841 appears embedded in a low density substructure extending farther south than Substructures 1 and 2. This southernmost extension (Southern Substructure 3; $\Delta$RA = $2^\circ$, $\Delta$Dec = $-17.5^\circ$) probably corresponds to the stream-like feature previously detected by \cite{Belokurov2019}. In our maps it has a high density in the old populations and its connection to Southern Substructure 1 is unclear.
Southern Substructures 1 and 3 characterise the maps obtained from LMC-based proper motions, whereas Southern Substructure 2 is also present in SMC-based proper motion maps, suggesting a possible connection between the LMC and the SMC. This connection, which is supported by an overall extension of the SMC populations towards the east, may be associated with the Old Bridge \citep{Belokurov2017}.

The east of the SMC shows several protuberances that emerge from its more regular elliptical body. We confirm the SMC East substructure \citep{Mackey2018} at ($\Delta$RA = $18^\circ$, $\Delta$Dec = $-14^\circ$) and the extension of the Wing towards the Magellanic Bridge at ($\Delta$RA = $9.5^\circ$, $\Delta$Dec = $-6.5^\circ$). The SMC East substructure together with the SMCNOD overdensity (\citealp{Pieres2017}) west of the SMC at ($\Delta$RA = $23^\circ$, $\Delta$Dec = $-10^\circ$) are most obvious in the old population, whereas the extension towards the Magellanic Bridge is better delineated in the young population. Between the Wing and SMCNOD features there is a clear overdensity extending outwards of the galaxy at ($\Delta$RA = $19.5^\circ$, $\Delta$Dec = $-7.5^\circ$) which is most prominent in the old population. This feature, which we refer to as Northern Substructure 2, appears also in previous studies (e.g. \citealp{Belokurov2019}), but it was not specifically mentioned. All of these SMC features are most prominent in maps created using SMC-based proper motions. Combining young and old populations enhances the level of substructure in the outskirts of the MCs with respect to the MW.

\subsection{Distance and proper motion of morphological features}
\begin{figure*}
	\centering
	\includegraphics[scale=0.25]{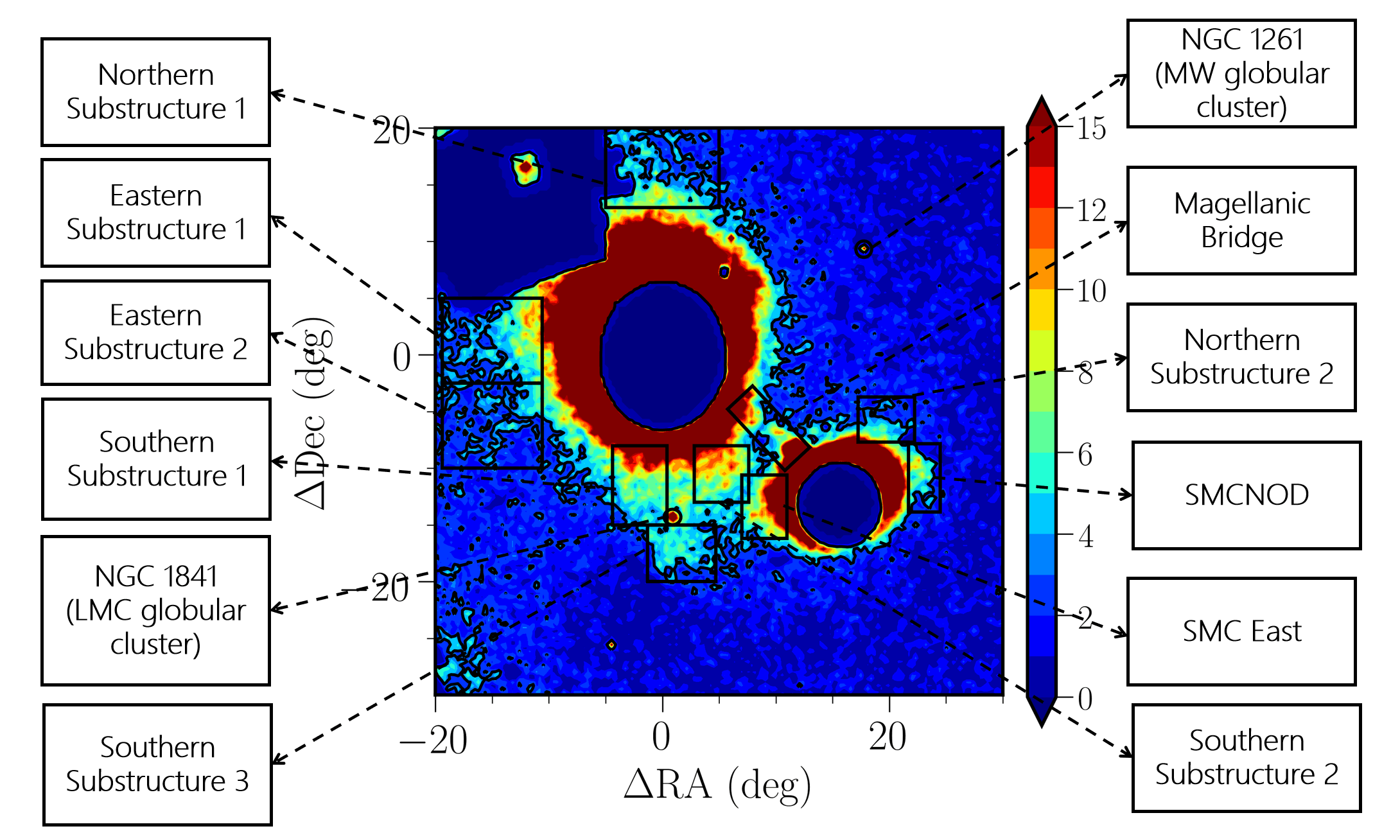}
	\caption{Morphological features in the outskirts of the MCs discussed in this paper. The underlying density is that of young and old stars, the bin size is $0.09$ deg$^2$, the colour bar shows the number of stars per bin whereas the map is centred at (RA$_0$, Dec$_0$) = ($81.24^\circ$, $-69.73^\circ$). We used  elliptical and circular masks of the central region of the LMC and SMC, respectively, to enhance the distribution of stars in the outer regions. Some of these morphological features have already been reported in previous studies: the Eastern Substructure 1 by \protect\cite{DeVaucouleurs1955}, the Northern Substructure 1 by \protect\cite{Mackey2016}, SMCNOD by \protect\cite{Pieres2017}, the Southern Substructure 3 by \protect\cite{Belokurov2019}, the Southern Substructures 1 and 2 as well as the SMC East by \protect\cite{Mackey2018}, and the Magellanic Bridge by \protect\cite{Hindman1963}. The extension of the LMC and SMC outskirts outlined by the SMC East and Southern Substructure 2 embody the Old Bridge of \protect\cite{Belokurov2017}, whereas the Eastern Substructure 2 and Northern Substructure 2 have been highlighted for the first time in this work.}
	\label{fig:features}
\end{figure*}

\begin{figure*}
	\includegraphics[scale=0.06]{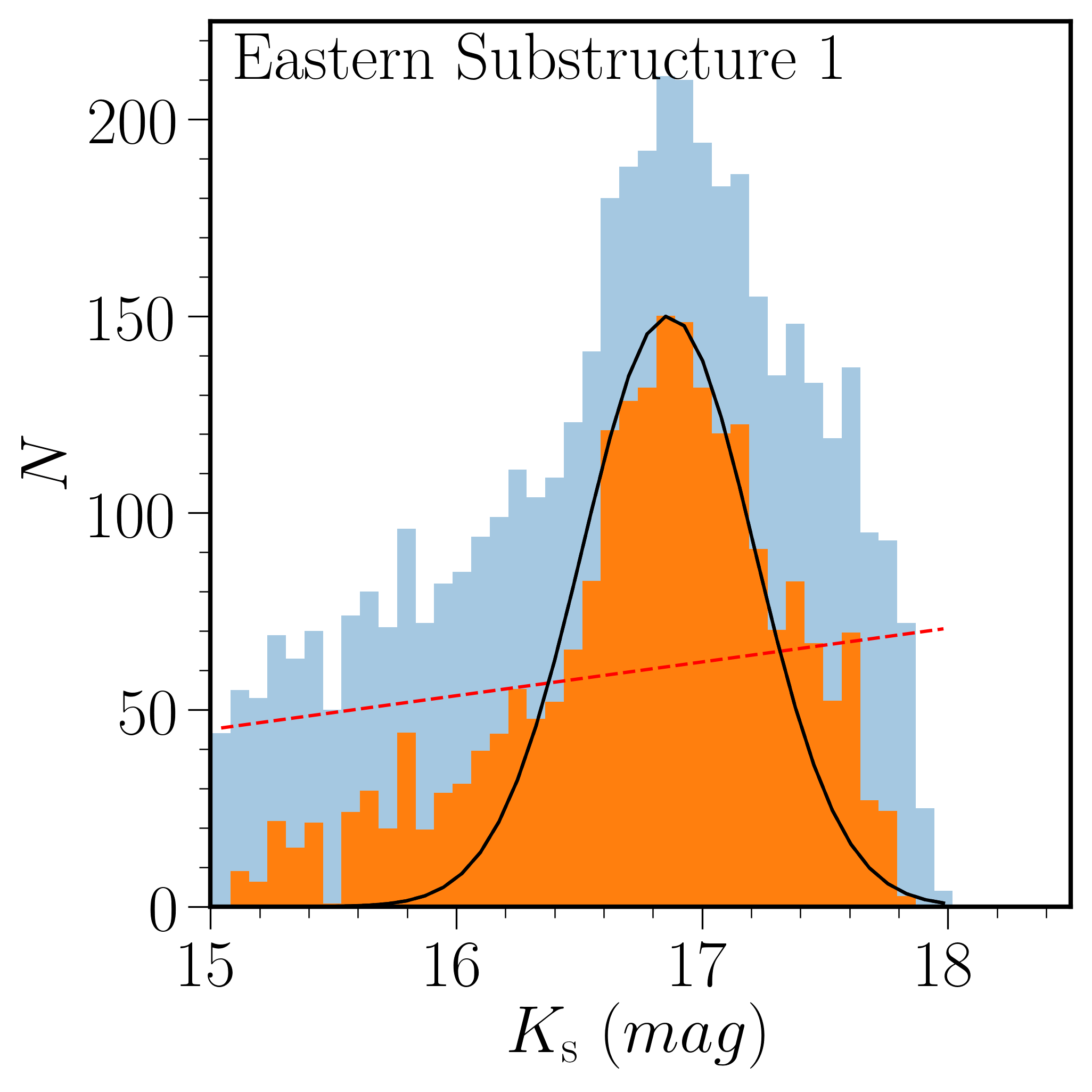}
	\includegraphics[scale=0.06]{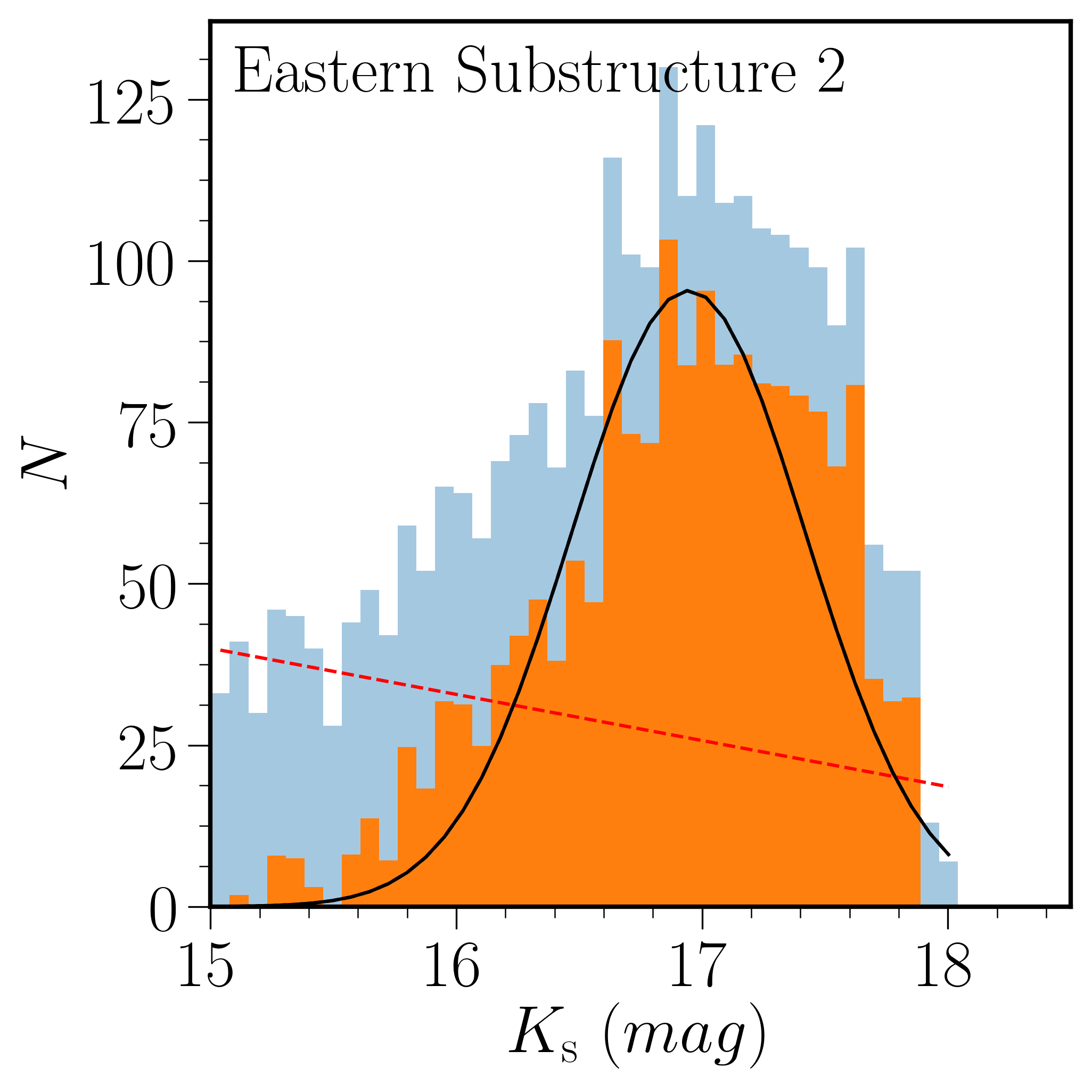}
	\includegraphics[scale=0.06]{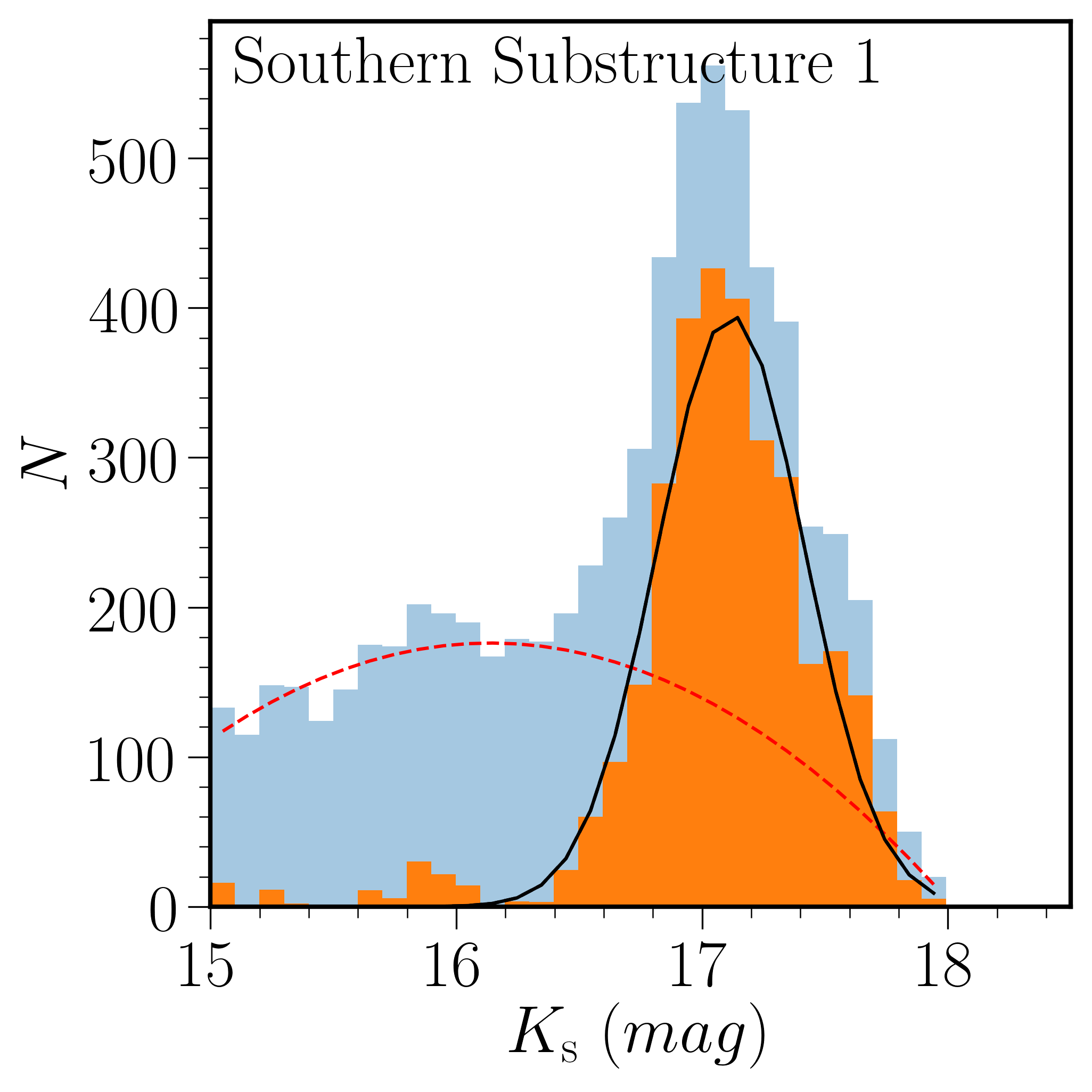}
	\includegraphics[scale=0.06]{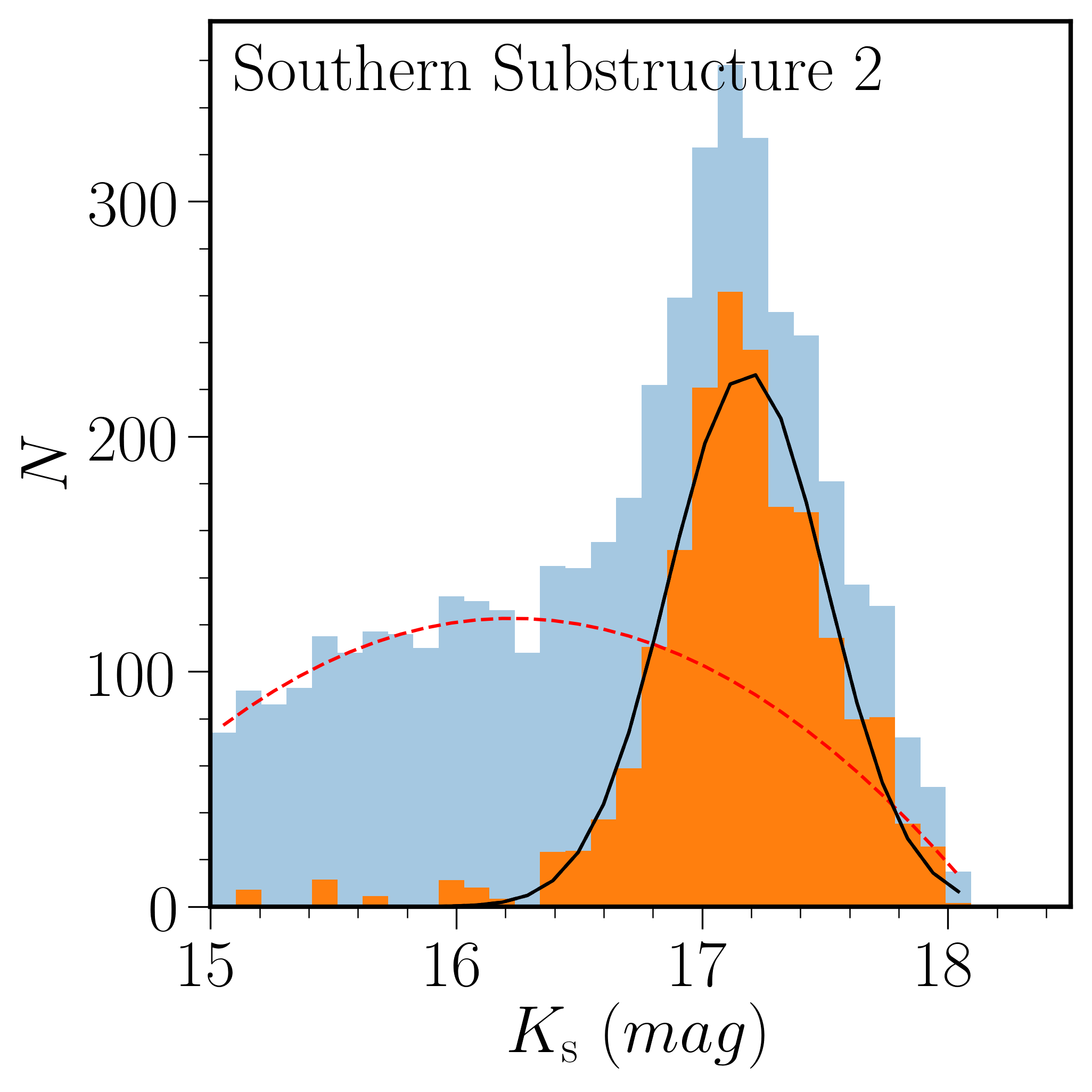}
	\includegraphics[scale=0.06]{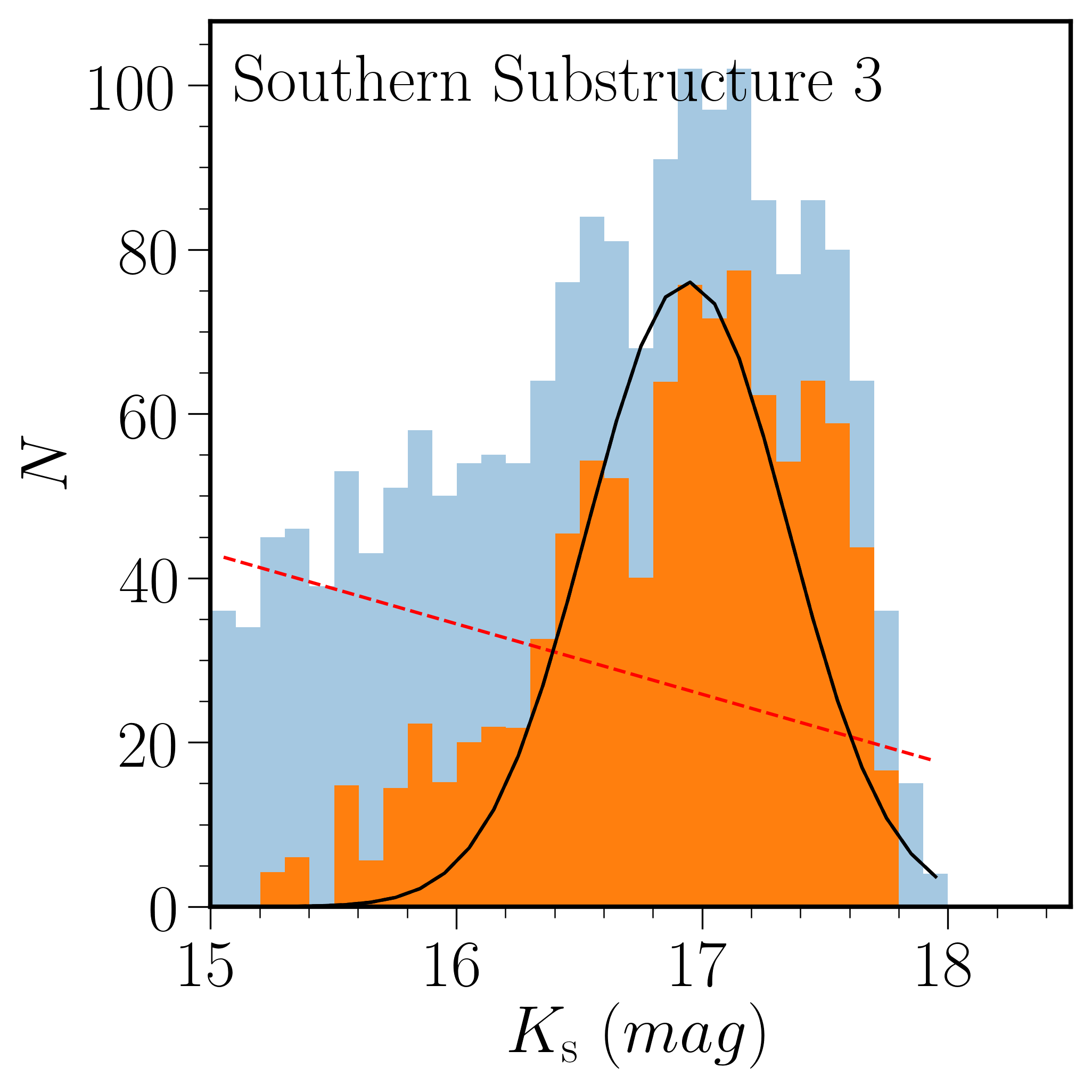}
	\includegraphics[scale=0.06]{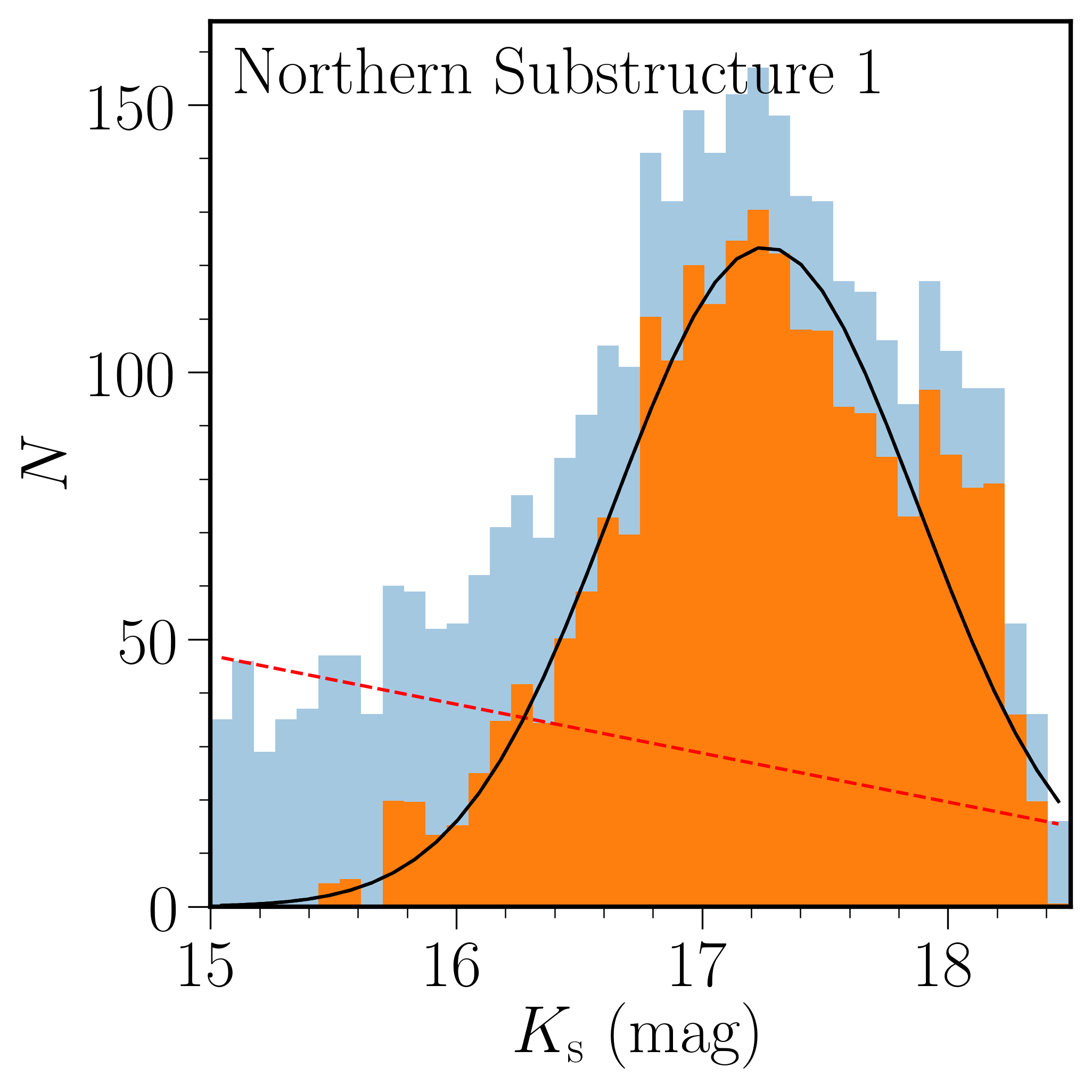}
	\includegraphics[scale=0.06]{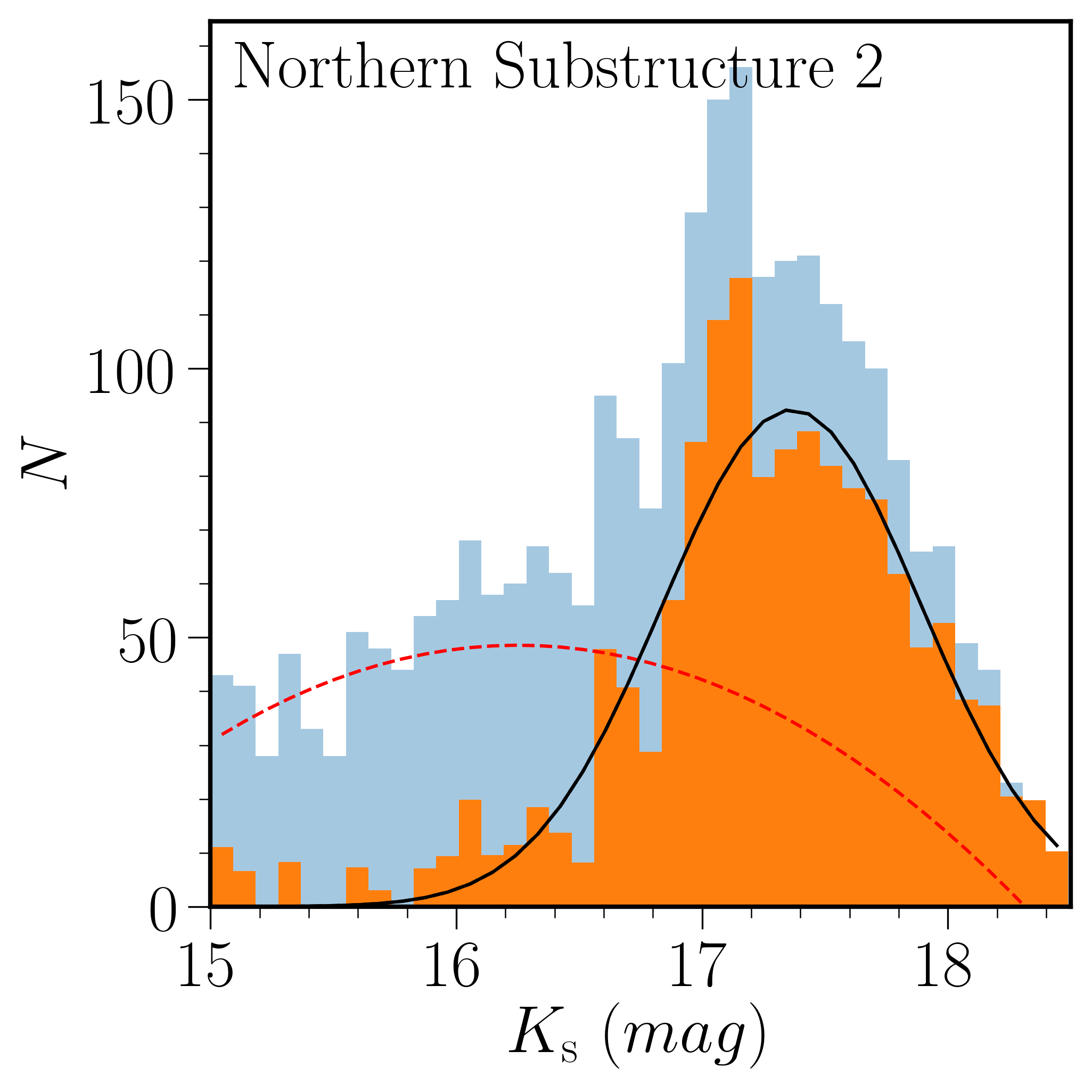}
	\includegraphics[scale=0.06]{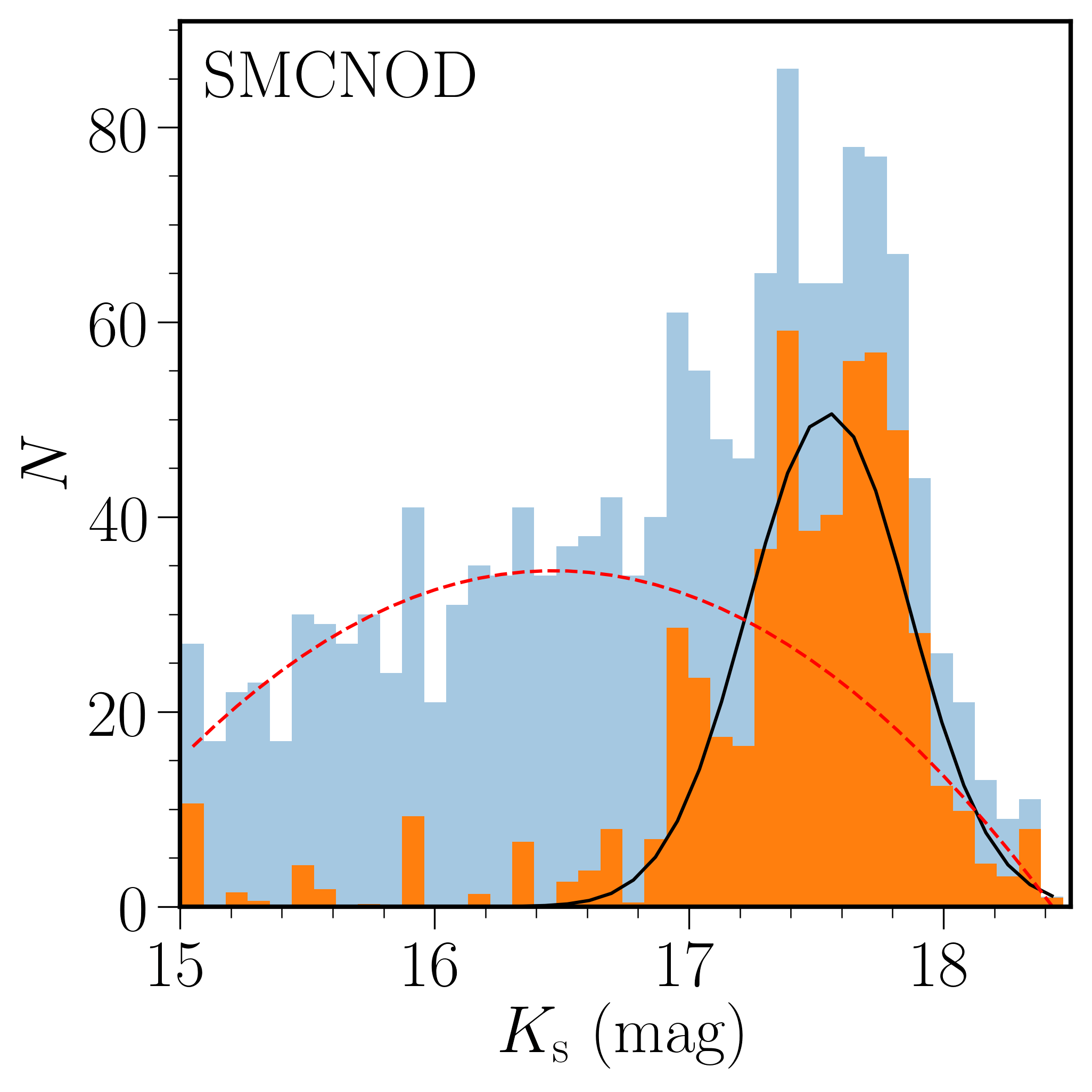}
	\includegraphics[scale=0.06]{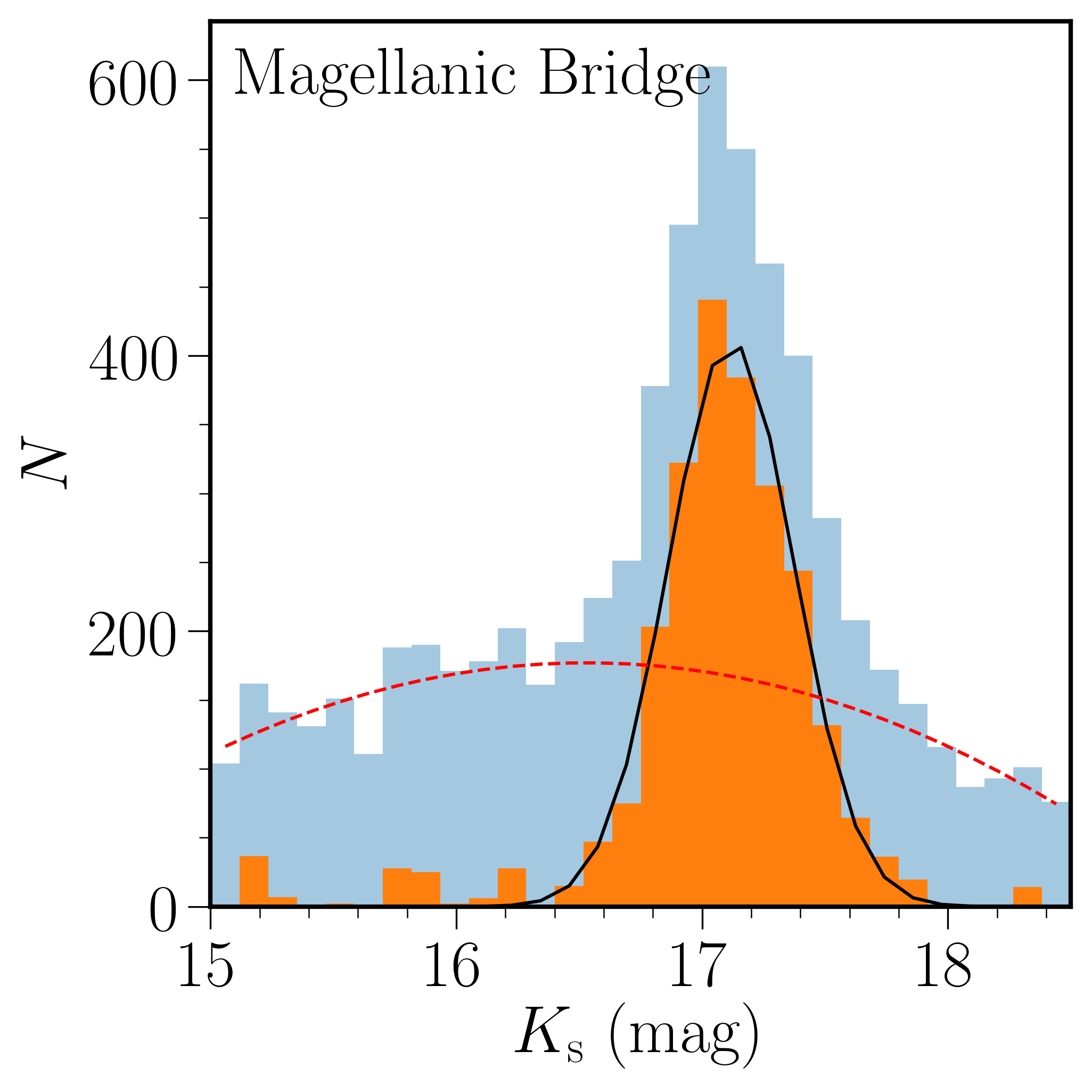}
	\includegraphics[scale=0.06]{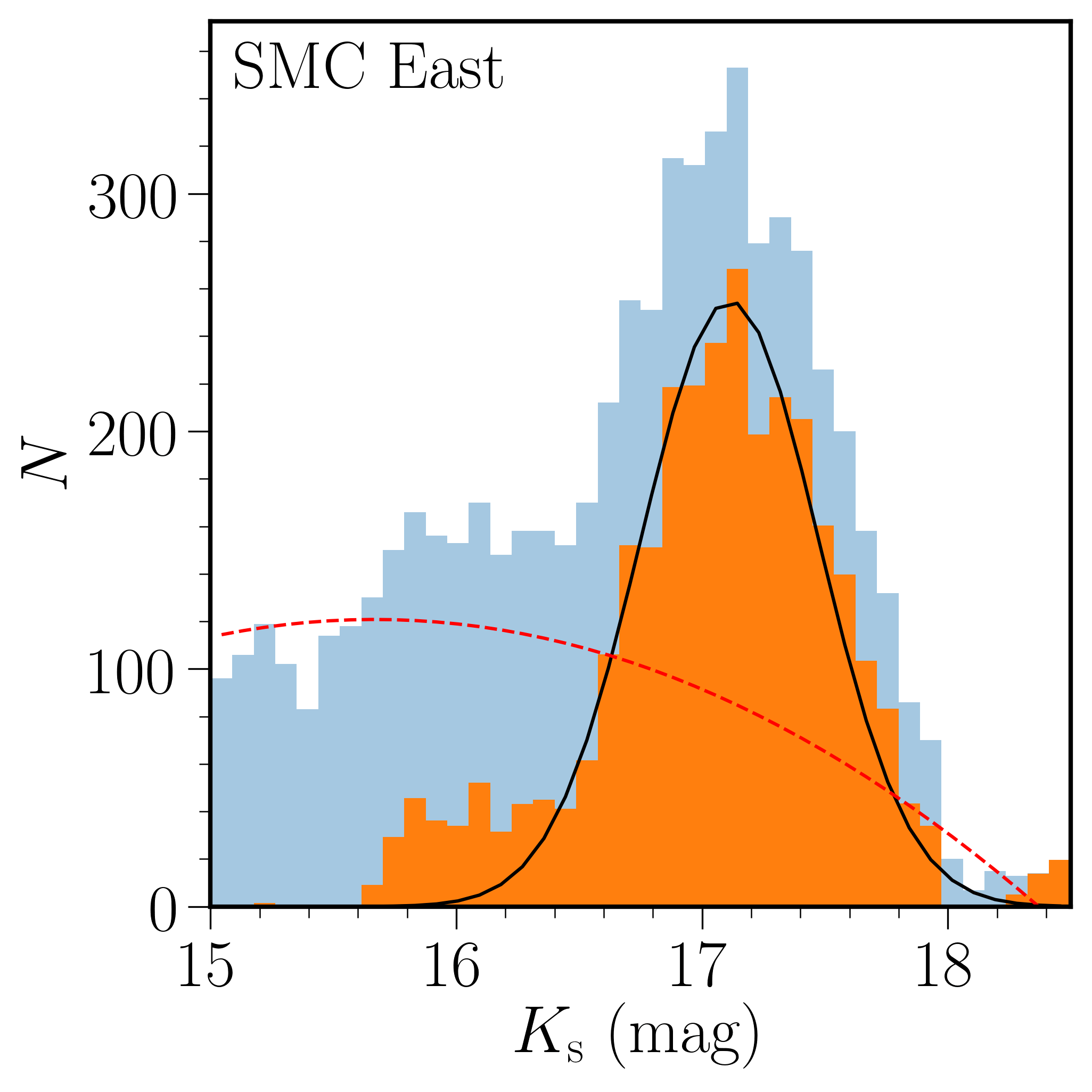}
	\caption{Luminosity function of the RC stars of the different morphological features outlined in Figure \ref{fig:features}. The blue histograms show the observed luminosity functions, whereas orange histograms show the distributions after subtracting the RGB components, the continuous lines show the total fits to the distributions whereas the dashed lines represent the separate components of the fits. The bin size is of 0.09 mag.}
	\label{fig:features_distance}
\end{figure*}

We determined distances to the morphological features by examining the luminosity function of RC stars within the areas defined in Figure \ref{fig:features} and using the CMD box outlined in Section \ref{section4}. We employed a multi--component non--linear least squares fitting technique to characterise the peak of the magnitude distribution produced by RC stars for each feature. Figure (\ref{fig:features_distance}) and Table \ref{table:rcfeatures} show the histograms and parameters, respectively, obtained for each morphological feature following the same methodology to study the double RC feature across the SMC (Section \ref{section4}). The stellar populations seen in Eastern substructures 1 and 2  features lie at distances similar to that of the LMC ($\sim$50 kpc, e.g. \citealp{DeGrijs2014b}). This strengthens the hypotheses that the two substructures are either owing to material being tidally stripped from the disc, or are overdensities in the structure of the disc itself, ruling out that they may be shreds of a dwarf galaxy. We found that Southern Substructures 1 and 2, as well as the SMC East Substructure, are located at distances similar to that of the Magellanic Bridge. If Southern Substructure 2 and SMC East were to be associated to an Old Bridge then this structure would be at the same distance of the Magellanic Bridge. We also found that Southern Substructure 3 is located somewhat closer than the other southern substructures, but this is only marginal in view of the uncertainties involved and the difficult RGB subtraction compared to the other regions. Northern Substructure 1 appears instead further away than the LMC main body suggesting it is made of material tidally stripped from the LMC disc as a result of an interaction with the SMC (\citealp{Mackey2016}). On the other hand, Northern Substructure 2 appears at the same distance of the SMC main body and it is probably a feature of its ellipsoidal structure. The feature might be undergoing tidal stripping as a consequence of ram pressure effects which dislodged material from the galaxy (e.g. \citealp{Tatton2020}). Additionally, we find that the SMCNOD substructure is the most distant feature. SMCNOD is located 8$^\circ$ away from the SMC centre and its stellar population is similar to that of the SMC main body which, in agreement with \cite{Pieres2017}, suggests that it was probably removed from the galaxy.

We have also examined the proper motions of the stars enclosed within each morphological feature and compared them with the proper motion ($\mu_{\mathrm{RA}}$, $\mu_{\mathrm{Dec}}$) of the LMC ($1.871$, $0.391$) mas~yr$^{-1}$ and SMC ($0.686$, $-1.237$) mas~yr$^{-1}$ obtained using \textit{Gaia} early DR3 (\citealp{Luri2020}). These values are consistent with the \textit{Gaia} DR2 values derived by \cite{Helmi2018} up to $6^\circ$ and $4^\circ$ from the LMC and SMC centres, respectively. Table \ref{table:pm} shows for each feature the number of stars (Y+O according to the CMD selection described in Sect.~\ref{section22}), the angular distance from the LMC/SMC centres, and the median proper motions with the respective spread.

The features around the LMC are located at angular distances of 11-18$^\circ$ from the centre, whereas those around the SMC are at $7-8^\circ$ from the centre. In RA the Southern and Northern substructures have higher proper motion values than the Easter substructures, all features show however proper motions lower than the LMC value. In Dec, Eastern substructures 1 and 2 as well as Southern Substructure 1 show higher proper motions than the LMC contrary to the other southern substructures and Northern Substructure 1 which show lower values. Assuming that the substructures belong to the LMC disc then the viewing perspective may account for proper motion differences of $0.15-0.31$ mas~yr$^{-1}$ (\citealp{Kallivayalil2006}). Except for Eastern Substructure 1 and 2, and for Southern Substructure 2 all other features have both proper motion components for which the difference with respect to the proper motion of the LMC can be explained by viewing perspective.

The proper motions of the features around the SMC appear higher in RA than in Dec compared to the values attributed to the galaxy. The structure of the SMC is complex and the influence of viewing perspective is difficult to estimate. The proper motion of the Magellanic Bridge is as expected the most discrepant (in RA) from the SMC because it connects both galaxies. \cite{Schmidt2020} derived ($\mu_\mathrm{RA}$ = $1.80\pm0.25$, $\mu_\mathrm{Dec}$ = $-0.72\pm0.13$) mas yr$^{-1}$ for the Bridge centre following a procedure that thoroughly removes the influence of MW stars.

\begin{figure*}
	\centering
	\includegraphics[scale=0.08]{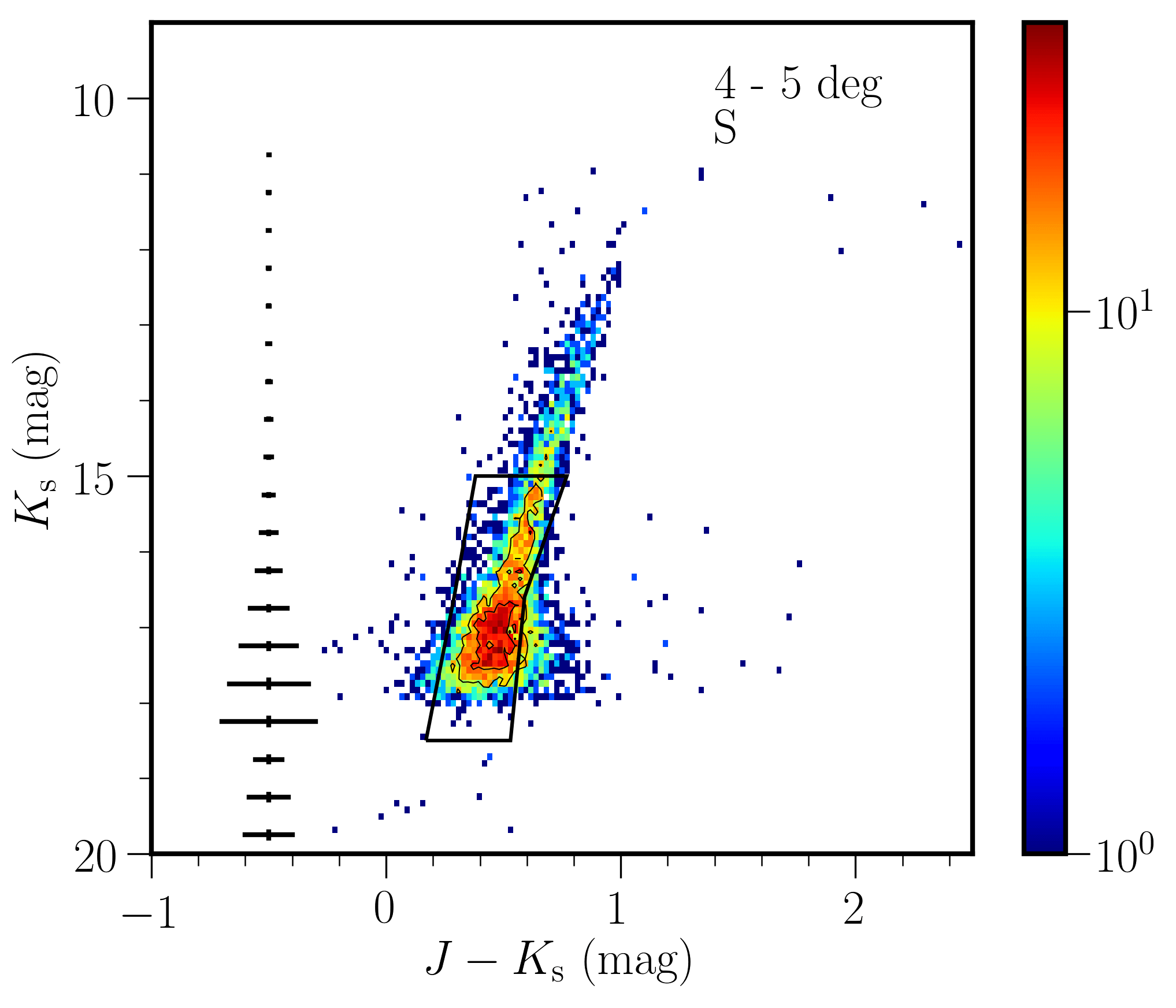}
	\includegraphics[scale=0.08]{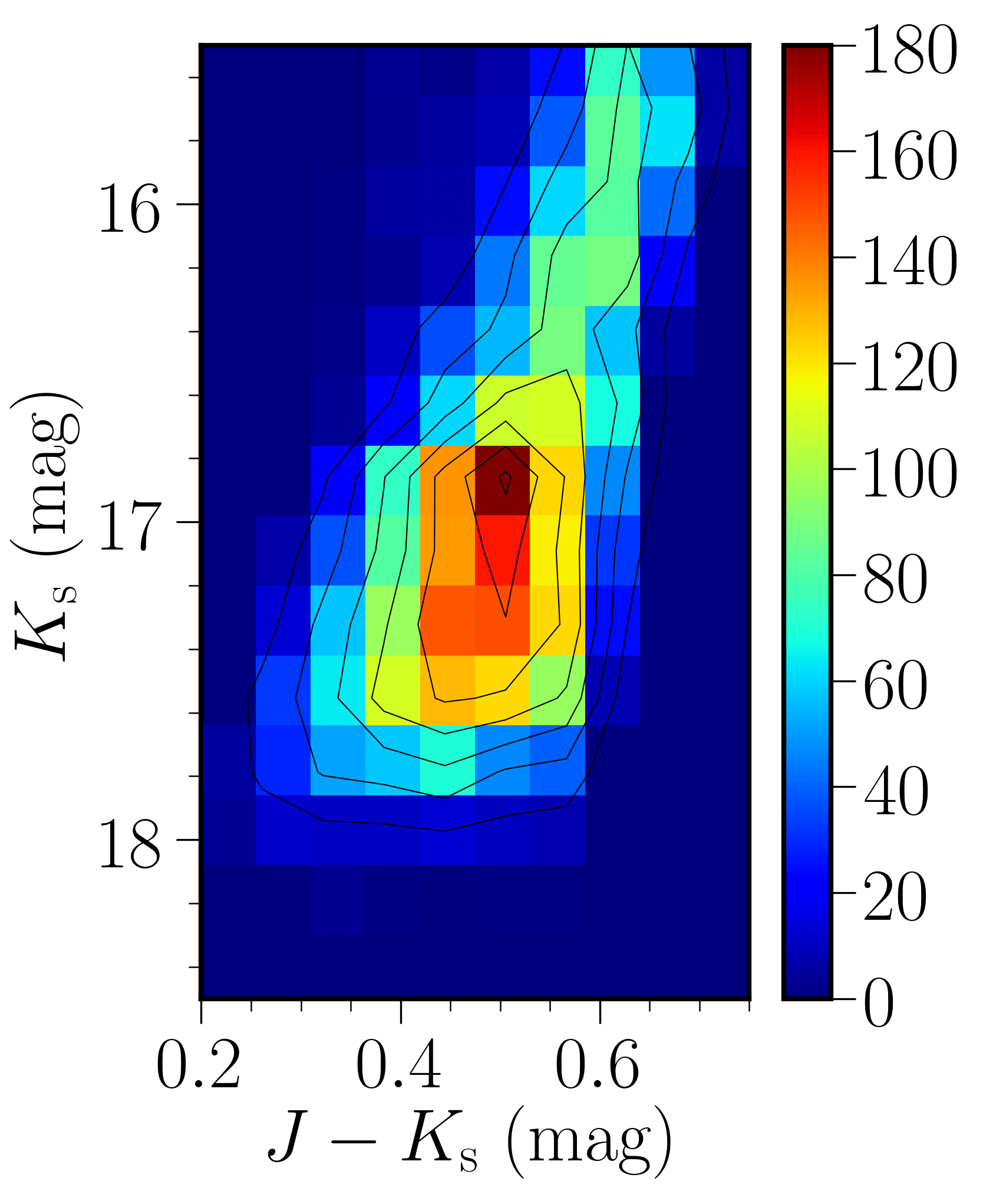}
	\includegraphics[scale=0.08]{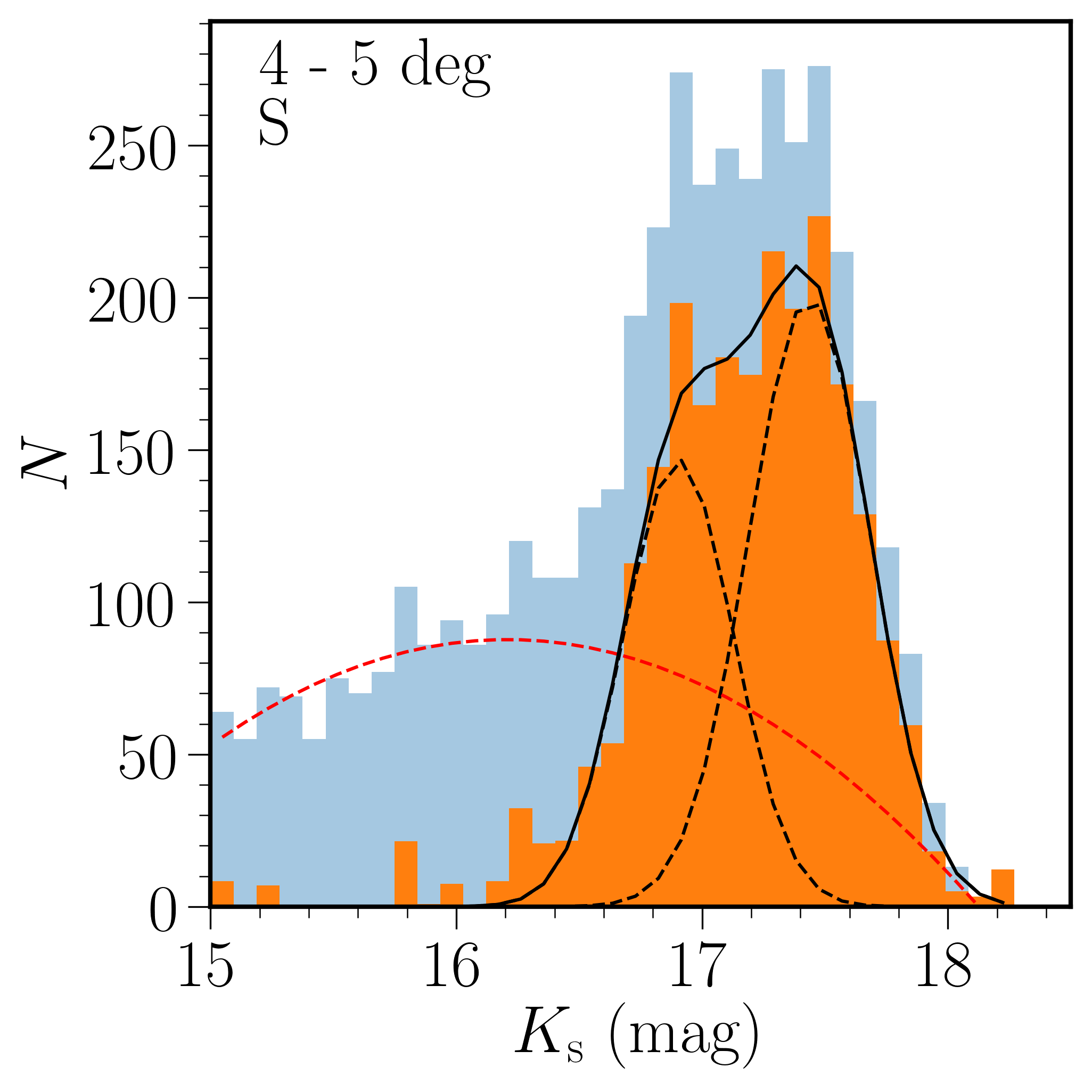}
	\caption{(left) Hess diagram (corrected for reddening) for SMC stars within the S sector and a 4--5$^\circ$ annulus. (middle) Zoom-in on the selected RC region. (right) Luminosity function of RC stars in the same sector and annulus. The blue histogram shows the distribution of all stars while the orange histogram shows the distribution after the subtraction of the RGB component. The continuous line shows their total fit, while dashed lines represent the separate Gaussian components of the fits (black) and the quadratic polynomial term used to account for RGB stars (red).}
	\label{fig:DRC}
\end{figure*}

\begin{table*}
	\caption{Gaussian parameters, distances and the reduced $\chi^2$ values of the profile fits to the luminosity function of the RC stars in different morphological features.}                        
	\label{table:rcfeatures}      
	
	\begin{tabular}{lrrrrrr}
		\hline
		Feature & $\mu$ & $\sigma$ &$N$ & $D$ & $\sigma_{\mathrm{D}}$ & $\chi^2$  \\
		             &  (mag) & (mag) & (counts) & (kpc) & (kpc) & \\
		\hline
		
		Eastern Substructure 1 & 16.86 $\pm$ 0.08 & 0.35 & 150 $\pm$ 35&47.9 $\pm$1.9 & 7.72 & 7.24\\
		Eastern Substructure 2 & 16.94 $\pm$ 0.04 & 0.47 & 95 $\pm$ 8 &49.7$\pm$1.0 & 10.76 & 4.85\\
		Southern Substructure 1 & 17.12 $\pm$ 0.01 & 0.29 & 395 $\pm$ 19 &53.9 $\pm$ 0.4 & 7.21 & 2.94\\
		Southern Substructure 2 & 17.18 $\pm$ 0.01 & 0.32 & 227 $\pm$ 10 &55.4 $\pm$ 0.5 & 8.18 & 1.30\\
		Southern Substructure 3 & 16.95 $\pm$ 0.06 & 0.40 & 531 $\pm$ 10 &50.0 $\pm$ 1.3 & 9.20 & 5.95\\
		Northern Substructure 1 & 17.26 $\pm$ 0.03 & 0.62 & 123 $\pm$ 6 &57.5 $\pm$  0.9 & 16.44 & 2.68\\
		Northern Substructure 2 & 17.36 $\pm$ 0.02 & 0.52 & 92 $\pm$ 5 & 60.2 $\pm$ 0.7 & 14.44 & 1.71\\
		SMCNOD & 17.54 $\pm$ 0.03 & 0.31 & 51 $\pm$ 5 & 65.4 $\pm$ 1.1 & 9.36 & 1.51\\
	    Magellanic Bridge & 17.12 $\pm$ 0.01 & 0.26 & 411 $\pm$ 23  & 53.9 $\pm$ 0.5 & 6.47 & 2.64\\	
	    SMC East & 17.11$\pm$ 0.03 & 0.36 & 255 $\pm$ 23  &53.7 $\pm$ 0.8 & 8.91 & 7.25\\			    

			\hline
	\end{tabular}
\end{table*}

\begin{table*}
	\caption{Proper motions of morphological features.}
	\label{table:pm}
	
	\begin{tabular}{lrrrrrrr}
		\hline
		Feature & $N$ & $\rho_{\mathrm{LMC}}$ & $\rho_{\mathrm{SMC}}$& $\mu_{\mathrm{RA}}$ &$\sigma_{\mu_\mathrm{RA}}$& $\mu_{\mathrm{Dec}}$& $\sigma_{\mu_{Dec}}$ \\
		&  (counts) & (deg) & (deg) &(mas~yr$^{-1}$) & (mas~yr$^{-1}$)&(mas~yr$^{-1}$)&(mas~yr$^{-1}$)\\
		\hline
		Eastern Substructure 1 & 4094 & 16& -- &1.33 &0.31&0.65&0.60 \\
		Eastern Substructure 2 & 3622 &  17&-- &1.31&0.33&0.49&0.57 \\
		Southern Substructure 1 & 4270 &13 &-- &1.61&0.33&0.59&0.47 \\
		Southern Substructure 2 & 3040 & 12 & -- &1.66&0.33&-0.12&0.43  \\
		Southern Substructure 3 & 2286 & 18  &-- &1.52&0.34&0.13& 0.65 \\
		Northern Substructure 1 & 3460 & 16  &--&1.58 &0.31&0.09& 0.53 \\
		Northern Substructure 2 &1720 &--  &7 &1.20&0.27&-1.05&0.20\\
		SMCNOD & 1116 &--  &7  &0.93&0.29& -1.13&0.15\\
		Magellanic Bridge &9604& 11 & 8 &1.43&0.19& -1.07&0.16 \\
		SMC East & 5516 &-- & 7&1.31  &0.22& -1.12&0.17\\
		\hline
	\end{tabular}
\end{table*}

\section{Double red clump feature in the SMC}\label{section4}
Figure \ref{fig:DRC_1} (left panel) shows the spatial distribution of stars around the SMC, after applying the proper motion selection criteria (Figure \ref{fig:pmsmc}), in sectors of annuli defined in Sect. \ref{section2}. The main motivation behind using this cardinal division, is that the tail/Wing as well as the Magellanic Bridge are completely encompassed by the E sector (Fig.~\ref{fig:B3}). Figure \ref{fig:DRC_1} (right panel) also shows a Hess diagram of the same stars outlining several stellar populations. A similar Hess diagram for stars within the S sector and the 4$-$5$^\circ$ annulus is shown in Figure \ref{fig:DRC}, whereas all Hess diagrams corresponding to all sectors in each annulus are shown in Appendix \ref{appendix_0} (Figure \ref{fig:RCCMD0}). These CMDs show that the vertical extent of the RC population is more prominent in the eastern regions out to 10$^\circ$ from the centre. 

\begin{figure*}
	\includegraphics[scale=0.039]{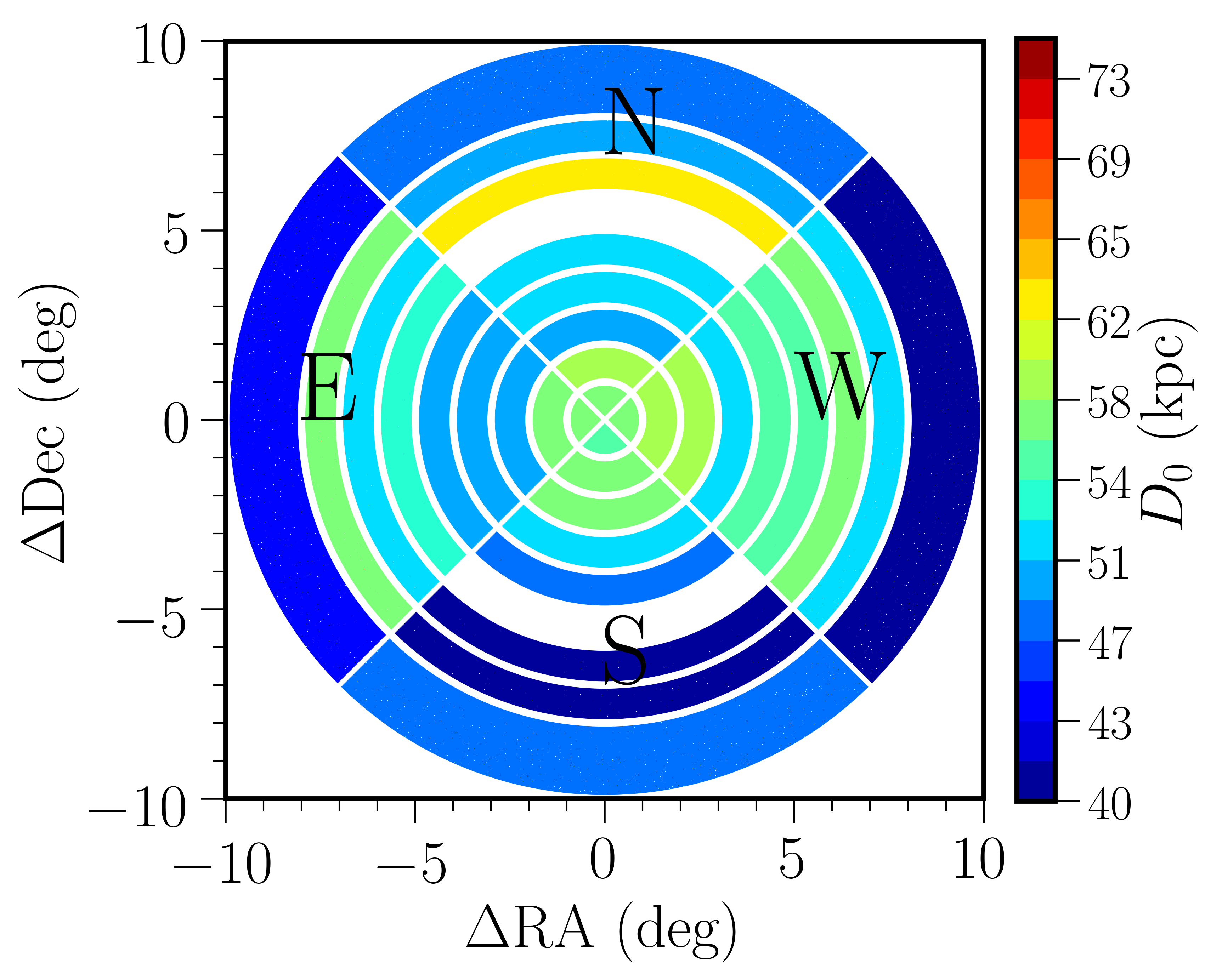}
	\includegraphics[scale=0.04]{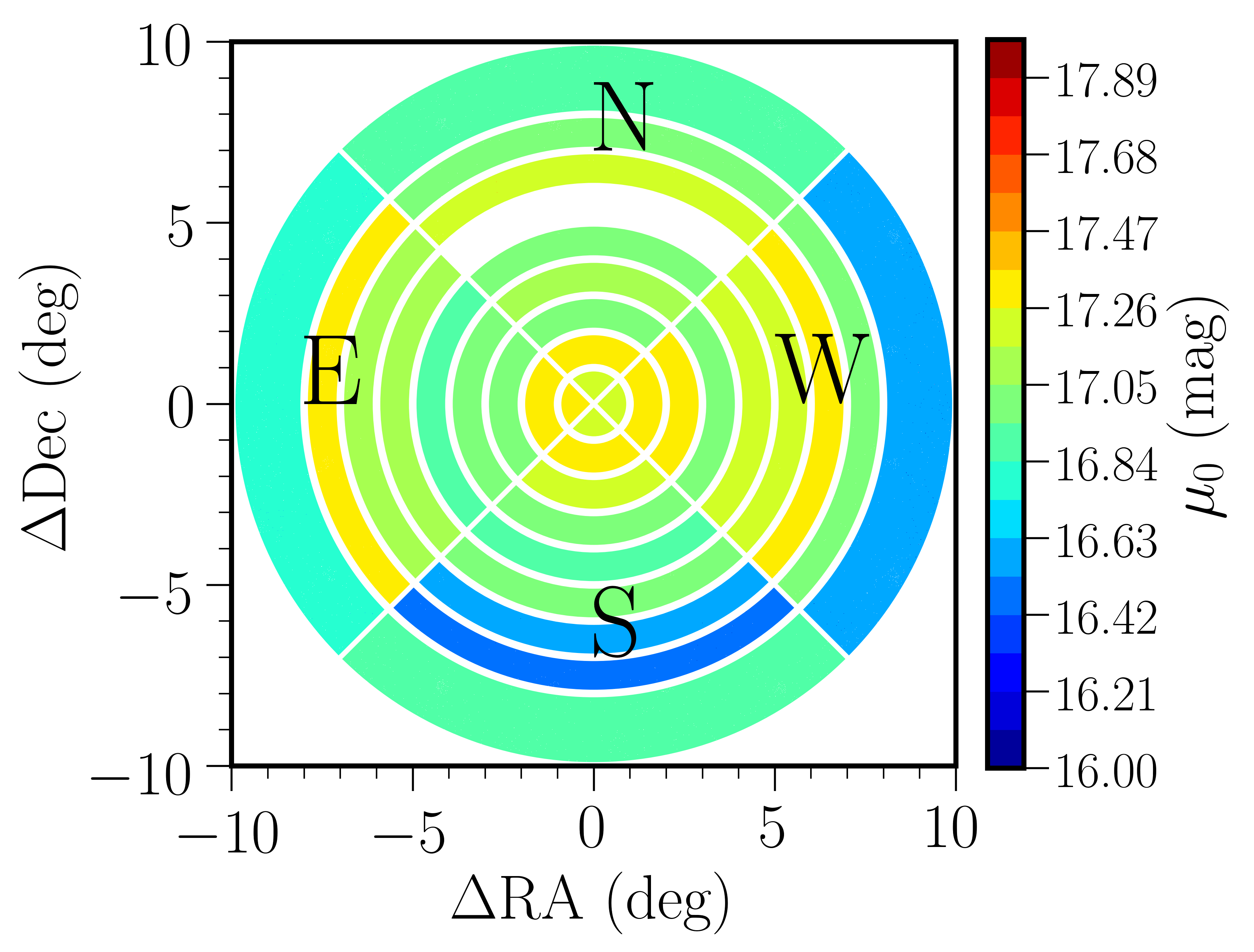}
	\includegraphics[scale=0.04]{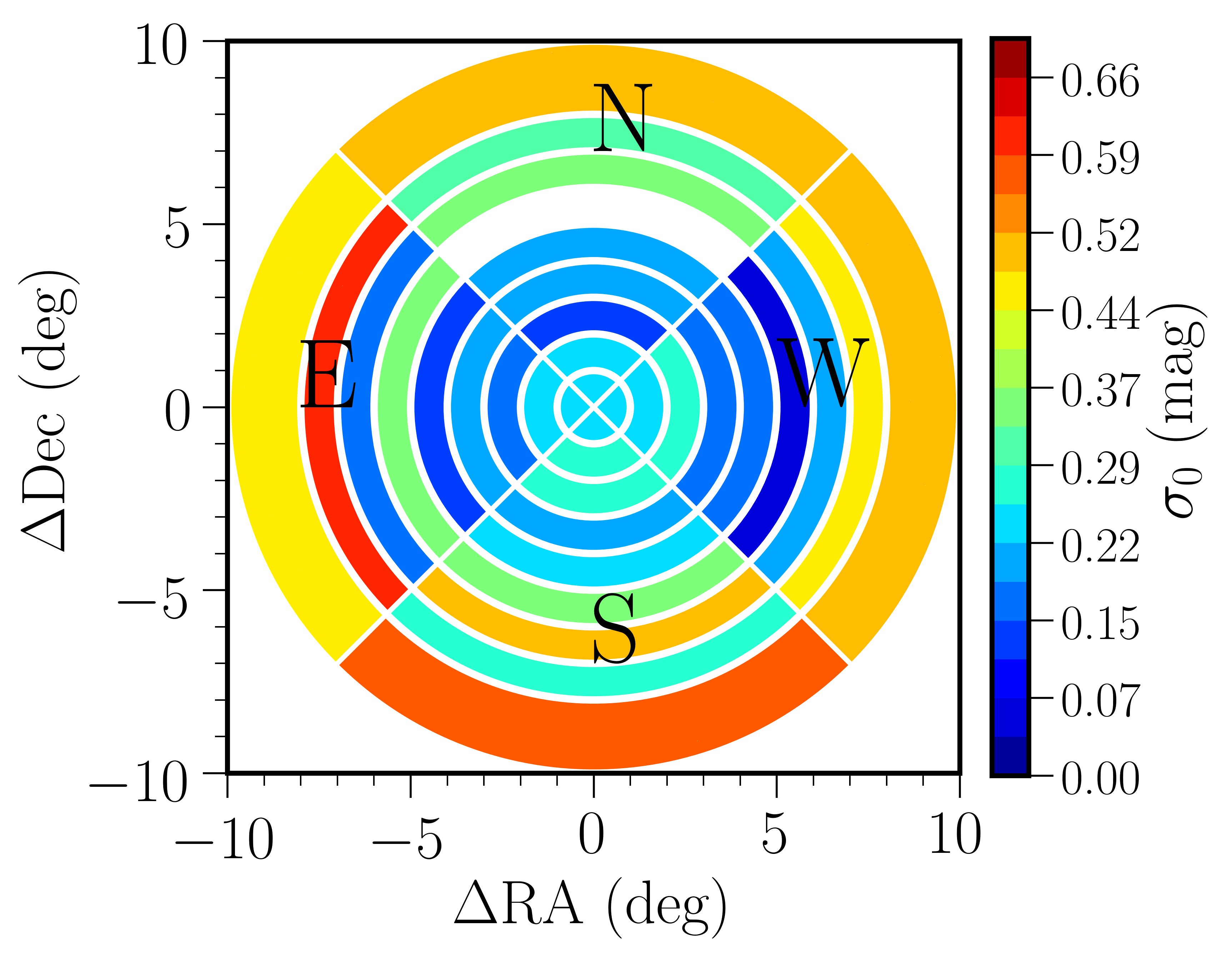}
	\includegraphics[scale=0.039]{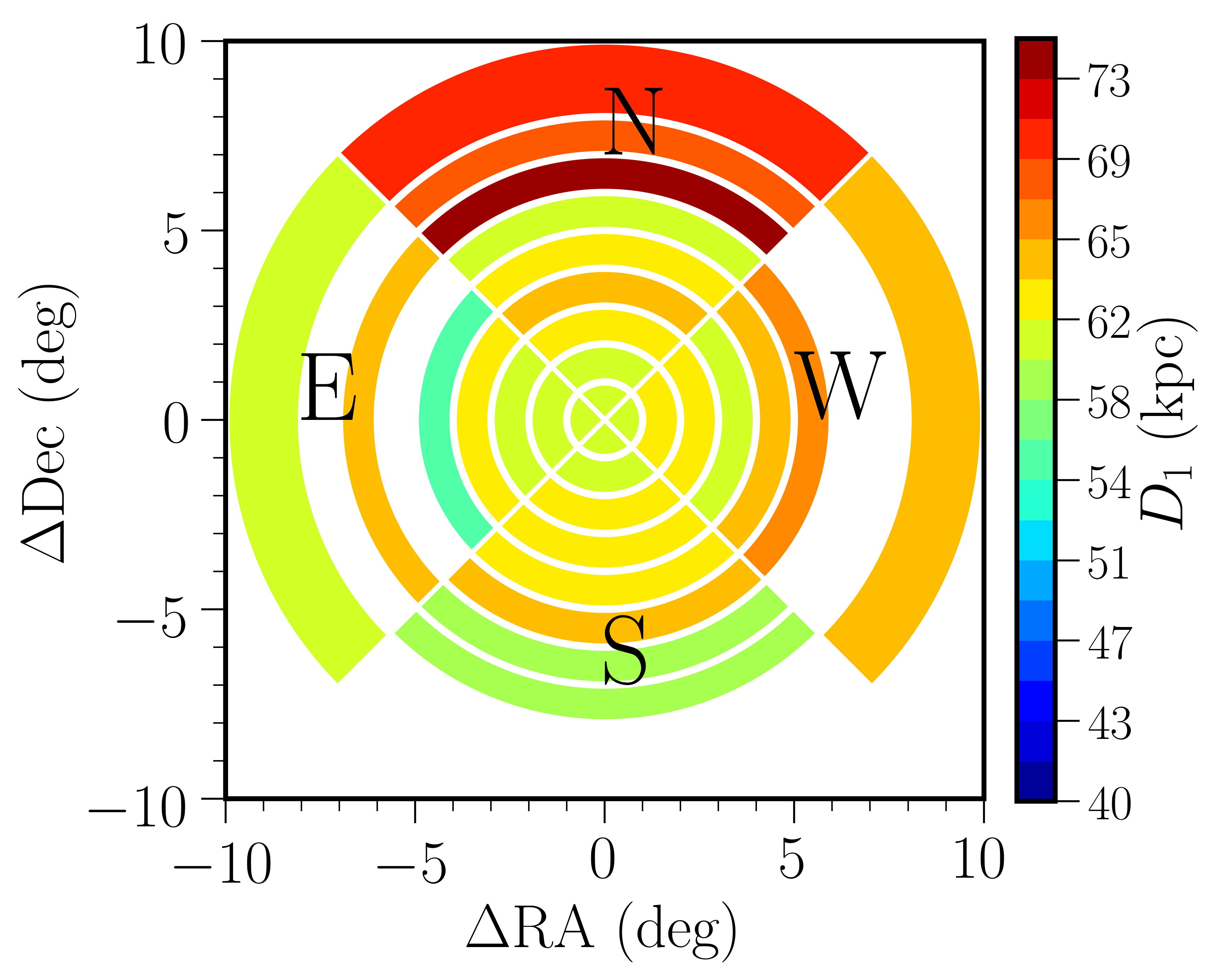}
	\includegraphics[scale=0.04]{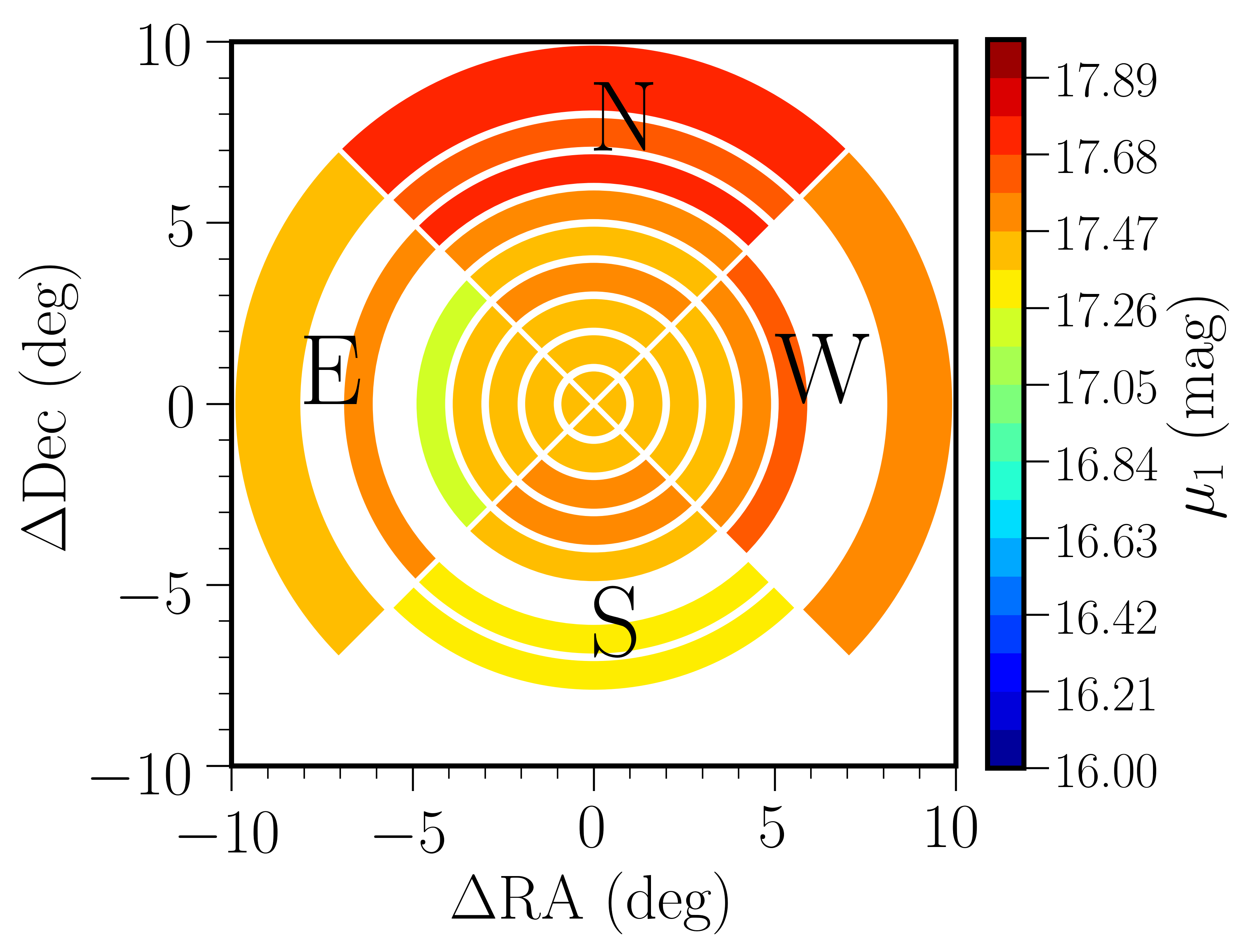}
	\includegraphics[scale=0.04]{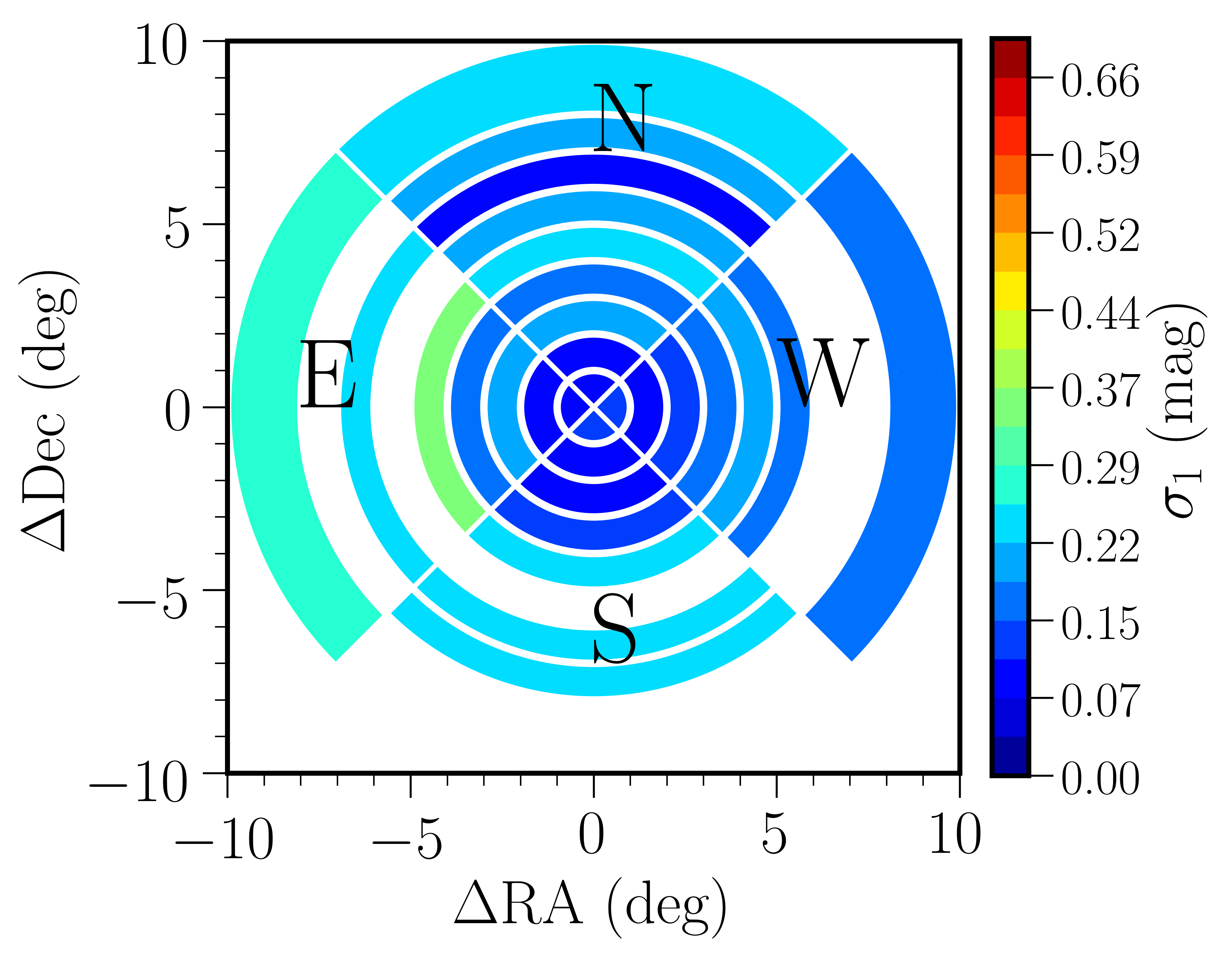}
	\caption{Two-dimensional maps of the distance (left), mean brightness (middle) and sigma (right) of the bright (top) and faint RC components (bottom) in all four sectors, N, E, S and W out to 10$^\circ$ from the centre of the SMC.}
	\label{fig:param}
\end{figure*}

To study the RC stars in more detail, we constructed the luminosity function of the population, see Figure \ref{fig:DRC} (right panel) for stars within the S sector and the 4$-$5$^\circ$ annulus and \ref{fig:RCHIST0} for all sectors of each annulus, and obtained distances.
Unlike \cite{Subramanian2017} and \cite{Tatton2020}, we used the $J$ instead of the $Y$ band  to its larger spatial coverage in the VHS survey. We constructed luminosity functions for RC stars using a bin size of 0.05~mag for regions with radius $R~\leq$4$^\circ$, and 0.1~mag for $R\geq$4$^\circ$ (the bin sizes are on the order of the errors in the $K_\mathrm{s}$ band). In order to evaluate the luminosity function, we employed a multi--component non--linear least squares fitting method. The CMD box, used to extract RC stars, also includes RGB stars, which we minimised by using a quadratic polynomial term. The main RC component is first modelled with one Gaussian and then, if the reduced $\chi^2$ parameter improves by more than 25 per cent when adding a second Gaussian, the second RC is deemed real. The parameters of the fits to the RC components and the reduced $\chi^2$ values are given in Table \ref{table:smc}, and the two-dimensional maps of the distances, means, and widths of the RC components are shown in Figure \ref{fig:param}. Modelling the RGB component proved challenging for some regions (5--6$^\circ$~S, 6--7$^\circ$~S, 7--8$^\circ$, 8--10$^\circ$), where only the brightest RGB component is present. This is owing to VHS observations being limited to just below the RC. In these cases, we modelled the RGB with a linear slope. 

$K_\mathrm{s}$ is an excellent photometric band for absolute RC magnitude determination as it is less affected by reddening and systematic dependence on metallicity than optical bands. Many studies established median RC absolute magnitudes in the $K_\mathrm{s}$ band, which mostly agree on $M_{K}$ = $-$1.61~mag. \cite{Alves2000} used $K$ magnitudes from the Caltech Two Micron Sky Survey (TMSS; \citealp{Neugebauer1969}) of 238 \textit{Hipparcos} RC stars and, under the assumption of no reddening, they found a value of $M_{K}$ = $-$1.61 $\pm$ 0.03~mag. Using revised \textit{Hipparcos} parallaxes \cite{Groenewegen2008} finds $M_{K}$ = $-$1.54 $\pm$ 0.04~mag, based on the 2MASS system. The difference in magnitude between the two studies was attributed to a selection bias due to a lack of data for bright nearby RC stars. This was confirmed by \cite{Laney2012} who found $M_{K}$ = $-$1.613 $\pm$ 0.015~mag from a sample of 226 bright RC stars using accurate $K$ band magnitudes. The drawback of the \cite{Alves2000} study was the quality of the available IR photometry. More recent studies such as \cite{Hawkins2017} and \cite{Ruiz-Dern2018} favour the brighter magnitude of $M_{K}$ = $-$1.61~mag. 

We converted the extinction corrected $K_\mathrm{s}$ magnitudes of the bright and faint RC components ($\mu_0$ and $\mu_1$) to distance moduli, using the high-precision observations of RC stars in the solar neighbourhood  by \citealp{Laney2012}. Their absolute magnitude in the 2MASS $K$ band corresponds to $M_{K}$ = $-$1.613 $\pm$ 0.015~mag whereas in the VISTA $K_\mathrm{s}$ band, it is $M_{K_\mathrm{s}}$ = $-$1.609 $\pm$ 0.015~mag \citep{Gonzalez-Fernandez2018}. However, since these stars are located in the solar neighbourhood, their absolute magnitude is expected to be different from those of RC stars in the SMC due to differences in metallicity, age and star formation rate. A correction for these population effects is needed and \cite{Salaris2002} estimated it to be $-$0.07~mag in the $K_\mathrm{s}$ band. They obtained this term by simulating RC stars using stellar population models and including the star formation rate and age--metallicity relation derived from observations and then compared the difference in the absolute magnitudes in the SMC and solar neighbourhood to quantify the corresponding population effects. Once we obtained distance moduli for the different peaks, we calculated distance moduli uncertainties which include both the uncertainty on the absolute magnitude estimate as well as the peak magnitude uncertainty. We converted them to distances and distance uncertainties in kpc, and the values are given in Table \ref{table:smc}. We followed the same methodology as \cite{Subramanian2017}.

Most regions show a bi--modality in their luminosity function distribution, it was first discovered by \cite{Nidever2013} and it can be due to several effects, including distance, line-of-sight depth as well population effects. Reddening has been ruled out as a significant contributor to the spread by \citealp{Subramanian2017}.
\begin{itemize}
\item 0$^\circ$$\leq$ $R$ $\leq$2$^\circ$: The bi--modality is less pronounced here than in the outer regions. The difference in magnitude between the bright and faint RC components is on the order of $\sim$ 0.15 mag and is generally less than the width of the bright Gaussian components. The difference between the two peaks is more pronounced in the southern regions. Note that these differences are ten times larger than the error on the peaks, but they are comparable with the width of the RC components. The width of the bright RC component is always larger than that of the faint component. The height of the faint component is always larger than the bright one, except for the northern and eastern regions in the 1$^\circ$$\leq$ $R$ $\leq$2$^\circ$ sector.

\item 2$^\circ$$\leq$ $R$ $\leq$5$^\circ$: A clear separation between the bright and faint RC components becomes obvious in most regions of 2$^\circ$$\leq$ $R$ $\leq$3$^\circ$. The difference in magnitude becomes on the order of $\sim$ 0.4~mag. \cite{Subramanian2017} found that this difference is too large to be attributed to any other effect but to different distances in the plane of the sky. The bi--modality has been detected by \cite {Nidever2013} at 4 $^\circ$ and is likely due to material stripped from the SMC $\sim$ 200 Myr ago. The bi-modality in the western regions was not seen by neither \cite{Nidever2013} or \cite{Subramanian2017}. 

\item 5$^\circ$$\leq$ $R$ $\leq$7$^\circ$: A bi--modality is only detected in a few regions. The dominant components are consistent with those in the inner SMC. In region 5--6$^\circ$~W we find a bright component, brighter than previously found, suggesting a foreground structure. In region 6--7$^\circ$~N we find a faint component, fainter than previously found which because of its large distance might belong to the structure known as the Counter Bridge \citep{Diaz2012}, whereas a bright component in region 6--7$^\circ$~S might be linked to the Old Bridge \citep{Belokurov2017}. These three features might be influenced by a poor RGB subtraction.

\item 7$^\circ$$\leq$ $R$ $\leq$10$^\circ$: In the northern regions we detect both a bright and a faint component where the difference in magnitude between the two components is on the order of $\sim$0.7. In the region 7--8$^\circ$~S we find a bright component as above, which is possibly linked to the Old Bridge and that at 8--10$^\circ$ appears instead to the East. This demonstrates that the double RC feature is traceable out to 10$^\circ$. A double component in the other regions is less well defined.
\end{itemize}
The fact that two components are present beyond the E sector at different radii demonstrates that the foreground structure in the form of a distance bi--modality is not solely owing to the Magellanic Bridge and Wing. To further investigate the presence of the double RC feature at these different locations, we explored an alternative cardinal division where we used four sectors (NE, SE, NW, SW) that are 45$^\circ$ offset counter-clockwise from the previous setup (N, E, S, W), see Appendix \ref{appendix_1}. We performed the same analysis for the luminosity functions of the stars within these sectors and found comparable results. This shows also that the faint component detected in the North, which might belong to the Counter Bridge, is present irrespective of the cardinal division scheme chosen. If the bi-modality in the luminosity function of RC stars is interpreted as a distance effect, it will correspond to a difference of $\sim$11 kpc consistent with the findings of \cite{Subramanian2017} and \cite{Omkumar2020}. The bright RC component which is detected throughout the N, S, and E sectors out to 10$^\circ$ might be owing to a foreground substructure in the periphery of the SMC (including the Magellanic Bridge and the Old Bridge), whereas a faint component detected in the northern outermost regions might belong to the Counter Bridge. Whilst the substructures discussed in Sect.~\ref{section3} are dominated by single components, the Northern Substructure 2 could also have two, in agreement with the analysis of the RC distribution across the galaxy.

\section{Conclusions}\label{section5}
In this paper, we have presented a study of the outer morphology of the MCs using NIR data from the VHS and ($J-K_\mathrm{s}$, $K_\mathrm{s}$) CMDs to distinguish between young (tip of MS and supergiant stars) and old (RGB, AGB and RC stars) stellar populations. We minimise the influence of MW stars by using proper motion and parallax selection criteria from \textit{Gaia} DR2. Our morphology maps demonstrate that it is possible to trace many of the diffuse stellar substructures present in the periphery of the MCs. Furthermore, we determine distances to these substructures using the luminosity function of RC stars. We confirm the detection of substructures discovered previously such as: the Northern and Southern features in the LMC disc by \cite{Mackey2016,Mackey2018} and \cite{Belokurov2019}. We also characterise for the first time two new substructures: one on the eastern side of the LMC and one on the northern side of the SMC. A substructure on the east of the LMC was already present in \cite{DeVaucouleurs1955} and it is possible that both eastern substructures are associated to the disc of the galaxy. The simulations described by \cite{Belokurov2019} showed that the tidal influence of the MW and SMC are responsible for creating different substructures in the LMC. The SMC is responsible for creating the northern arm and a more prominent southern arm, while the MW's tidal field can also be responsible for creating, bending and extending the northern arm. The eastern substructures appear consistent with the influence of the MW deflecting the LMC disc. The substructure on the northern side of the SMC is probably associated to the ellipsoidal structure of the galaxy.
Overall, we have been able to uncover a wealth of substructures in the periphery of the MCs, showcasing that their outskirts are susceptible to tidal interactions. Mapping these morphological features with fainter stellar populations and comparing their morphology to simulations can provide further constraints on the orbits of past encounters and the mass of the LMC.

We also present a study of the luminosity function of the RC stars from the centre of the SMC to a radius of 10$^\circ$, using NIR data from the VHS and VMC surveys. Previous studies such as \cite{Nidever2013}, \cite{Subramanian2017}, \cite{TattonThesis} and \cite{Tatton2020} were limited to a radius of 4$^\circ$, therefore only covering the main body of the galaxy. \cite{Omkumar2020} used \textit{Gaia} DR2 data to examine the luminosity function of RC stars up to 10$^\circ$, however, they only detect a double red clump feature from 2.5$^\circ$ to 5--6$^\circ$ in the eastern regions. In this study we find that a bimodal RC feature is present in the eastern, northern and southern regions at different radii suggesting that tidal interactions have affected the entire inner eastern parts of the galaxy and not only its outskirts.  In addition, our results reveal that RC stars form a background structure in the North of the SMC which might trace the Counter Bridge \citep{Diaz2012, Wang2019}, and a foreground structure in the South which might trace the Old Bridge \citep{Belokurov2017}. This is consistent with previous studies based on metal-poor red giants \citep{Dobbie2014}, Cepheids and RR Lyrae stars \citep{Ripepi2017,Jacyszyn-Dobrzeniecka2020}.



\begin{figure*}
	\centering
	
	\includegraphics[scale=0.06]{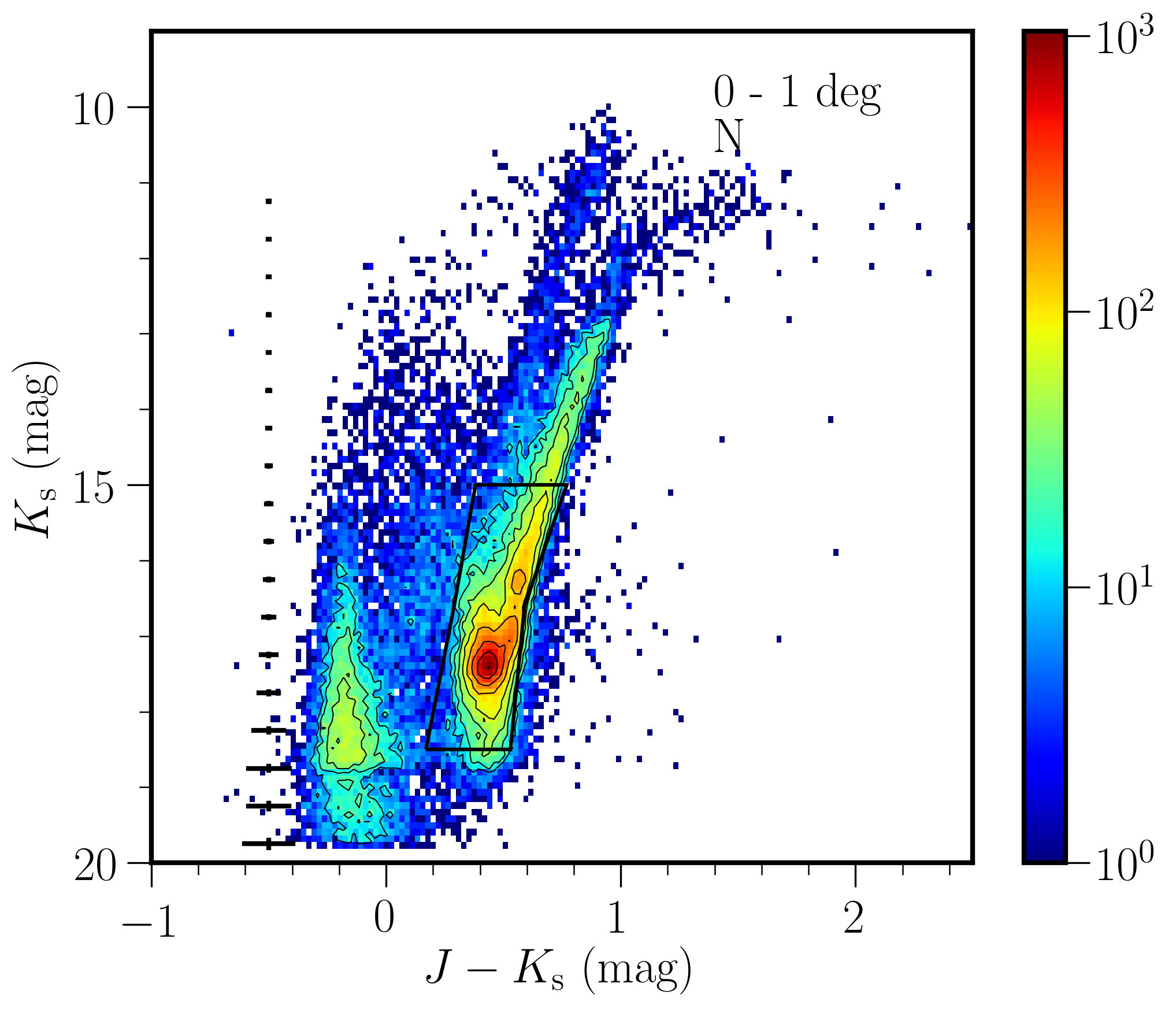}
	\includegraphics[scale=0.06]{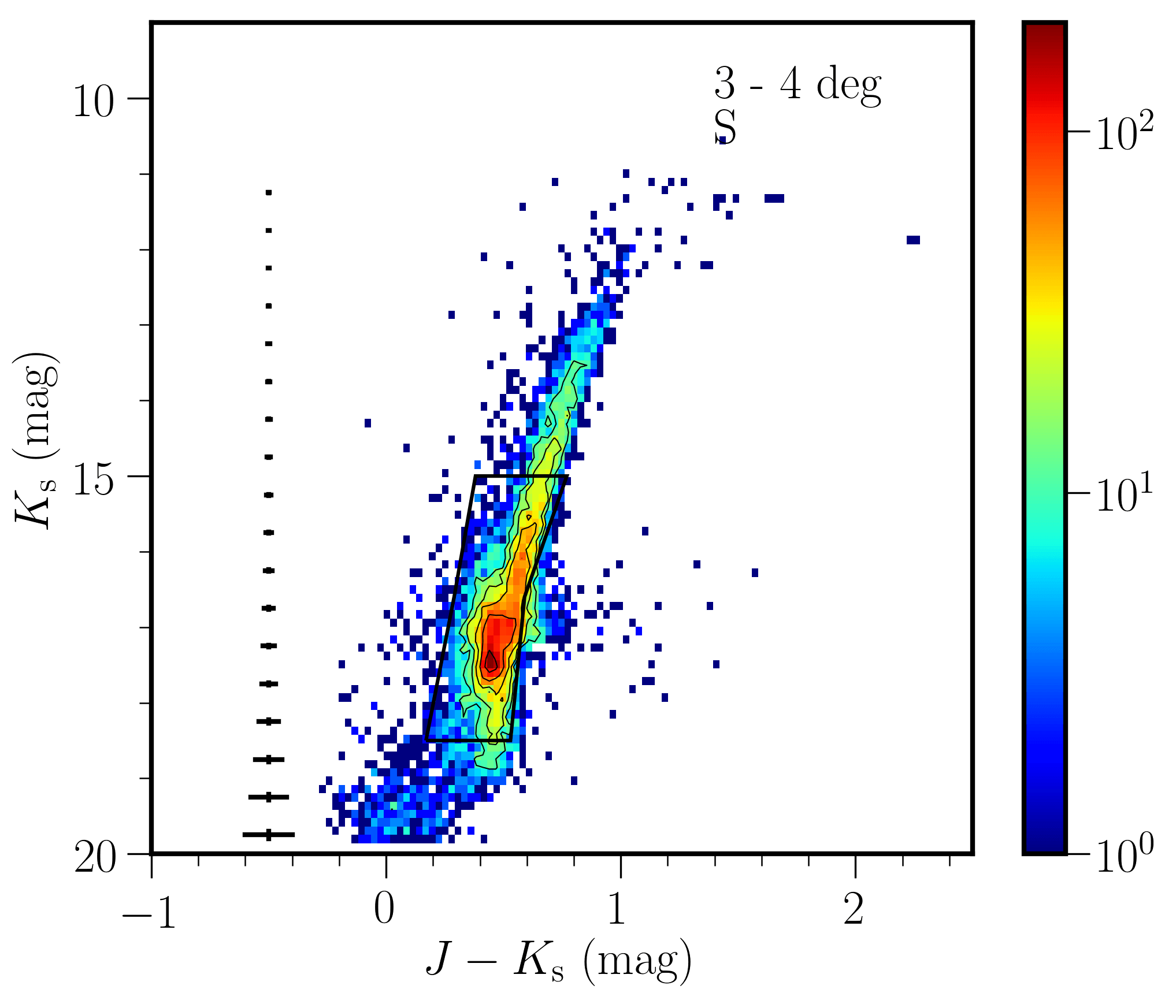}
	\includegraphics[scale=0.06]{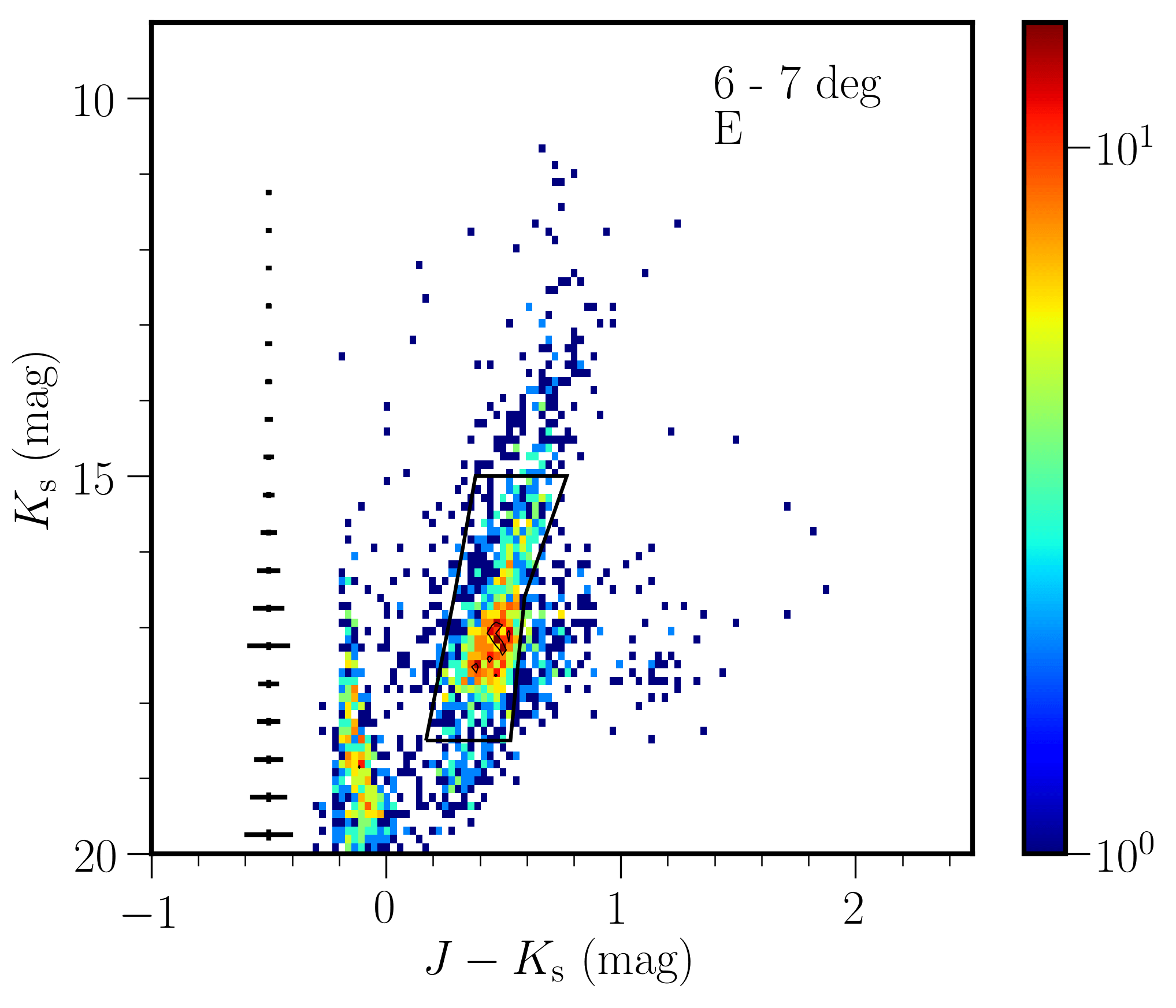}
	\includegraphics[scale=0.06]{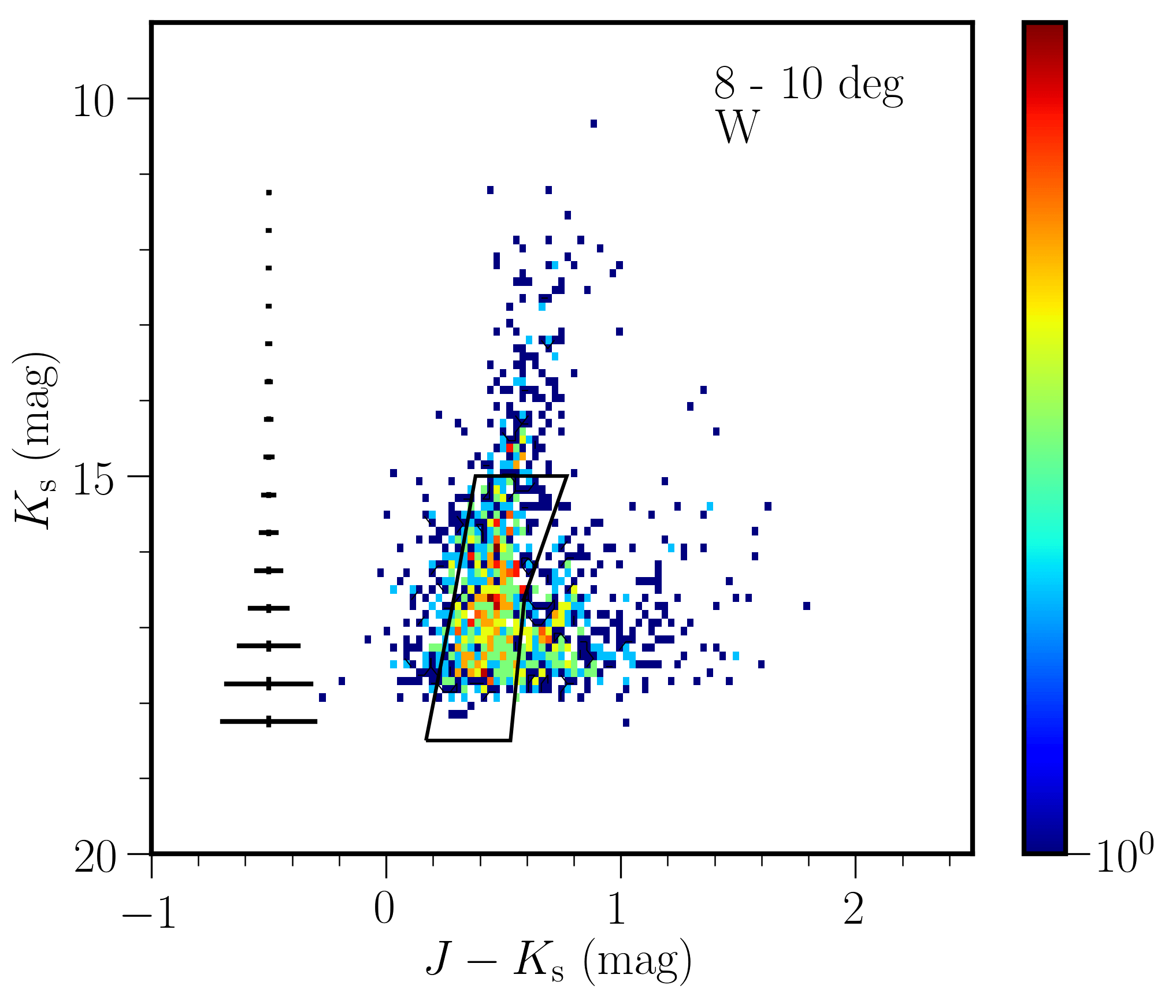}
	
	\caption{NIR ($J-K_\mathrm{s}$, $K_\mathrm{s}$) Hess diagrams of a few of the SMC regions discussed in Sect. \ref{section4} and presented in full in Appendix \ref{appendix_0}. The colour scale indicates the stellar density on a logarithmic scale per bin. The bin size is 0.027$\times$0.110 deg$^2$ and the black box limits the region defined to study the distribution of RC stars.}
	\label{fig:RCCMD}
\end{figure*}
\begin{figure*}
\includegraphics[scale=0.06]{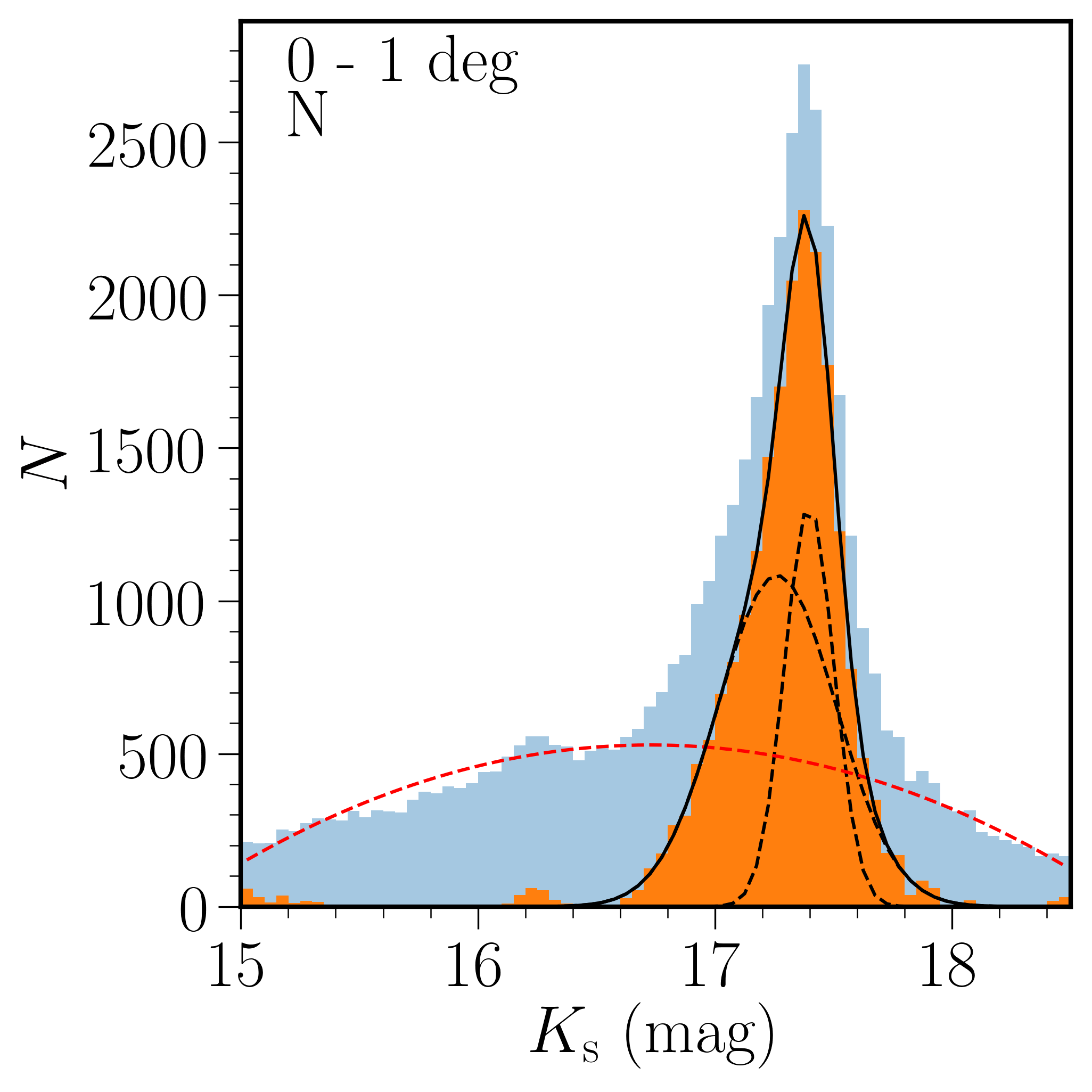}
\includegraphics[scale=0.06]{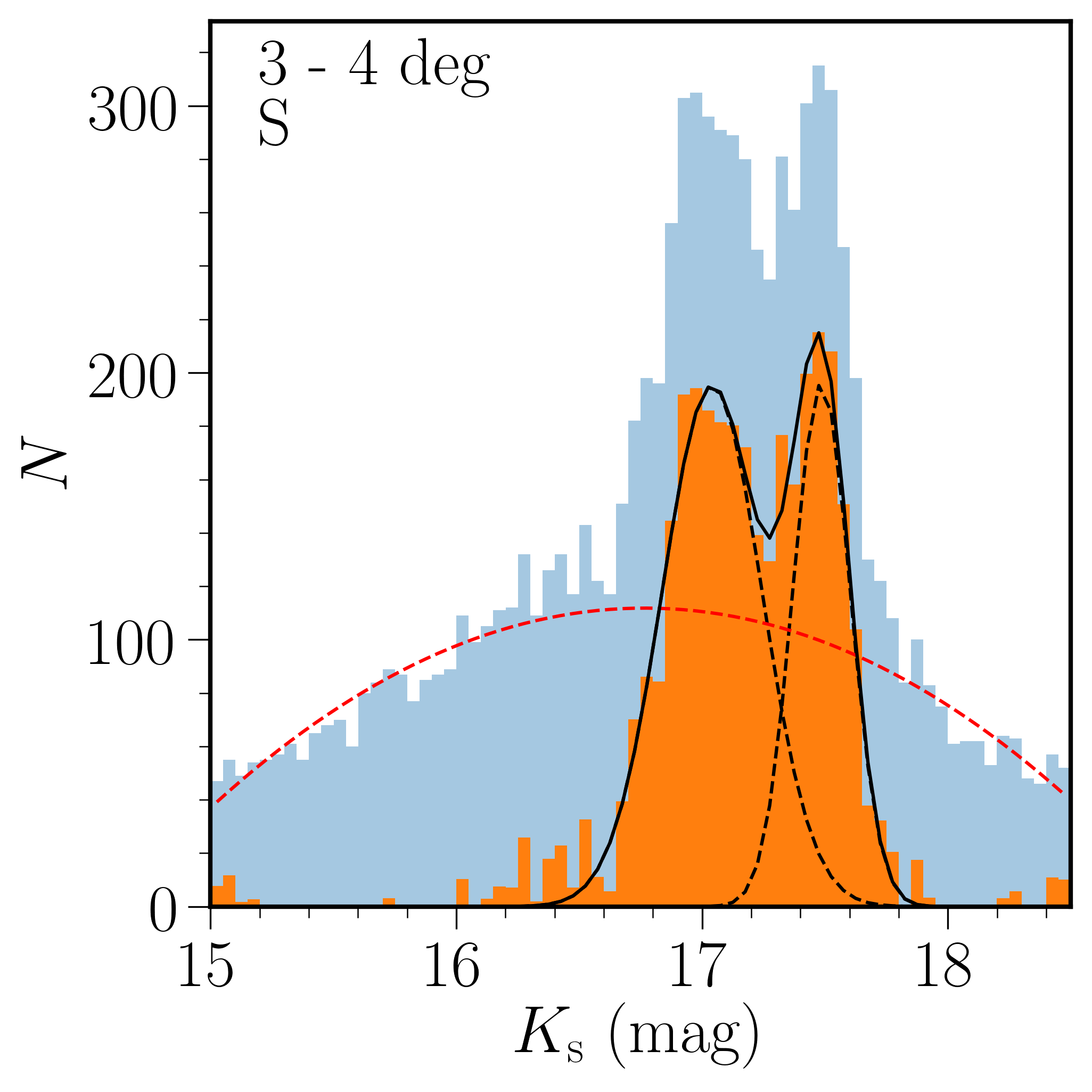}
\includegraphics[scale=0.06]{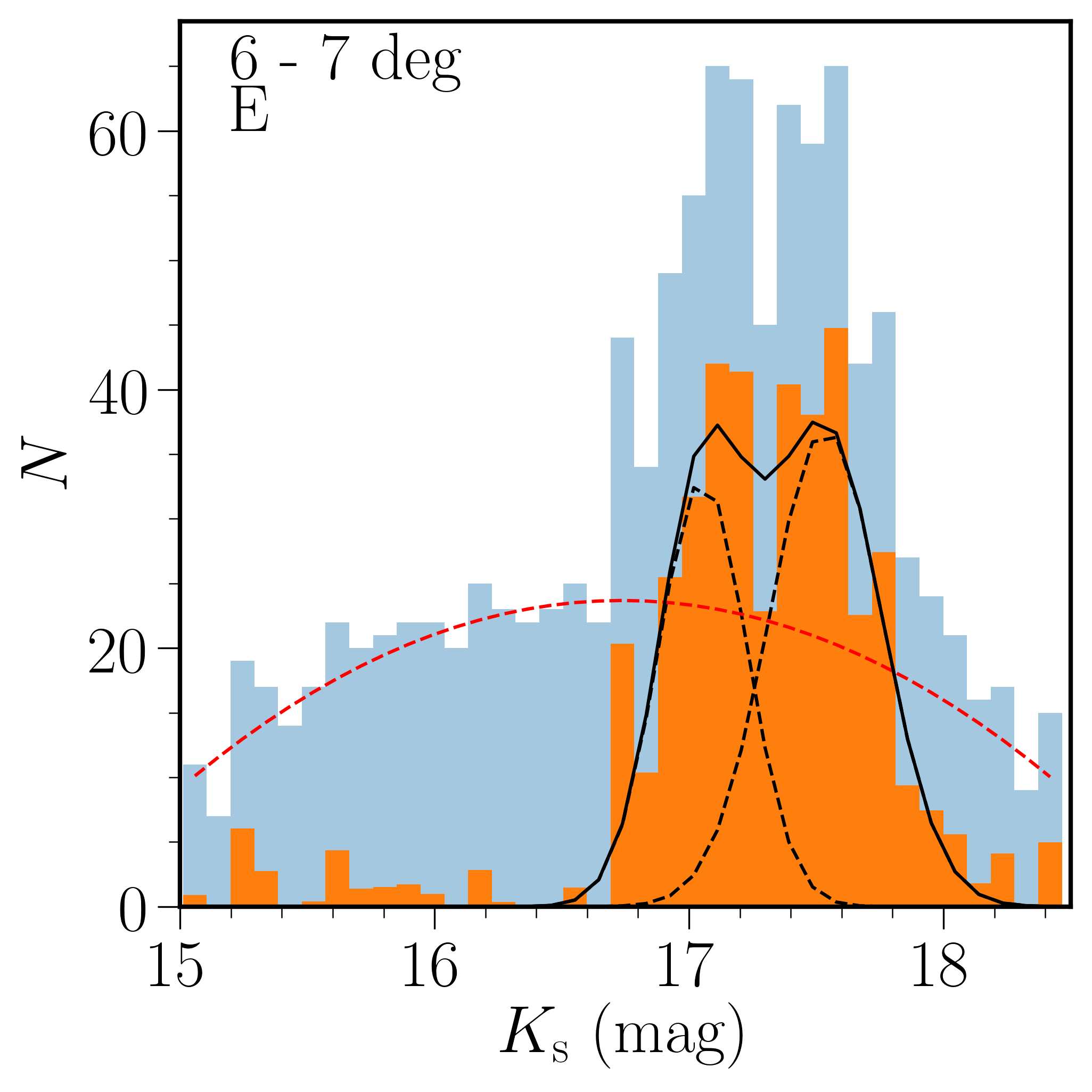}
\includegraphics[scale=0.06]{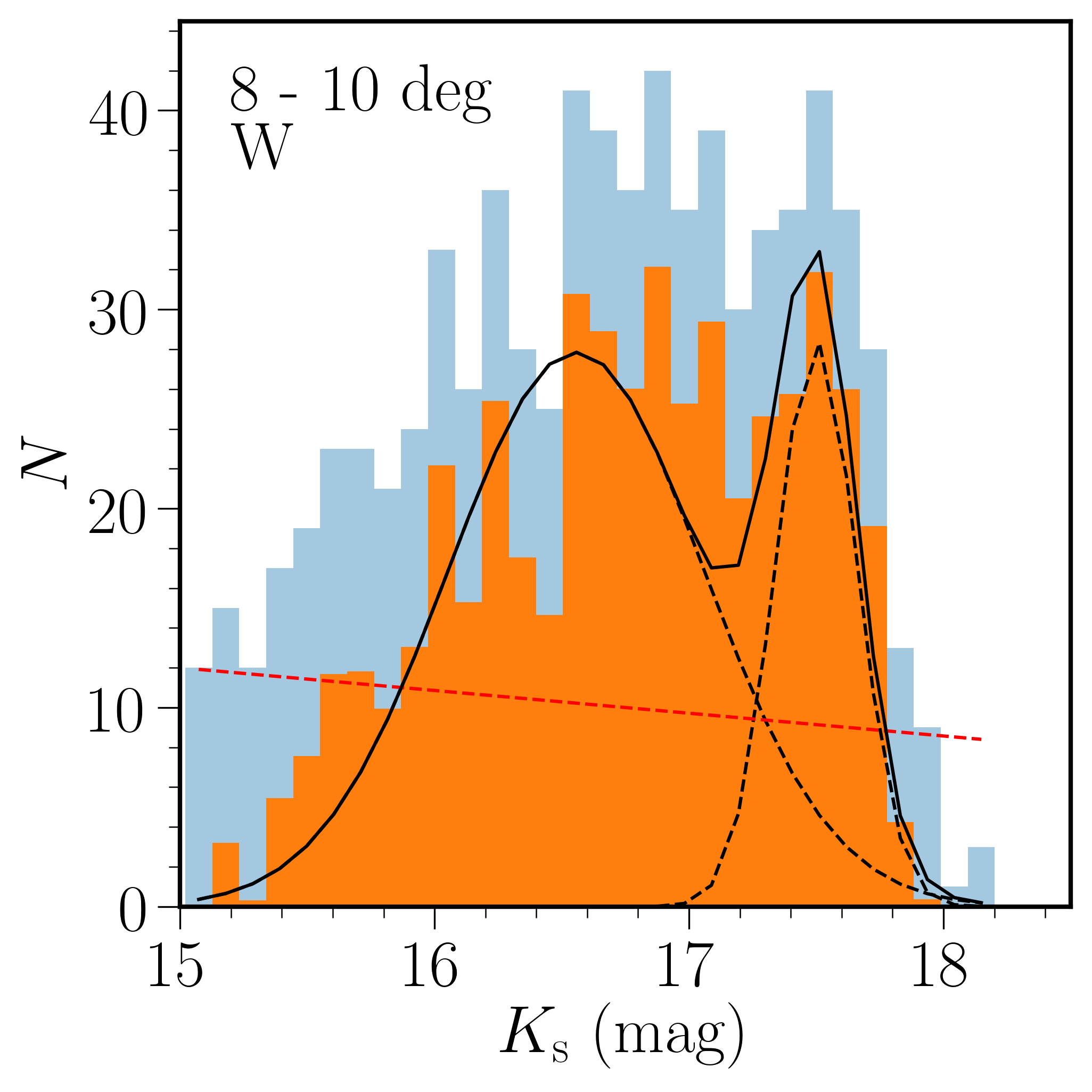}

\caption{Luminosity function of the RC stars in a few different SMC regions discussed in Sect. \ref{section4} and presented in full in Appendix \ref{appendix_0}. The blue histograms show the observed luminosity functions, whereas orange histograms show the distributions after subtracting the RGB components and the continuous lines show the total fits to these distributions whereas the dashed lines represent the separate components of the fits.}
\label{fig:RCHIST}	
\end{figure*}

\begin{table*}
		\caption{Gaussian parameters and the reduced $\chi^2$ values of the profile fits to the luminosity function of the RC stars in different areas.}                        
		\label{table:smc}      
		
		\begin{tabular}{lrrrrrrrrrrr}
			\hline
			Region & $\mu_0$ & $\sigma_0$ &$N_0$ &$D_0$&$\sigma_{D0}$& $\mu_1$ & $\sigma_1$& $N_1$&$D_1$&$\sigma_{D1}$ & $\chi^2$\\
			            &  (mag) & (mag) & (counts) & (kpc) & (kpc)& (mag) & (mag) & (counts) & (kpc) & (kpc) & \\
			\hline

			0--1$^\circ$~N& 17.26 $\pm$ 0.01 & 0.25  & 1083 $\pm$ 83  &57.5 $\pm$  0.5 & 6.62 & 17.40 $\pm$ 0.01 & 0.10  & 1312 $\pm$ 90 &61.3 $\pm$ 0.4 & 2.97 &2.93\\
			0--1$^\circ$~E& 17.27 $\pm$ 0.01 & 0.25  & 1282 $\pm$ 146 &57.9 $\pm$  0.6 & 6.77 & 17.40 $\pm$ 0.01 & 0.10  & 1460 $\pm$ 155 & 61.3 $\pm$ 0.5 & 2.97 &6.93\\
			0--1$^\circ$~S& 17.21 $\pm$ 0.03 & 0.24  & 947 $\pm$ 147 & 56.3 $\pm$  1.0 & 6.30 & 17.40 $\pm$ 0.01 & 0.12  & 1607 $\pm$ 177 & 61.5 $\pm$ 0.5 & 3.34 &6.62\\
			0--1$^\circ$~W& 17.25 $\pm$ 0.02 & 0.24  & 1094 $\pm$ 147 &57.2 $\pm$  0.7 & 6.45 & 17.40 $\pm$ 0.01 & 0.12  & 1444 $\pm$ 155 & 61.3 $\pm$ 0.5 & 3.34 &5.22\\	
			&&&&&&&&&&&\\
			
			1--2$^\circ$~N& 17.31 $\pm$ 0.01 & 0.25  &1242 $\pm$ 86 &58.7 $\pm$ 0.5 & 6.89 & 17.41 $\pm$ 0.01 & 0.10 & 761 $\pm$ 87  & 61.5 $\pm$ 0.5 & 2.98 &2.48\\
			1--2$^\circ$~E& 17.26 $\pm$ 0.01 & 0.25  &1959 $\pm$ 136 &57.6 $\pm$ 0.5 & 6.58 & 17.40 $\pm$ 0.01& 0.10 & 1691 $\pm$ 132 &61.5 $\pm$ 0.5 & 2.89 &4.83\\
			1--2$^\circ$~S& 17.26 $\pm$ 0.01 & 0.27  &1703 $\pm$ 92 &57.6 $\pm$ 0.5 & 7.16 & 17.43 $\pm$ 0.01 & 0.10 & 2448 $\pm$ 163 & 62.1 $\pm$ 0.4 & 2.87 &6.51\\
			1--2$^\circ$~W& 17.32 $\pm$ 0.01 & 0.24 &1658 $\pm$ 127 &59.2 $\pm$ 0.5 & 6.64 & 17.44 $\pm$ 0.01 & 0.10 & 1768 $\pm$ 134 & 62.5 $\pm$ 0.4 & 2.94 &4.24\\		
			&&&&&&&&&&&\\

			2--3$^\circ$~N& 16.98 $\pm$ 0.02 & 0.14  & 168 $\pm$ 20 &50.6 $\pm$ 0.7 & 3.19 & 17.44 $\pm$ 0.01 & 0.21 & 650 $\pm$ 15 &62.5 $\pm$ 0.5 & 6.15 &1.42\\
			2--3$^\circ$~E& 16.95 $\pm$ 0.03 & 0.18  & 418 $\pm$ 64 &49.8 $\pm$ 0.9 & 4.08 & 17.38 $\pm$ 0.02 & 0.22 & 904 $\pm$ 238 & 60.7 $\pm$ 0.7 & 6.14 &1.65\\
			2--3$^\circ$~S& 17.26 $\pm$ 0.01 & 0.29  & 692 $\pm$ 29 &57.4 $\pm$ 0.5 & 7.66 & 17.47 $\pm$ 0.01 & 0.11 & 844 $\pm$ 40 & 63.5 $\pm$ 0.4 & 3.19 &1.49\\
			2--3$^\circ$~W& 17.35 $\pm$ 0.01 & 0.28 & 582 $\pm$ 54 &60.1 $\pm$ 0.5 & 7.88 & 17.46 $\pm$ 0.01 & 0.11 & 829 $\pm$ 57 & 63.2 $\pm$ 0.5 & 3.32 &2.13\\				
			&&&&&&&&&&&\\

			3--4$^\circ$~N& 17.08 $\pm$ 0.04&0.19 &153 $\pm$ 14 &52.9 $\pm$ 1.1 & 4.72 & 17.52 $\pm$ 0.04 & 0.18 & 143 $\pm$ 17 &64.8 $\pm$ 1.3 & 5.31 &1.36\\
			3--4$^\circ$~E& 16.99 $\pm$ 0.01&0.19 &520 $\pm$ 17 &50.8 $\pm$ 0.5 & 4.44 & 17.45 $\pm$ 0.02 & 0.16 & 356 $\pm$ 21 & 62.7 $\pm$ 0.6 & 4.56 &1.57\\	
			3--4$^\circ$~S& 17.04 $\pm$ 0.01&0.20 &195 $\pm$ 8 &52.0 $\pm$ 0.5 & 4.88 & 17.49 $\pm$ 0.01 & 0.12 & 196 $\pm$ 12 & 63.8 $\pm$ 0.5& 3.44 &1.08\\
			3--4$^\circ$~W& 17.01 $\pm$ 0.08&0.16 & 48 $\pm$ 12 &51.4 $\pm$ 2.0 & 3.78 & 17.42 $\pm$ 0.02 & 0.17 & 240 $\pm$ 12 & 62.0 $\pm$ 0.7 & 4.79 &1.38\\					
			&&&&&&&&&&&\\

			4--5$^\circ$~N&17.02 $\pm$ 0.12& 0.19 &111 $\pm$ 73 &51.6 $\pm$ 2.9 & 4.46 & 17.45 $\pm$ 0.06 &0.24 & 377 $\pm$ 42 &62.7 $\pm$ 1.7 & 7.10 & 1.60\\
			4--5$^\circ$~E&16.94 $\pm$ 0.01& 0.13 &350 $\pm$ 37 &49.6 $\pm$ 0.4 & 3.08 & 17.21 $\pm$ 0.02 &0.35 & 435 $\pm$ 28 &56.2 $\pm$ 0.7 & 9.05 & 1.31\\
			4--5$^\circ$~S&16.90 $\pm$ 0.08 & 0.22 &147 $\pm$ 29 &48.8 $\pm$ 1.9 & 5.05 & 17.44 $\pm$ 0.07 &0.25 & 200 $\pm$ 28 &62.4 $\pm$ 2.0 & 7.15 & 2.75\\
			4--5$^\circ$~W&17.16 $\pm$ 0.14 & 0.16 &53  $\pm$ 42 &55.0 $\pm$ 3.6 & 4.12 & 17.53 $\pm$ 0.10 &0.20 &117 $\pm$ 26 & 65.3 $\pm$ 2.9 & 6.16 & 1.66\\			
			&&&&&&&&&&&\\
			
			5--6$^\circ$~N&--&--&--&--&--& 17.40 $\pm$ 0.01 & 0.32 & 200 $\pm$ 7 & 61.3 $\pm$ 0.5 & 9.14 &0.99 \\
			5--6$^\circ$~E& 17.15 $\pm$ 0.02 & 0.35 &146 $\pm$ 7 &54.7 $\pm$ 0.6 & 8.88 &--&--& -- &--&--&1.37\\
			5--6$^\circ$~S& 17.02 $\pm$ 0.09 & 0.35 &101 $\pm$ 9 &51.5 $\pm$ 2.3 & 8.22 &17.54 $\pm$ 0.07 & 0.19 & 47 $\pm$ 27 & 65.5 $\pm$ 2.1 & 5.70 & 1.42\\
			5--6$^\circ$~W& 16.83 $\pm$ 0.19 & 0.38 & 23 $\pm$ 5 &47.2 $\pm$ 4.1 & 8.27 &17.52 $\pm$ 0.06 & 0.16 & 35 $\pm$ 12 & 64.8 $\pm$ 1.5 & 4.80 &4.62\\
			&&&&&&&&&&&\\
			
			6--7$^\circ$~N&17.47 $\pm$ 0.03 & 0.35 & 65 $\pm$ 5 &63.5 $\pm$ 0.9 & 10.13 &17.79 $\pm$ 0.04 &0.10 & 23 $\pm$ 9 &73.4 $\pm$ 1.4& 3.29 &0.99\\
			6--7$^\circ$~E&17.05 $\pm$ 0.07 & 0.17 & 33 $\pm$ 8 &52.3 $\pm$ 1.7 &  4.19 &17.54 $\pm$ 0.08 &0.22 & 37 $\pm$ 5 &65.3 $\pm$ 2.4 & 6.70 &0.79\\
			6--7$^\circ$~S&16.54 $\pm$ 0.28 & 0.50 & 19 $\pm$ 3 &41.3 $\pm$ 5.3 &  9.43 &17.32 $\pm$ 0.04 &0.26 & 50 $\pm$ 11 &59.0 $\pm$ 1.2 & 6.98 &1.29\\
			6--7$^\circ$~W&17.30 $\pm$ 0.04 & 0.19 & 14 $\pm$ 3 &47.9  $\pm$ 1.2&--&--&--&--&--& 5.11 &1.15\\
			&&&&&&&&&&&\\ 
			
			7--8$^\circ$~N&17.00 $\pm$ 0.32&0.32 & 14 $\pm$ 5 &51.1 $\pm$ 7.5 & 7.64 &17.67 $\pm$ 0.07 & 0.22 & 44 $\pm$ 11 &69.5 $\pm$ 2.4 & 6.97 &2.85\\
			7--8$^\circ$~E&17.27 $\pm$ 0.07 &0.62 & 17 $\pm$ 2 &57.8 $\pm$ 1.9 & 16.53 &--&--&--&--&--&1.25\\
			7--8$^\circ$~S&16.52 $\pm$ 0.07 &0.27 & 19 $\pm$ 3 &40.9 $\pm$ 1.4 & 5.16 &17.34 $\pm$  0.04 & 0.23 & 38 $\pm$ 5 &59.8 $\pm$ 1.1 & 6.43& 1.87\\
			7--8$^\circ$~W&17.03 $\pm$ 0.08 &0.45 &  7 $\pm$ 1 &51.8 $\pm$ 2.1 & 10.77&--&--&--&--&--&1.06\\
			&&&&&&&&&&&\\
			
			8--10$^\circ$~N&16.92 $\pm$ 0.39 & 0.50 & 21 $\pm$ 5 & 49.3 $\pm$ 8.8 & 11.47 & 17.70 $\pm$ 0.05 & 0.22 & 53 $\pm$ 17 &70.3 $\pm$ 1.8 & 7.24 &3.21\\
			8--10$^\circ$~E&16.74 $\pm$ 0.62 & 0.47 & 21 $\pm$ 11 & 45.3 $\pm$ 13.0 & 9.74 & 17.41 $\pm$ 0.12 & 0.27 & 39 $\pm$ 28 &61.8 $\pm$ 3.1 & 7.81 &1.81\\
			8--10$^\circ$~S&16.91 $\pm$ 0.06 & 0.56 & 39 $\pm$ 4 & 48.9 $\pm$ 1.4 & 12.71 &--&--&--&--&--&2.43\\
			8--10$^\circ$~W&16.56 $\pm$ 0.13 & 0.50 & 28 $\pm$ 5 & 41.6 $\pm$ 2.6 & 9.65 & 17.50 $\pm$ 0.06 &0.16 & 28 $\pm$ 10 & 64.2 $\pm$ 1.8 & 4.8 &4.18\\

			\hline
		\end{tabular} 
	\end{table*}

\section*{acknowledgements}
We are grateful to the anonymous referee who helped us shape the content and format of the paper with valuable suggestions. We thank the Cambridge Astronomy Survey Unit (CASU) and the Wide Field Astronomy Unit (WFAU) in Edinburgh for providing calibrated data products under the support of the Science and Technology Facility Council (STFC). This project has received funding from the European Research Council (ERC) under European Union's Horizon 2020 research and innovation programme (project INTERCLOUDS, grant agreement no. 682115). We thank Vasiliev Belokurov for the fruitful exchange concerning this work on multiple occasions. This research was supported by the DAAD with funds from the German Federal Ministry of Education and Research within an exchange grant between the Leibniz Institute of Astrophysics Potsdam (AIP, Germany) and the Indian Institute of Astrophysics (IIA, Bangalore). This study is based on observations obtained with VISTA at the Paranal Observatory under programmes 179.B-2003 and 179.A-2010. This work has made use of data from the European Space Agency (ESA)
mission {\it Gaia} (\url{https://www.cosmos.esa.int/gaia}), processed by
the {\it Gaia} Data Processing and Analysis Consortium (DPAC,
\url{https://www.cosmos.esa.int/web/gaia/dpac/consortium}). Funding
for the DPAC has been provided by national institutions, in particular
the institutions participating in the {\it Gaia} Multilateral Agreement. SS acknowledges support from the Science and Engineering Research Board, India, through a Ramanujan Fellowship.
This project has made extensive use of the Tool for OPerations on Catalogues And Tables (TOPCAT) software package \citep{Taylor2005} as well as the following open-source Python packages: matplotlib \citep{Hunter2007}, NumPy \citep{Oliphant2015}, pandas \citep{McKinney2010}, SciPy \citep{Jones2001}.

\section*{Data Availability}
The mean magnitudes, spread in brightness and distances of RC stars in different
sectors are provided in Table \ref{table:smc}. The VMC data were released as part of Data Release 5 (DR5) of the VMC survey (see \url{https://www.eso.org/sci/publications/announcements/sciann17232.html}). VHS observations obtained until 2017 March 30 were released as part of DR5 of the VHS survey (see \url{https://www.eso.org/sci/publications/announcements/sciann17290.html}), whereas VHS observations obtained until 2017 September 30 will be released soon.

\bibliographystyle{mnras} 
\bibliography{ref-VHS}

\appendix
\section{Hess diagrams and luminosity functions of RC stars}\label{appendix_0}
\begin{figure*}
	\centering
	
	\includegraphics[scale=0.06]{VHS-SMC-CMD-R-01-N_der_new_box.png}
	\includegraphics[scale=0.06]{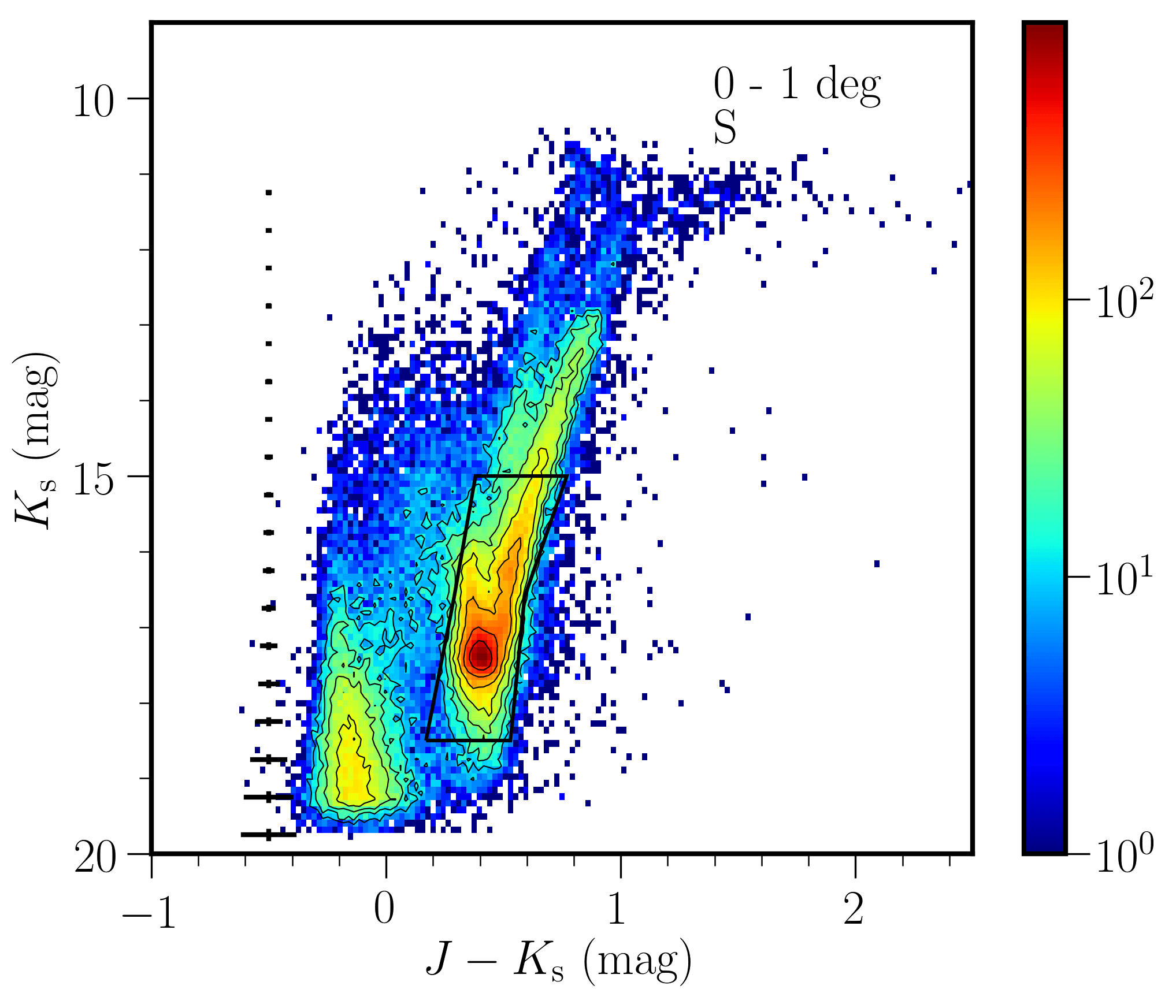}
	\includegraphics[scale=0.06]{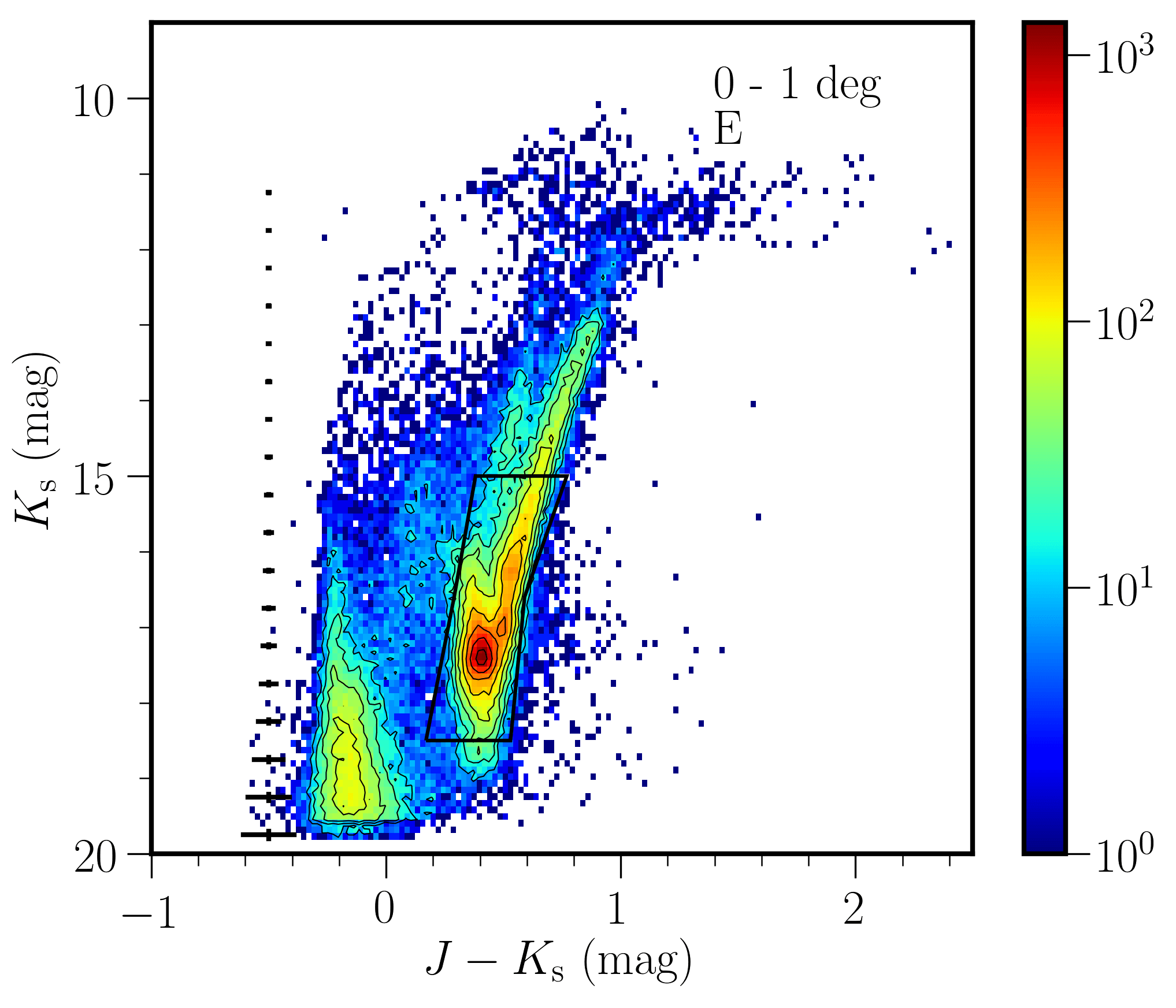}
	\includegraphics[scale=0.06]{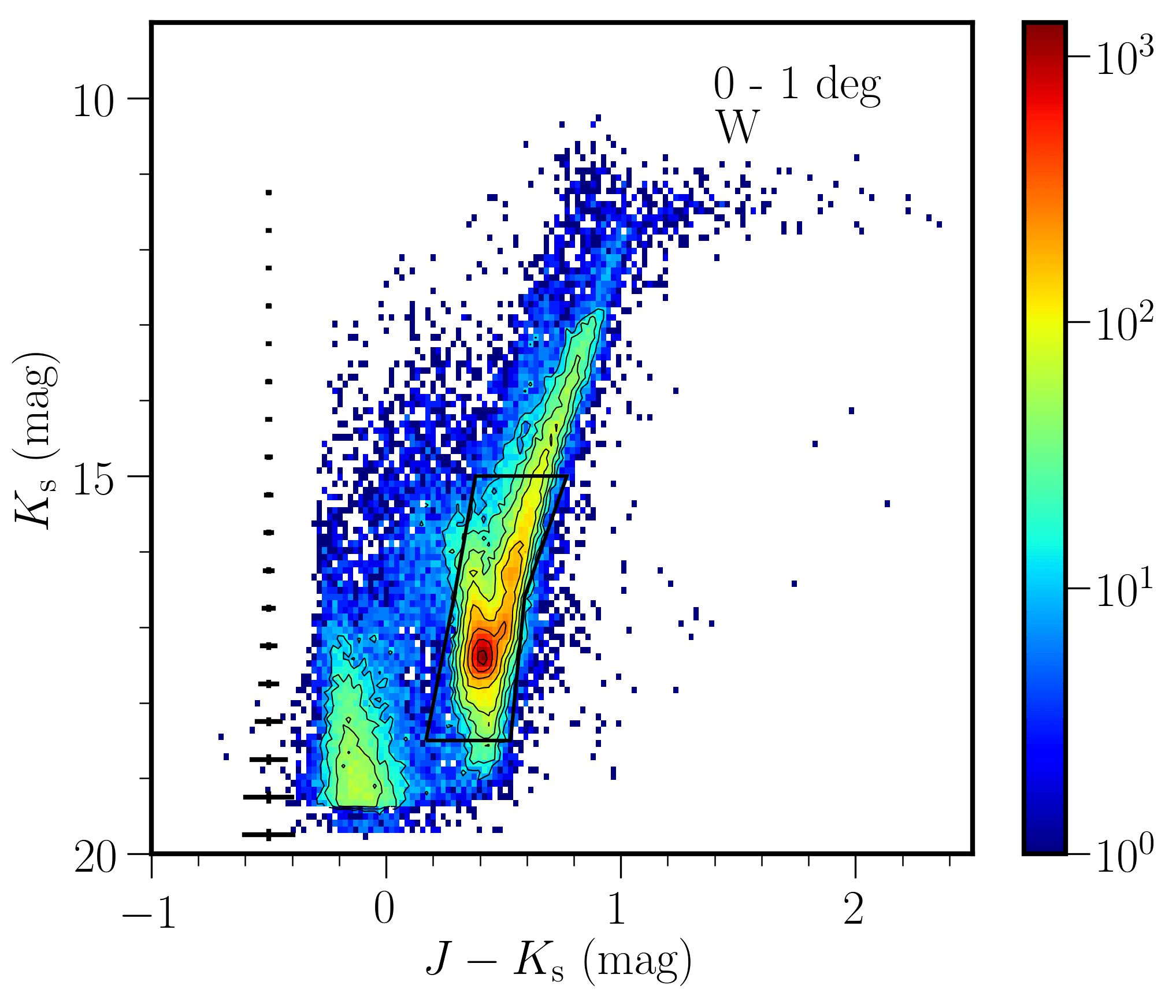}
	
	\includegraphics[scale=0.06]{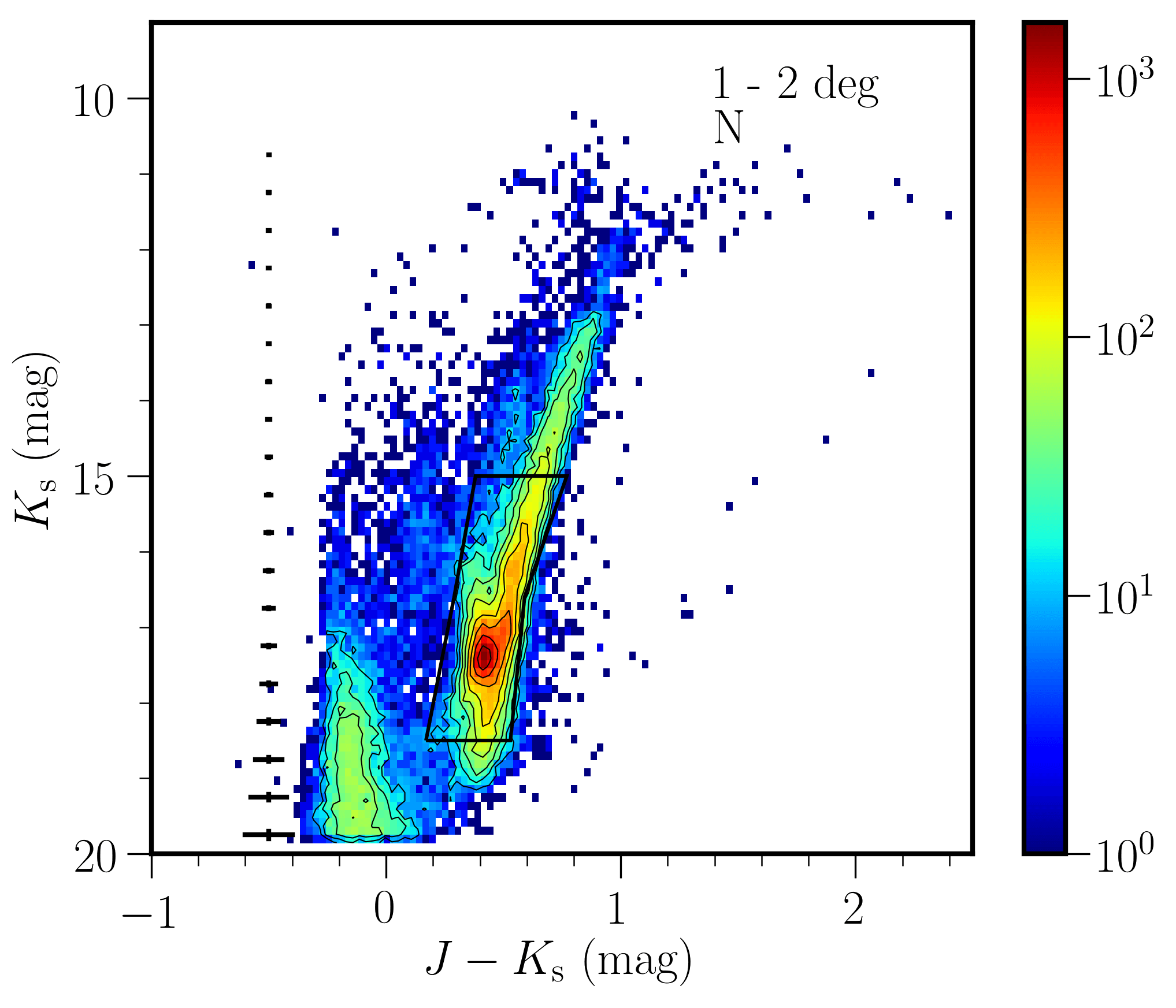}
	\includegraphics[scale=0.06]{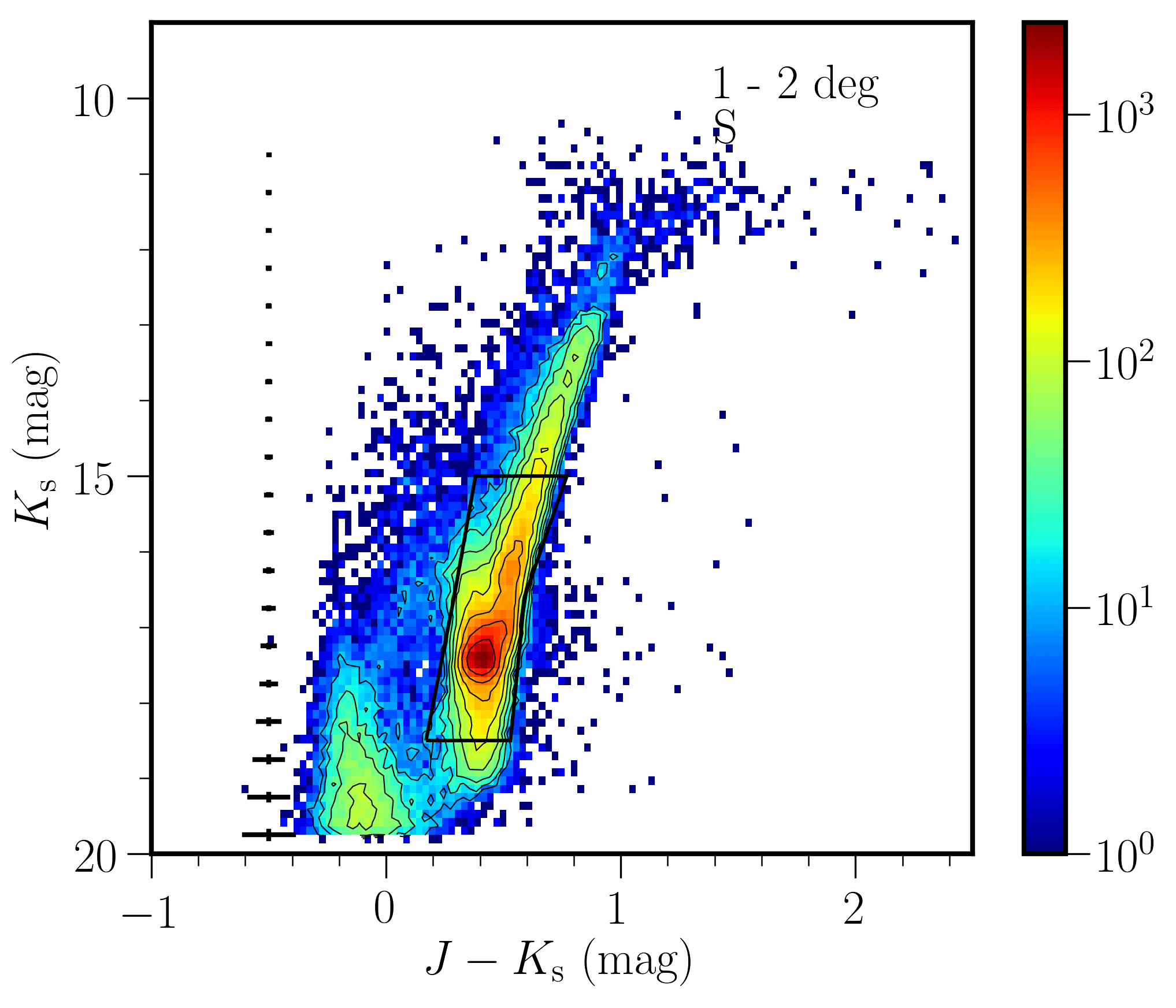}
	\includegraphics[scale=0.06]{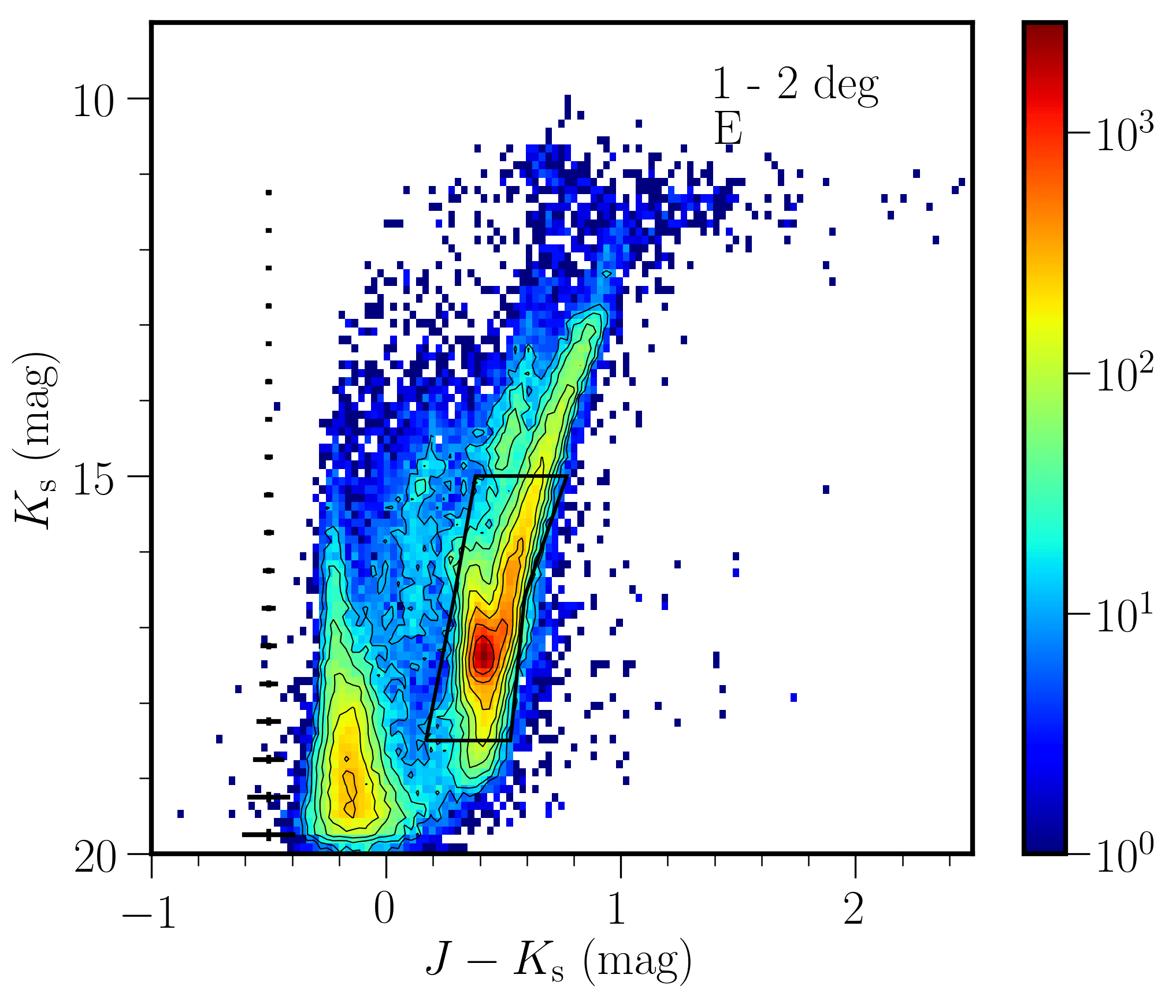}
	\includegraphics[scale=0.06]{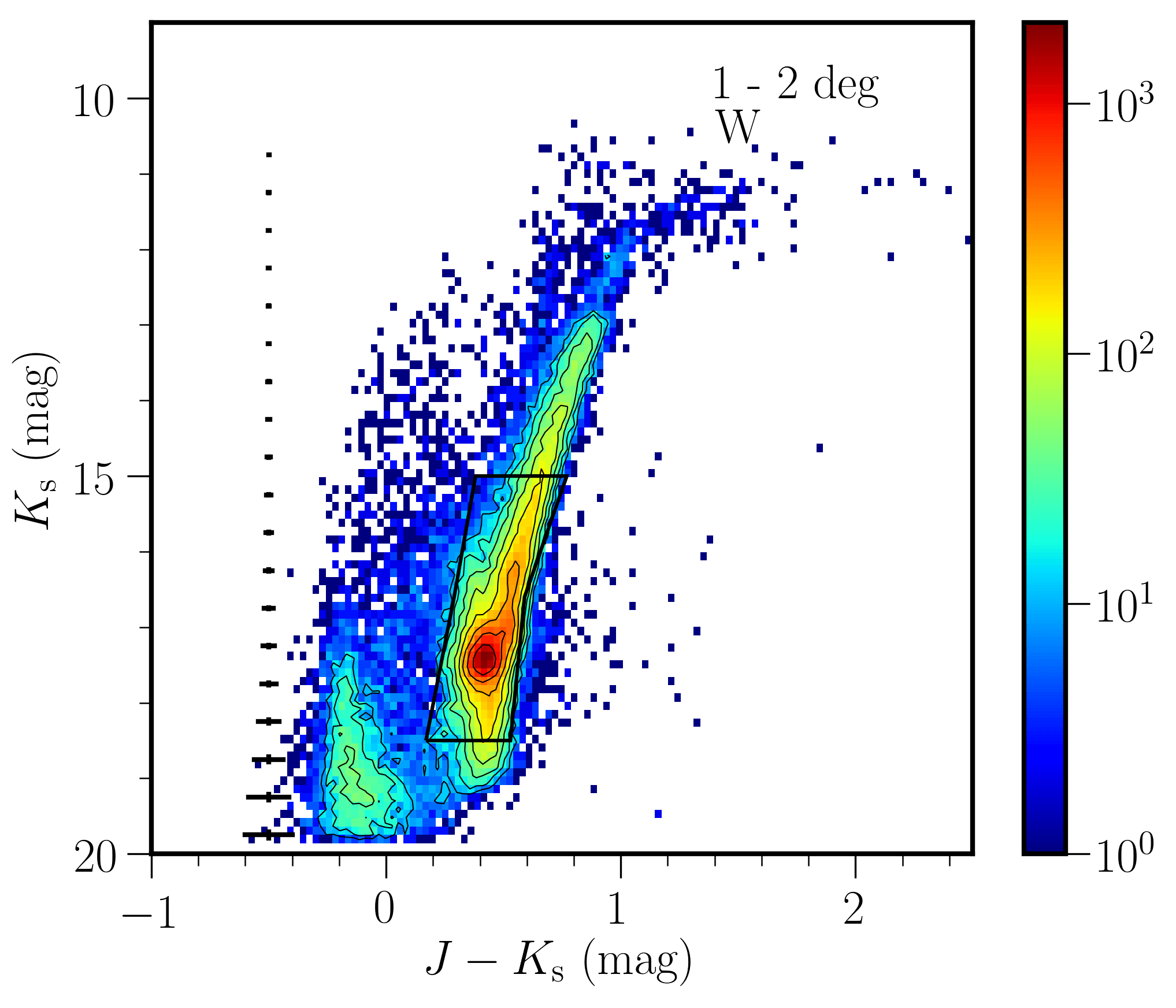}
	
	\includegraphics[scale=0.06]{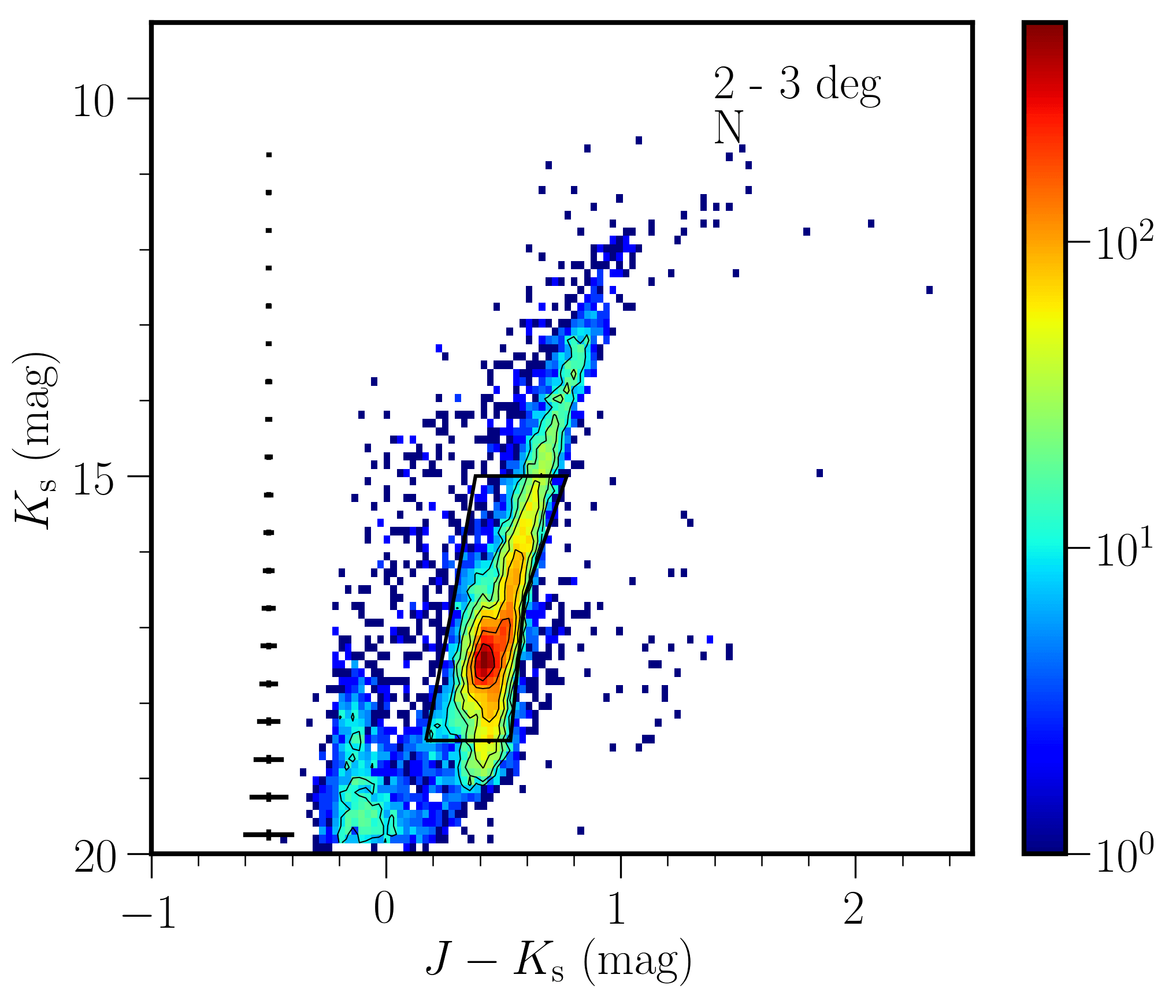}
	\includegraphics[scale=0.06]{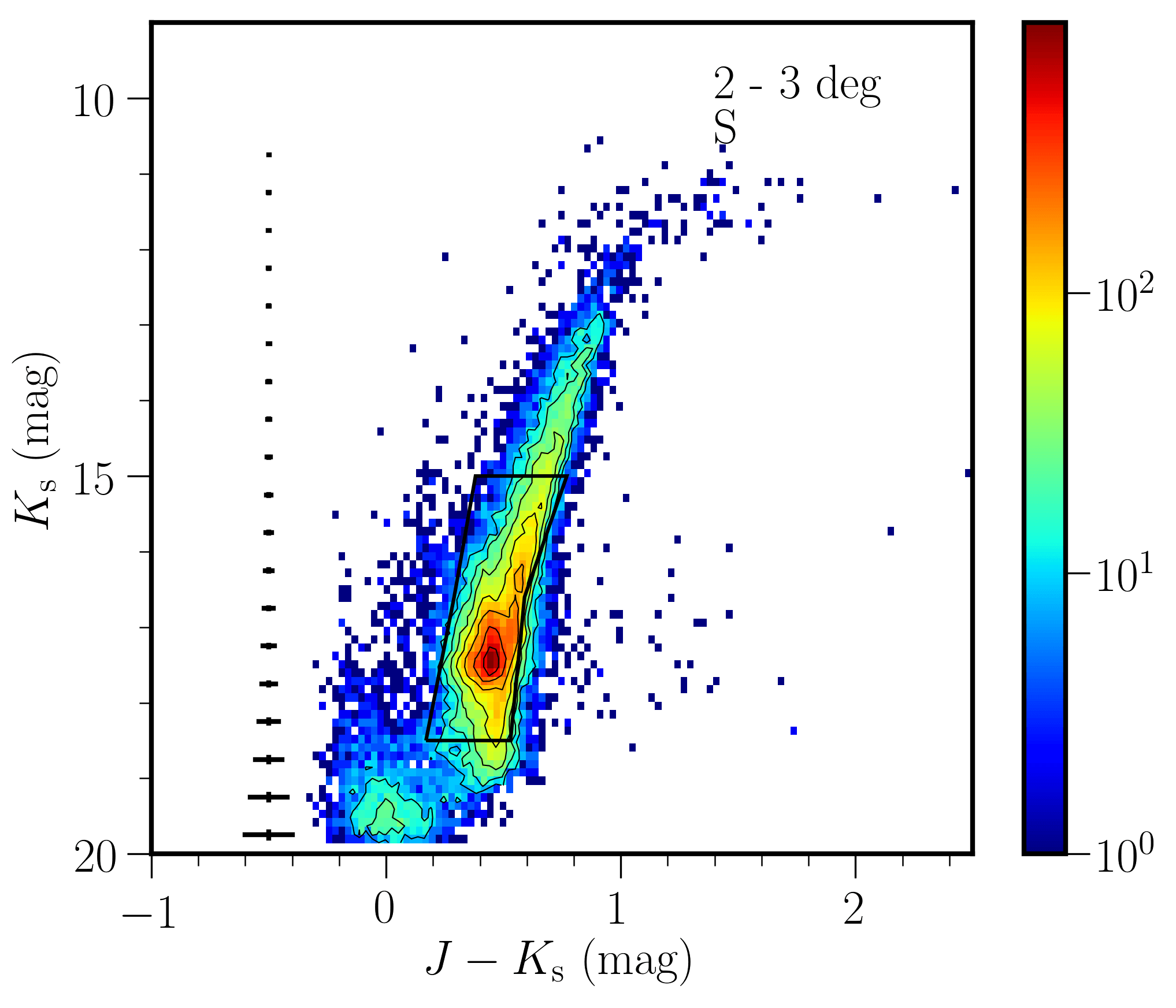}
	\includegraphics[scale=0.06]{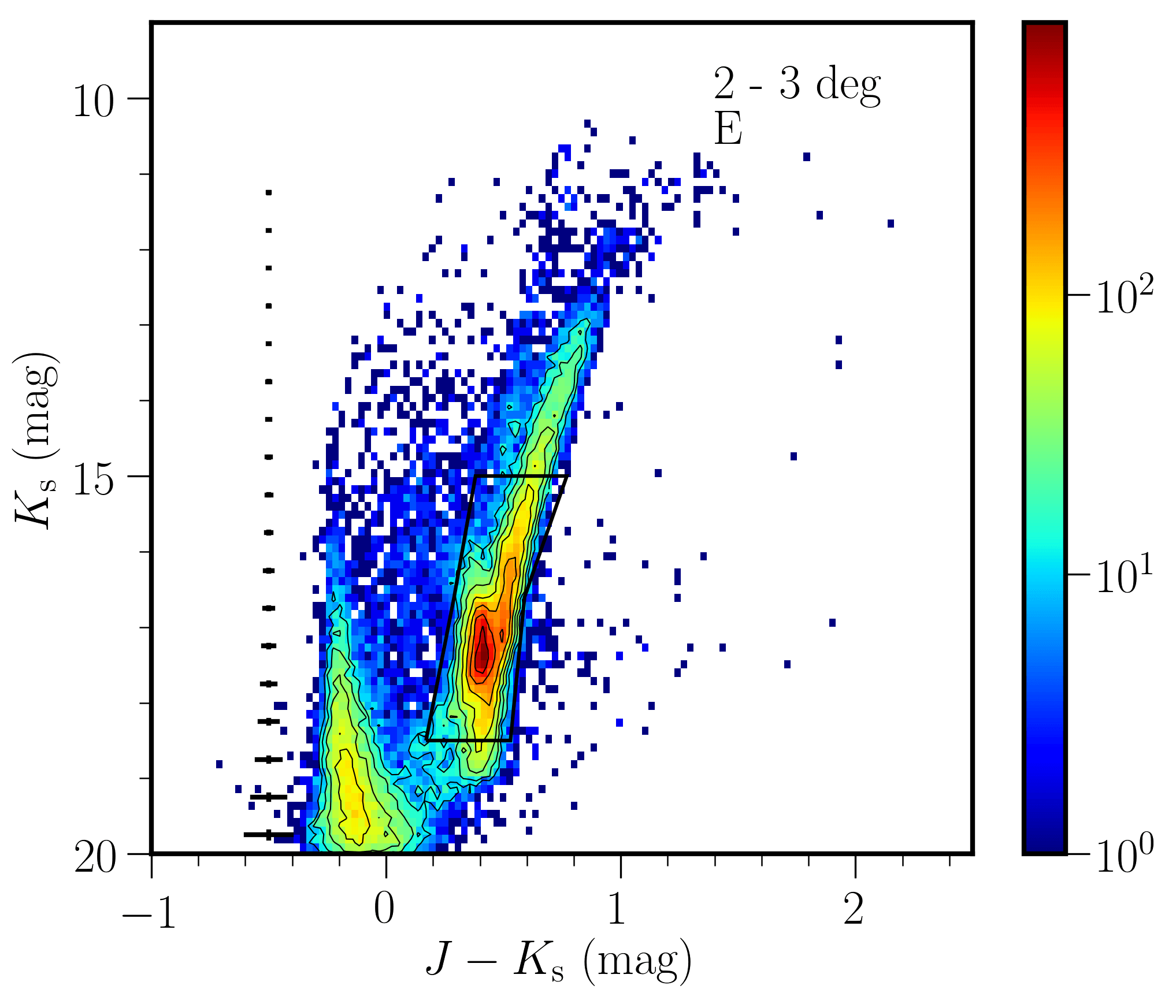}
	\includegraphics[scale=0.06]{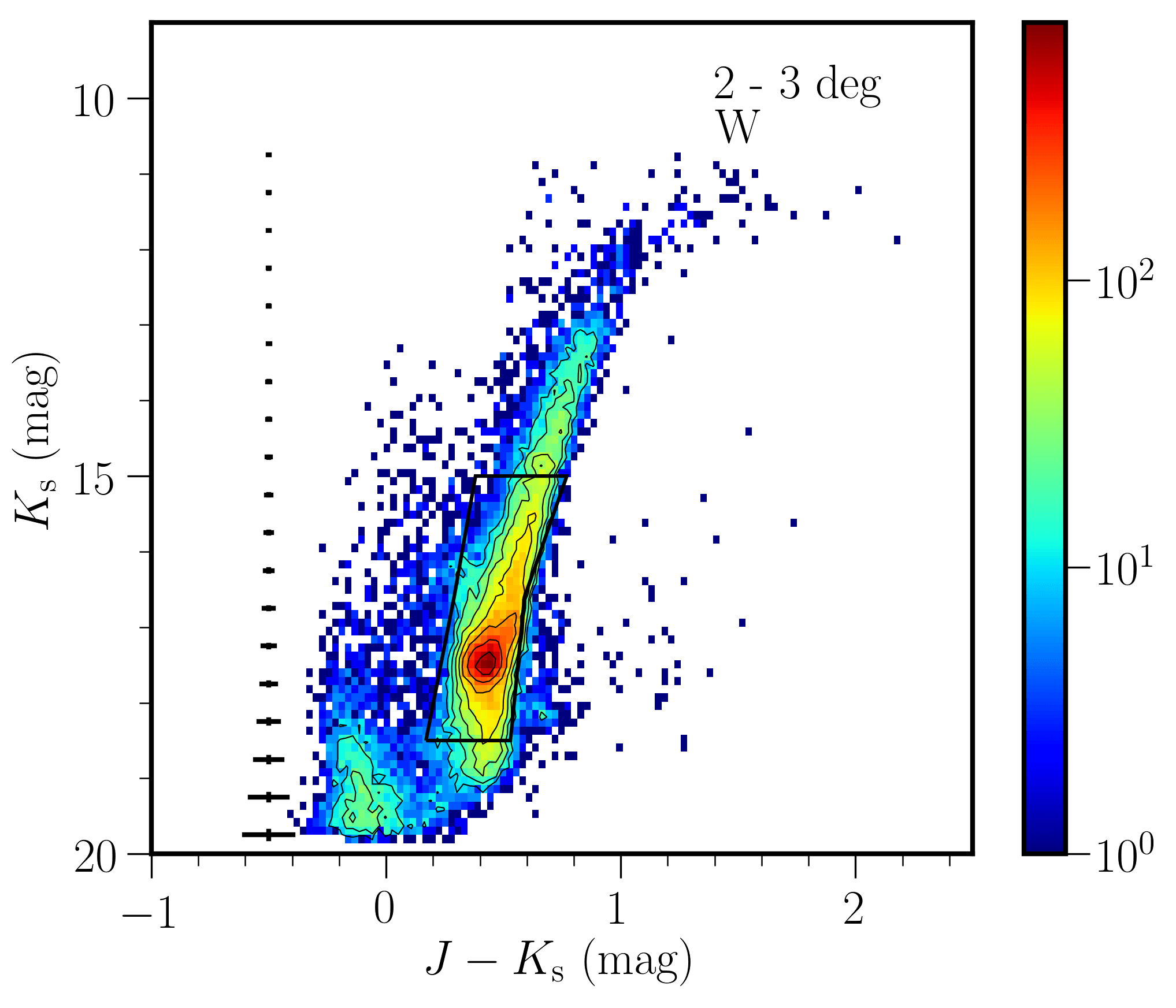}

	\includegraphics[scale=0.06]{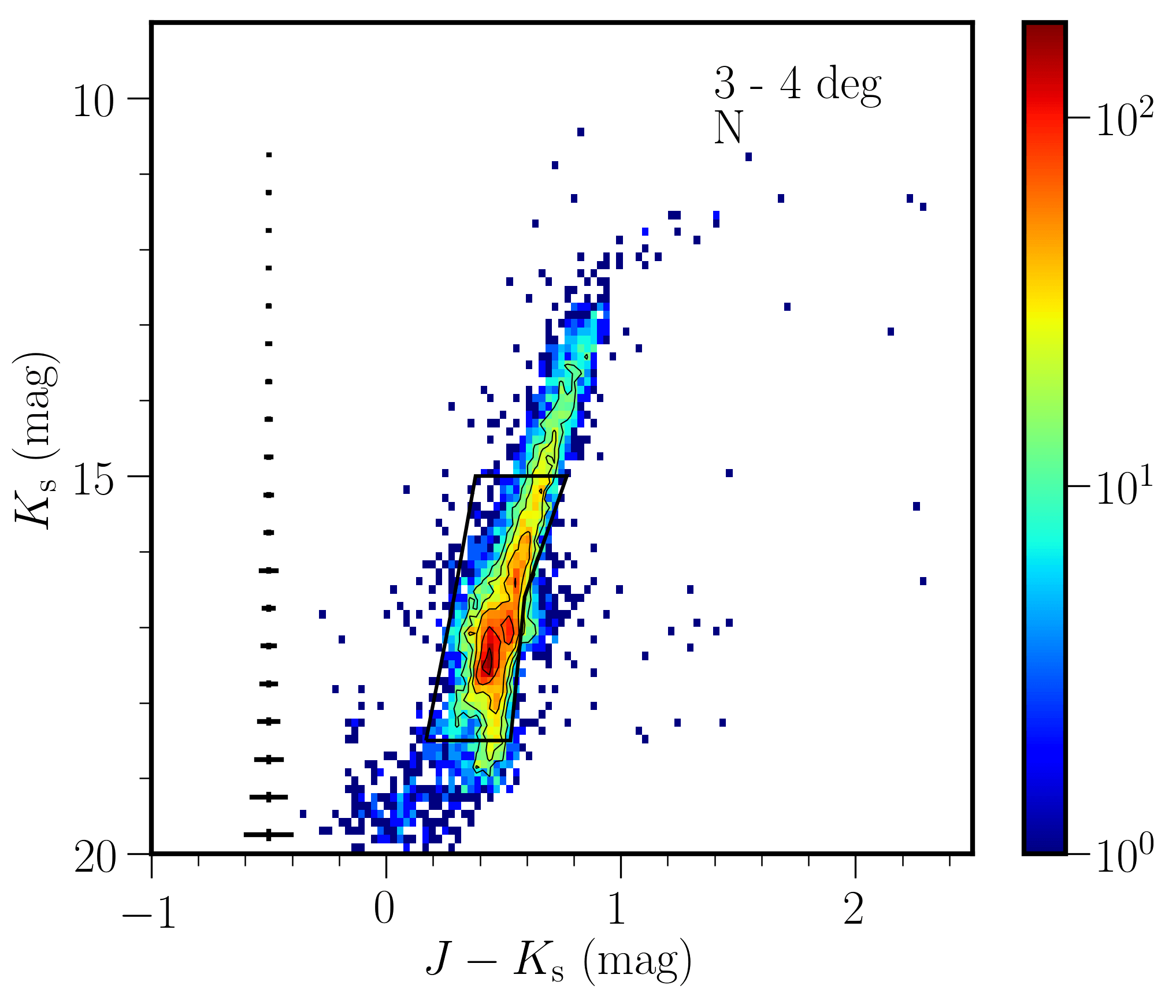}
	\includegraphics[scale=0.06]{VHS-SMC-CMD-R-34-S_der_new_box.png}
	\includegraphics[scale=0.06]{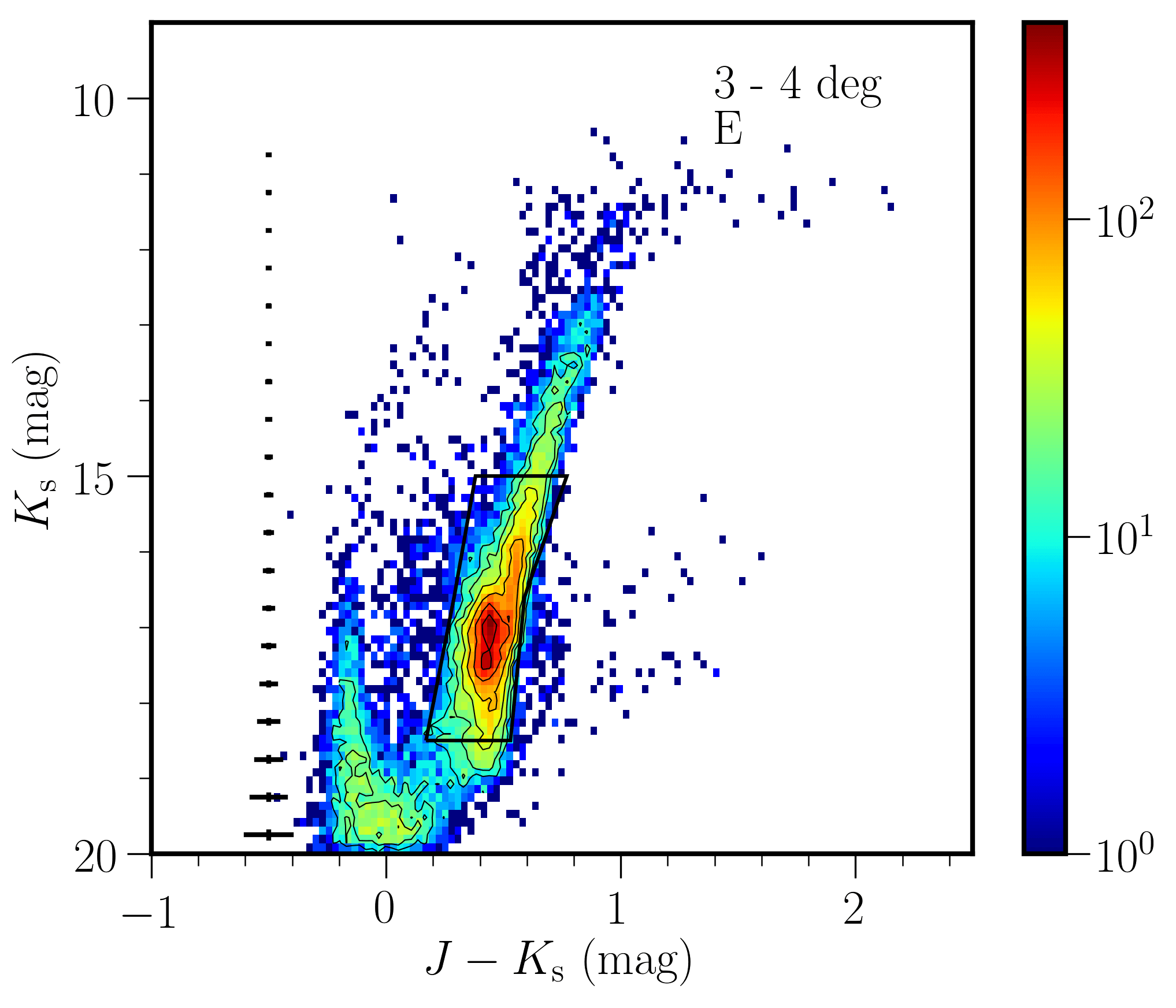}
	\includegraphics[scale=0.06]{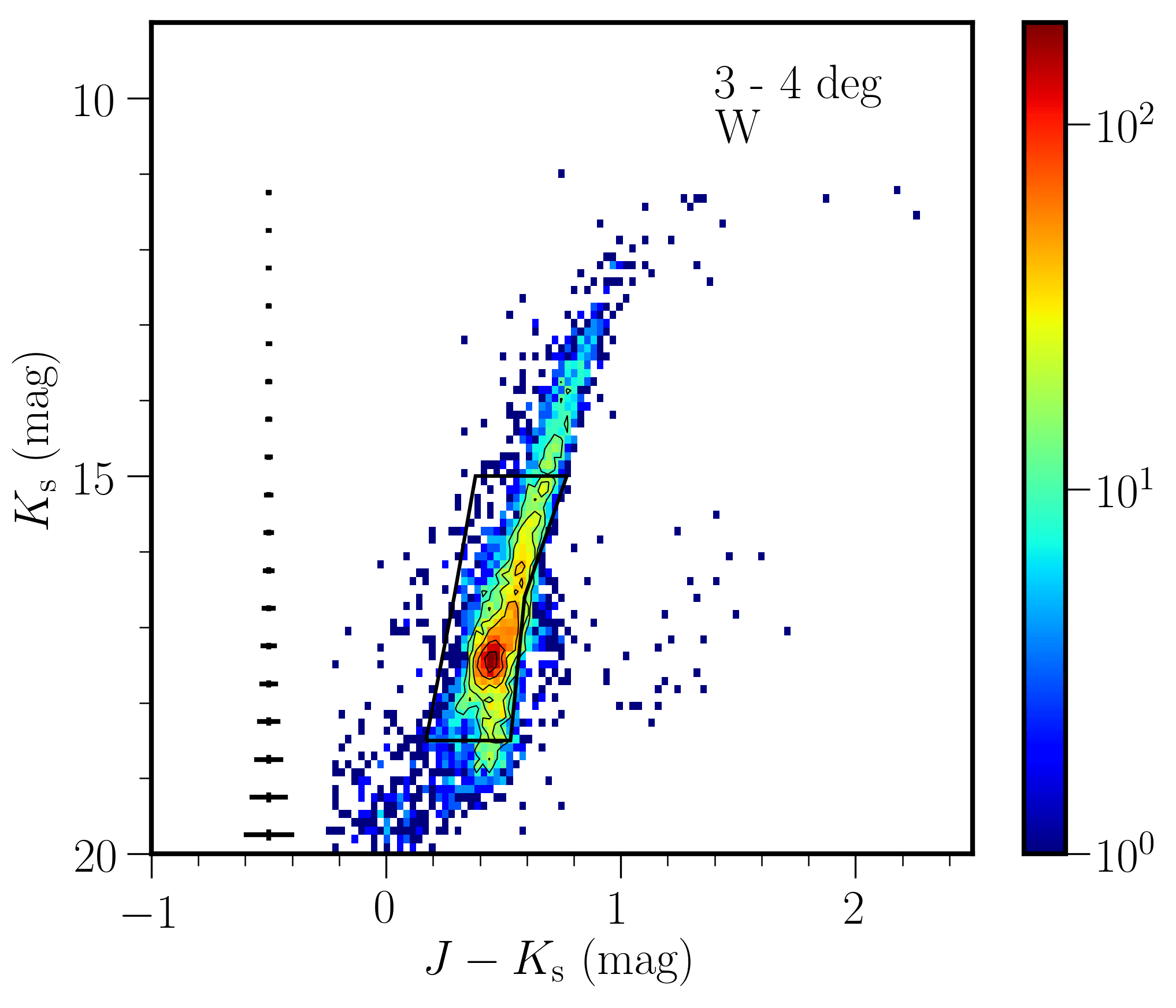}

	\includegraphics[scale=0.06]{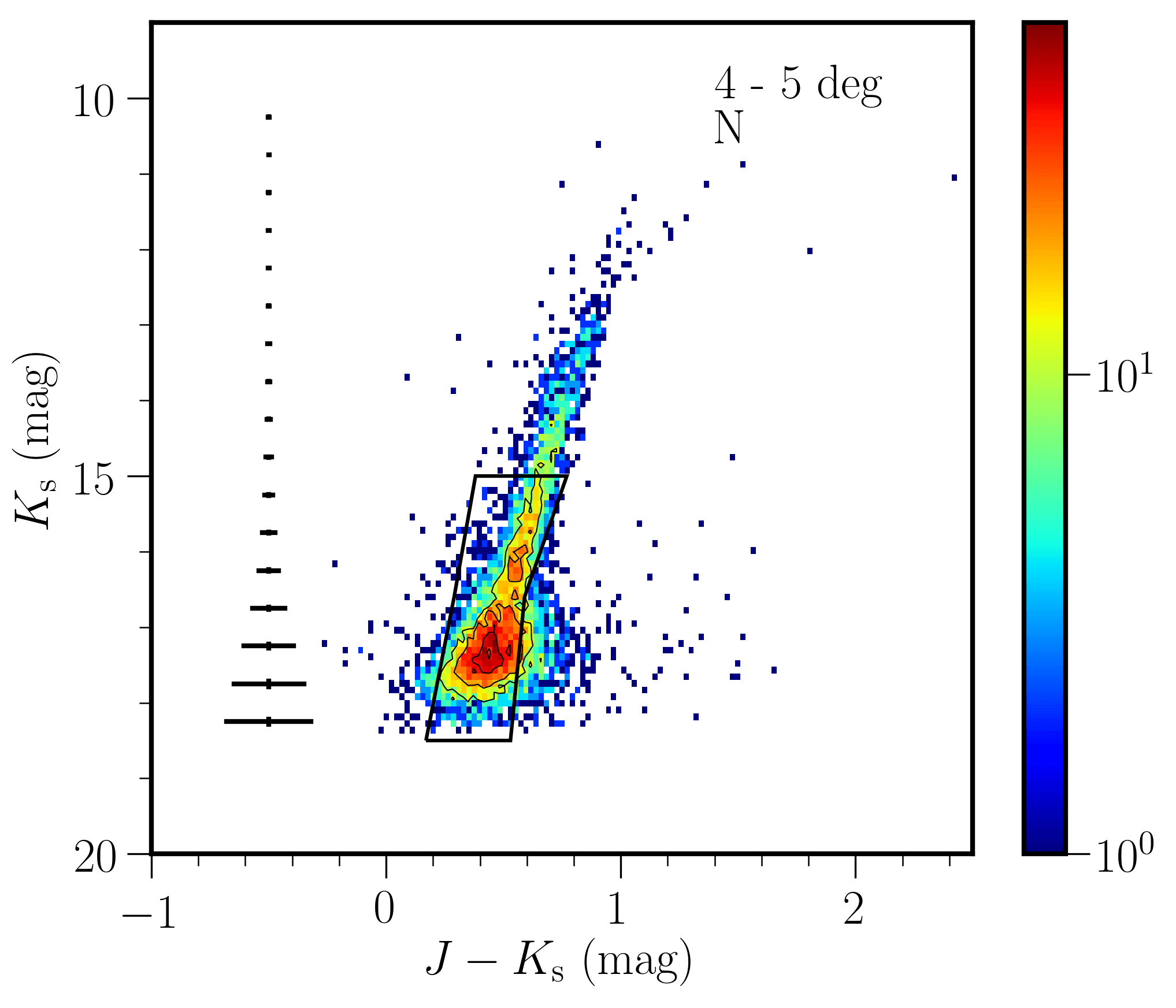}
	\includegraphics[scale=0.06]{VHS-SMC-CMD-R-45-S_der_new_box.png}
	\includegraphics[scale=0.06]{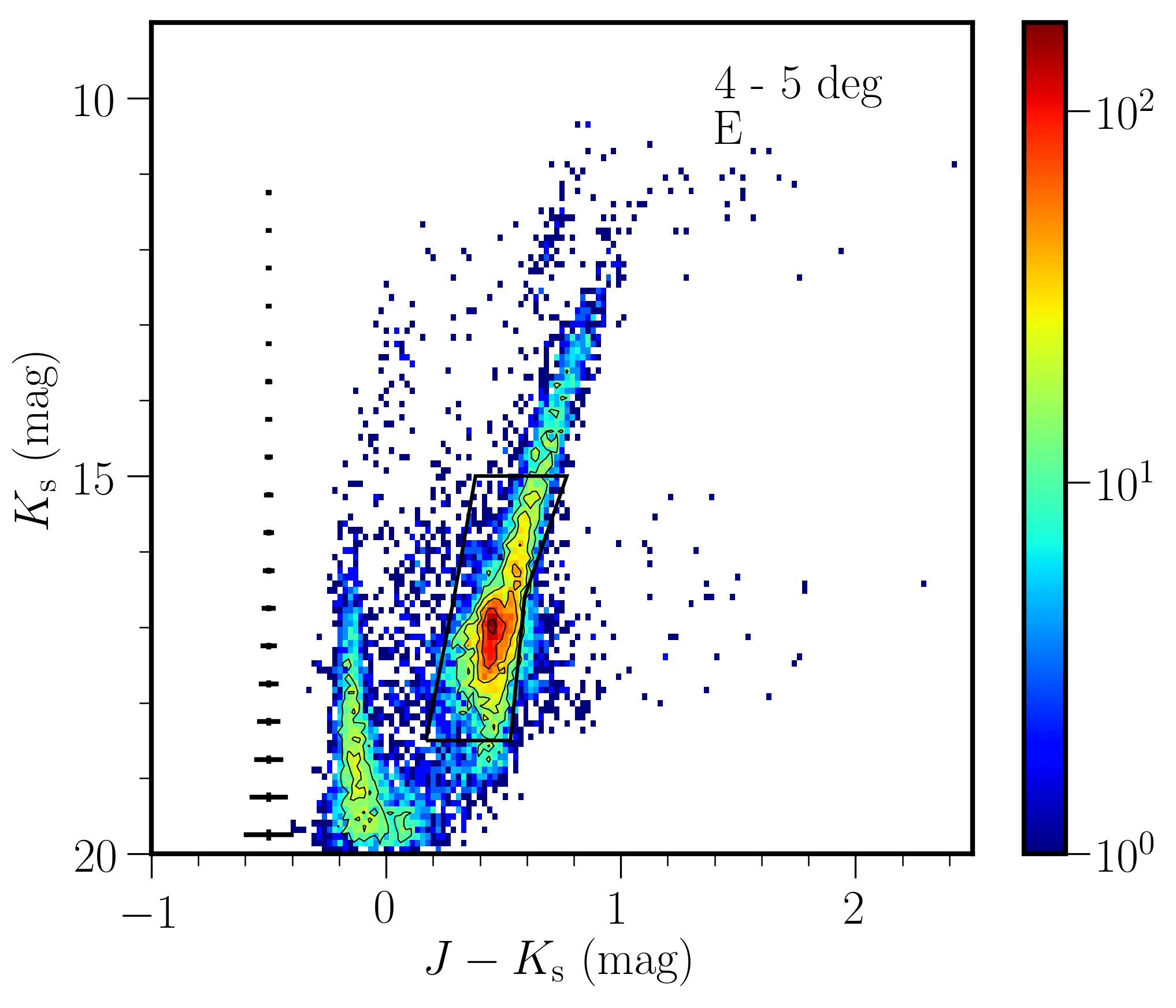}
	\includegraphics[scale=0.06]{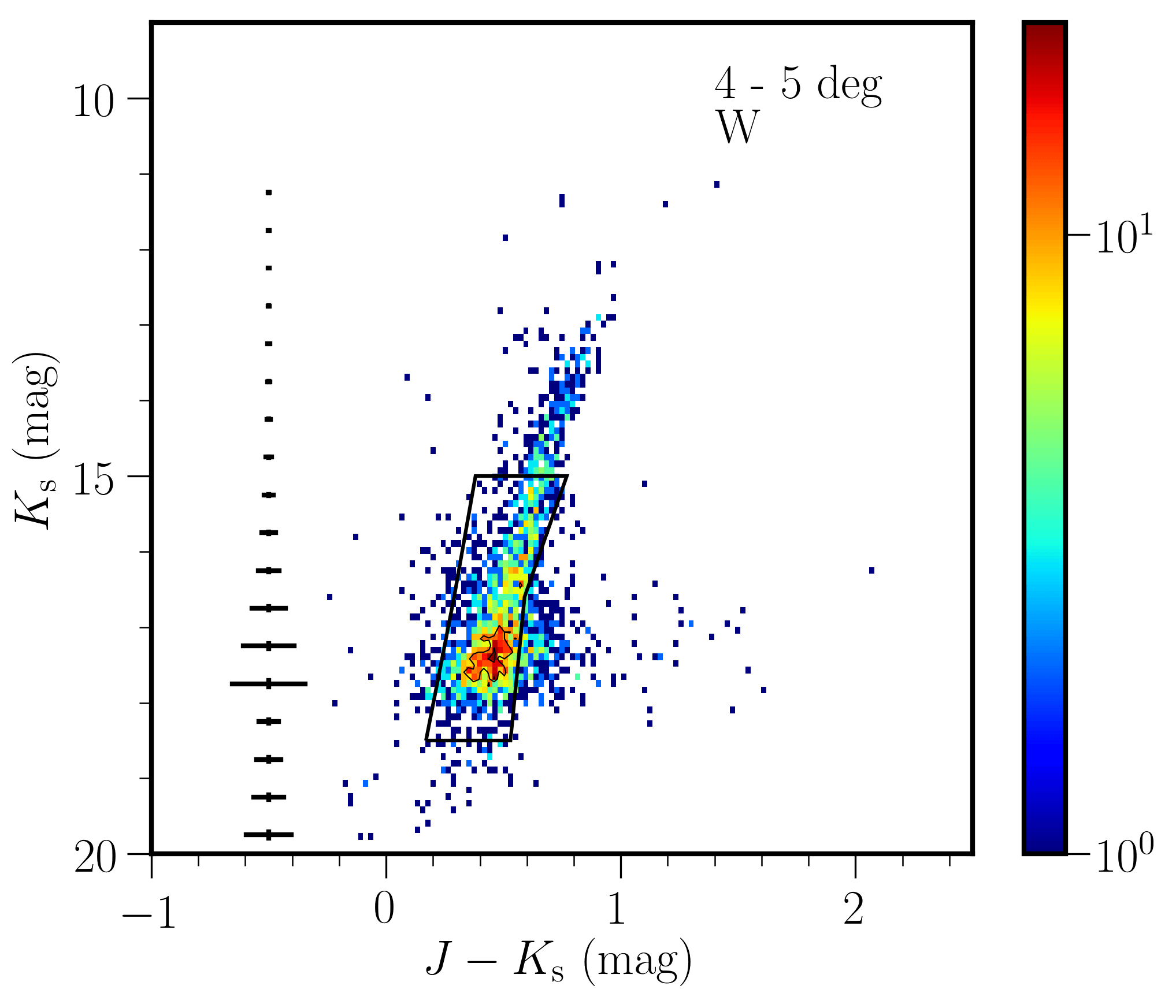}
	
	\includegraphics[scale=0.06]{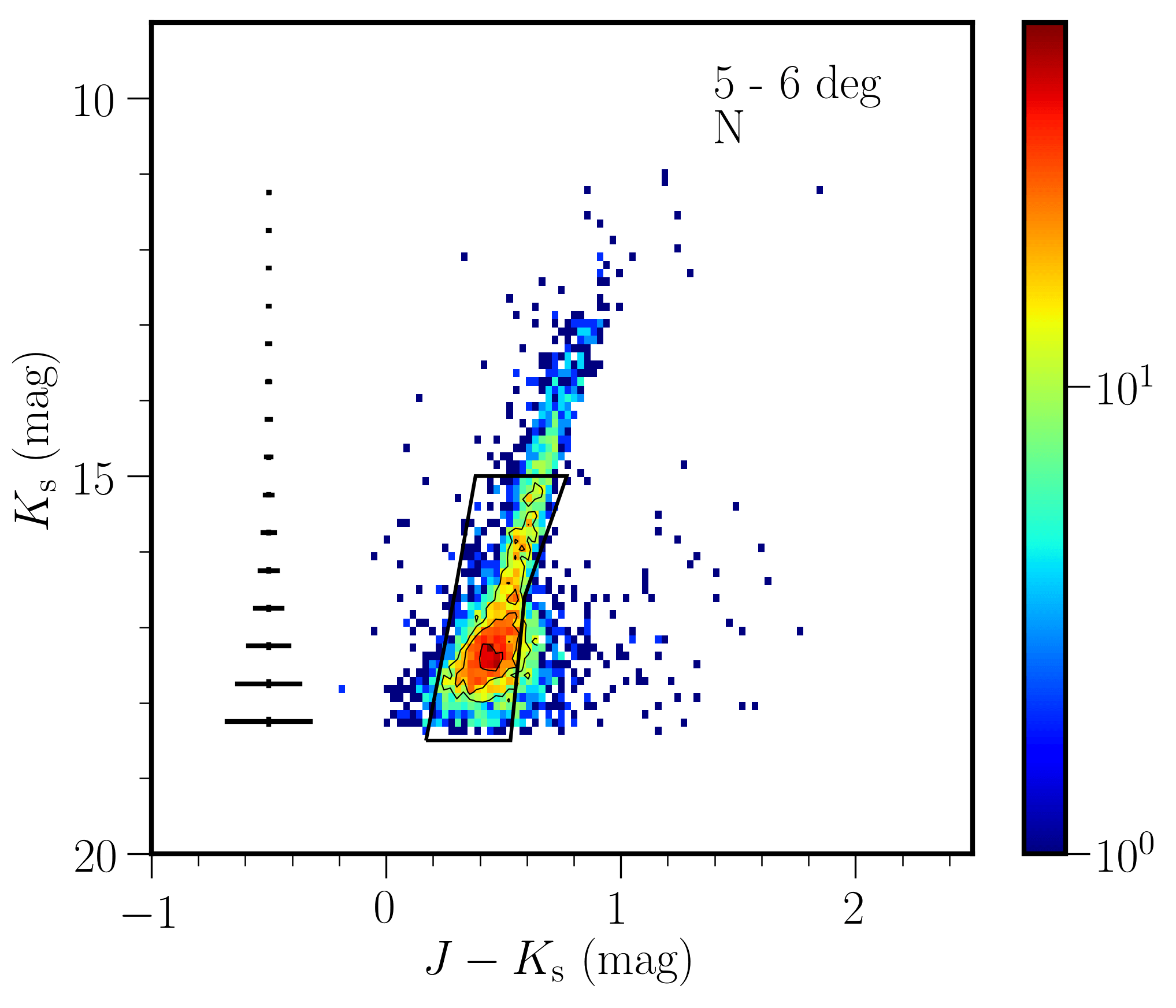}
	\includegraphics[scale=0.06]{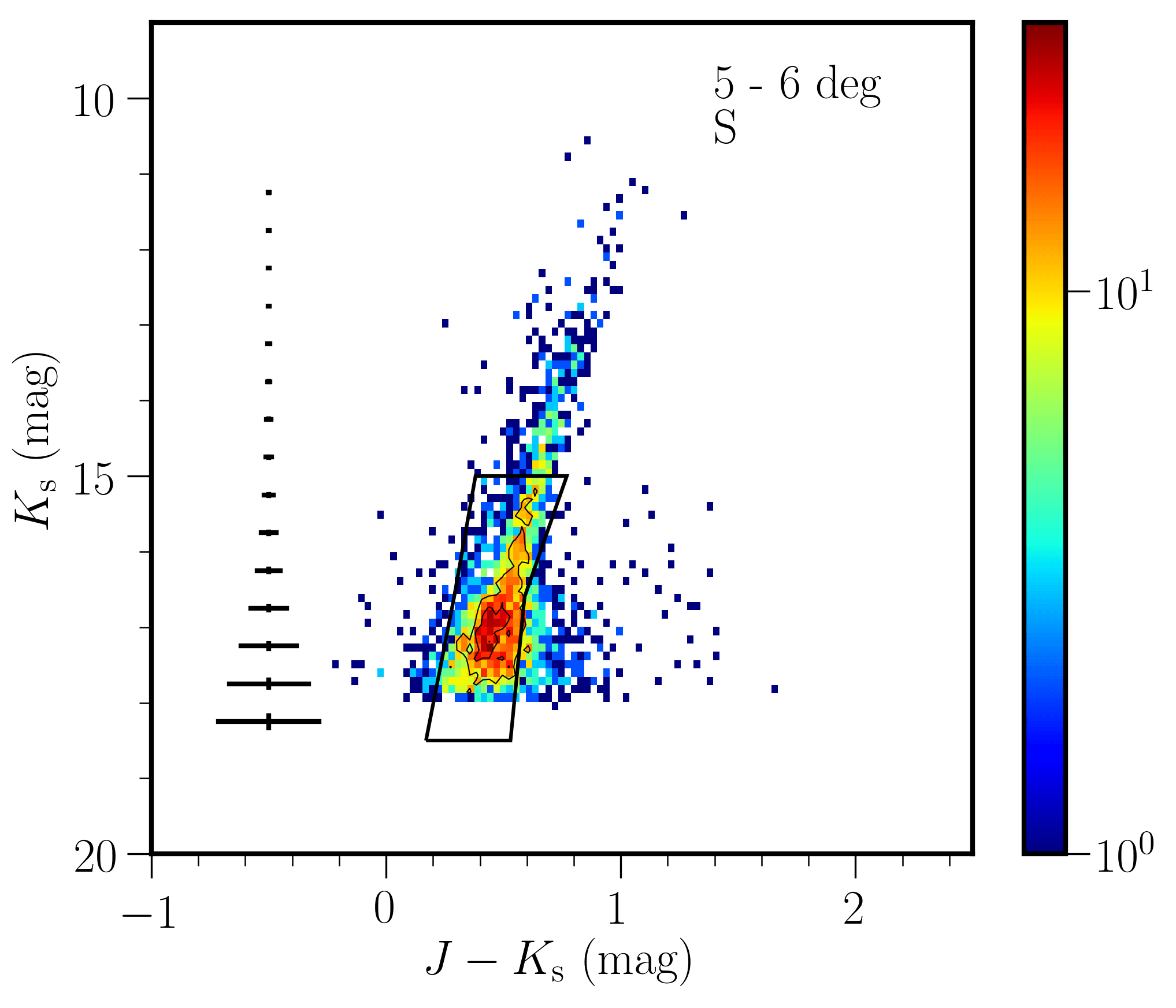}
	\includegraphics[scale=0.06]{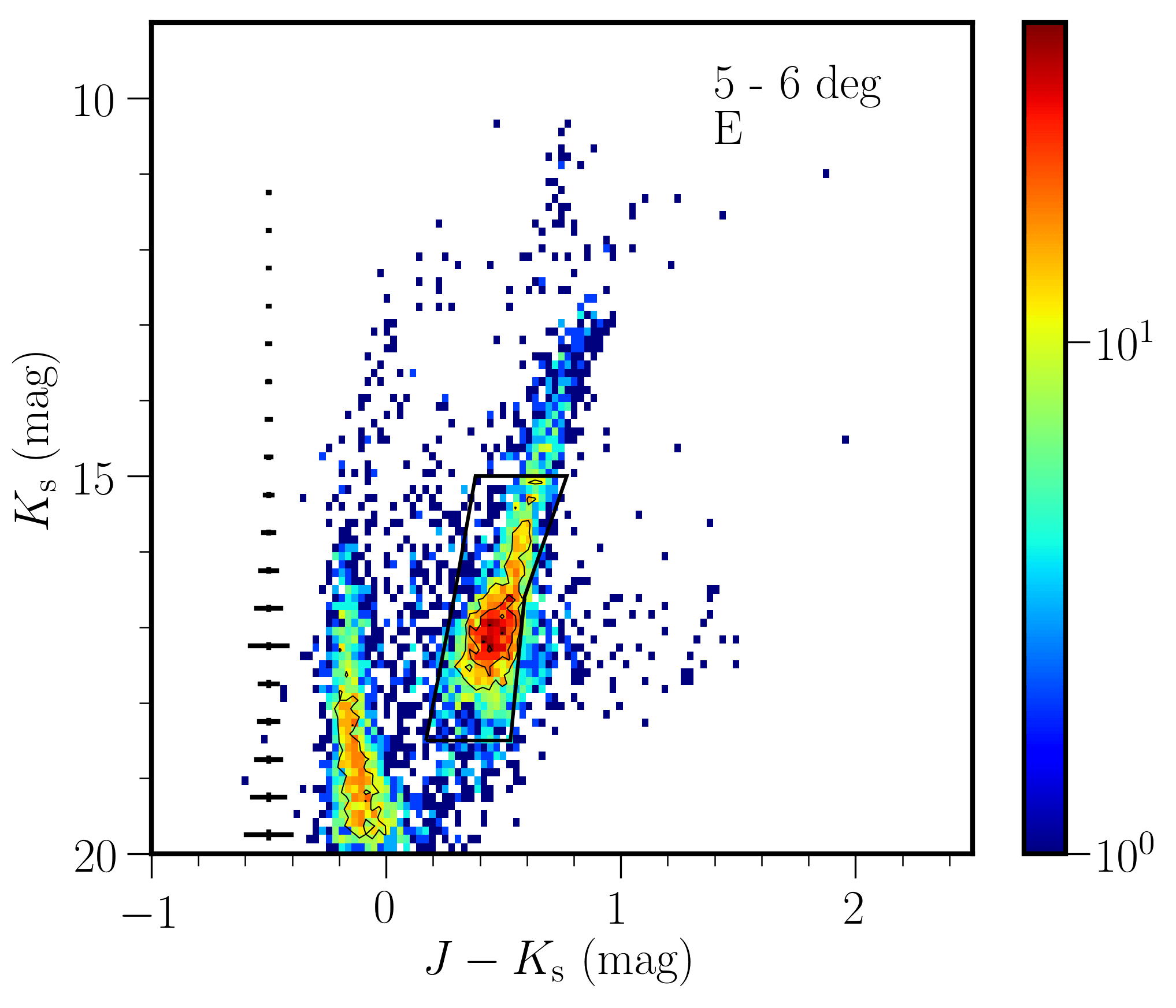}
	\includegraphics[scale=0.06]{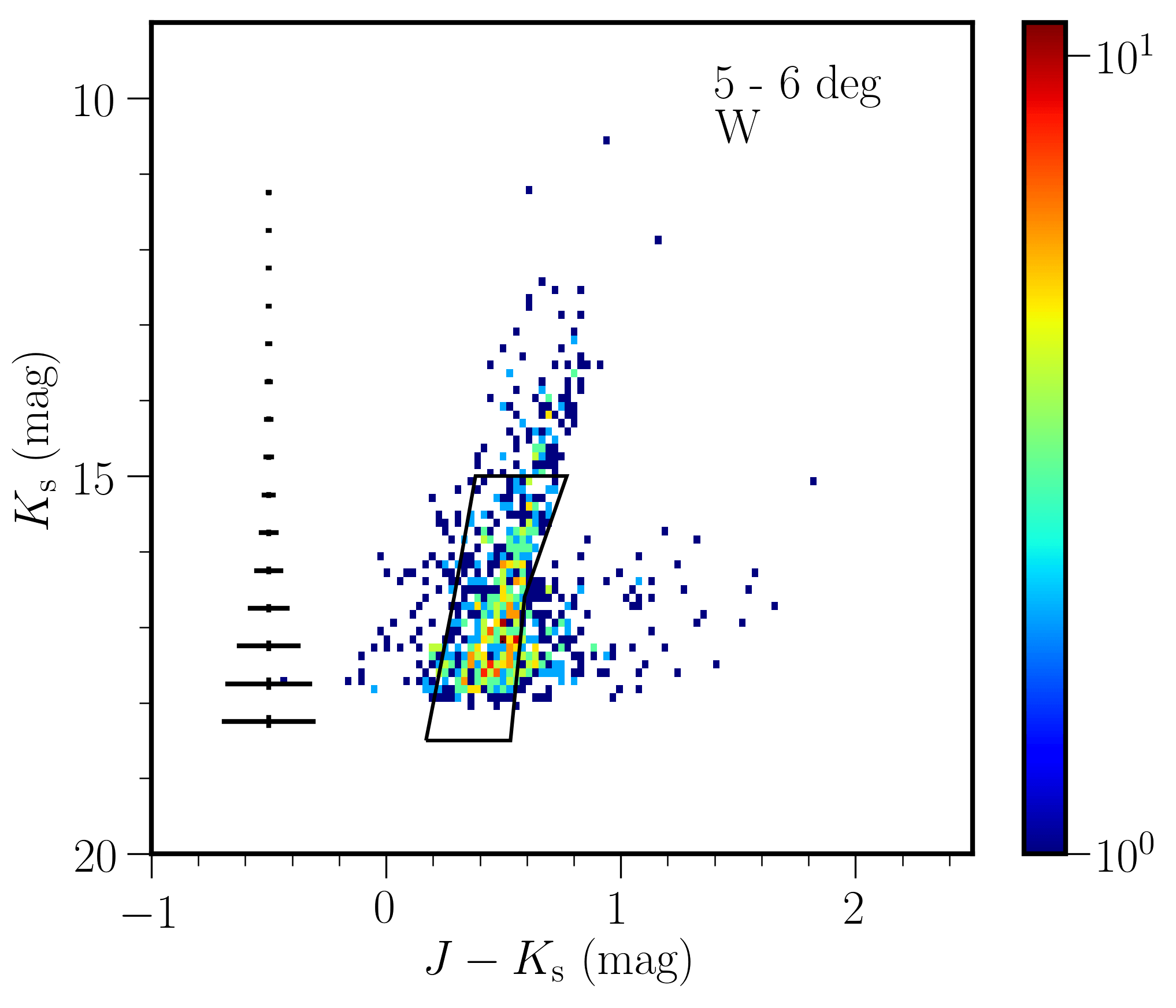}
	
	\caption{NIR ($J-K_\mathrm{s}$, $K_\mathrm{s}$) Hess diagrams of different SMC regions. The colour scale indicates the stellar density on a logarithmic scale per bin. The bin size is 0.027$\times$0.110 deg$^2$ and the black box limits the region defined to study the distribution of RC stars.}
	\label{fig:RCCMD0}
\end{figure*}

\begin{figure*}
	\setcounter{figure}{0}

	\includegraphics[scale=0.06]{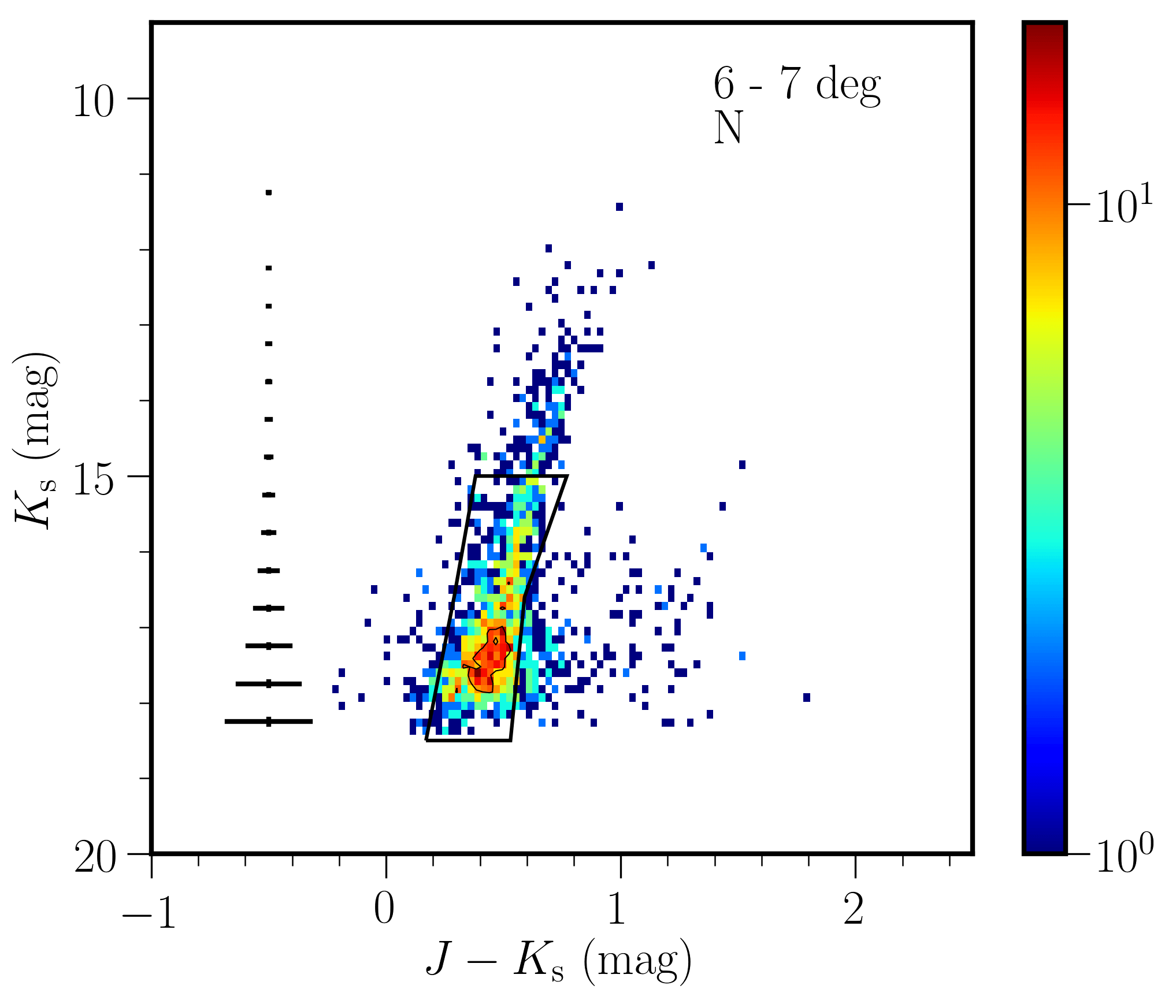}
	\includegraphics[scale=0.06]{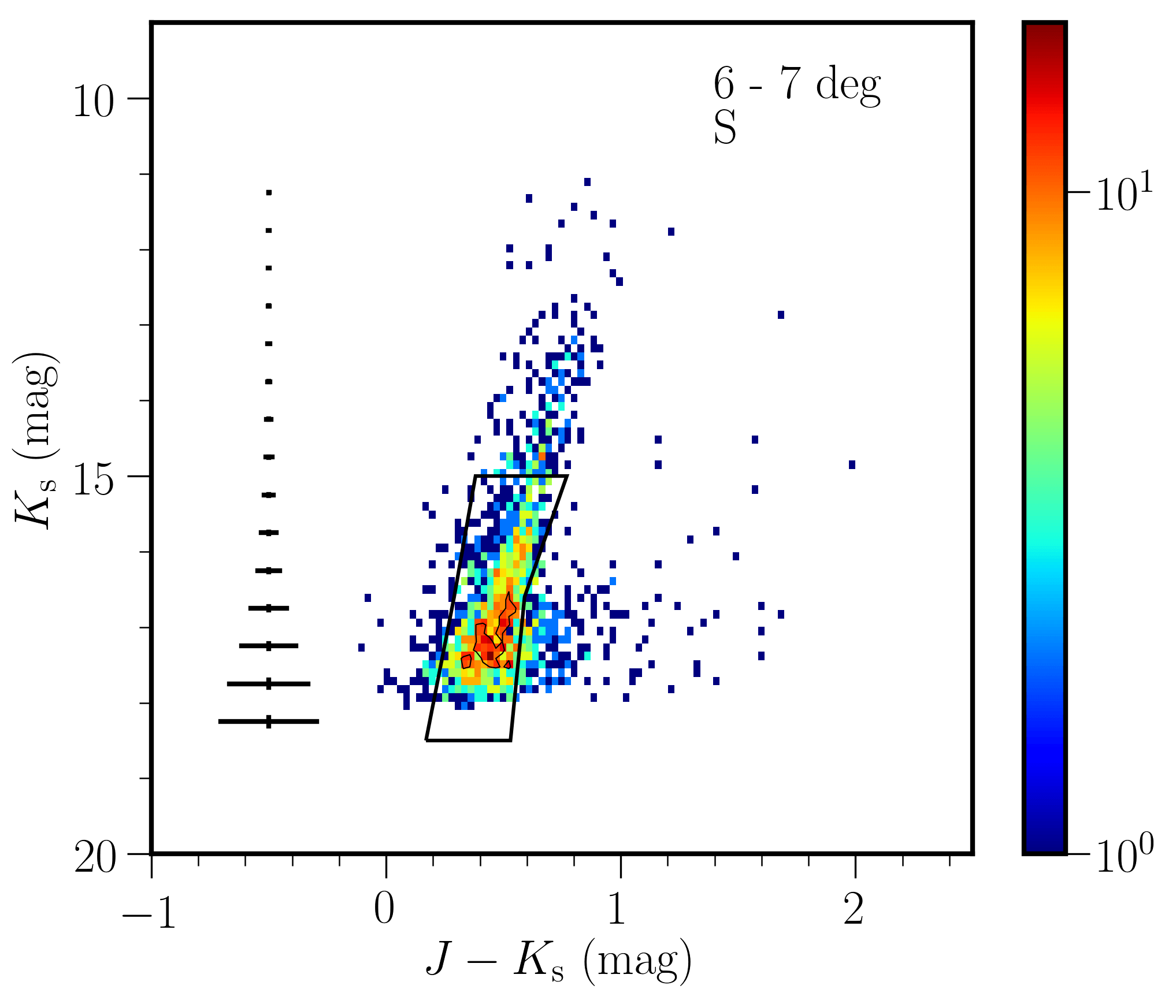}
	\includegraphics[scale=0.06]{VHS-SMC-CMD-R-67-E_der_new_box.png}
	\includegraphics[scale=0.06]{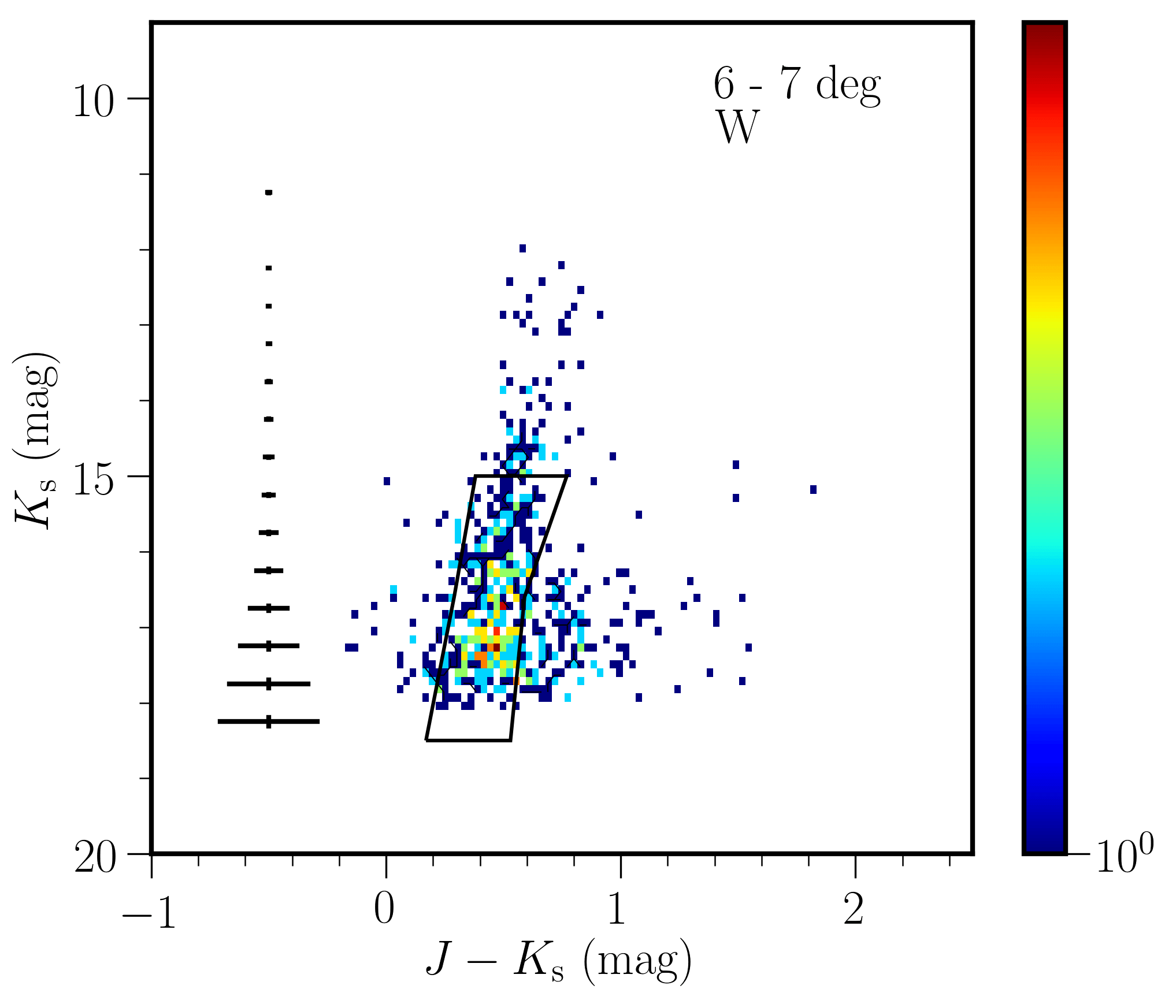}
	
	\includegraphics[scale=0.06]{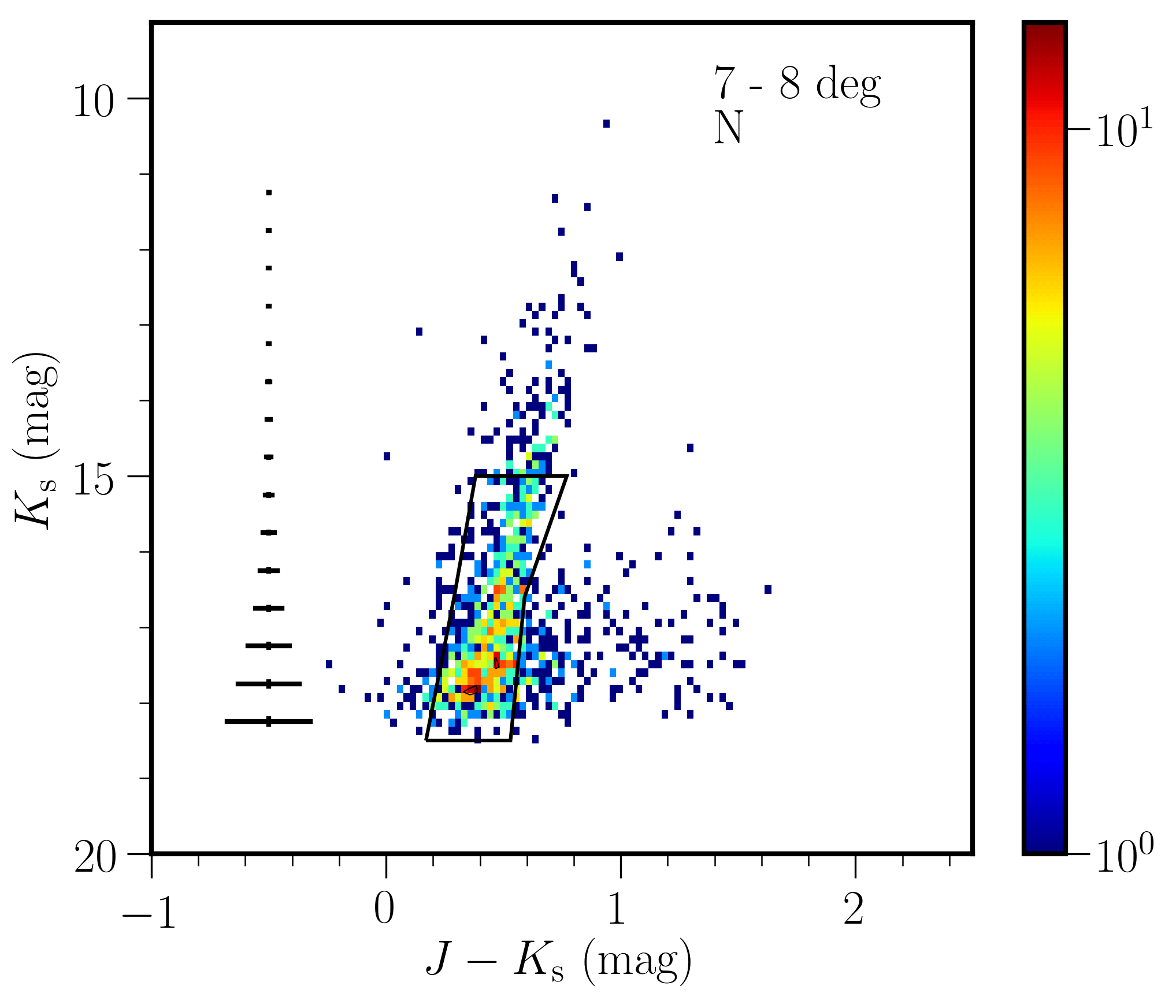}
	\includegraphics[scale=0.06]{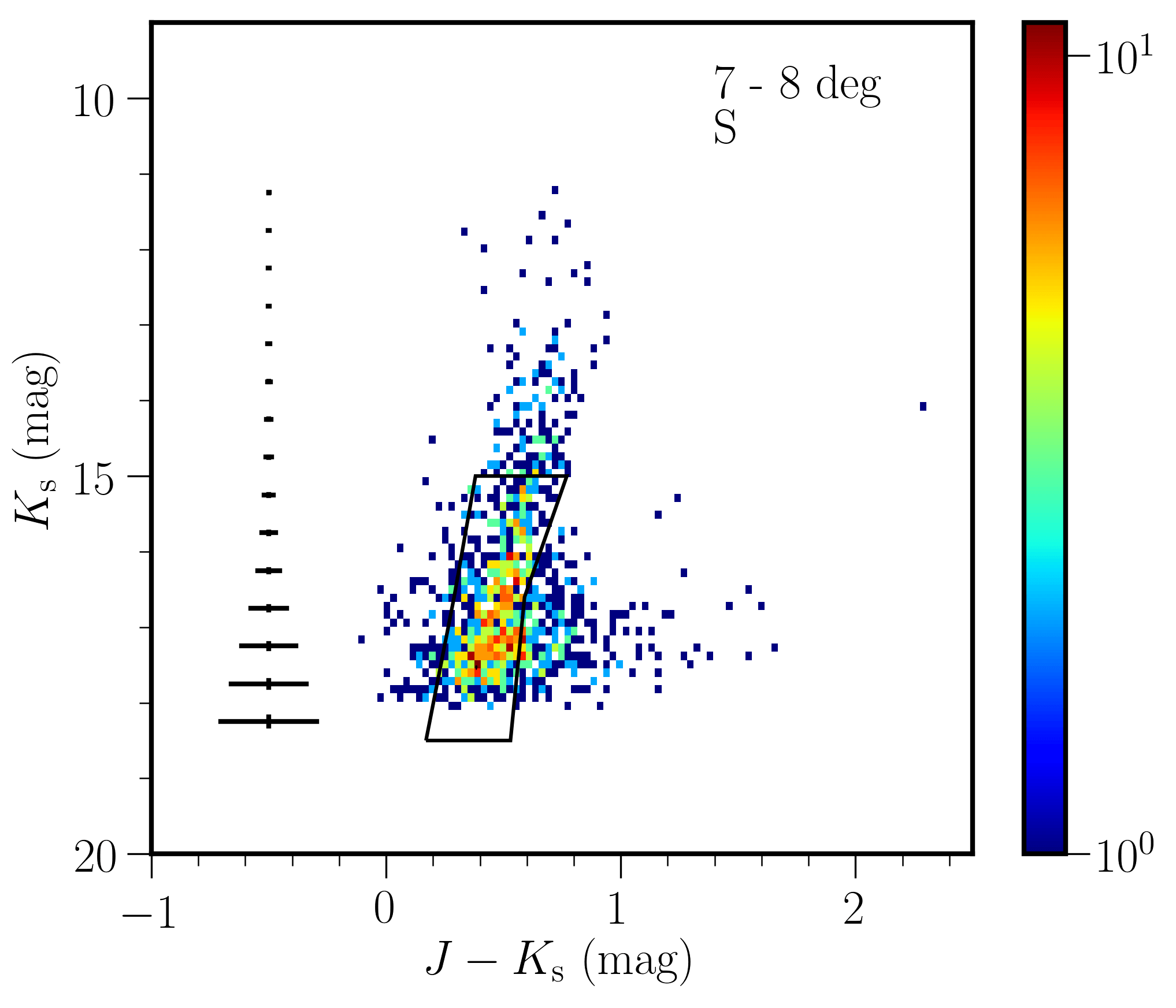}
	\includegraphics[scale=0.06]{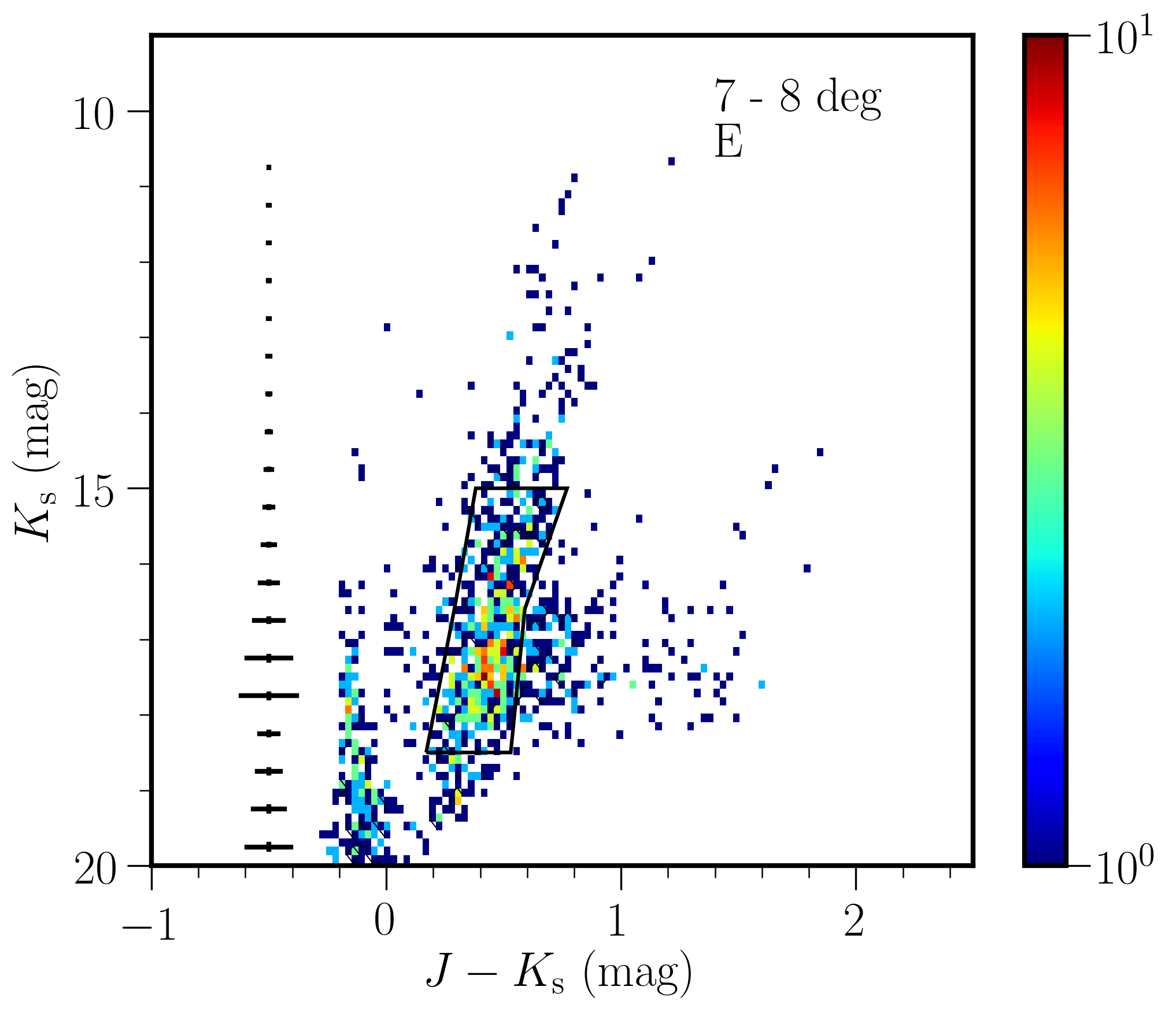}
	\includegraphics[scale=0.06]{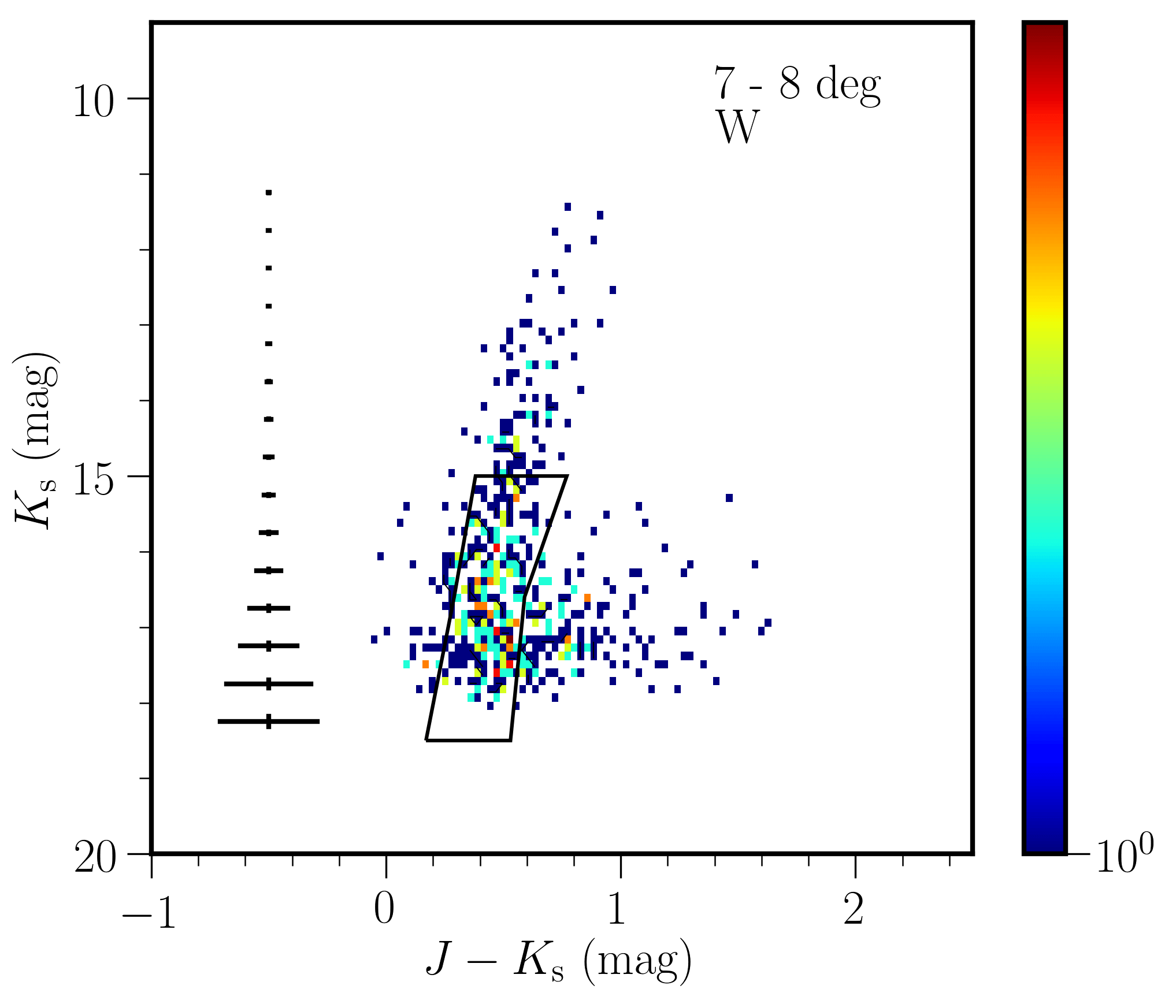}
	\includegraphics[scale=0.06]{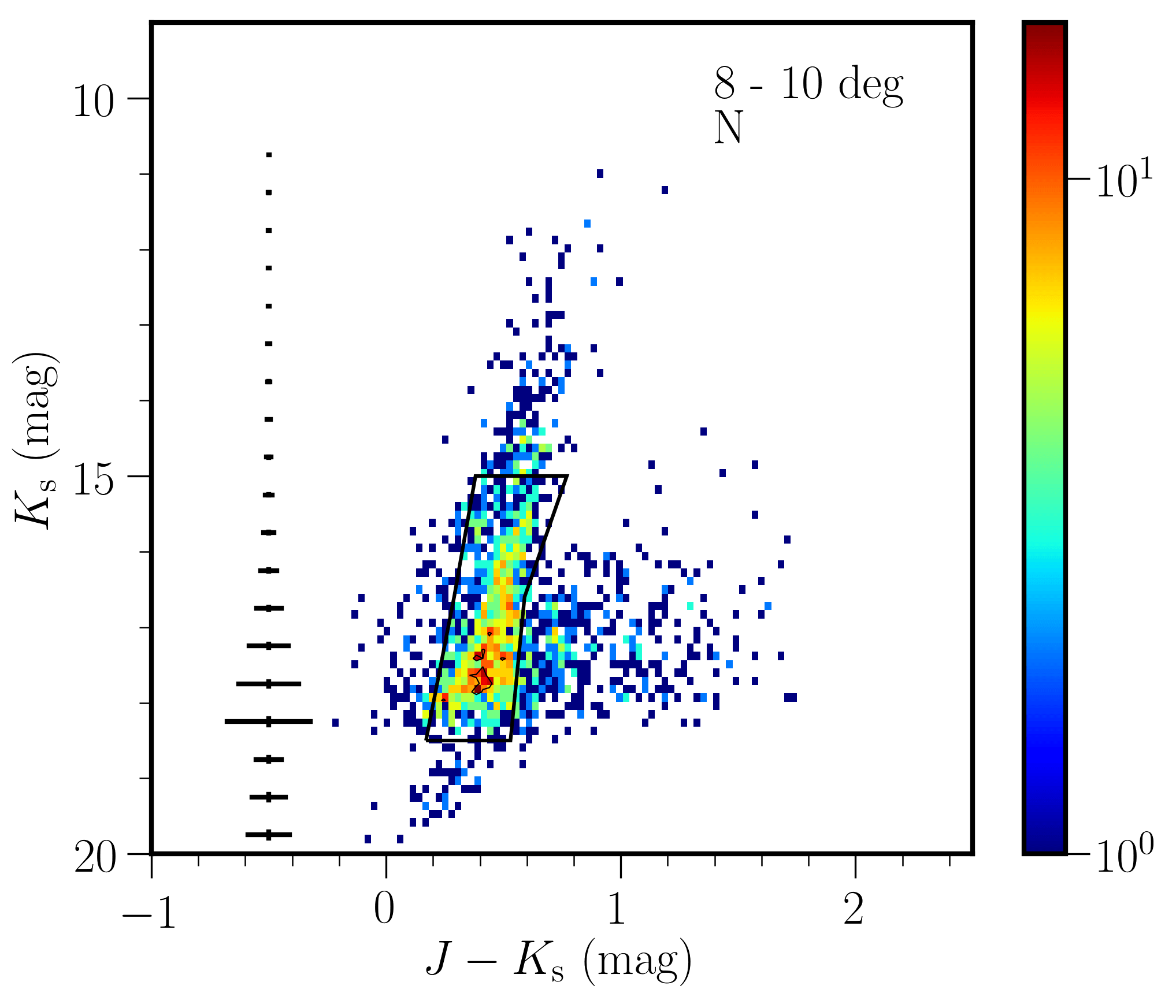}
	\includegraphics[scale=0.06]{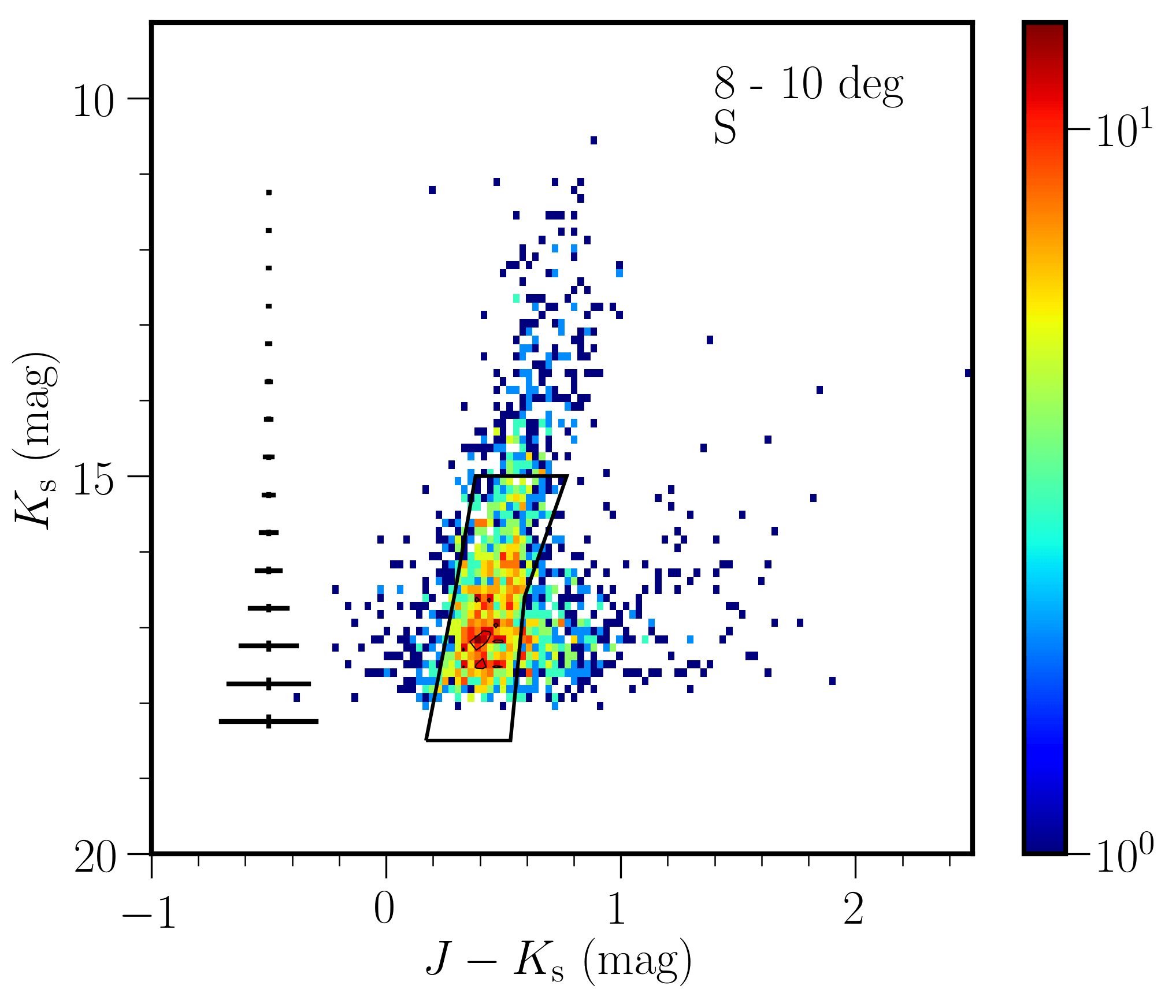}
	\includegraphics[scale=0.06]{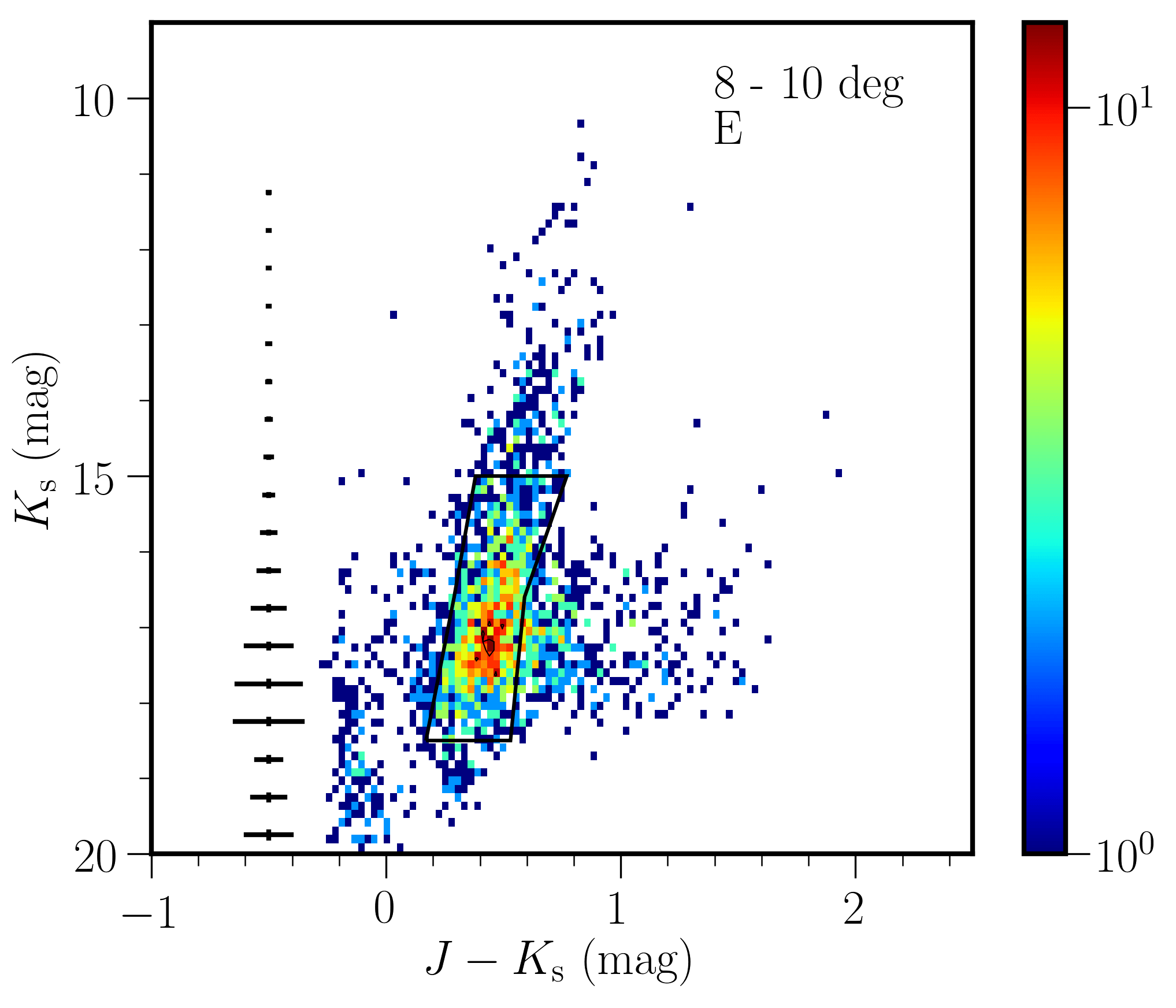}
	\includegraphics[scale=0.06]{VHS-SMC-CMD-R-810-W_der_new_box.png}

	\caption{(continued)}	
\end{figure*}

\begin{figure*}	
	\centering
	
	\includegraphics[scale=0.05]{Hist_RC_0_1deg_N_der.png}
	\includegraphics[scale=0.05]{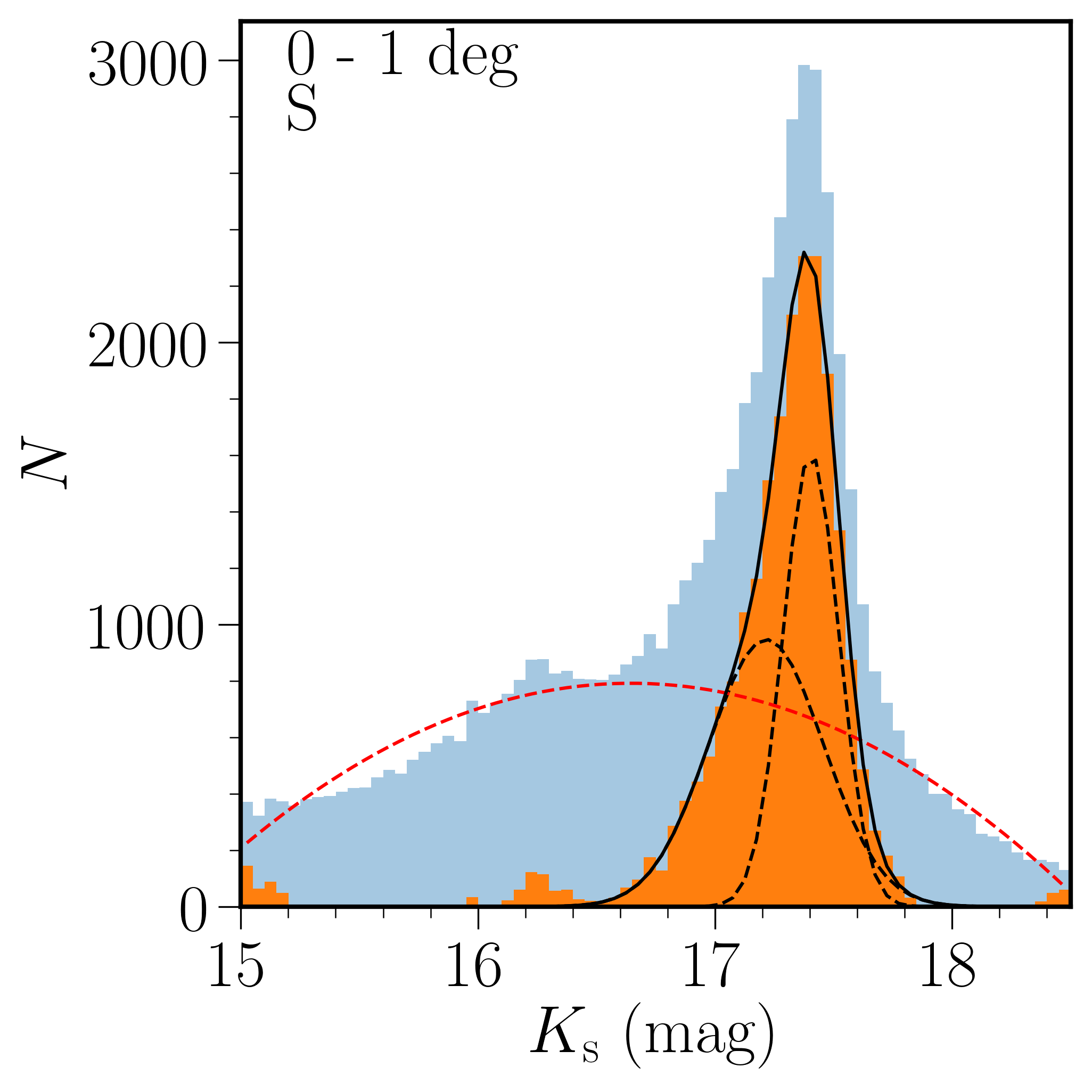}
	\includegraphics[scale=0.05]{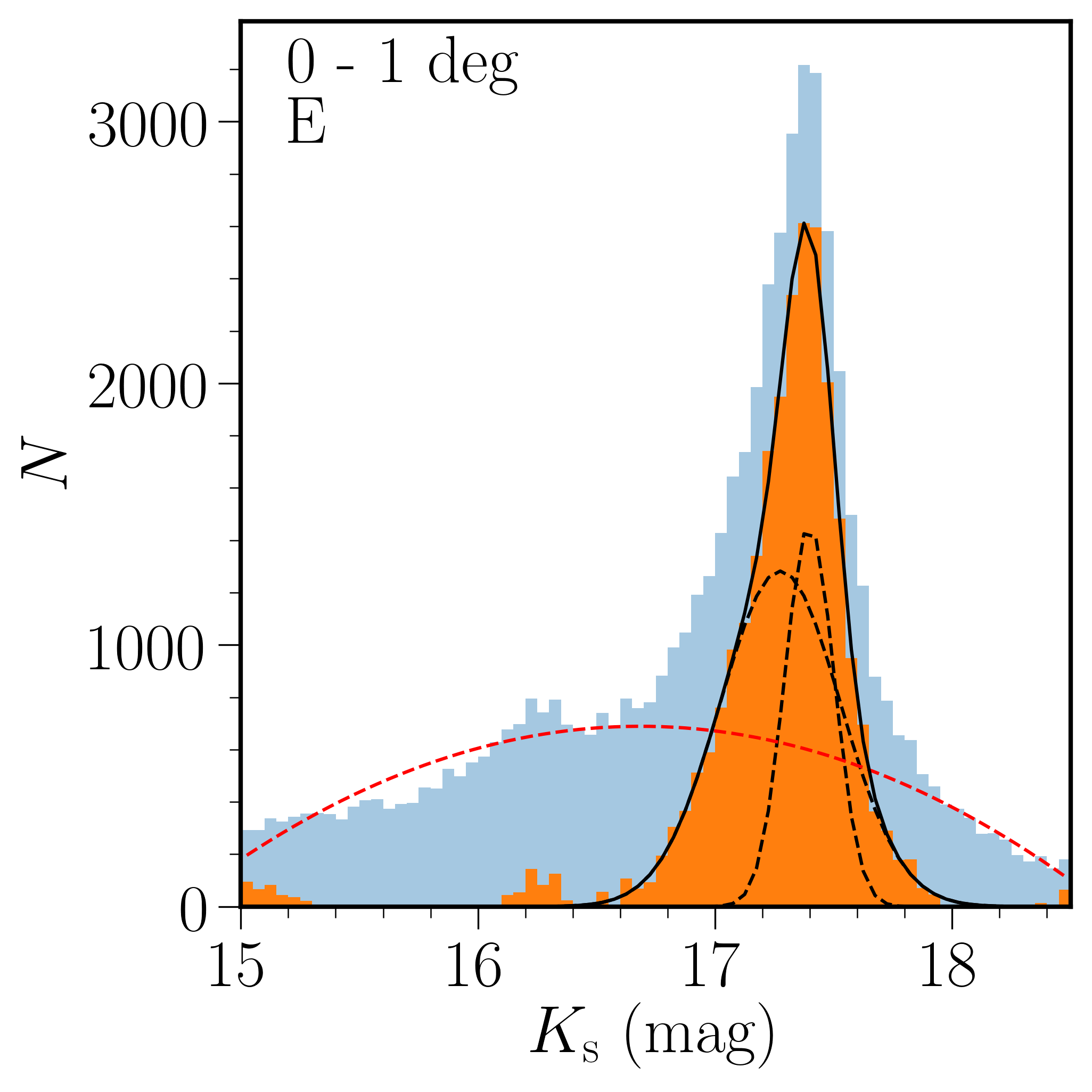}
	\includegraphics[scale=0.05]{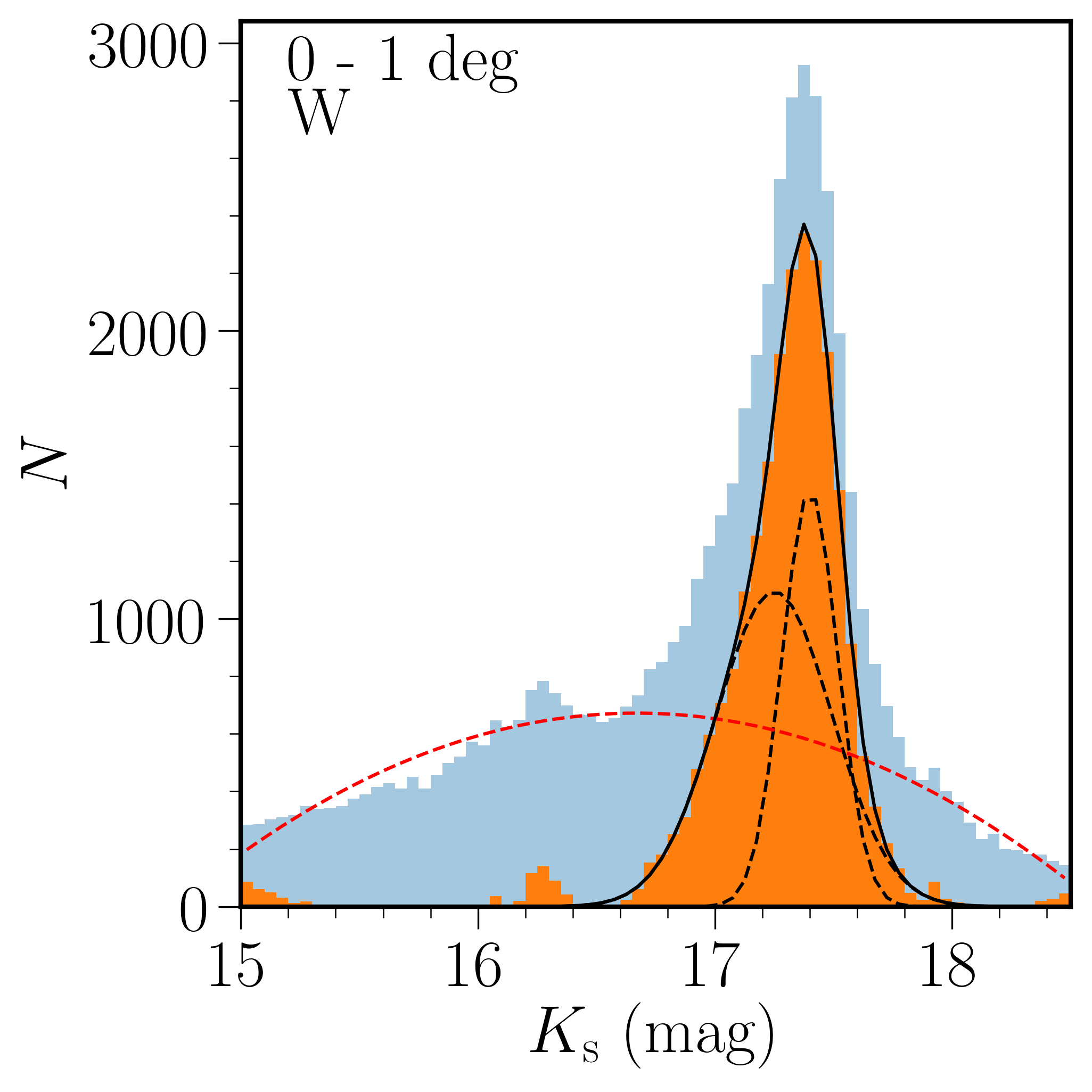}
	
	\includegraphics[scale=0.05]{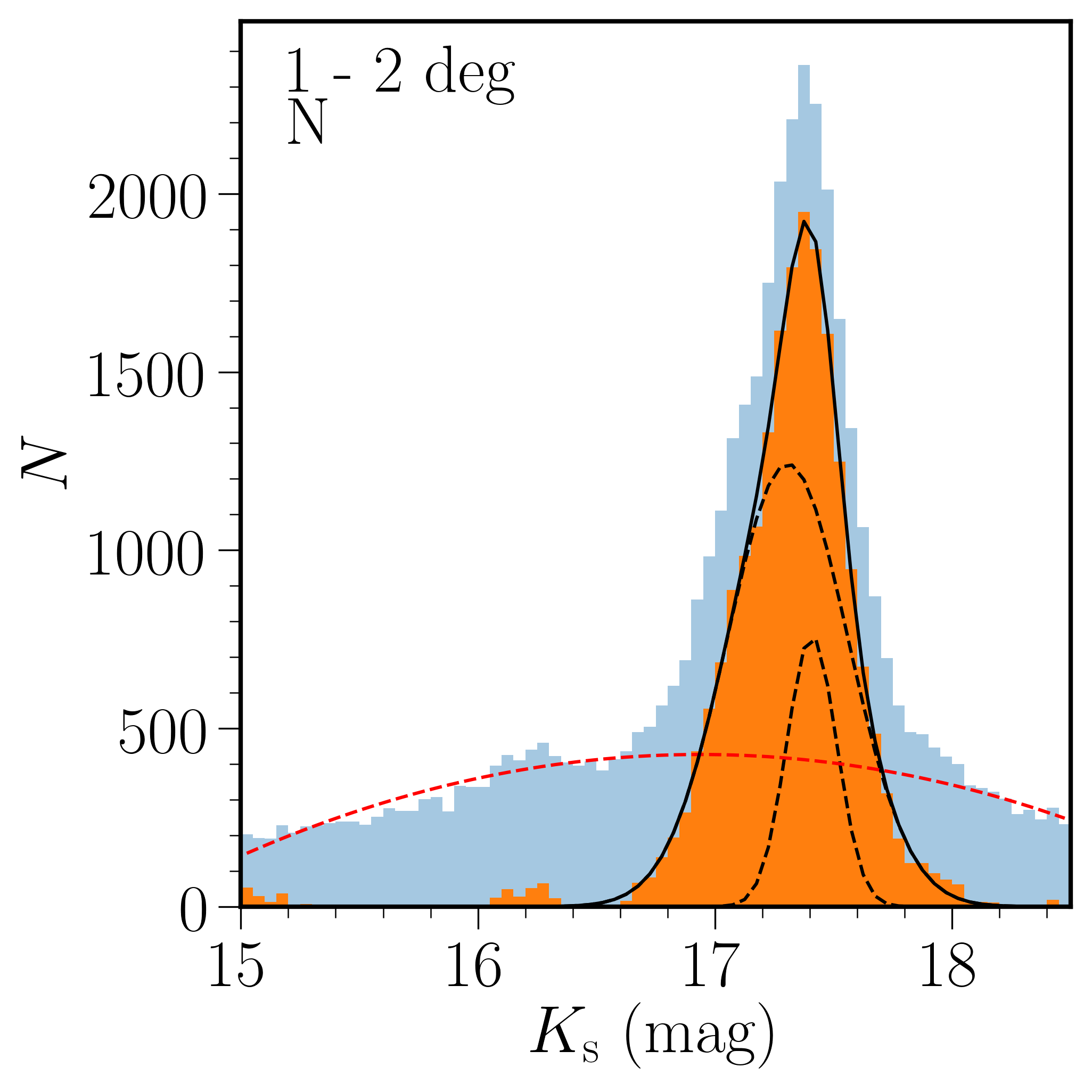}
	\includegraphics[scale=0.05]{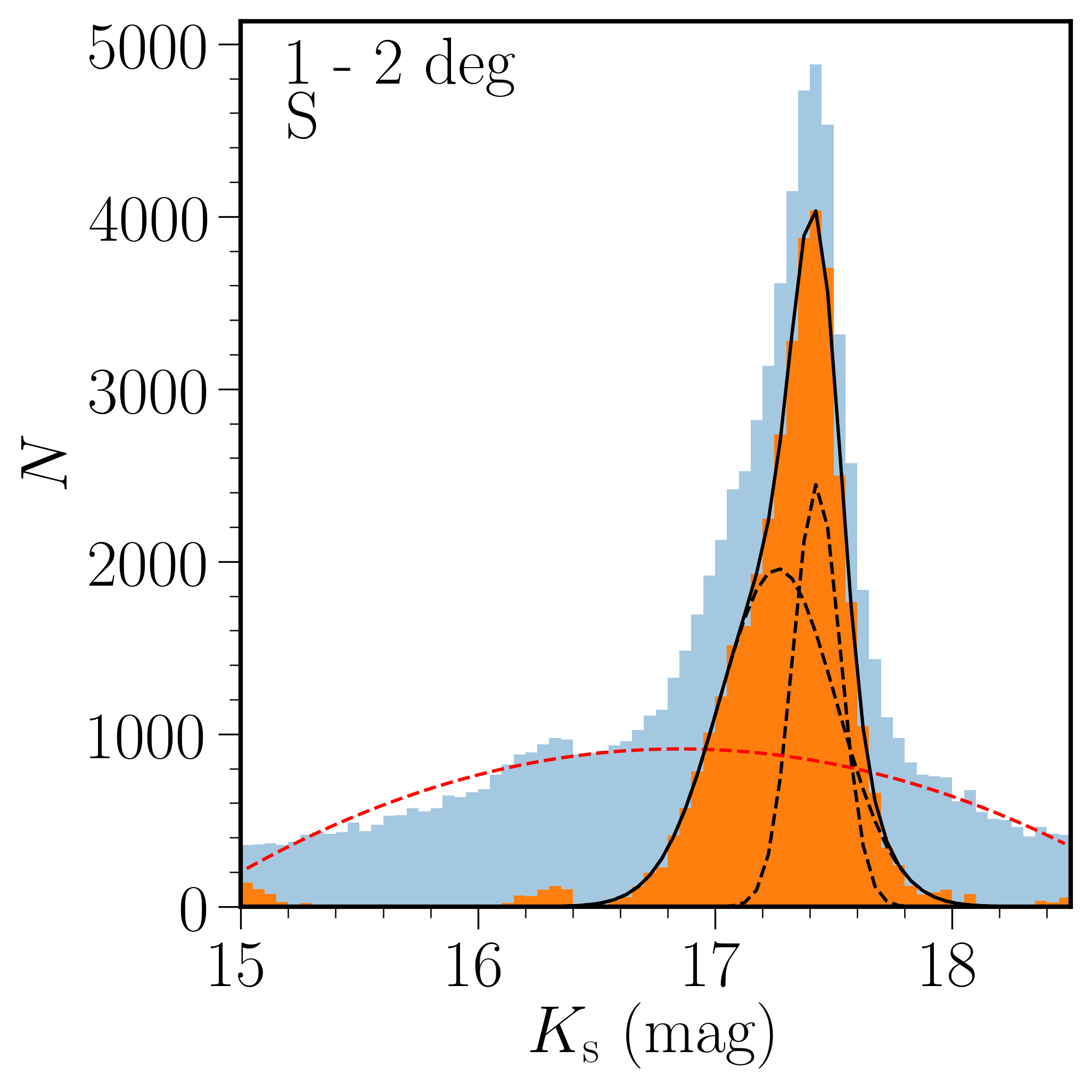}
	\includegraphics[scale=0.05]{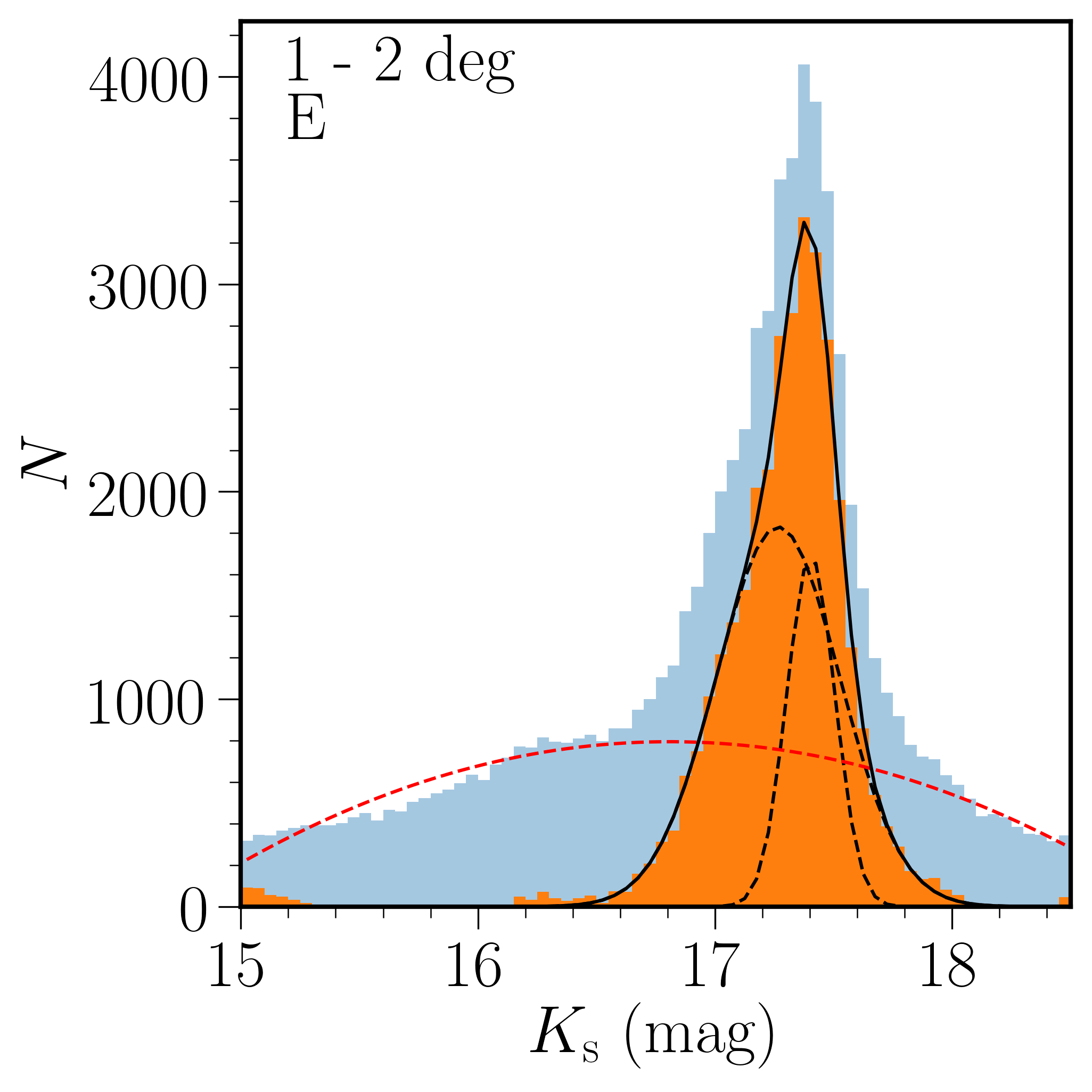}
	\includegraphics[scale=0.05]{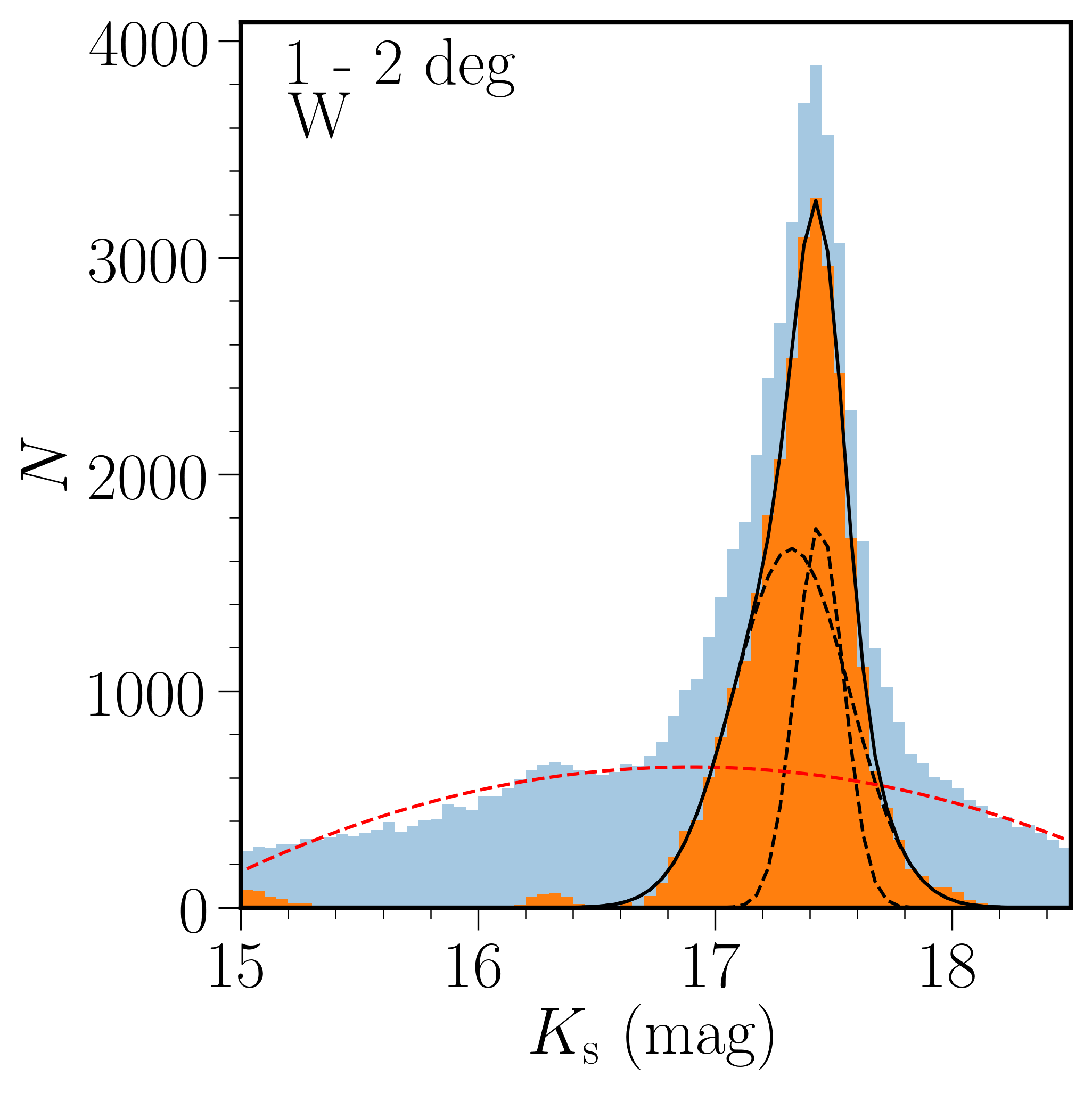}
	
	\includegraphics[scale=0.05]{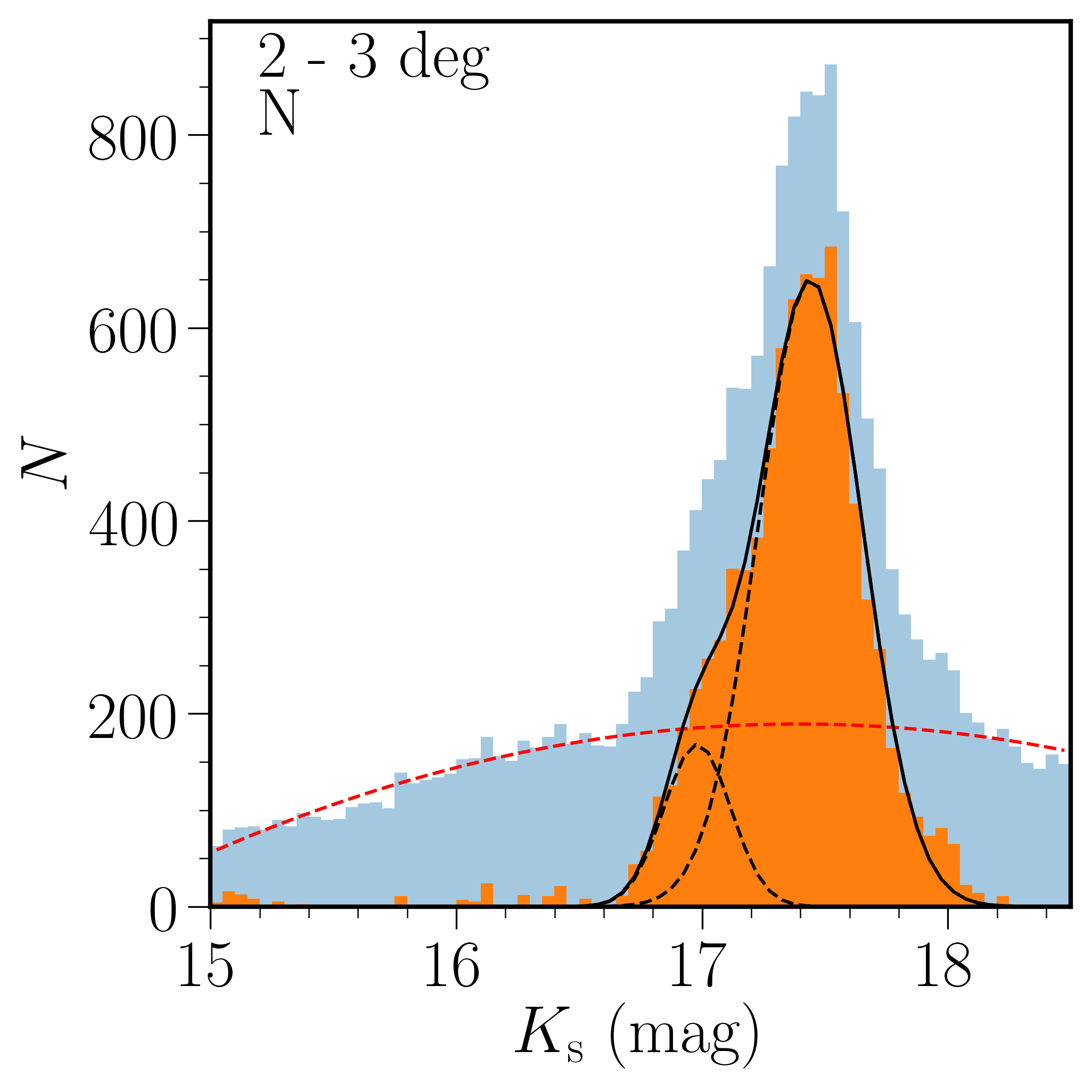}
	\includegraphics[scale=0.05]{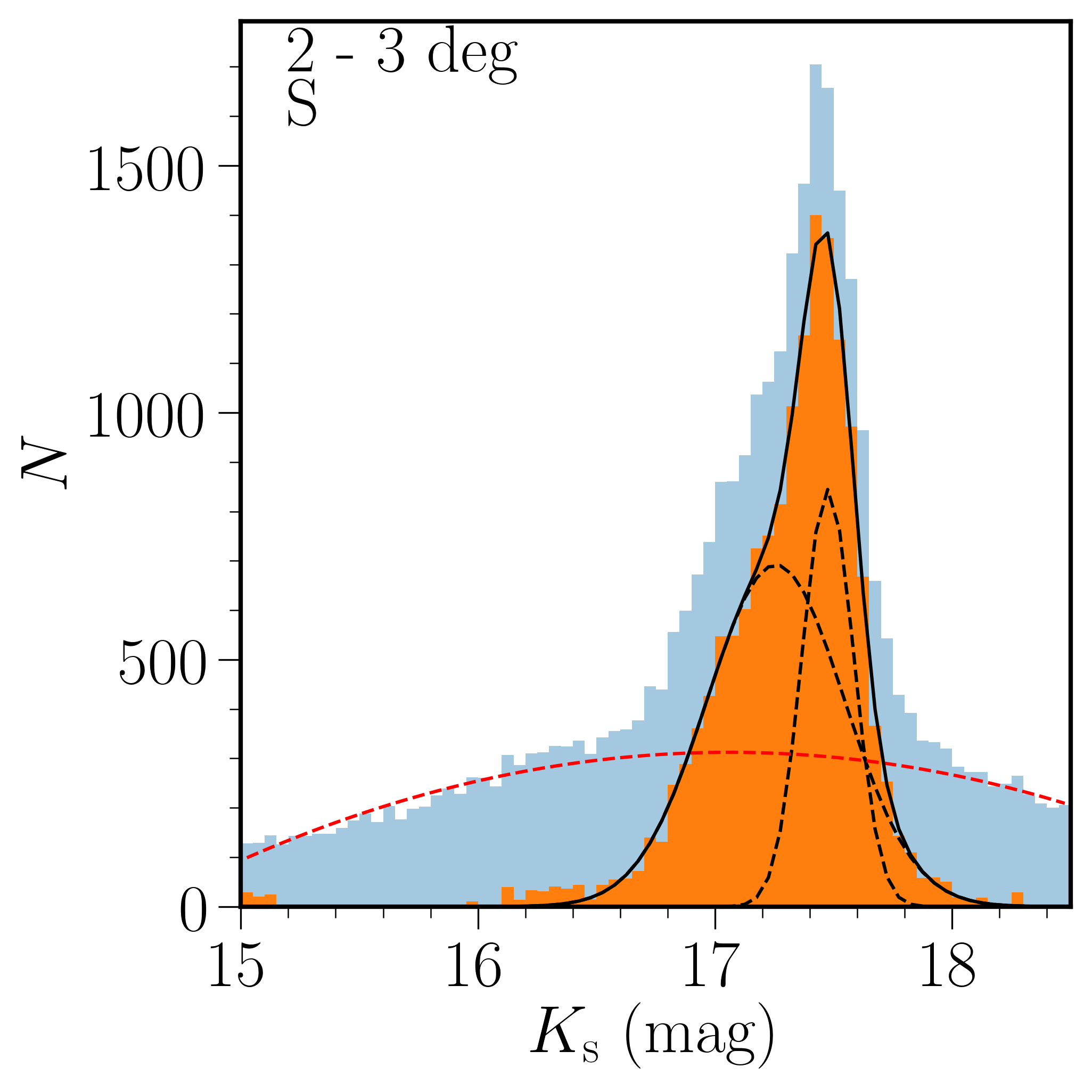}
	\includegraphics[scale=0.05]{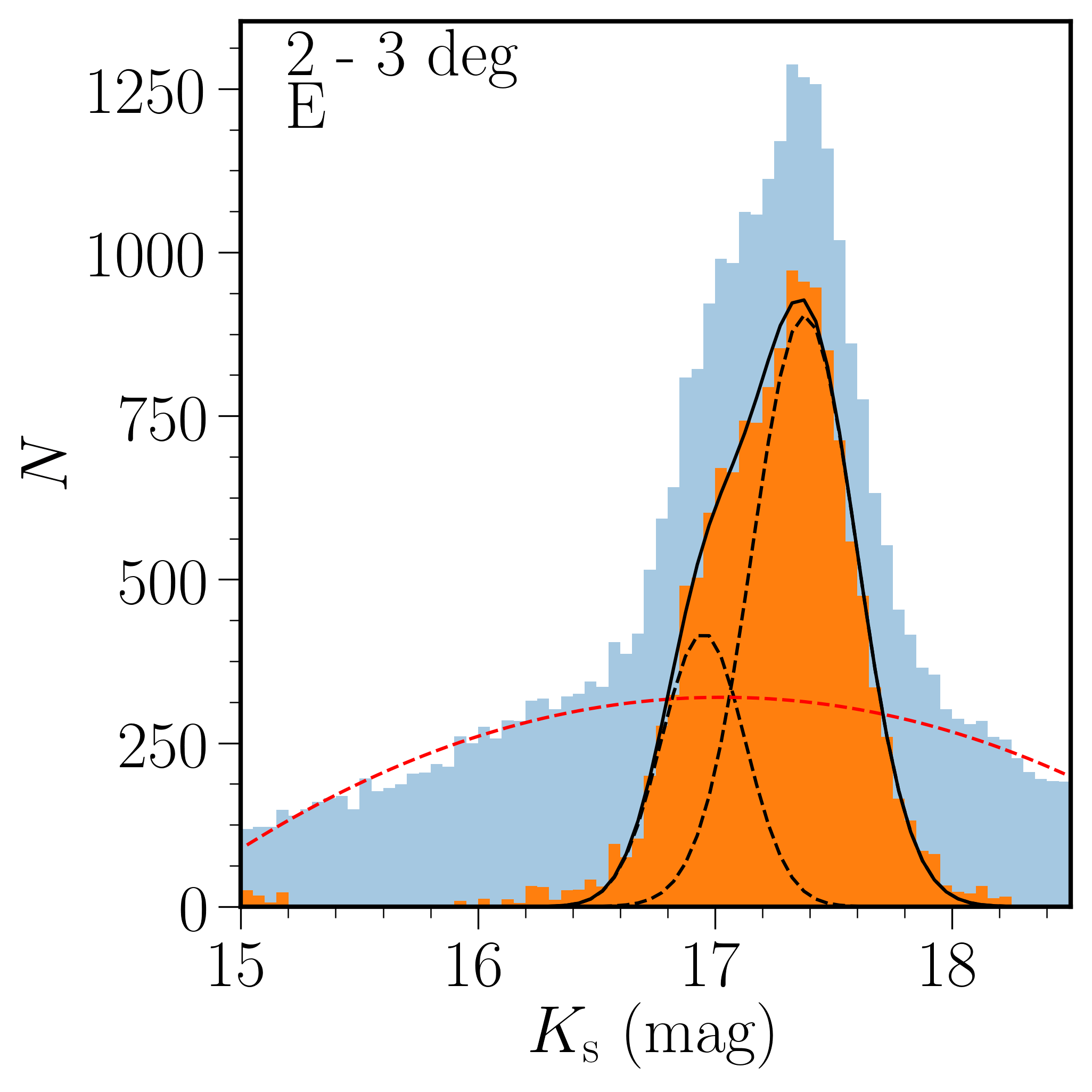}
	\includegraphics[scale=0.05]{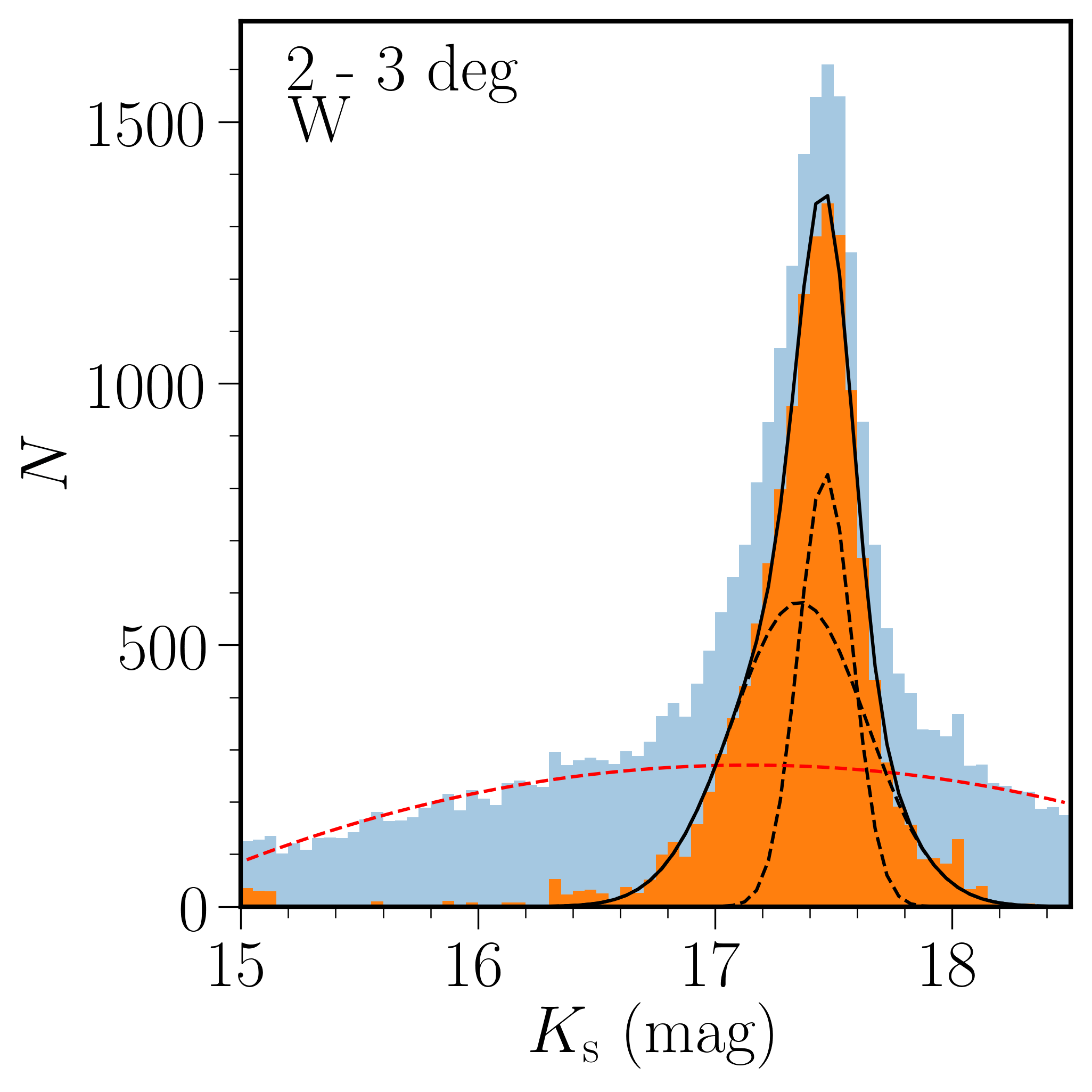}
	
	\includegraphics[scale=0.05]{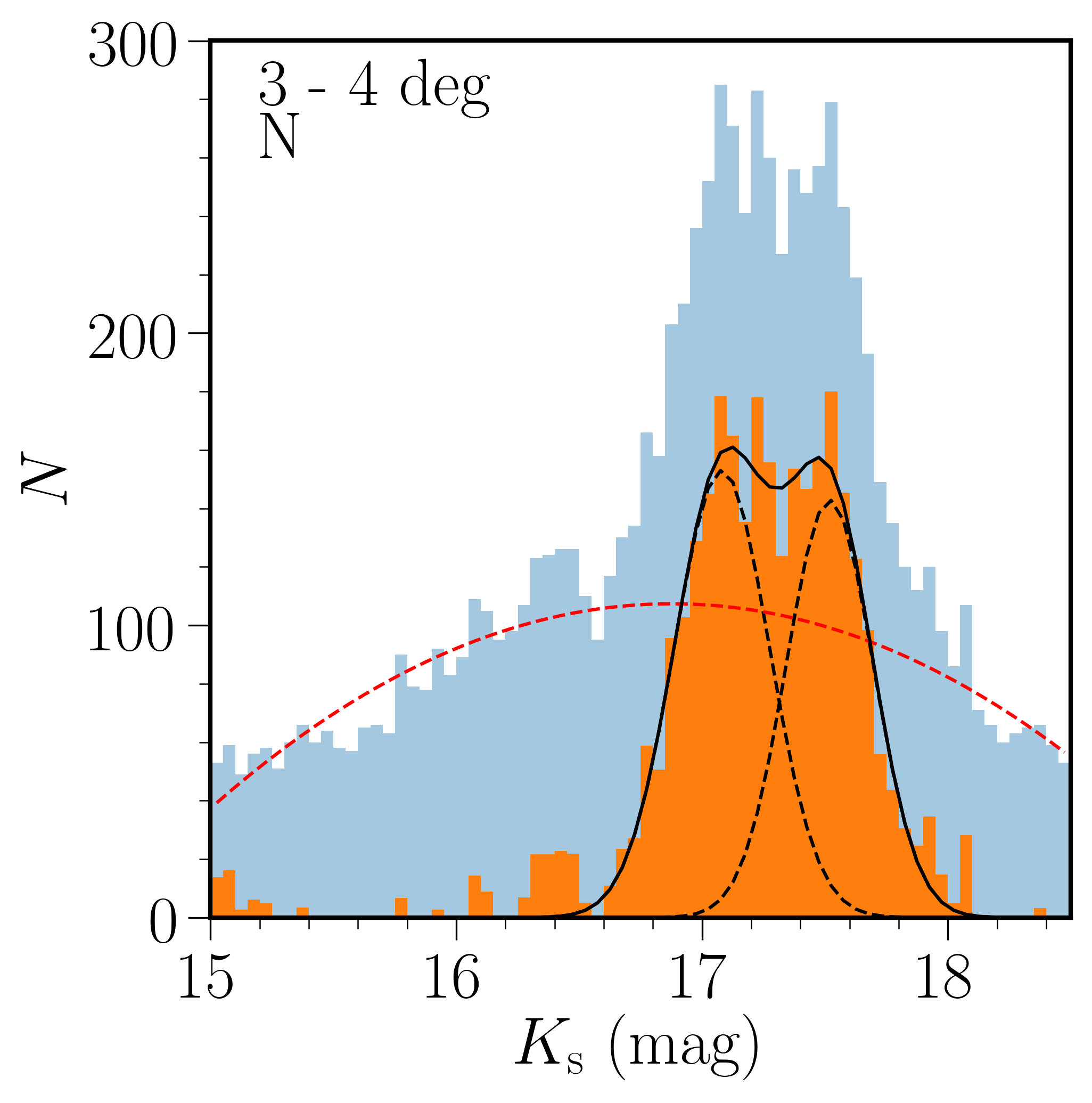}
	\includegraphics[scale=0.05]{Hist_RC_3_4deg_S_der.png}
	\includegraphics[scale=0.05]{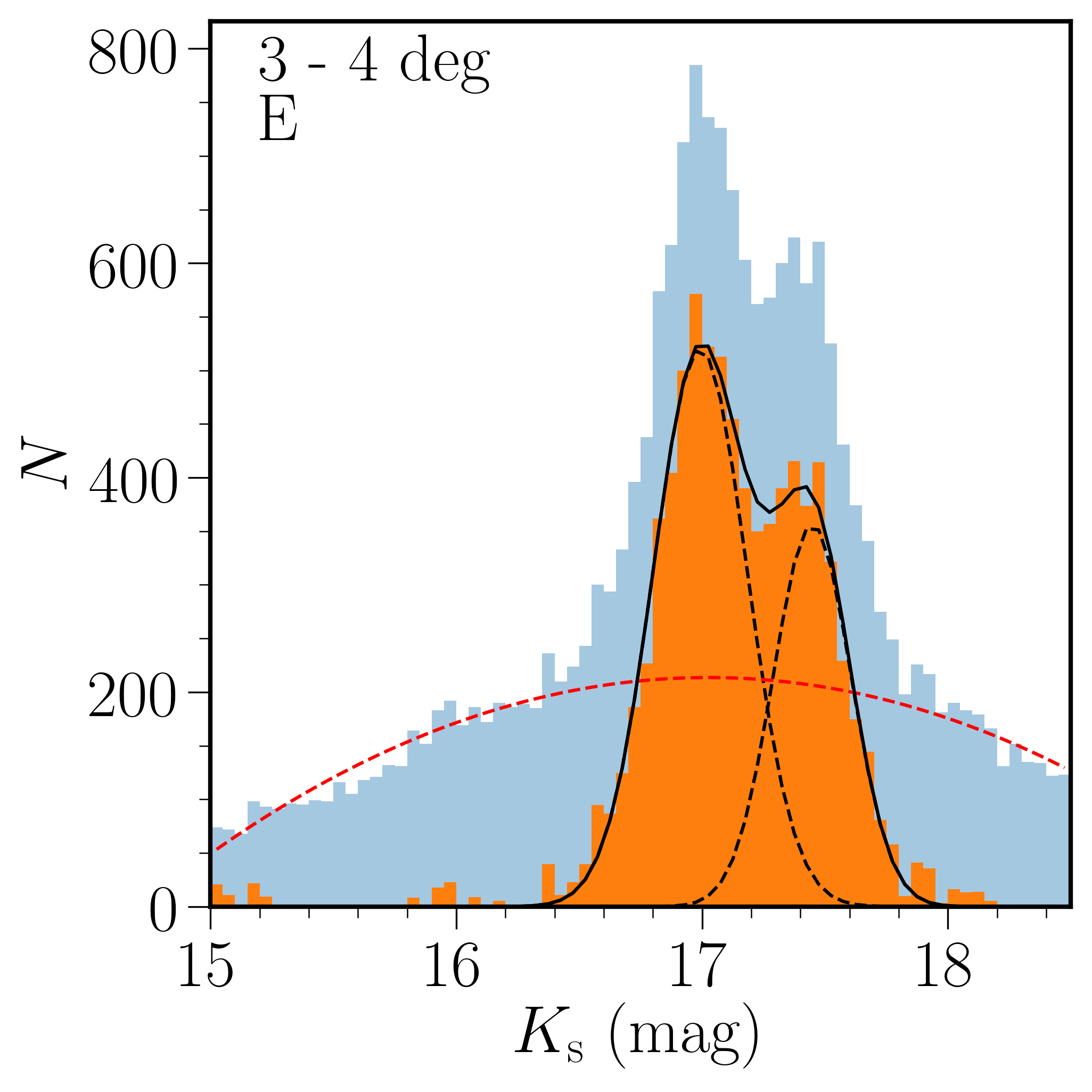}
	\includegraphics[scale=0.05]{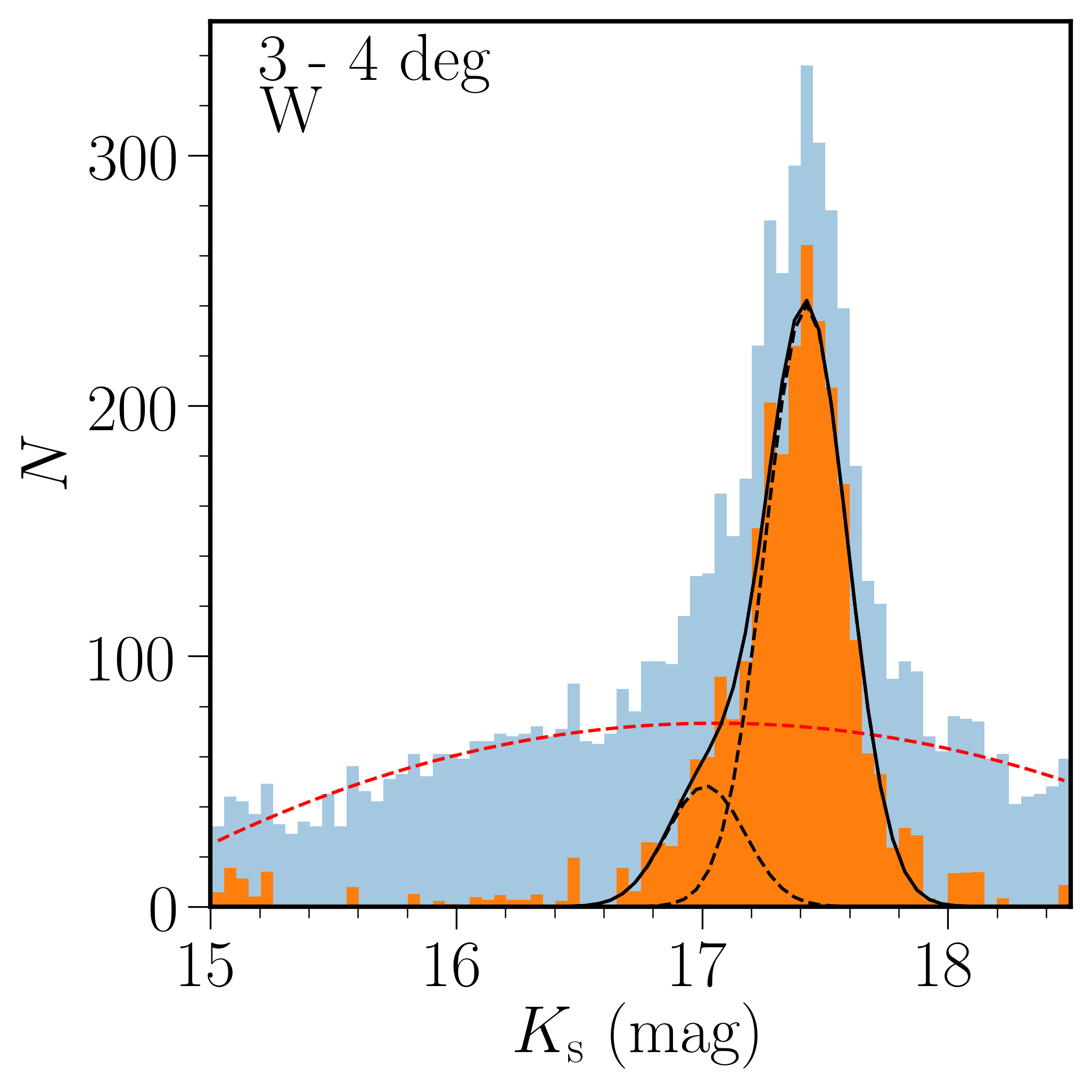}
	
	\includegraphics[scale=0.05]{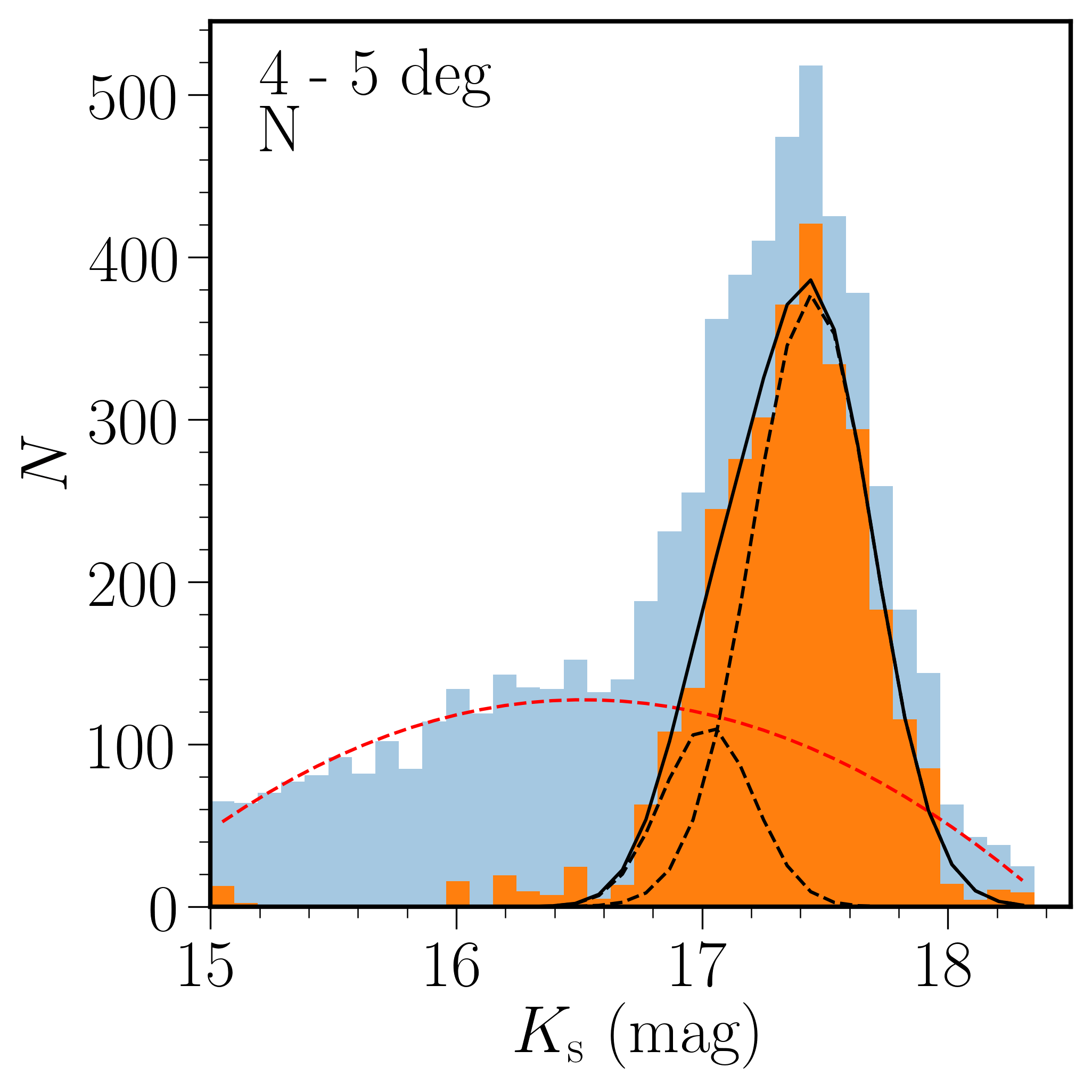}
	\includegraphics[scale=0.05]{Hist_RC_4_5deg_S_der.png}
	\includegraphics[scale=0.05]{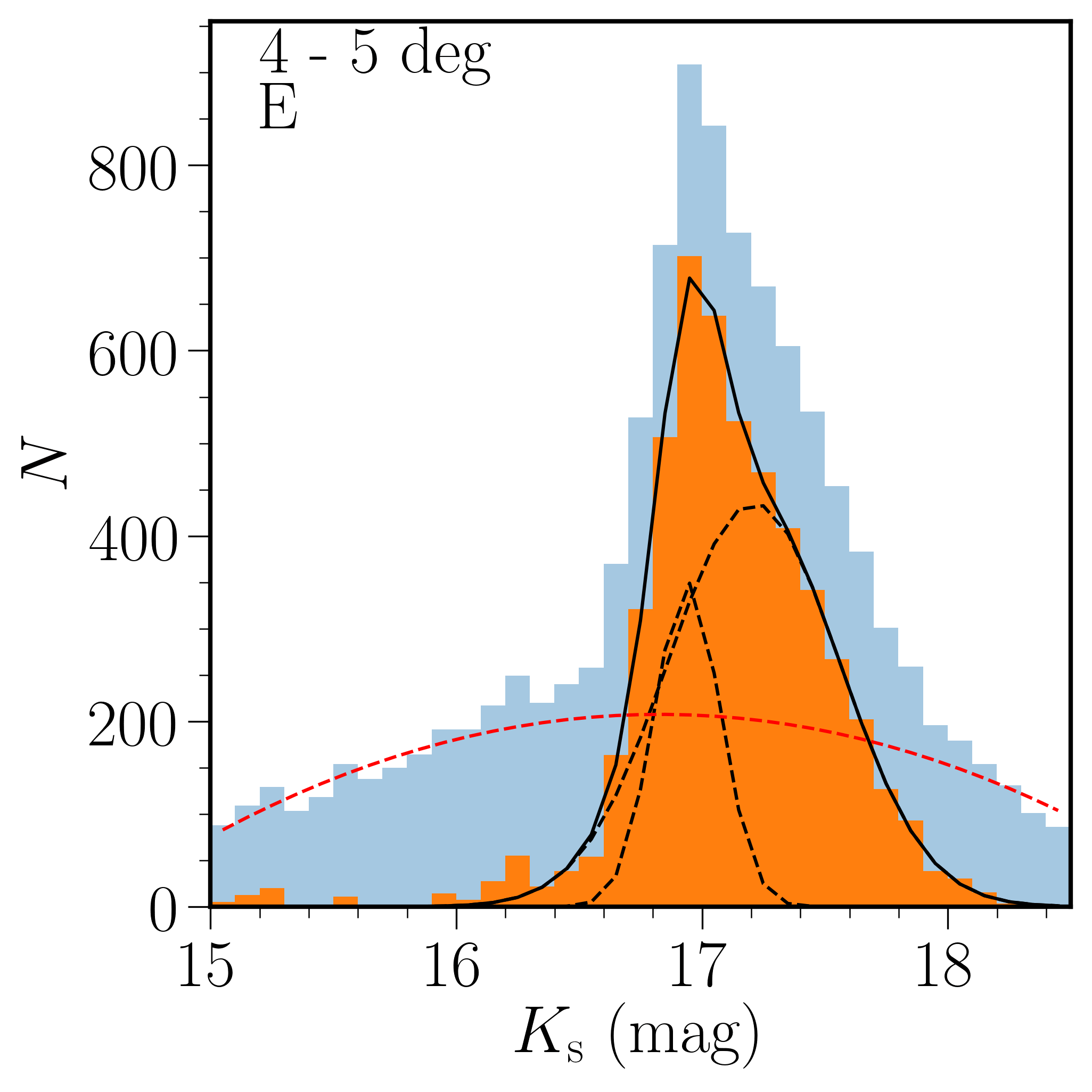}
	\includegraphics[scale=0.05]{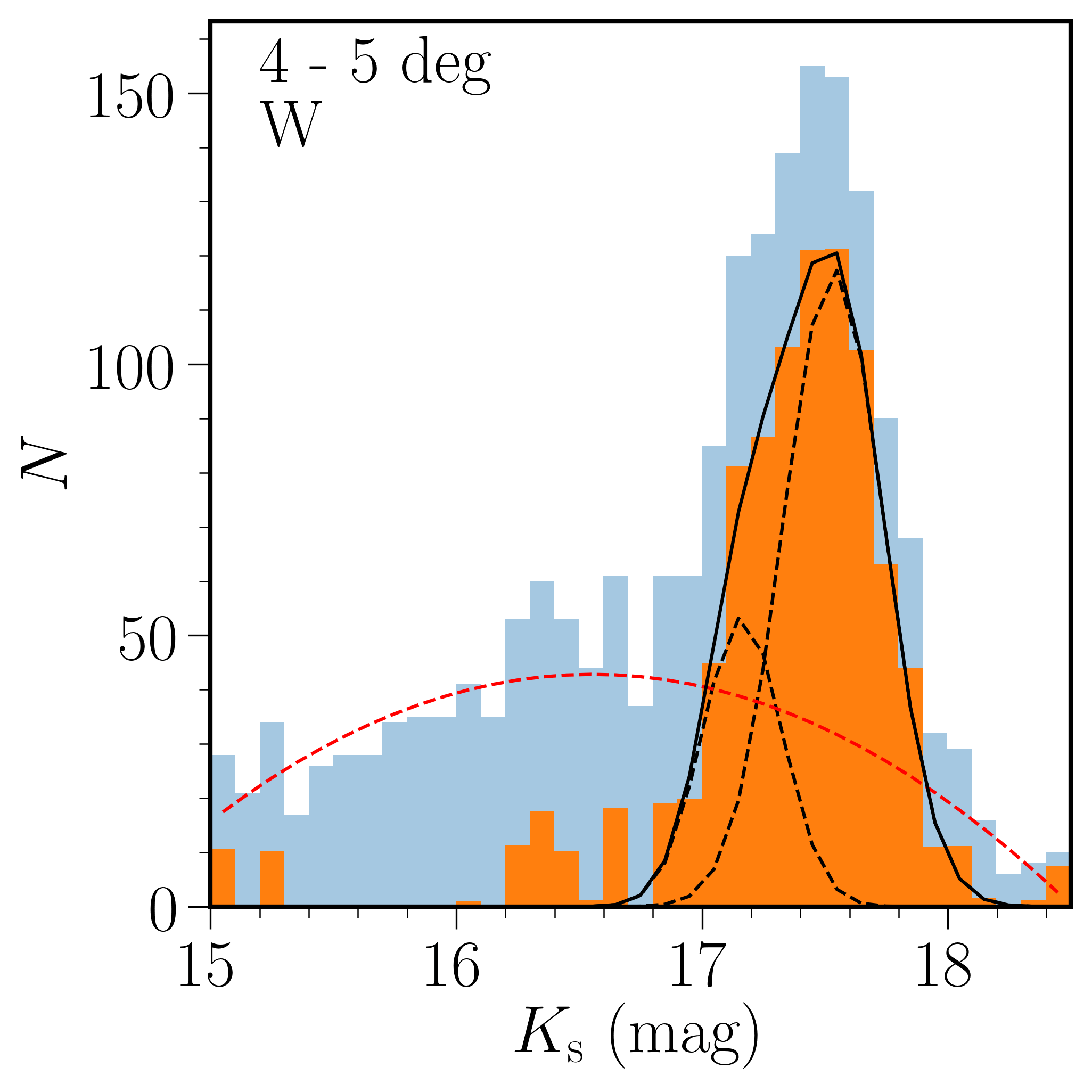}
	
	\caption{Luminosity function of the RC stars in different SMC regions. The blue histograms show the observed luminosity functions, whereas orange histograms show the distributions after subtracting the RGB components and the continuous lines show the total fits to these distributions while the dashed lines represent the separate components of the fits.}
	\label{fig:RCHIST0}	
\end{figure*}

\begin{figure*}
	\setcounter{figure}{1}

	\includegraphics[scale=0.05]{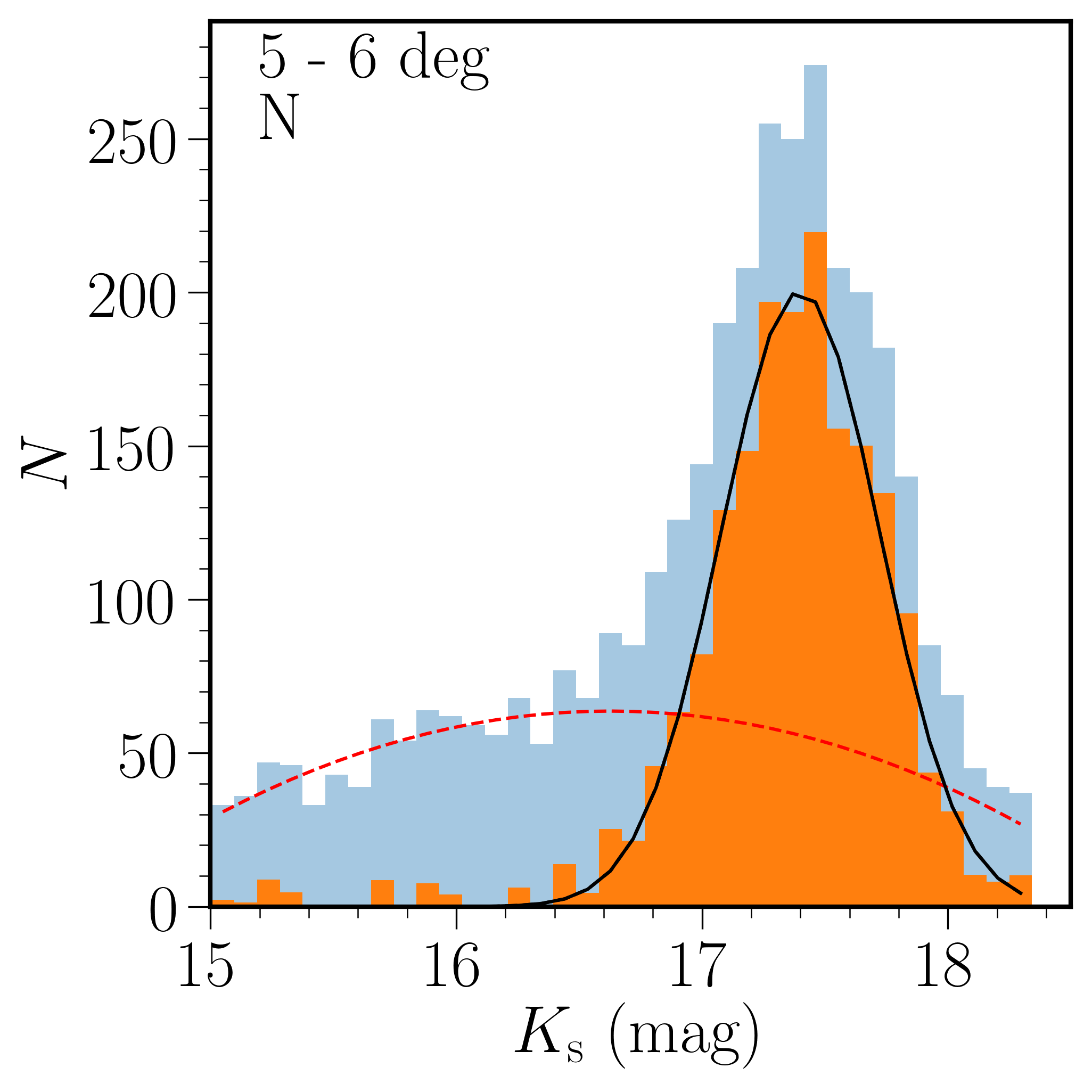}
	\includegraphics[scale=0.05]{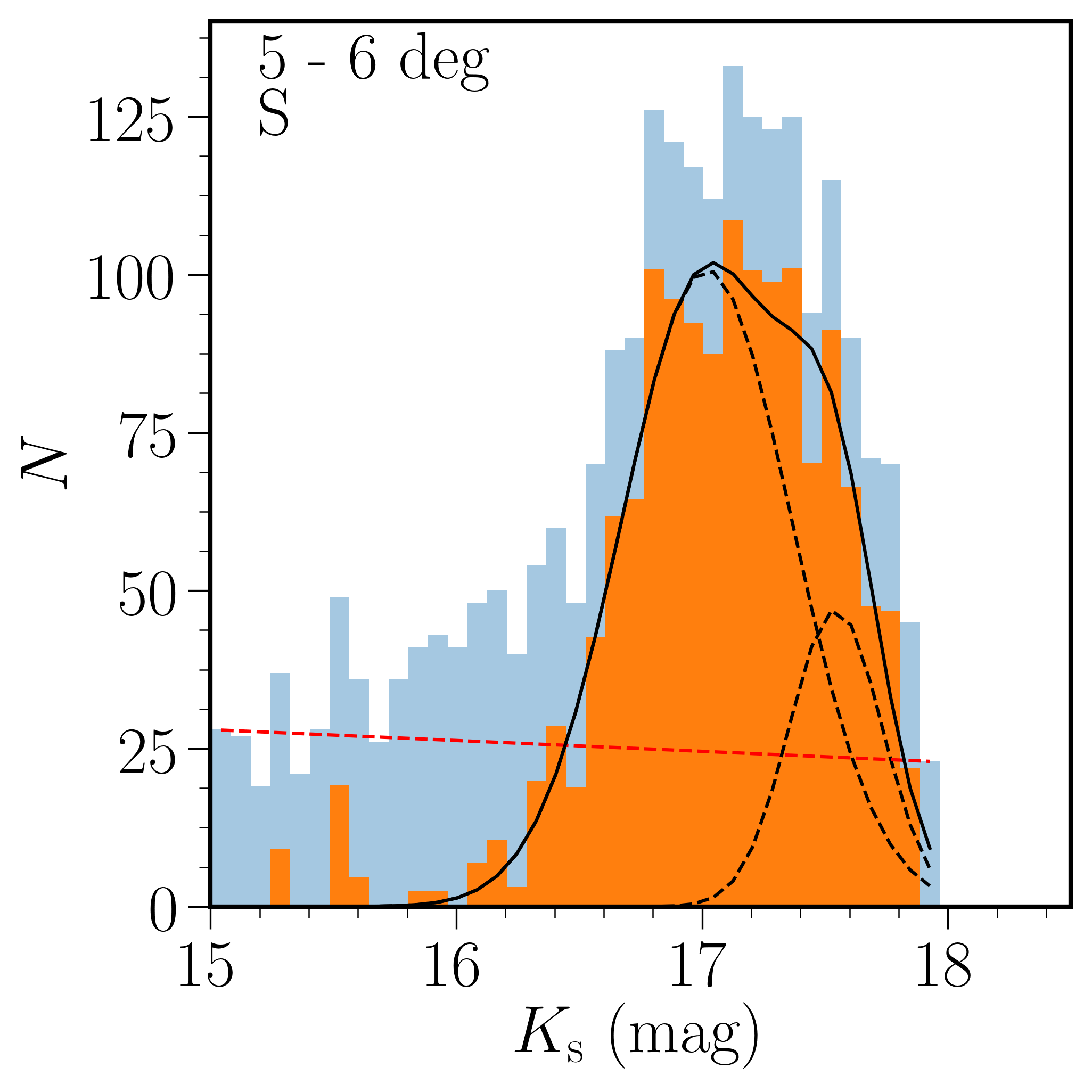}
	\includegraphics[scale=0.05]{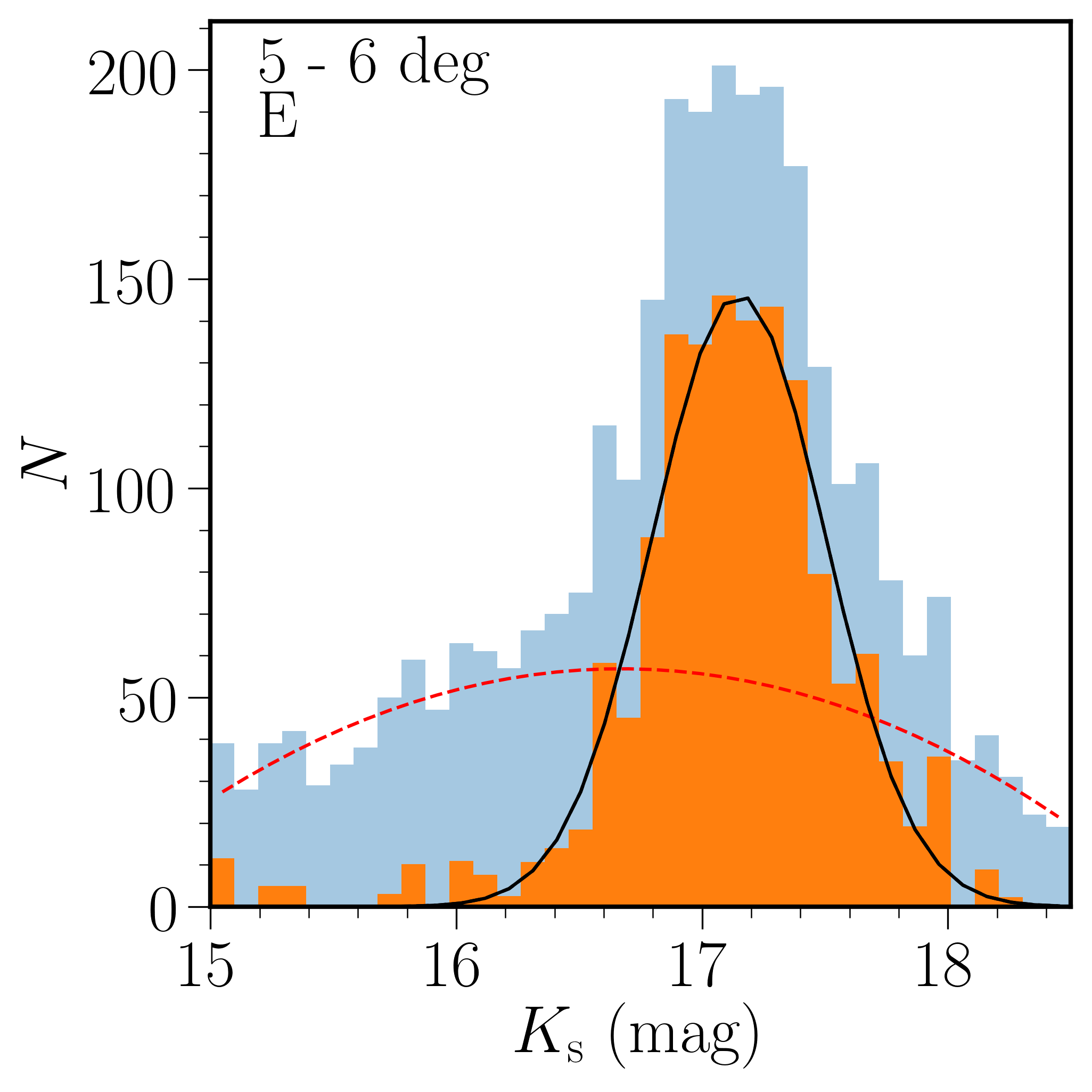}
	\includegraphics[scale=0.05]{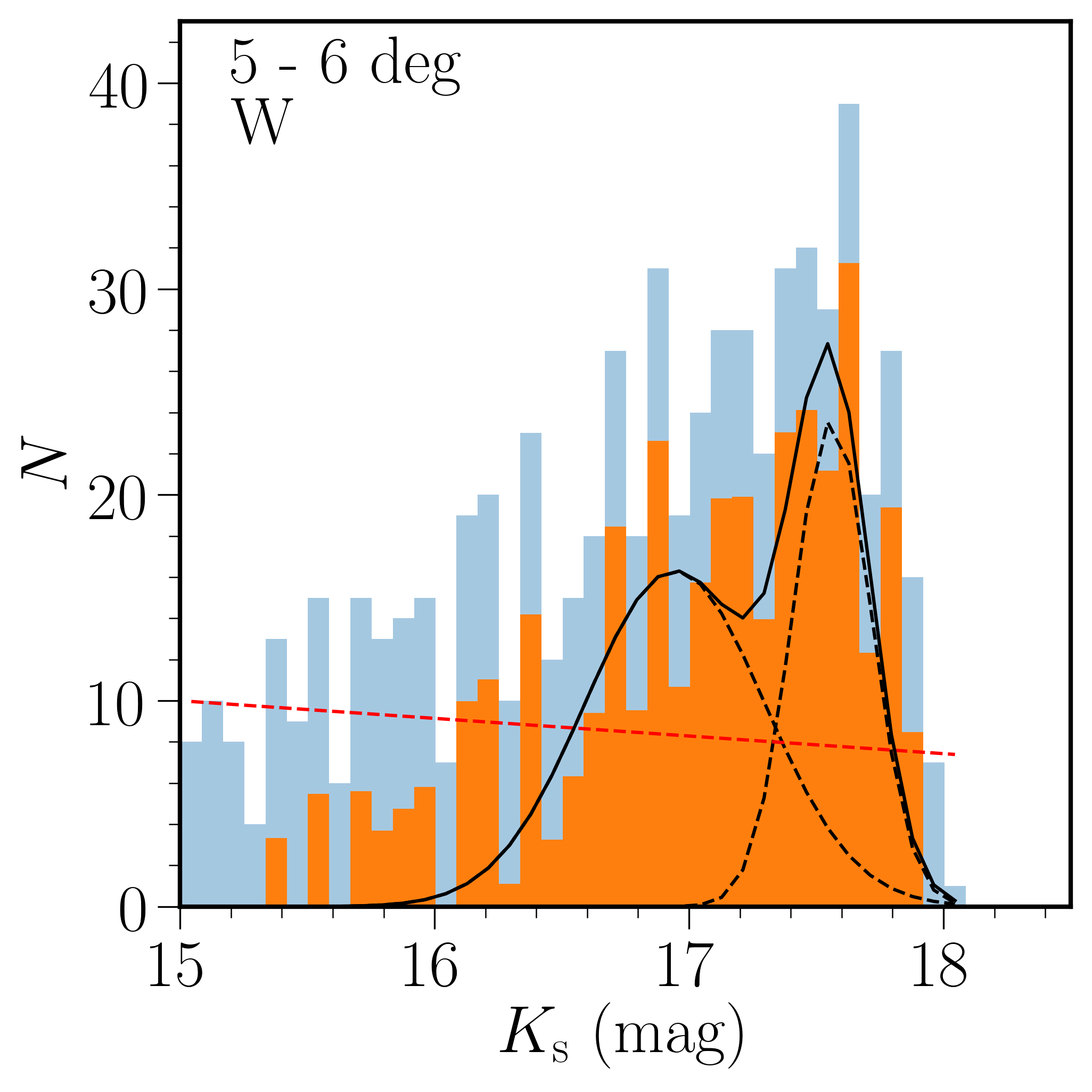}
	
	\includegraphics[scale=0.05]{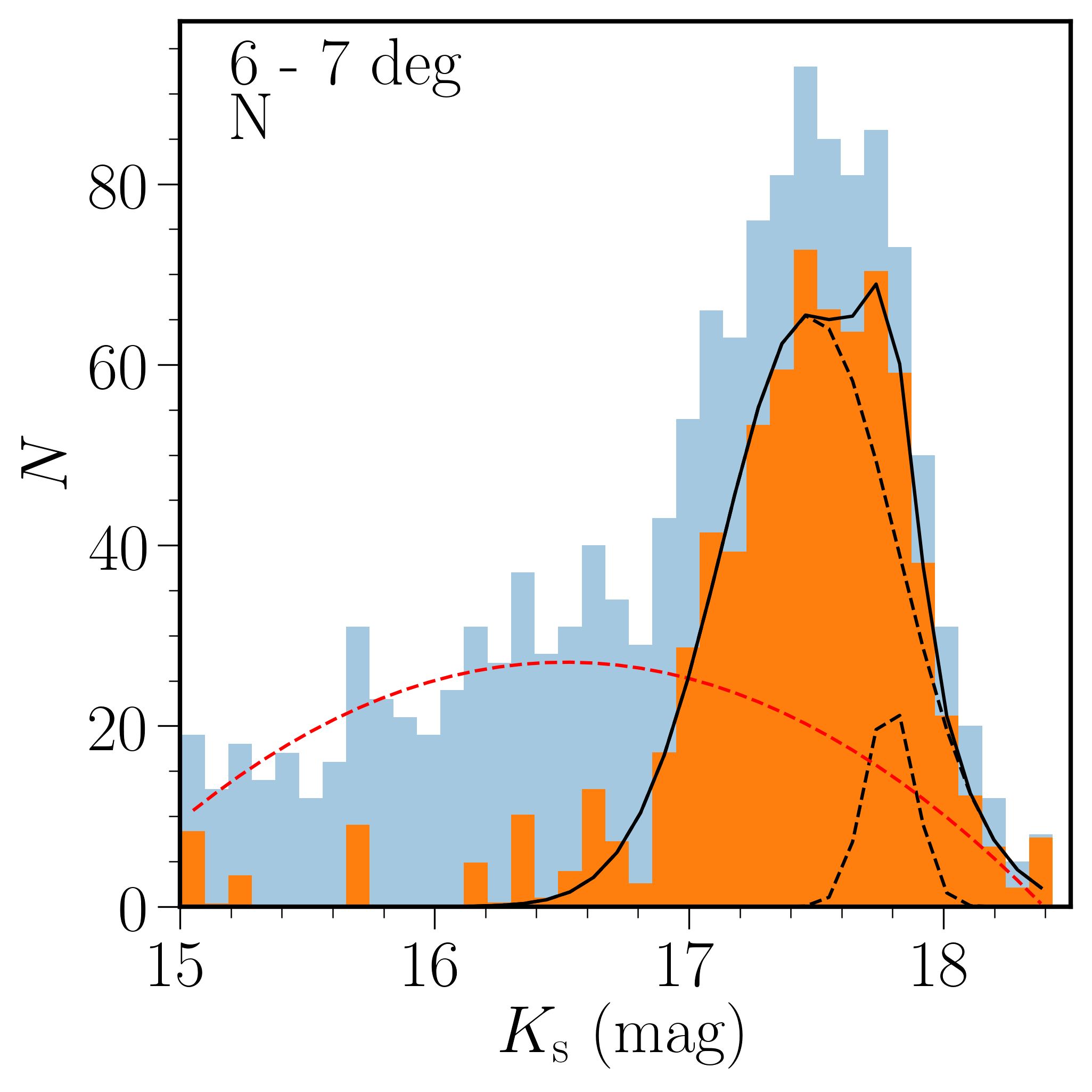}
	\includegraphics[scale=0.05]{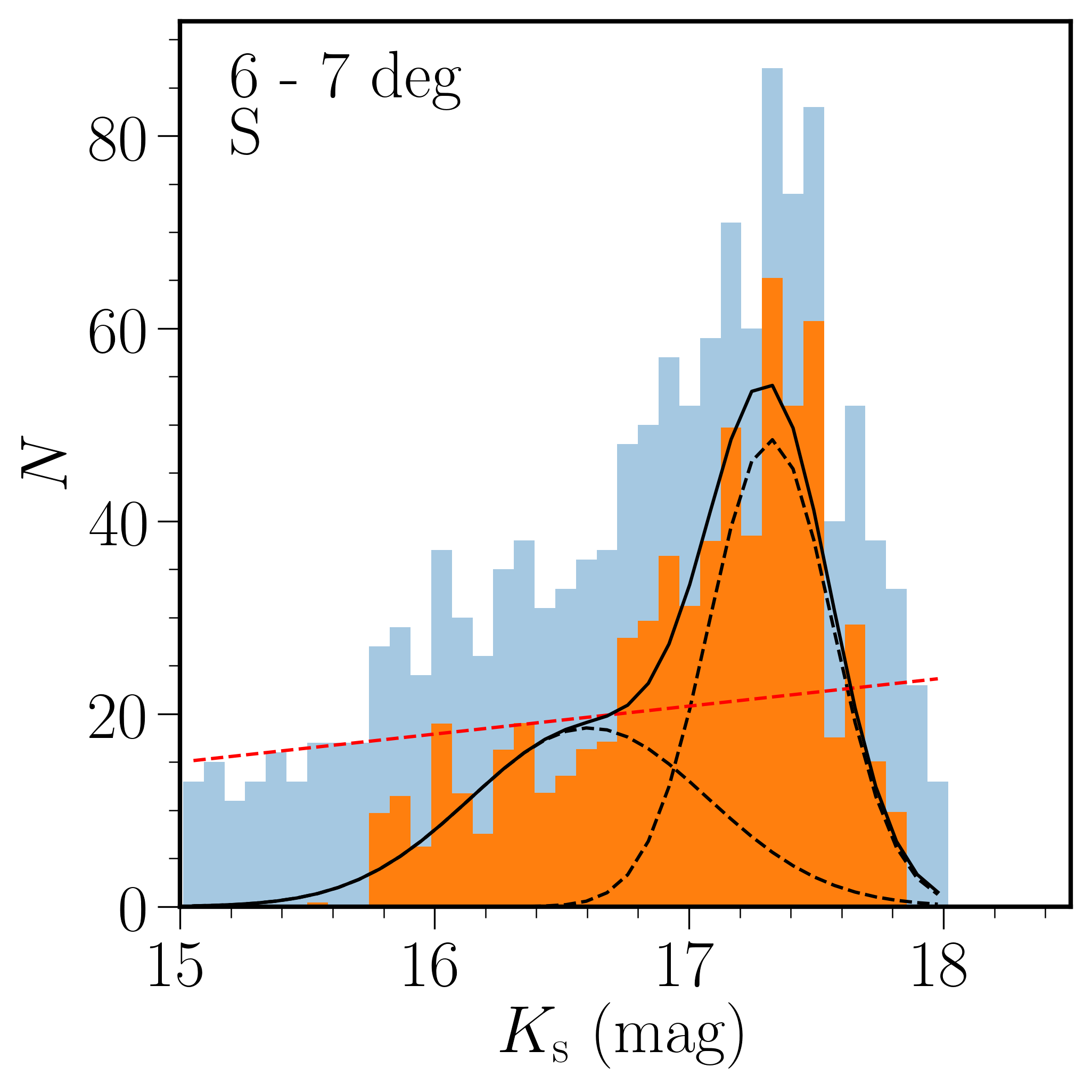}
	\includegraphics[scale=0.05]{Hist_RC_6_7deg_E_der.png}
	\includegraphics[scale=0.05]{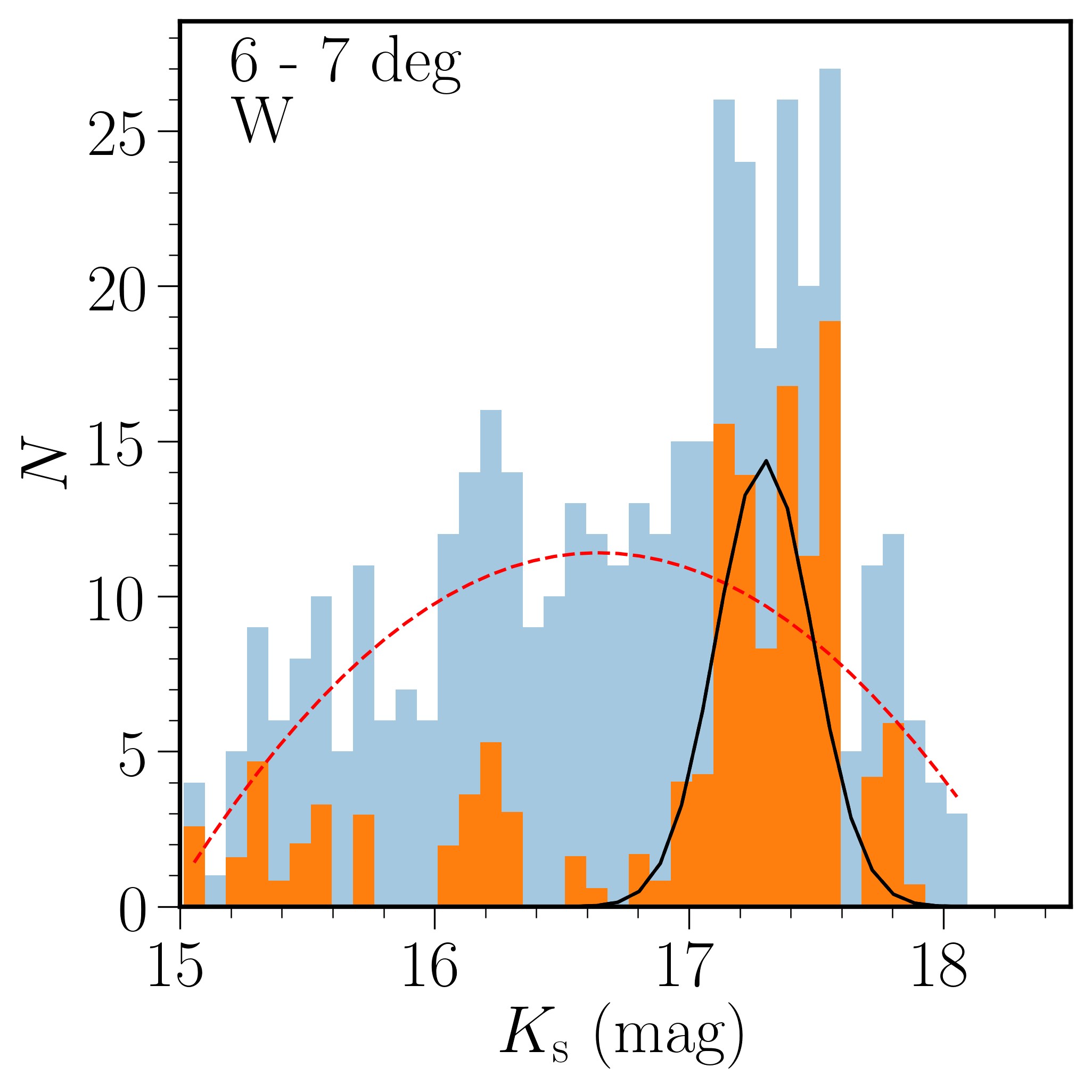}
	
	\includegraphics[scale=0.05]{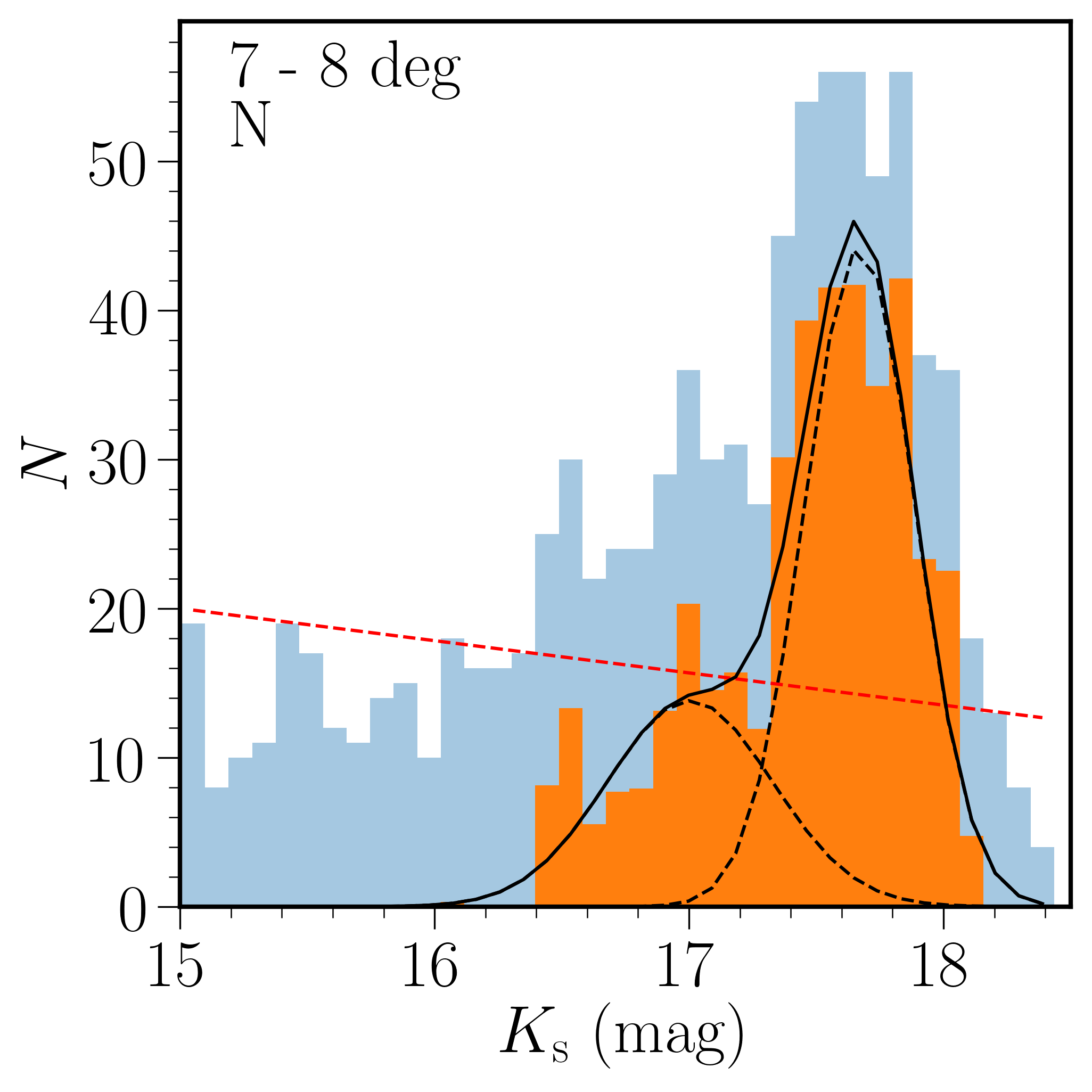}
	\includegraphics[scale=0.05]{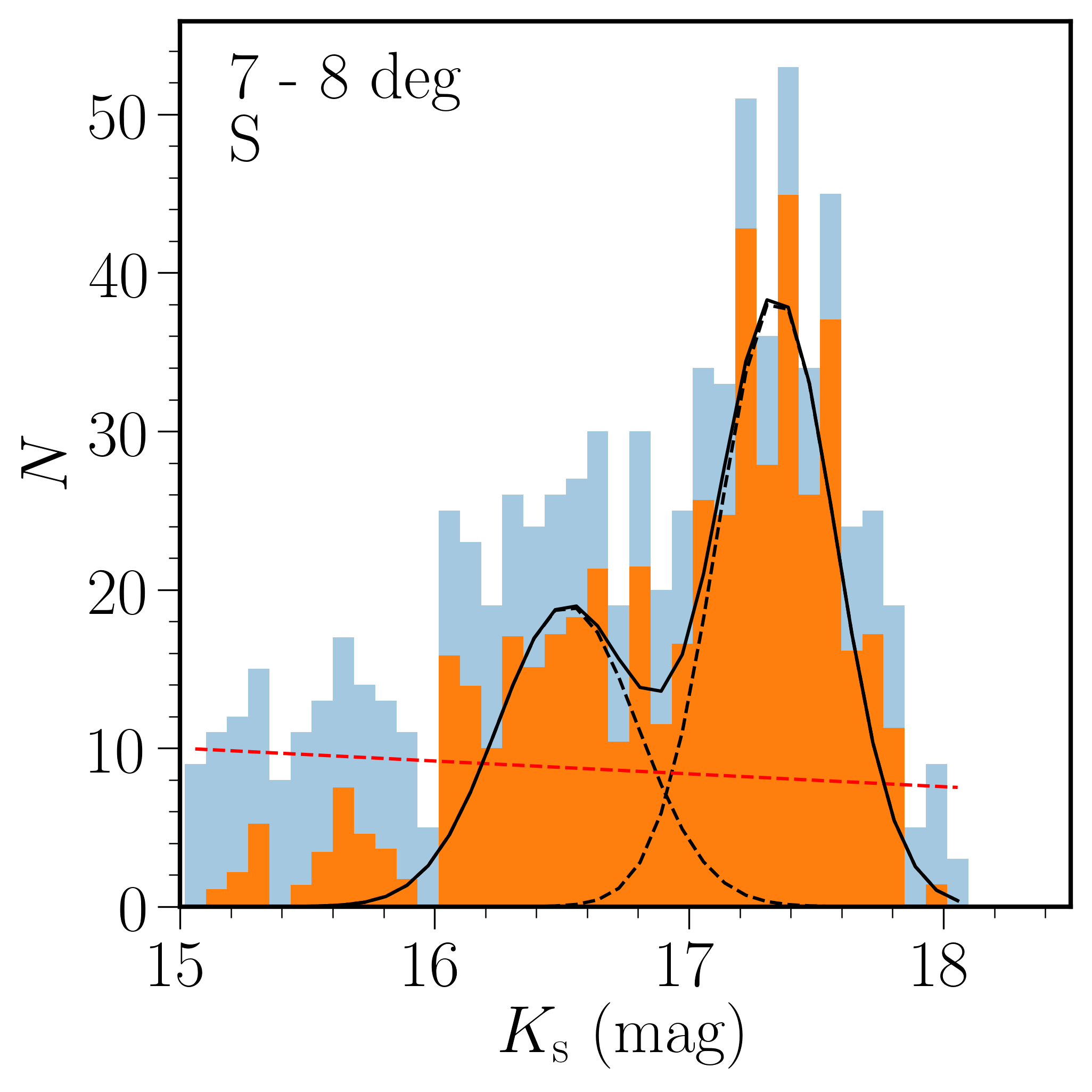}
	\includegraphics[scale=0.05]{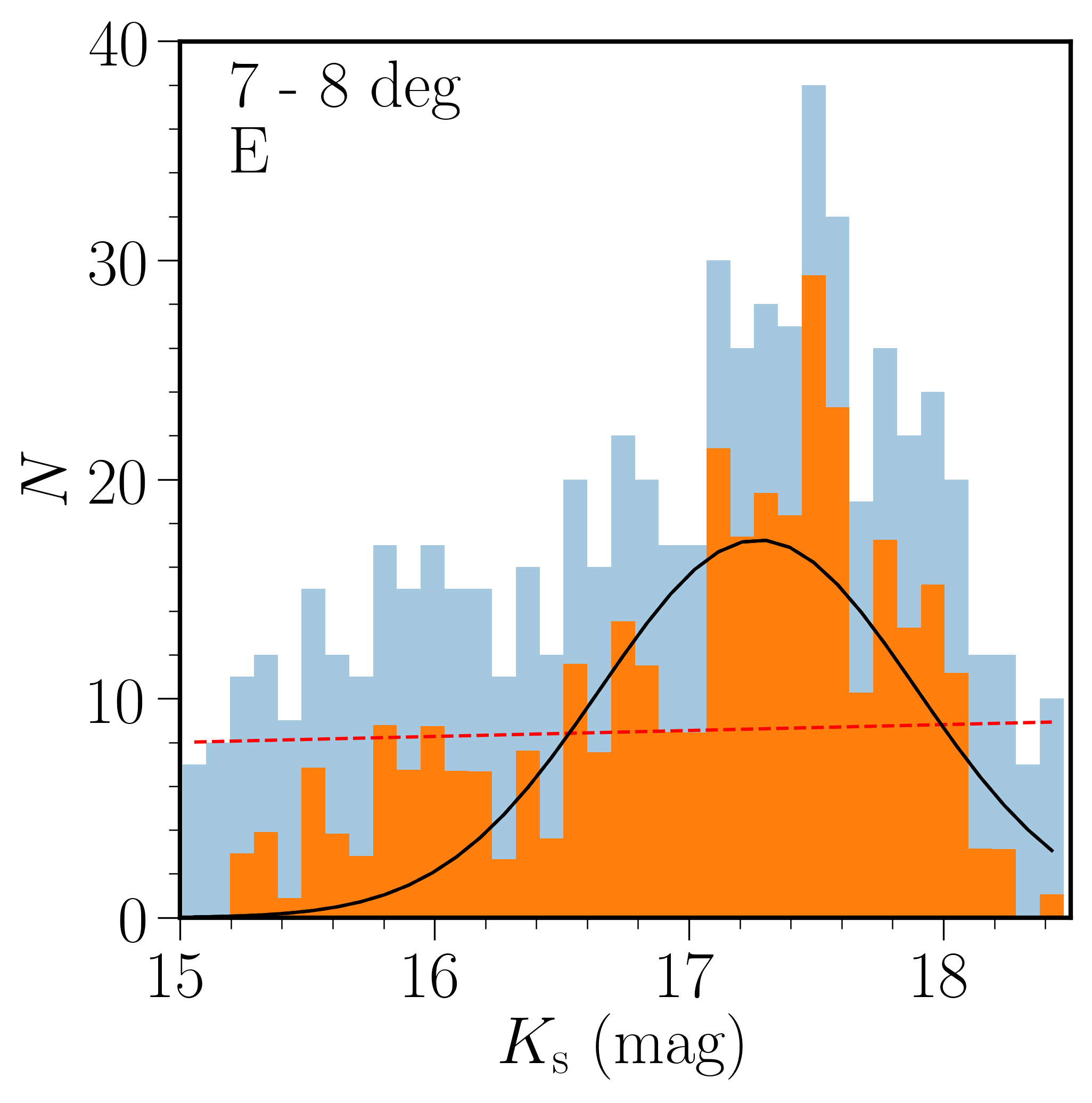}
	\includegraphics[scale=0.05]{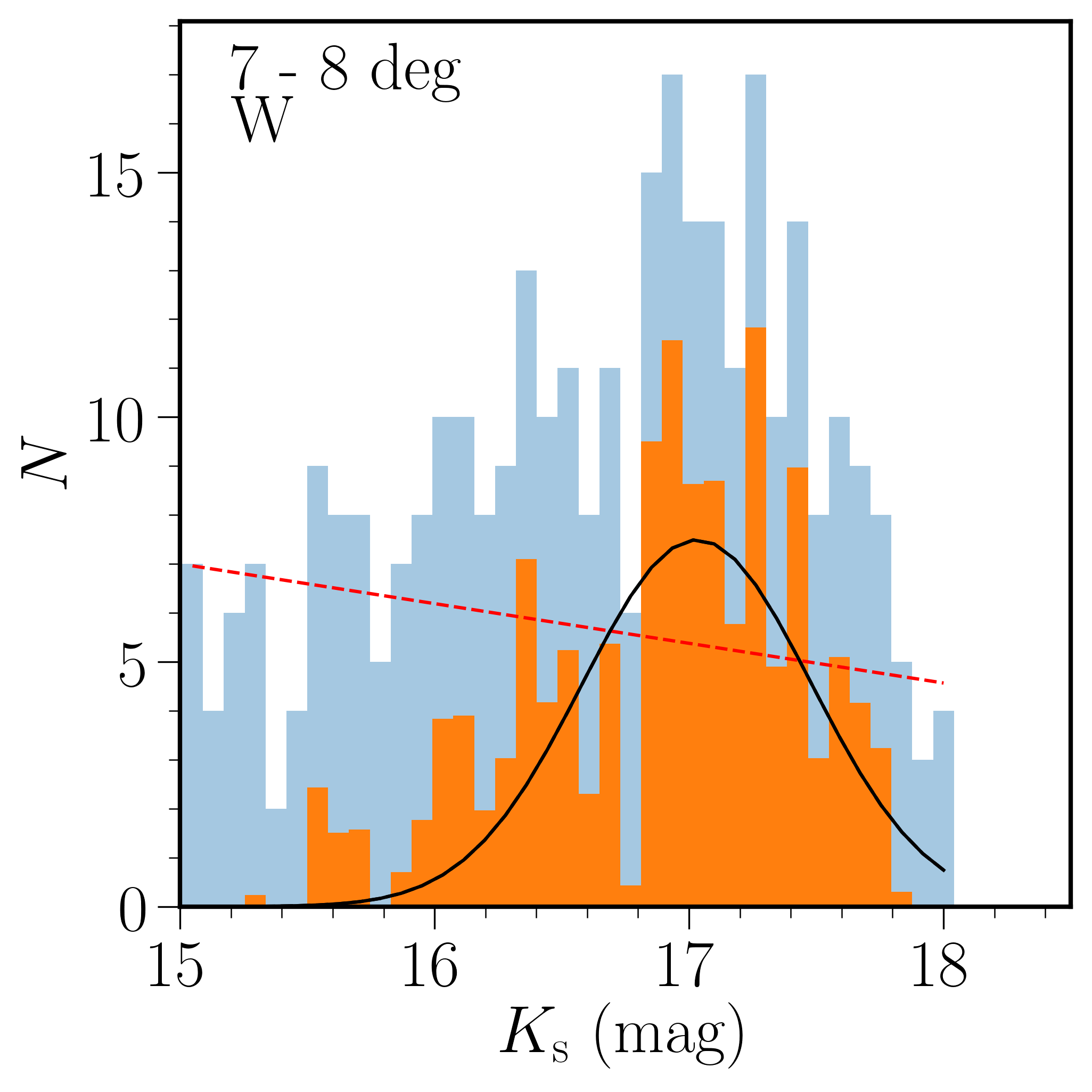}
	
	\includegraphics[scale=0.05]{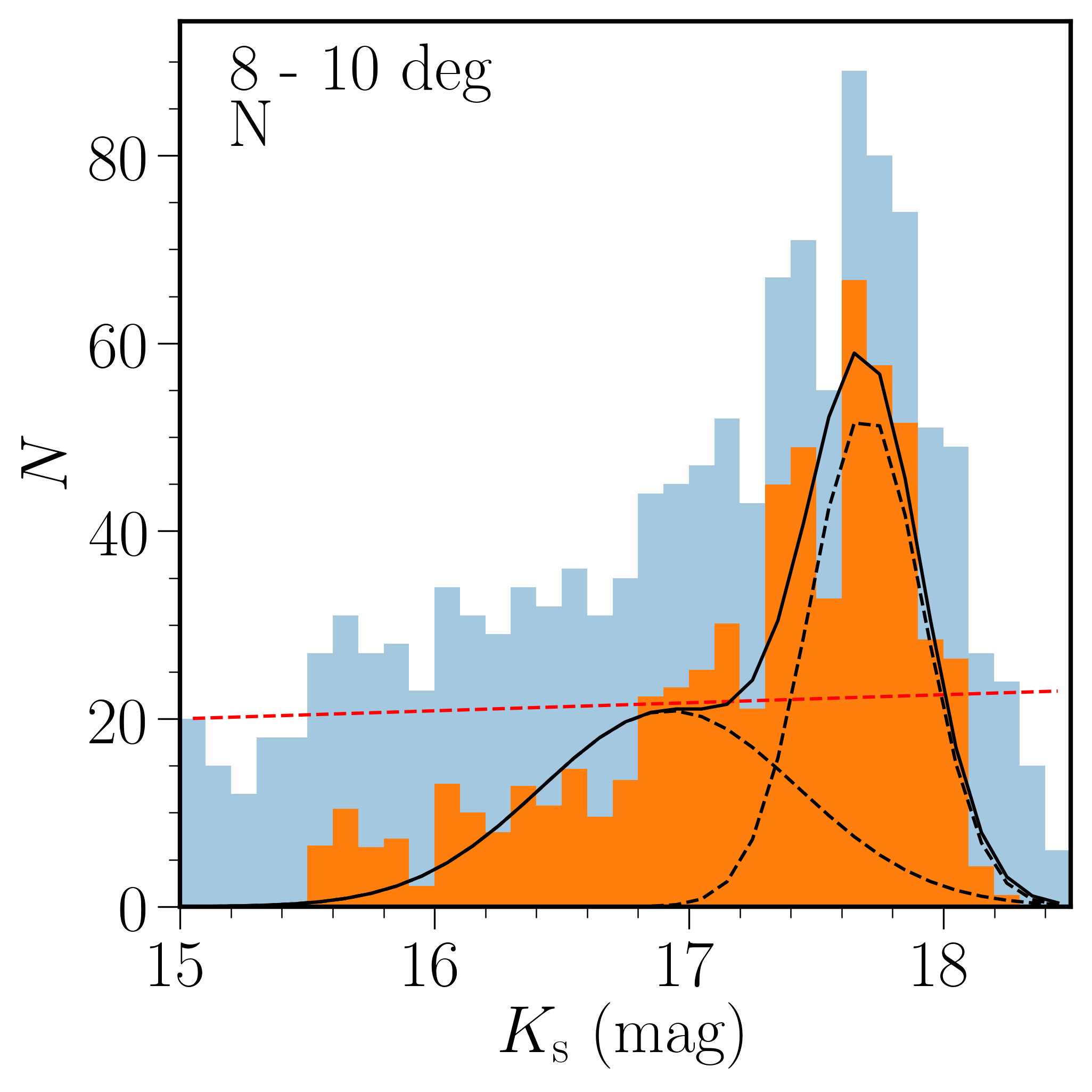}
	\includegraphics[scale=0.05]{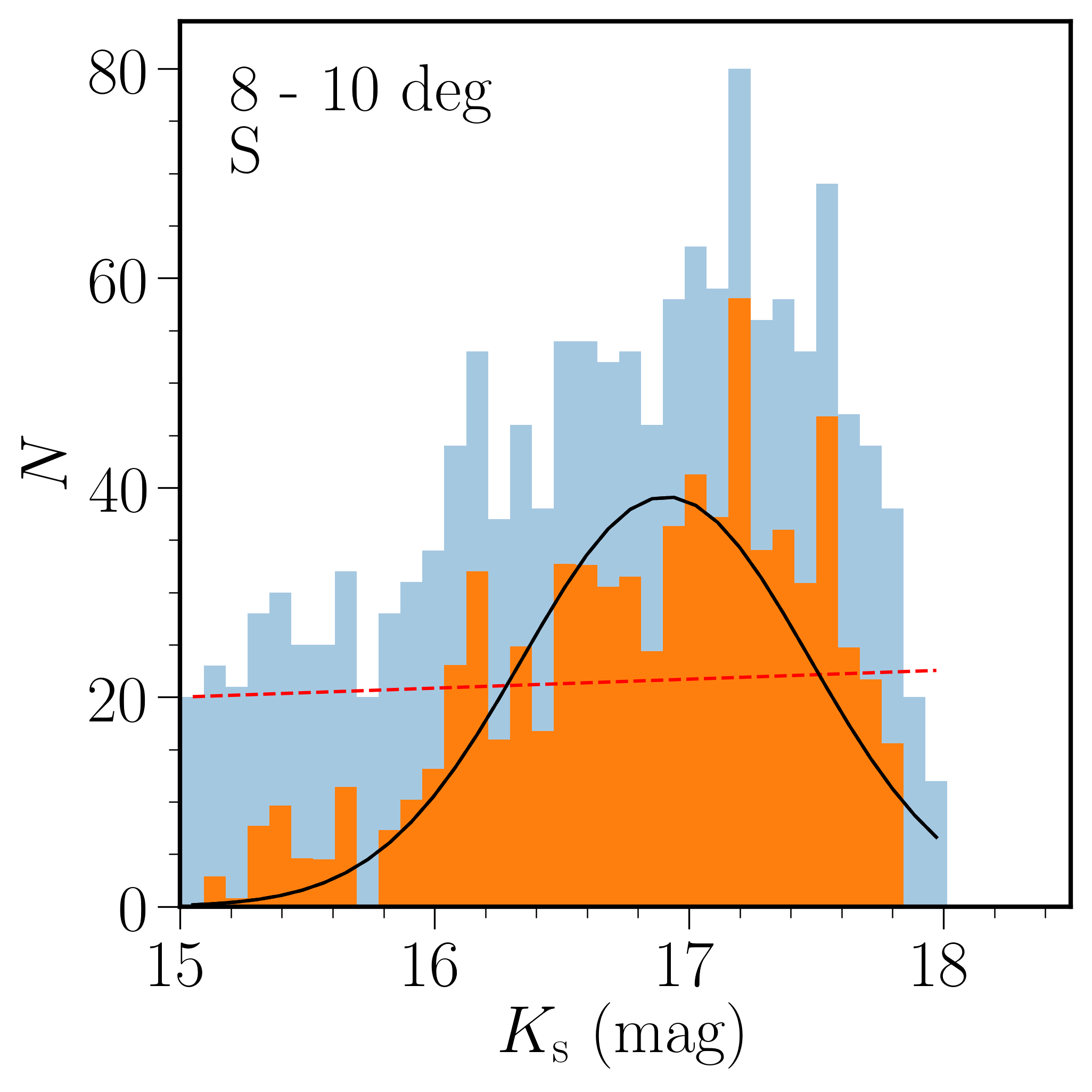}
	\includegraphics[scale=0.05]{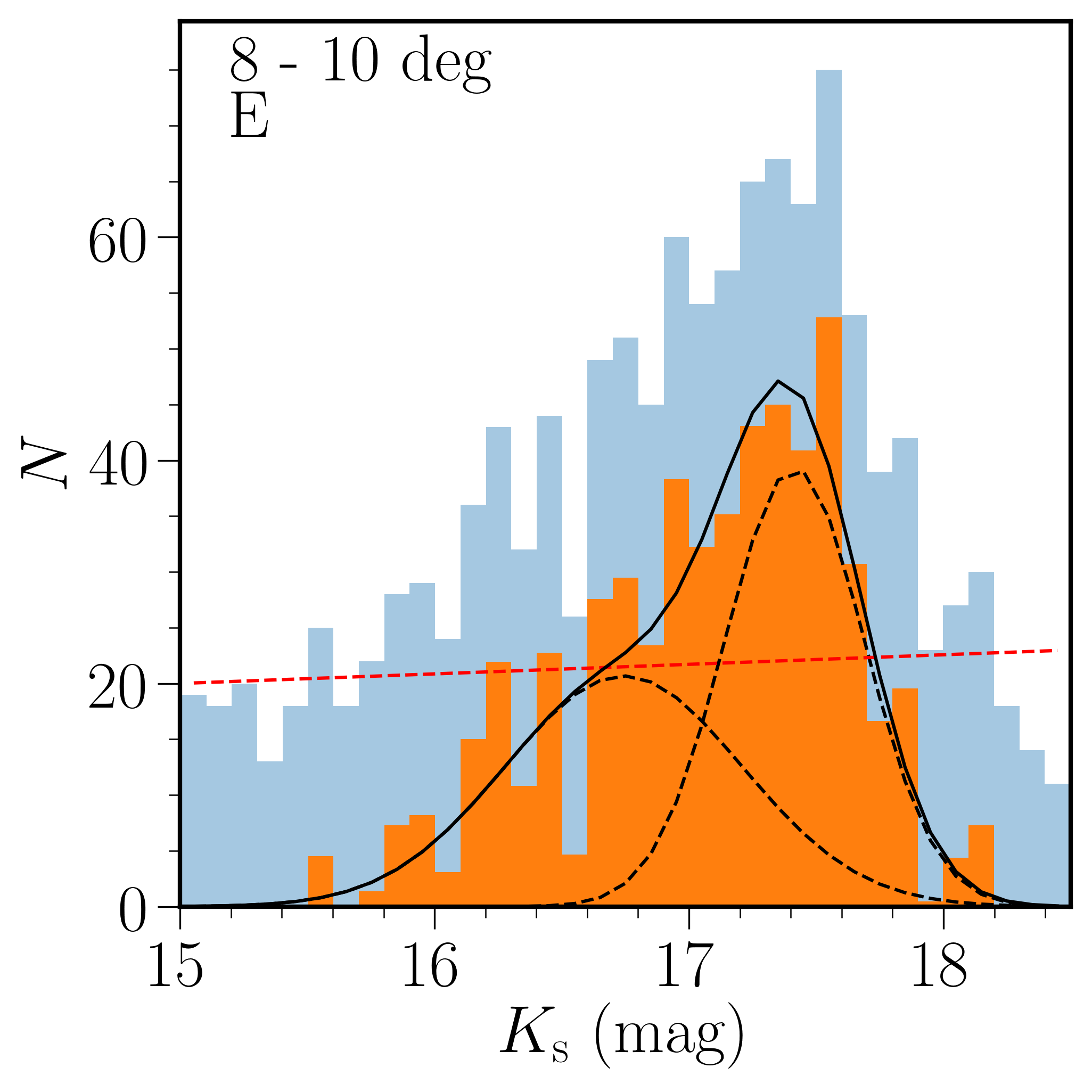}
	\includegraphics[scale=0.05]{Hist_RC_8_10deg_W_der.png}
	
	\caption{(continued)}
\end{figure*}

\section{Alternative cardinal division for RC stars}\label{appendix_1}
\begin{figure}
		\includegraphics[scale=0.12]{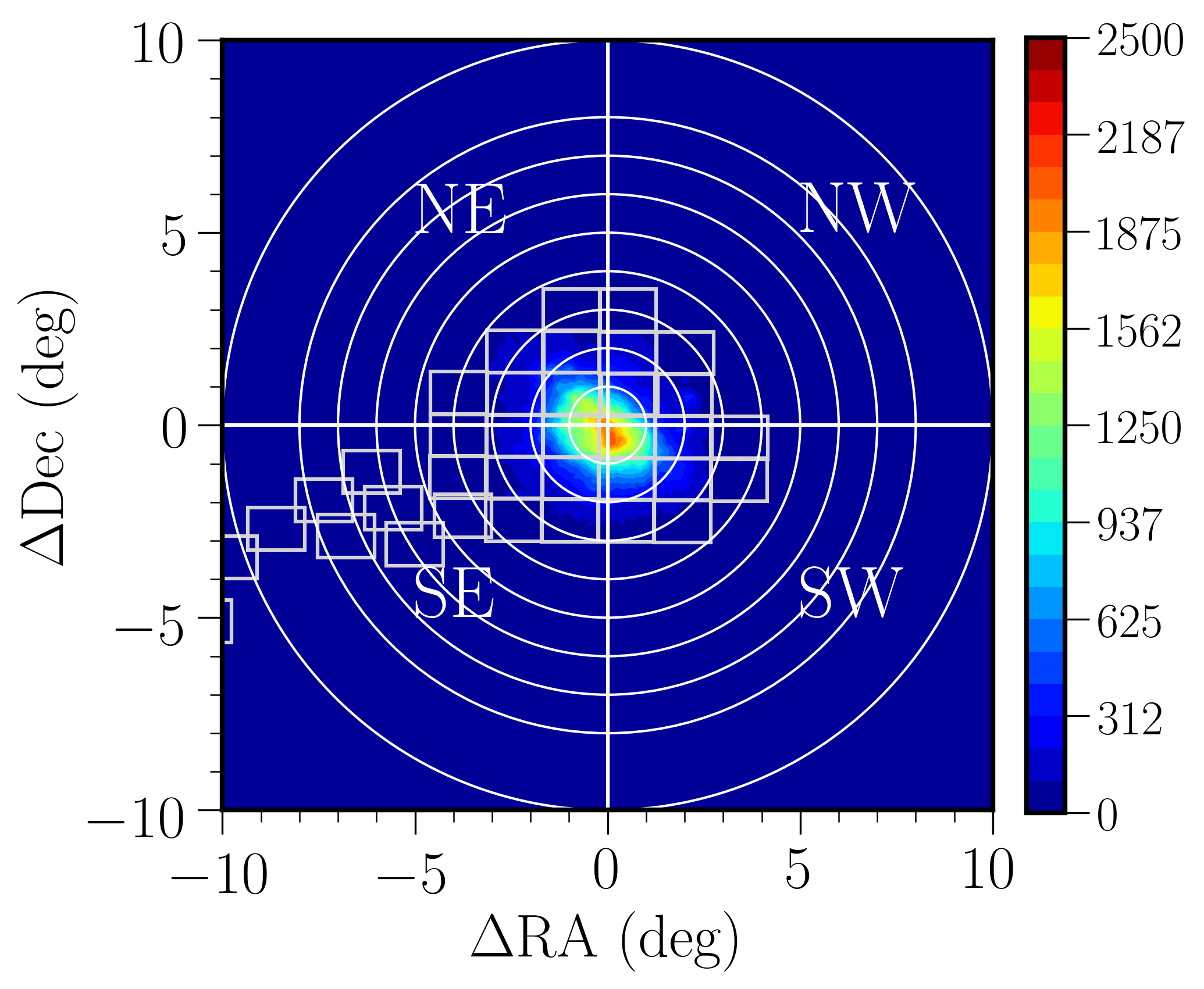}
		\caption{Spatial distribution of stars around the SMC and plotted annuli from 0$^\circ$ to 10$^\circ$ further divided into 4 regions: NE, NW, SE and SW. VMC tiles are also shown.}
		\label{fig:FOV_1}
\end{figure}

\begin{figure}
	\centering
\includegraphics[scale=0.05]{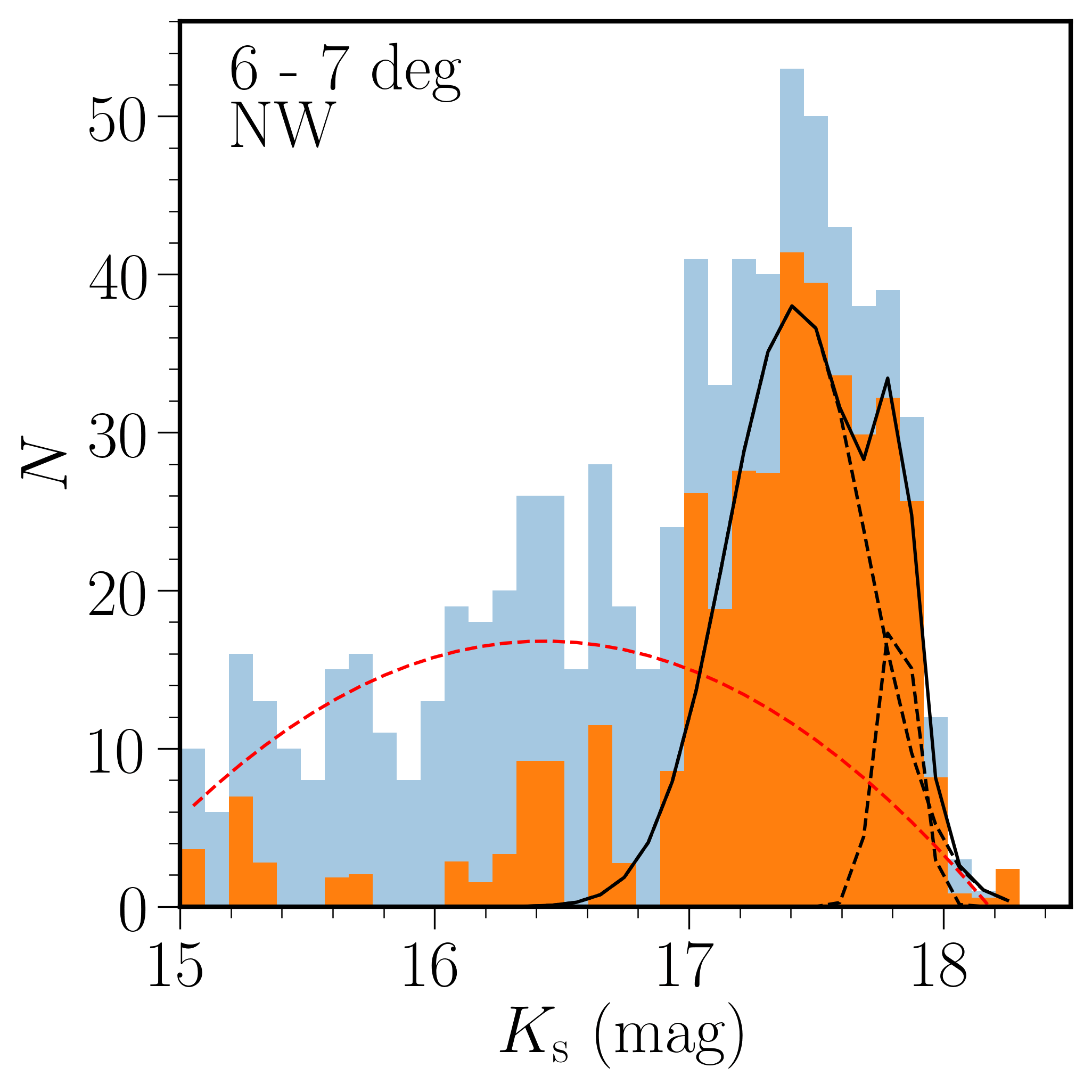}
\includegraphics[scale=0.05]{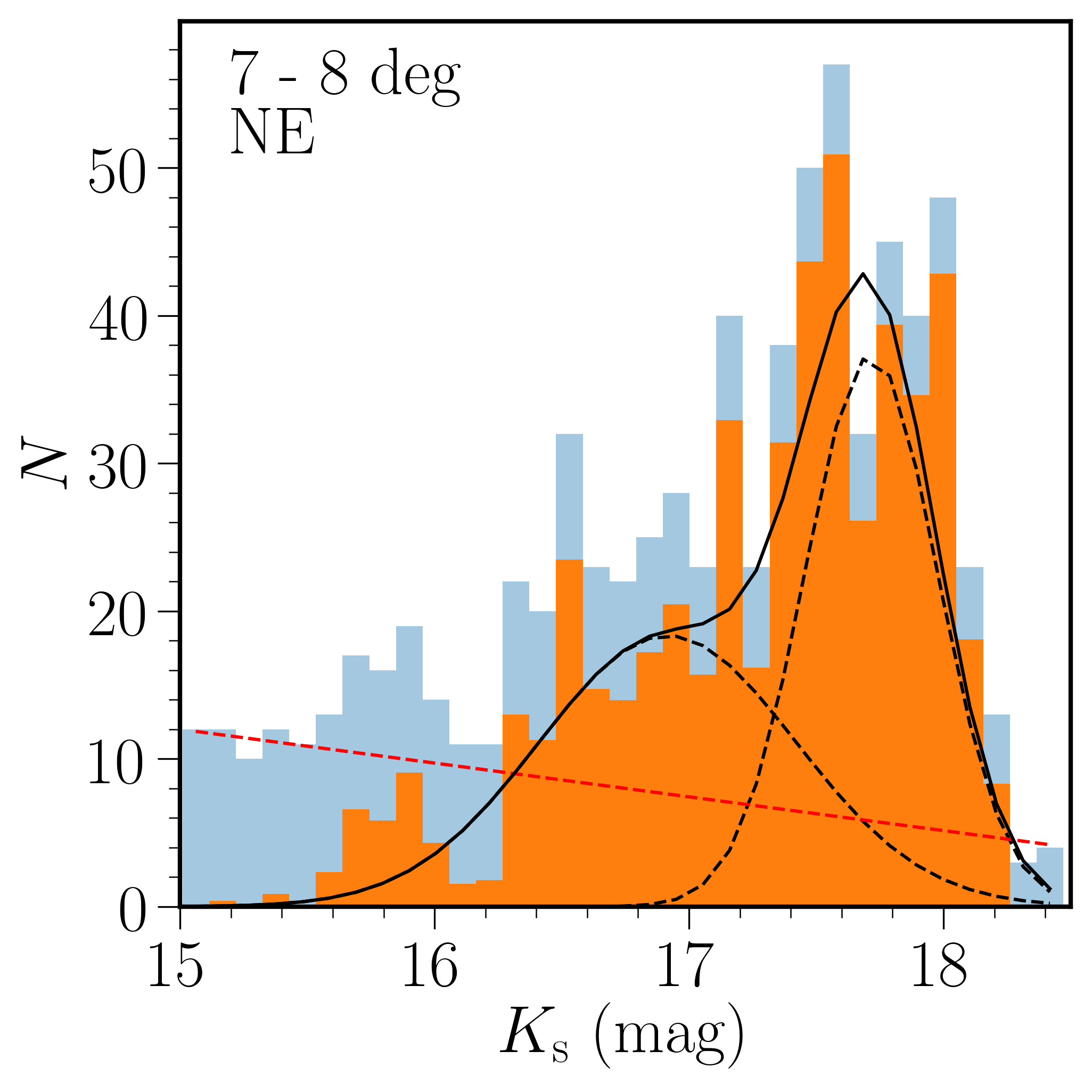}
\includegraphics[scale=0.05]{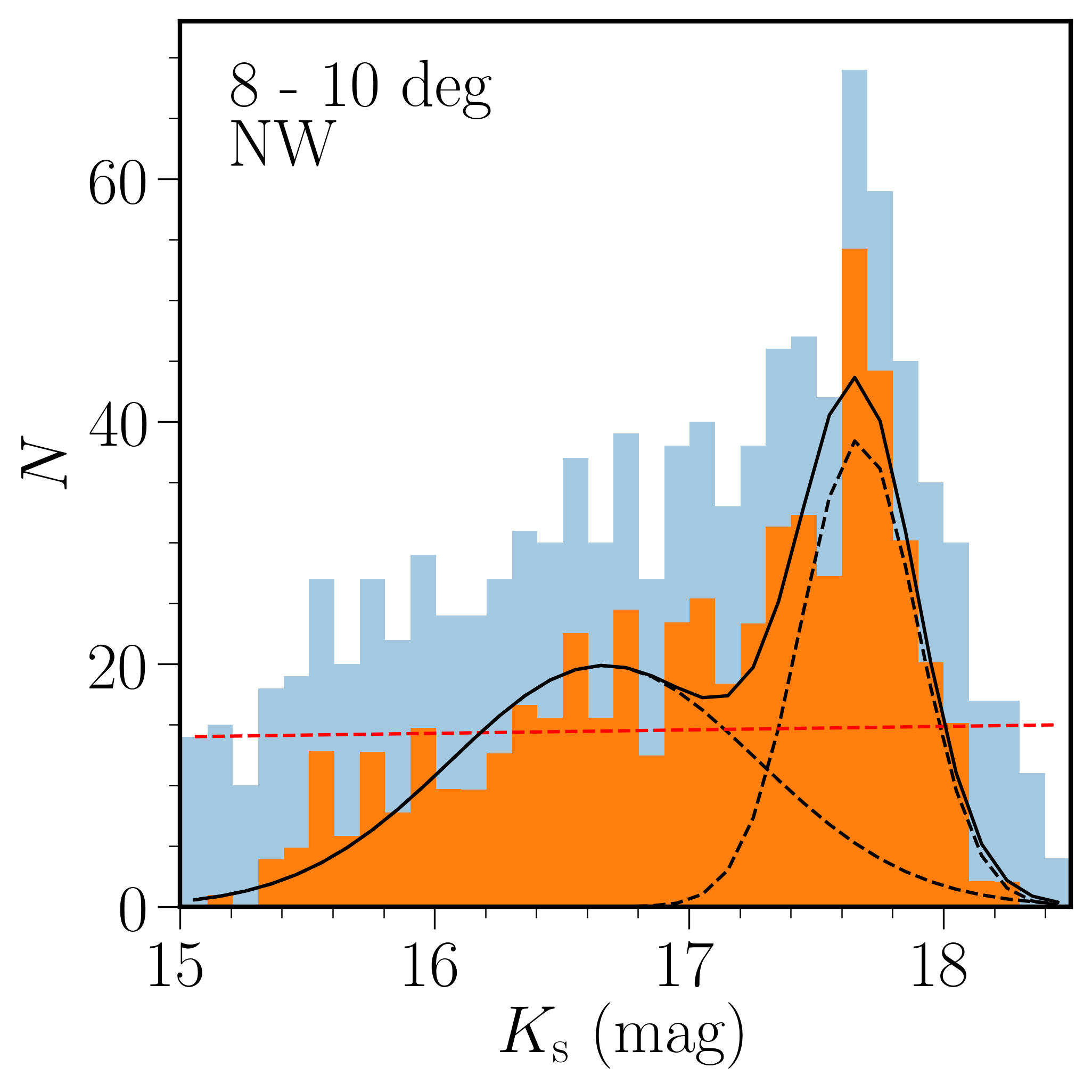}
\caption{Luminosity function of the RC stars in different regions. Each panel shows the luminosity function of the regions discussed in Sect.~\ref{section4} and Appendix \ref{appendix_1}. The blue histograms show the observed luminosity functions, whereas orange histograms show the distributions after subtracting the RGB components and the continuous lines show the total fits to these distributions whereas the dashed lines represent the separate components of the fits.}
\label{fig:RCHIST1}
\end{figure}

\begin{figure}
\includegraphics[scale=0.11]{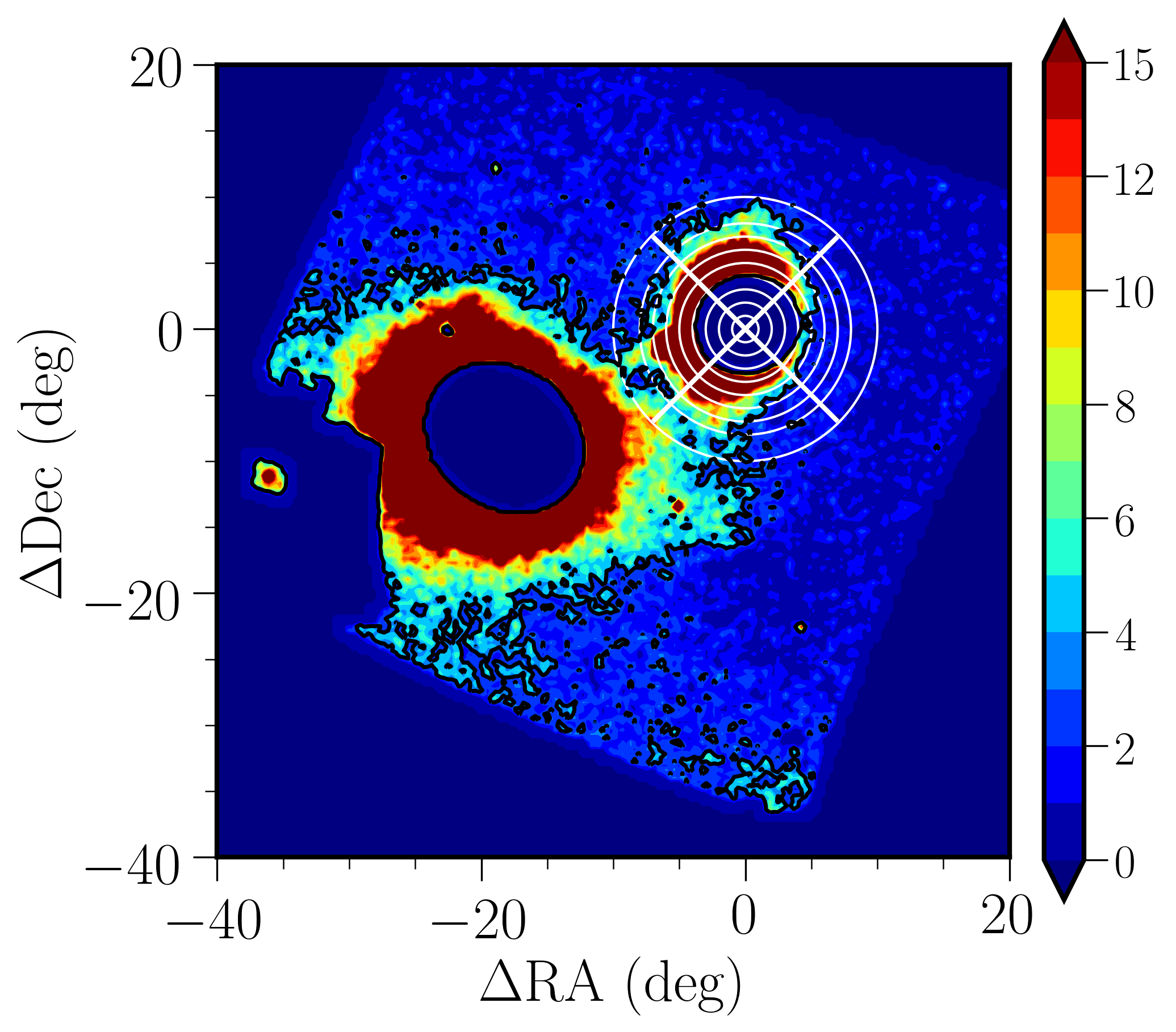}
\includegraphics[scale=0.11]{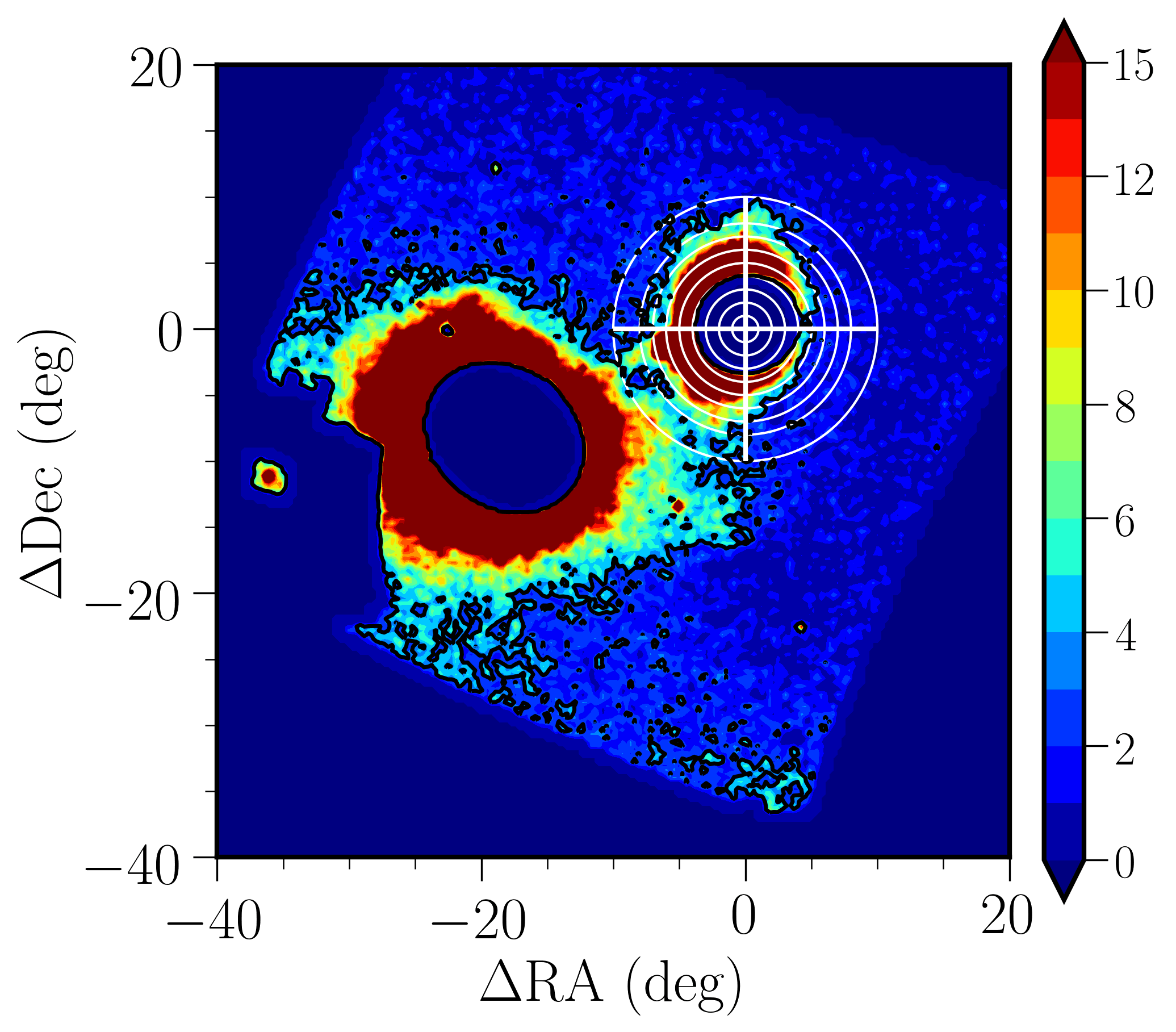}
\caption{Spatial division of the SMC area into sectors and rings as in Fig.~\ref{fig:DRC_1} (top) and Fig.~\ref{fig:FOV_1} (bottom) superimposed on the density of young and old stars as in Fig.~\ref{fig:features}, depicting the morphological features analysed in this study.}
\label{fig:B3}
\end{figure}

We explore an alternative cardinal division of the spatial distribution of stars around the SMC. The galaxy is still divided into circular annuli of radii 1$^\circ$, 2$^\circ$, 3$^\circ$, 4$^\circ$, 5$^\circ$, 6$^\circ$, 7$^\circ$, 8$^\circ$ and 10$^\circ$ from the centre and each annulus into four sectors: North--East (NE), North--West (NW), South--East (E) and South--West (W), see Figure \ref{fig:FOV_1}. These four regions are 45$^\circ$ offset counter-clockwise from the sectors adopted in Sect. \ref{section4}. As in the previous analysis, we study their NIR ($J-K_\mathrm{s}$, $K_\mathrm{s}$) CMD diagrams and RC luminosity functions (Figure \ref{fig:RCHIST1}). We only show sectors where a faint component which might be related to the Counter Bridge is detected. Figure \ref{fig:B3} shows how the cardinal divisions explored in this study map onto the morphological features.

\end{document}